\documentclass[useAMS,usenatbib]{mnras}

\usepackage{graphicx, bm, amssymb, xcolor}
\usepackage{verbatim}
\usepackage[all]{hypcap}
\usepackage{xspace}
\usepackage{multirow,bigdelim}
\usepackage[fleqn]{amsmath}
\usepackage{mathtools}

\usepackage{epsfig}
\usepackage{aas_macros}
\usepackage{natbib}
\usepackage{times}

\newcommand{\msun}{\mbox{M$_\odot$}}
\newcommand{\yr}{\mbox{${\rm yr}$}}
\newcommand{\myr}{\mbox{${\rm Myr}$}}
\newcommand{\gyr}{\mbox{${\rm Gyr}$}}
\newcommand{\pc}{\mbox{${\rm pc}$}}
\newcommand{\kpc}{\mbox{${\rm kpc}$}}
\newcommand{\mpc}{\mbox{${\rm Mpc}$}}
\newcommand{\kms}{\mbox{${\rm km}~{\rm s}^{-1}$}}
\newcommand{\cmc}{\mbox{${\rm cm}^{-3}$}}
\newcommand{\mh}{\mbox{${\rm m}_{\rm H}$}}

\newcommand{\dex}{\mbox{${\rm dex}$}}
\newcommand{\pix}{\mbox{${\rm pix}$}}
\newcommand{\sfr}{\mbox{${\rm SFR}$}}

\newcommand{\lap}{\mbox{$l_{\rm ap}$}}
\newcommand{\lapeff}{\mbox{$l_{\rm ap,eff}$}}
\newcommand{\lapstar}{\mbox{$l_{\rm ap,star}$}}
\newcommand{\lapgas}{\mbox{$l_{\rm ap,gas}$}}
\newcommand{\lapmin}{l_{\rm ap,min}}
\newcommand{\lapminlr}{l_{\rm ap,min,LR}}
\newcommand{\lapminhr}{l_{\rm ap,min,HR}}
\newcommand{\lapmax}{\mbox{$l_{\rm ap,max}$}}
\newcommand{\nap}{\mbox{$N_{\rm ap}$}}
\newcommand{\lambdamin}{\mbox{$\lambda_{\rm min}$}}
\newcommand{\lambdamax}{\mbox{$\lambda_{\rm max}$}}
\newcommand{\lambdamc}{\mbox{$\lambda_{\rm mc}$}}
\newcommand{\tgas}{\mbox{$t_{\rm gas}$}}
\newcommand{\tgasmin}{\mbox{$t_{\rm gas,min}$}}
\newcommand{\tgasmax}{\mbox{$t_{\rm gas,max}$}}
\newcommand{\tgasin}{\mbox{$t_{\rm gas,in}$}}
\newcommand{\tgasout}{\mbox{$t_{\rm gas,out}$}}
\newcommand{\tgasmc}{\mbox{$t_{\rm gas,mc}$}}
\newcommand{\tstar}{\mbox{$t_{\rm star}$}}
\newcommand{\tover}{\mbox{$t_{\rm over}$}}
\newcommand{\tovermin}{\mbox{$t_{\rm over,min}$}}
\newcommand{\tovermax}{\mbox{$t_{\rm over,max}$}}
\newcommand{\toverin}{\mbox{$t_{\rm over,in}$}}
\newcommand{\toverout}{\mbox{$t_{\rm over,out}$}}
\newcommand{\tovermc}{\mbox{$t_{\rm over,mc}$}}
\newcommand{\tauin}{\mbox{$\tau_{\rm in}$}}
\newcommand{\tauout}{\mbox{$\tau_{\rm out}$}}
\newcommand{\toff}{\mbox{$t_{\rm off}$}}

\newcommand{\tstarref}{\mbox{$t_{\rm star,ref}$}}
\newcommand{\rhomin}{\rho_{\rm min}}
\newcommand{\rhominlr}{\rho_{\rm min,LR}}
\newcommand{\rhominhr}{\rho_{\rm min,HR}}
\newcommand{\tgasexp}{\mbox{$t_{\rm gas,exp}$}}
\newcommand{\toverexp}{\mbox{$t_{\rm over,exp}$}}
\newcommand{\lambdaexp}{\mbox{$\lambda_{\rm exp}$}}
\newcommand{\rhog}{\mbox{$\rho_{\rm g}$}}
\newcommand{\rhocrit}{\mbox{$\rho_{\rm crit}$}}
\newcommand{\tdyn}{\mbox{$t_{\rm dyn}$}}
\newcommand{\tsn}{\mbox{$t_{\rm SN}$}}
\newcommand{\tdepl}{\mbox{$t_{\rm depl}$}}

\newcommand{\betastar}{\beta_{\rm star}}
\newcommand{\betagas}{\beta_{\rm gas}}

\newcommand{\rgas}{r_{\rm gas}}
\newcommand{\rstar}{r_{\rm star}}
\newcommand{\zetastar}{\mbox{$\zeta_{\rm star}$}}
\newcommand{\zetagas}{\mbox{$\zeta_{\rm gas}$}}
\newcommand{\zetacrit}{\mbox{$\zeta_{\rm crit}$}}
\newcommand{\sfe}{\mbox{$\epsilon_{\rm sf}$}}
\newcommand{\mdotsf}{\mbox{$\dot{M}_{\rm sf}$}}
\newcommand{\mdotfb}{\mbox{$\dot{M}_{\rm fb}$}}
\newcommand{\vfb}{v_{\rm fb}}
\newcommand{\vfbr}{v_{{\rm fb},r}}
\newcommand{\etafb}{\mbox{$\eta_{\rm fb}$}}
\newcommand{\etaavgfb}{\mbox{$\overline{\eta}_{\rm fb}$}}
\newcommand{\chifbe}{\mbox{$\chi_{{\rm fb},E}$}}
\newcommand{\chifber}{\mbox{$\chi_{{\rm fb},E,r}$}}
\newcommand{\chifbp}{\mbox{$\chi_{{\rm fb},p}$}}
\newcommand{\chifbpr}{\mbox{$\chi_{{\rm fb},p,r}$}}
\newcommand{\halpha}{\mbox{${\rm H}\alpha$}\xspace}
\newcommand{\chired}{\chi_{\rm red}^2}
\newcommand{\chimin}{\chi_{\rm red,min}^2}
\newcommand{\f}{{\cal F}}
\newcommand{\bias}{{\cal B}}
\newcommand{\rat}{{\cal R}}
\newcommand{\exc}{{\cal E}}
\newcommand{\nsamp}{N_{\rm samp}}
\newcommand{\nbins}{N_{\rm bins}}
\newcommand{\npixmin}{N_{\rm pix,min}}
\newcommand{\nlinstar}{N_{\rm lin,star}}
\newcommand{\nlingas}{N_{\rm lin,gas}}
\newcommand{\nmcpeak}{N_{\rm mc,peak}}
\newcommand{\ndepth}{N_{\rm depth}}
\newcommand{\ntry}{N_{\rm try}}
\newcommand{\nmcphys}{N_{\rm mc,phys}}
\newcommand{\klprinciple}{\citetalias{kruijssen14} principle\xspace}
\newcommand{\code}{{\sc Heisenberg}\xspace}
\newcommand{\idl}{{\sc idl}\xspace}
\newcommand{\python}{{\sc Python}\xspace}
\newcommand{\clfind}{{\sc Clumpfind}\xspace}
\newcommand{\ds}{{\sc DS9}\xspace}
\newcommand{\be}{\begin{equation}}
\newcommand{\ee}{\end{equation}}
\newcommand{\bea}{\begin{eqnarray}}
\newcommand{\eea}{\end{eqnarray}}

\defcitealias{kruijssen14}{KL14}

\title{\vspace{-2mm}An uncertainty principle for star formation -- II. A new method for characterising the cloud-scale physics of star formation and feedback across cosmic history\vspace{-2mm}}
\author{J.~M.~Diederik Kruijssen,$^{1,2,3}$\thanks{kruijssen@uni-heidelberg.de}  Andreas Schruba,$^4$ Alexander~P.~S.~Hygate,$^{2,1}$ 
\newauthor Chia-Yu Hu,$^{3,5}$ Daniel~T.~Haydon$^1$ and Steven~N.~Longmore$^6$\\
$^1$Astronomisches Rechen-Institut, Zentrum f\"{u}r Astronomie der Universit\"{a}t Heidelberg, M\"{o}nchhofstra\ss e 12-14, 69120 Heidelberg, Germany\\
$^2$Max-Planck Institut f\"{u}r Astronomie, K\"{o}nigstuhl 17, 69117 Heidelberg, Germany\\
$^3$Max-Planck Institut f\"{u}r Astrophysik, Karl-Schwarzschild-Stra\ss e 1, 85748 Garching, Germany\\
$^4$Max-Planck Institut f\"{u}r Extraterrestrische Physik, Giessenbachstra\ss e 1, 85748 Garching, Germany\\
$^5$Center for Computational Astrophysics, 160 Fifth Avenue, New York, NY 10010, USA\\
$^6$Astrophysics Research Institute, Liverpool John Moores University, IC2, Liverpool Science Park, 146 Brownlow Hill,
Liverpool L3 5RF, United Kingdom\vspace{-1mm}}

\begin{document}

\date{Accepted 2018 April 27. Received 2018 April 23; in original form 2017 October 5.\vspace{-1mm}}

\pagerange{\pageref{firstpage}--\pageref{lastpage}} \pubyear{2018}

\maketitle

\label{firstpage}

\begin{abstract}
The cloud-scale physics of star formation and feedback represent the main uncertainty in galaxy formation studies. Progress is hampered by the limited empirical constraints outside the restricted environment of the Local Group. In particular, the poorly-quantified time evolution of the molecular cloud lifecycle, star formation, and feedback obstructs robust predictions on the scales smaller than the disc scale height that are resolved in modern galaxy formation simulations. We present a new statistical method to derive the evolutionary timeline of molecular clouds and star-forming regions. By quantifying the excess or deficit of the gas-to-stellar flux ratio around peaks of gas or star formation tracer emission, we directly measure the relative rarity of these peaks, which allows us to derive their lifetimes. We present a step-by-step, quantitative description of the method and demonstrate its practical application. The method's accuracy is tested in nearly 300 experiments using simulated galaxy maps, showing that it is capable of constraining the molecular cloud lifetime and feedback time-scale to $<0.1$~dex precision. Access to the evolutionary timeline provides a variety of additional physical quantities, such as the cloud-scale star formation efficiency, the feedback outflow velocity, the mass loading factor, and the feedback energy or momentum coupling efficiencies to the ambient medium. We show that the results are robust for a wide variety of gas and star formation tracers, spatial resolutions, galaxy inclinations, and galaxy sizes. Finally, we demonstrate that our method can be applied out to high redshift ($z\la4$) with a feasible time investment on current large-scale observatories. This is a major shift from previous studies that constrained the physics of star formation and feedback in the immediate vicinity of the Sun.
\end{abstract}

\begin{keywords}
stars: formation -- ISM: evolution -- galaxies: evolution -- galaxies: formation -- galaxies: ISM -- galaxies: stellar content\vspace{-1mm}
\end{keywords}

\section{Introduction} \label{sec:intro}
Current theoretical and numerical models aiming to reproduce the observed galaxy population are strongly limited by uncertainties in the baryonic physics \citep[e.g.][]{hopkins11,haas13,vogelsberger14,schaye15}. While the boundary conditions provided by cosmology and the hierarchical growth of the dark matter haloes in which galaxies reside are relatively well-constrained \citep[e.g.][]{springel05d,planck14}, connecting these haloes to the visible galaxy population hinges critically on the unknown physics of star formation and feedback \citep{mckee07,kennicutt12}.

Galaxy formation simulations generally describe star formation using the phenomenological, galaxy-scale relations between the gas mass and the star formation rate (SFR), observed from nearby spiral galaxies \citep[e.g.][]{bigiel08,schruba11,kennicutt12,leroy13} out to high redshift \citep[e.g.][]{daddi10b,genzel10,tacconi13}. Broadly, these galactic `star formation relations' represent variations of the form
\be
\label{eq:sfrelation}
{\rm SFR}=\frac{\sfe}{\tau_{\rm sf}}M_{\rm gas} ,
\ee
where $\sfe$ represents the star formation efficiency, $\tau_{\rm sf}$ is the star formation time-scale, and $M_{\rm gas}$ indicates the gas mass. One of the main reasons that the physics underpinning the galactic star formation relation have been so challenging to constrain \citep[see e.g.][among many others]{krumholz05,hennebelle11,ostriker11,padoan11,krumholz12} is that the star formation efficiency and time-scale represent degenerate quantities in the above expression. It is impossible to determine from the star formation relation alone whether star formation is rapid and inefficient or slow and efficient. This problem is exacerbated by the fact that both quantities are notoriously hard to measure -- star formation takes place on time-scales much longer than a human lifetime, implying that $\tau_{\rm sf}$ cannot be measured directly and $\sfe$ cannot be obtained by simply considering the initial and final states of a single system. However, knowing either $\sfe$ or $\tau_{\rm sf}$ would allow one to immediately constrain the other. Analogously to equation~(\ref{eq:sfrelation}), the commonly-used star formation `recipes' in galaxy formation simulations generally assume that gas is turned into stars on some time-scale $\tau_{\rm sf}$ with some efficiency $\sfe$. Both quantities are unknown on the small ($10$--$100~\pc$) scales of the individual gas clouds resolved in modern galaxy formation simulations, but strongly affect the structure of simulated galaxies \citep[e.g.][]{hopkins13b,braun15,semenov16}. It is therefore a priority to obtain empirical constraints on these quantities and their variation with the galactic environment.

Likewise, the deposition of mass, momentum and energy by feedback is often based on a simple representation of supernova (SN) feedback, i.e.~by injecting an amount of energy appropriate for the expected number of SNe into the interstellar medium (ISM) surrounding each young stellar population \citep[e.g.][]{springel03} or by defining an empirical mass outflow rate in units of the SFR that may depend on the local galactic environmental conditions \citep[e.g.][]{oppenheimer06,oppenheimer08}. However, it is shown by theory \citep[e.g.][]{murray10,dale13,agertz13} and observations \citep[e.g.][]{pellegrini10,lopez11,lopez14} of nearby star-forming regions that other feedback mechanisms (e.g.~photoionization, stellar winds, radiation pressure) can be at least equally effective as SNe \citep[see e.g.~the recent reviews by][]{kruijssen13,krumholz14b,dale15}. Obtaining an empirically-motivated prescription for stellar feedback in galaxy formation simulations is crucial, because most discrepancies between the observed and simulated galaxy populations can be alleviated with a suitable (but possibly ad-hoc) choice of feedback model \citep[e.g.][]{hopkins14,schaye15}.

In \citet[hereafter \citetalias{kruijssen14}]{kruijssen14}, we presented a new statistical model, named the `uncertainty principle for star formation', which explains how the galaxy-scale star formation properties of a galaxy emerge by the summation over its constituent population of independent star-forming regions. In particular, the `\klprinciple' predicts that the star formation relation changes form towards small ($\la\kpc$) spatial scales in a way that depends sensitively on the evolutionary timeline of star-forming regions. We showed that it can therefore be used to directly measure the efficiencies and time-, size- and velocity-scales of cloud collapse, star formation, and feedback without requiring individual clouds to be resolved. By comparing the flux ratios between tracers of gas and star formation on $0.1$--$1~\kpc$ scales across a galaxy, the method effectively counts the relative occurrence of these phases without resolving them spatially, thereby directly constraining their characteristic lifetimes and spatial separations. The method is easy to apply and could potentially be used to characterize cloud-scale star formation and feedback across cosmic history.

The \klprinciple does not provide the first attempt of characterising cloud-scale star formation and feedback. For instance, molecular cloud lifetimes have been estimated in the Local Group for the Milky Way, the Magellanic Clouds, and M33 \citep{engargiola03,kawamura09,murray11}, as well as for M51 \citep{meidt15}, stellar feedback energy and momentum deposition rates have been determined for the Magellanic Clouds \citep{pellegrini10,lopez11,lopez14}, and the rapidity and efficiency of star formation have been quantified for star-forming clouds in the Milky Way, albeit with a broad range of outcomes \citep[e.g.][]{elmegreen00,krumholz07,evans09,kruijssen15,barnes17}. The common shortcoming of the methods used in these studies is that they can only be applied to the limited range of galactic environments present in Local Group galaxies or to exceptionally high-resolution observations of external galaxies, because they require individual clouds to be resolved. The main improvement of the \klprinciple is that it can be applied out to much larger distances because only the typical {\it separation} length between clouds must be (marginally) resolved. In addition, the robust statistical basis of the method enables its application to large galaxy samples with little computational effort or human intervention.

In this paper, a general framework for applying the \klprinciple is presented and the method is validated using numerical simulations of isolated disc galaxies. We first summarise the basic concepts behind the \klprinciple (Section~\ref{sec:kl14}). We then present a step-by-step description of the method's application to integrated intensity maps of gas and star formation tracers across galaxies (Section~\ref{sec:method}). We validate the process by applying it to numerical simulations of disc galaxies and verifying the accuracy of the extracted quantities (Section~\ref{sec:valid}), the results of which are summarised as a set of guidelines for the reliable application of the method (Section~\ref{sec:guide}). The extracted quantities are then used to calculate several derived quantities that are demonstrated to accurately describe cloud-scale star formation and feedback (Section~\ref{sec:derivephys}). We then use the several performed tests of the method to determine out to which distances reliable constraints on cloud-scale star formation and feedback can be obtained (Section~\ref{sec:dist}). We include an extensive discussion of the method's caveats and limitations, as well as of its planned future improvements (Section~\ref{sec:disc}). The paper is finished with a summary of our conclusions and a discussion of the method's potential for future applications (Section~\ref{sec:concl}). Finally, Appendices~\ref{sec:appcircseg}--\ref{sec:appexp} provide the necessary background for a number of technical considerations in this paper, as well as the complete set of test results used in Section~\ref{sec:valid}.

The reader less interested in the quantitative details of the method and its validation can refer to parts of the paper for a qualitative summary. For a cursory overview of the results, we recommend consulting only Sections~\ref{sec:intro}, \ref{sec:kl14}, \ref{sec:qualdes}, \ref{sec:guide}, \ref{sec:dist}, and~\ref{sec:concl}. This effectively shortens the paper to a little over ten pages.

\section{Summary of the formalism} \label{sec:kl14}
The fundamental concept underpinning the \klprinciple is that a galaxy consists of some number of independent star-forming regions (hereafter `independent regions') separated by some length-scale $\lambda$, which each reside in an evolutionary phase independently of their neighbours. \autoref{fig:tschem} shows the evolutionary timeline for a single independent region. The physical meaning of the phases depends on the adopted tracers. While this example shows the transition from gas to stars, a similar timeline can be constructed for pairs of gas tracers or star formation tracers. By combining many such tracer pairs, a long timeline of several different, partially-overlapping phases is obtained.

\subsection{The star formation relation must break down on small spatial scales} \label{sec:form1}
\begin{figure}
\includegraphics[width=\hsize]{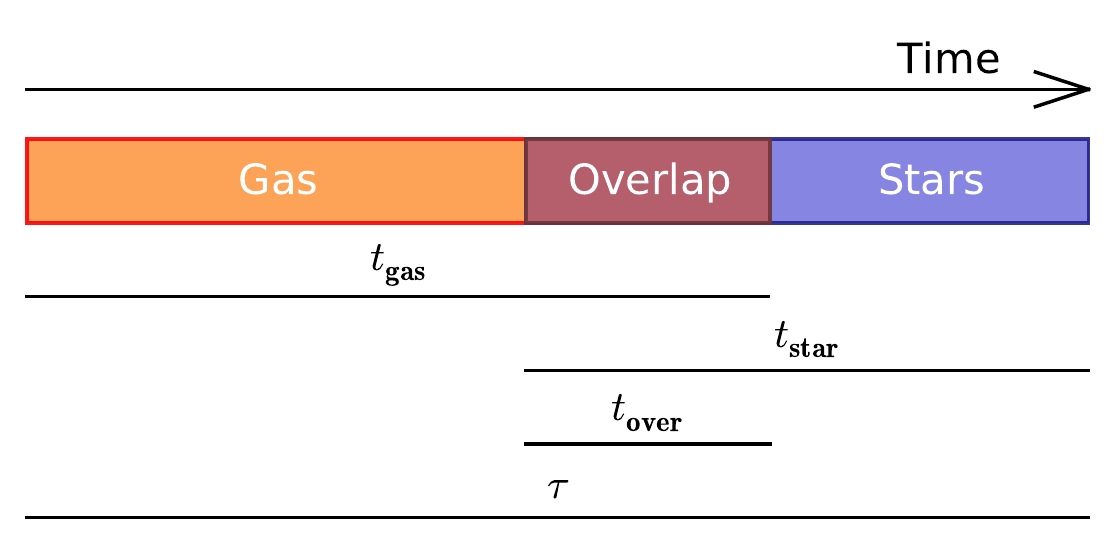}%
\vspace{-1mm}\caption{
\label{fig:tschem}
Schematic representation of the timeline of an individual star-forming region, which starts out being visible in gas tracer emission (e.g.~H{\sc i}, CO or HCN), and ends up being visible in young stellar tracer emission (e.g.~$\halpha$ or FUV). In between, there is an overlap phase during which both tracers are visible. The duration of each corresponding time-scale is indicated below the timeline.\vspace{-1mm}
}
\end{figure}

In its most general form, the first result of \citetalias{kruijssen14} is that if a macroscopic correlation (e.g.~the star formation relation between the gas mass $M_{\rm gas}$ and the SFR) is caused by a time evolution (e.g.~the conversion from gas to stars), then it {\it must} break down on small spatial scales because the subsequent phases of that time evolution are being resolved. When treating galaxies as a whole, this is generally not a concern, because most often they consist of a number of independent regions sufficiently large to ensure that the timeline of \autoref{fig:tschem} is well-sampled.\footnote{Certain dwarf galaxies, high-redshift galaxies, and galaxy mergers may represent exceptions to this idea if they host a limited number of star-forming regions, or their star-forming regions are subject to a large-scale synchronisation of their evolutionary states.} However, there must be some spatial scale below which this is no longer true. Given some typical separation length $\lambda$ between independent regions, the \klprinciple predicts that the relative uncertainty or Poisson error on the star formation relation is only smaller than unity (i.e.~the relation is well-defined and does not `break down') when the condition is satisfied that
\be
\label{eq:klprinciple}
\Delta x\Delta t^{1/2} \geq \lambda\tau^{1/2} .
\ee
Here, $\Delta x$ is the size scale over which the star formation relation is evaluated (the `aperture size'), $\Delta t$ is the shortest\footnote{This minimization only draws from the gas and young stellar phases ($\tgas$ and $\tstar$) and does not include the `overlap' phase $\tover$.} evolutionary phase from \autoref{fig:tschem}, $\lambda$ is the separation length of independent regions, and $\tau$ is the total duration of the evolutionary timeline as in \autoref{fig:tschem}. In \citetalias{kruijssen14}, we showed that the observed scale dependence of the scatter on the star formation relation \citep[e.g.][]{bigiel08,schruba10,onodera10,liu11,leroy13} is quantitatively reproduced by this simple model.\footnote{While \autoref{fig:tschem} shows a Lagrangian model that follows a single idealised region in time, a similar conclusion was reached by \citet{feldmann11,feldmann12} using a Eulerian model describing instantaneous snapshots of populations of independent regions.}

The key insight drawn from reproducing the observed, scale-dependent scatter with the simple schematic model of \autoref{fig:tschem}, is that the observed relation on 0.1--1~kpc scales provides information on processes taking place on much smaller scales. Another important implication is that the galactic-scale star formation relation results from taking an ensemble average over the cloud-scale physics. This is particularly relevant for interpreting the recent work showing that the cloud-scale star formation relations of actively star-forming regions have gas depletion times ($t_{\rm depl}\equiv M_{\rm gas}/{\rm SFR}$) shorter by a factor of 5--50 than the galactic star formation relation \citep[e.g.][]{heiderman10,lada10,lada12}.

In the context of the \klprinciple, the difference between the star formation relations on cloud and galaxy scales is not surprising. In order to place a single region in the same $M_{\rm gas}$--${\rm SFR}$ diagram as galaxies, it must both contain gas and star formation tracer emission. For the timeline of \autoref{fig:tschem}, this is equivalent to requiring the region to reside in the overlap phase. As a result, cloud-scale star formation relations cannot consider gaseous regions on the part of the timeline preceding the overlap phase and they therefore omit the gas emission from all regions outside the overlap phase. Assuming that the gas flux of a gaseous region (i.e.~a region residing in the phase covered by $\tgas$) does not strongly depend on the evolutionary phase, this means that the gas depletion times of individual regions should be a factor of $\tover/\tgas$ shorter than on galactic scales. For fiducial values of $\tover\sim3~\myr$ and $\tgas\sim30~\myr$, this implies a bias of roughly one order of magnitude, which is consistent with the large offset observed between $\tdepl\sim100~\myr$ on the cloud-scale \citep[using CO,][]{heiderman10,lada12} and $\tdepl\sim\gyr$ on the scales of entire galaxies \citep{bigiel08,schruba11}.\footnote{This depletion time bias can be increased further by differences in the adopted tracers. For instance, the quoted example adopts CO-based gas masses and depletion times, but cloud-scale studies often use a dense gas tracer like HCN or a minimum extinction contour containing even less mass than traced with CO. Such a choice results in an even shorter gas depletion time. Differences in SFR tracers are less important, because these are generally calibrated to translate the observed flux to a rate (which implicitly involves division by a reference time-scale). This is why the cloud-scale SFR inferred from young stellar object counts and galaxy-scale SFR measurements using \halpha do not necessarily differ by orders of magnitude, even though they technically trace different phases in \autoref{fig:tschem}. These differences and similarities between SFR and gas tracers are physical in nature and reflect the time evolution of the collapse and star formation process. No empirical star formation relation is therefore intrinsically incorrect. However, unless the appropriate care is taken to compare the right quantities, it may be misleading to compare different star formation relations.} An example of how to account for the region selection bias in targeted star formation studies is given in \citet{schruba17}.

Next to setting the absolute value of the gas depletion time, the part of the evolutionary timeline that is being traced should also influence the scatter of the star formation relation. Selecting higher-density gas (tracers) as in \citet{lada10} places even stronger limits on the part of the evolutionary timeline that is being probed than when using low-density gas (defined by using CO or a certain level of extinction), thus limiting the sample to regions at even more similar evolutionary stages. As a result, the scatter on the star formation relation should decrease towards higher gas densities, as is indeed observed by \citet{lada10}.
\begin{figure*}
\includegraphics[width=\hsize]{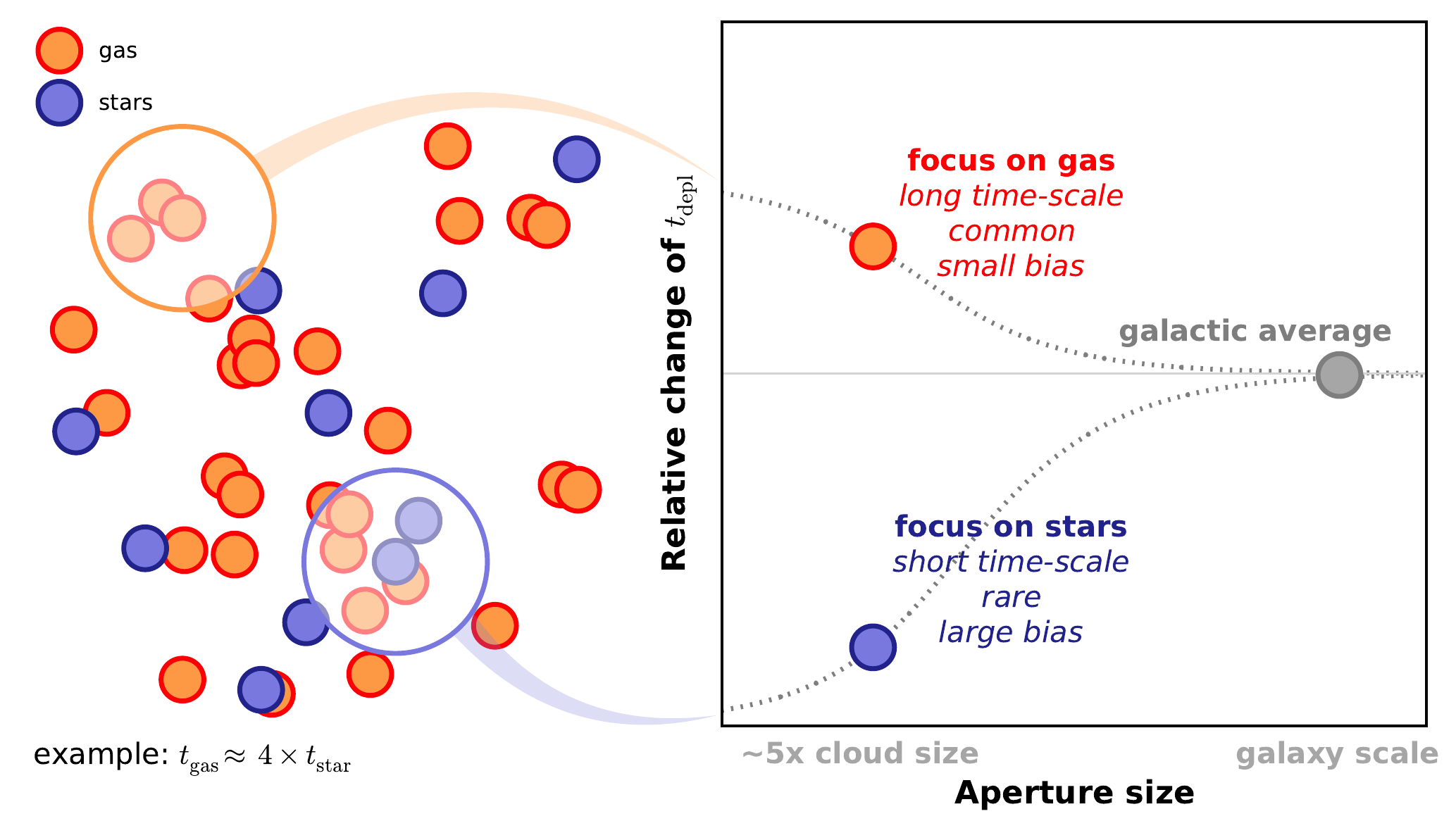}%
\vspace{-1mm}\caption{
\label{fig:tuningfork}
Schematic example of how the \klprinciple can be used to characterize the cloud-scale evolutionary timeline of star formation and feedback. Left panel: cartoon of a `galaxy' consisting of a random distribution of independent regions (circles), which are situated on the timeline of \autoref{fig:tschem} in a way that is uncorrelated to their neighbours. Orange circles indicate regions in the gas phase, whereas blue circles indicate those in the young stellar phase. In this example, the duration of the gas phase is 4 times that of the young stellar phase. The large circles represent apertures focused on a gas peak (orange) or a stellar peak (blue). Right panel: relative change of the gas depletion time (or the gas-to-stellar flux ratio) when focusing apertures on gas peaks (top branch) or young stellar peaks (bottom branch), as a function of the aperture size. On large scales, the galactic average is retrieved and the relative change is unity. However, on small scales (corresponding to the separation length $\lambda$, which is typically several times the cloud size), the excess or deficit of the gas depletion time in this `tuning fork diagram' is a non-degenerate, direct probe of the time and size scales governing the timeline of \autoref{fig:tschem}.\vspace{-1mm}
}
\end{figure*}

\subsection{The scale dependence of the star formation relation reveals cloud-scale physics} \label{sec:form2}
The second result presented in \citetalias{kruijssen14} is that the {\it way in which} star formation relations depend on the spatial scale is a direct probe of the physics of star formation and feedback on the cloud scale. It exploits the aforementioned notion that the observed star formation relation on 0.1--1~kpc scales provides information on processes taking place on much smaller scales. However, contrary to phrasing the scale-dependence in terms of the scatter of the star formation relation to access this information, this feature of the \klprinciple uses the absolute change (or `bias') of the gas depletion time when focusing a small aperture on gas or young stellar peaks to determine how rare or common the central peak is. That way, the \klprinciple can be used to constrain the time-scales governing the evolutionary timeline of \autoref{fig:tschem}.

We illustrate how the \klprinciple is applied to characterize cloud-scale star formation with an idealised example in \autoref{fig:tuningfork}. Imagine a two-dimensional, random distribution of points representing independent regions. Some part of these are dominated by gas, whereas the other part is dominated by young stellar emission. These regions represent two successive phases in the star formation process that do not overlap in time. Let us assume that time spent by a star-forming region in the `gas' phase is 4 times longer than the time spent in the `stellar' phase. This means that gas-dominated regions (`gas peaks') will be 4 times more numerous than stellar-dominated regions (`stellar peaks'). If we then consider a small region around a gas peak (defining an aperture of some size smaller than the galaxy), the local gas depletion time (i.e.~the gas-to-stellar flux ratio) in that region will be elevated compared to the galaxy-wide average, because focusing on a gas peak guarantees some excess gas flux to be present, whereas the rest of the aperture is randomly filled with gas or stellar peaks according to the galactic average. However, the depletion time excess will be minor -- the gas peaks are 4 times more common than the stellar peaks, hence the relative effect of guaranteeing the already-ubiquitous gas flux to be present is small. By contrast, if we focus an aperture on one of the rare stellar peaks (with a duration 4 times shorter than the gas phase), the corresponding decrease of the local gas depletion time is large compared to the galactic average, because a very rare phase is guaranteed to be present in the aperture.

This idealised example illustrates the fundamental thought behind the \klprinciple: the relative rarity of the subsequent phases in the cloud-scale star formation process is set by their relative durations and can be constrained from variations of the gas depletion time on small spatial scales. The \klprinciple provides a way of `counting' (and assigning relative time-scales to) these phases even if the two-dimensional structure of their emission is continuous and hard to quantise. If the duration of one of the two phases is known, then the relative time-scales translate to an absolute evolutionary timeline. In practice, this known `reference time-scale' will often refer to the stellar phase in the examples of Figures~\ref{fig:tschem} and~\ref{fig:tuningfork}, because stellar population synthesis models provide the age range over which a coeval stellar population is bright in commonly-used SFR tracers such as \halpha, FUV, and NUV \citep{haydon18}. As demonstrated in \citetalias{kruijssen14}, the small-scale excess or deficit of the depletion time when focusing apertures on gas or young stellar peaks (with a known lifetime) is then set by the three free parameters $\tgas$, $\tover$, and $\lambda$. Most importantly, we showed that these free parameters describe the `tuning fork diagram' of \autoref{fig:tuningfork} in a non-degenerate way. The absolute timeline of \autoref{fig:tschem} can therefore be uniquely constrained by fitting a model tuning fork to observations.

The sampling effects described by the \klprinciple manifest themselves on spatial scales much larger than the sizes of individual clouds or star-forming regions \citep{schruba10}, because they emerge on size scales comparable to the typical separation between such regions. This new method is therefore highly suitable for application to observations with physical resolutions of $10$--$10^3~\pc$, depending on the properties of the target galaxy. This provides a strong contrast with respect to previous methods, which required individual clouds to be resolved and could therefore only be applied to galaxies in (or around) the Local Group. Here, we present the step-by-step process for applying the \klprinciple to observations and carry out detailed numerical tests using numerical simulations of isolated disc galaxies.

\section{Step-by-step description of applying the method to galaxy maps} \label{sec:method}
In this section, we first describe the method in qualitative terms, before turning to a detailed description. The reader interested in obtaining a cursory understanding of the method is referred to Section~\ref{sec:qualdes}. Anyone planning practical applications of the method to observations or to simulated galaxy maps is encouraged to also give Section~\ref{sec:detdes} a close read.

\subsection{Qualitative description of the method} \label{sec:qualdes}
We have named the machinery for applying the method the `\code' code. \code has been developed in the Interactive Data Language (\idl),\footnote{\url{http://www.harrisgeospatial.com/ProductsandSolutions/GeospatialProducts/IDL.aspx}} which is commonly used for handling and analysing astrophysical data sets and is well-supported through several public libraries. It is compatible with \idl versions 7.0 and later. A translation of the code into \python is planned for the near future. \code currently has a couple of dependences on publicly available routines, most notably from the \idl Astronomy User's Library,\footnote{\url{https://idlastro.gsfc.nasa.gov/}} the \idl Coyote Library,\footnote{\url{http://idlcoyote.com/}} and \clfind \citep{williams94}.\footnote{\url{http://www.ifa.hawaii.edu/users/jpw/clumpfind.shtml}} Several of these routines have required modifications to ensure or optimise their compatibility with \code. The public release of \code will therefore include these modified routines.\footnote{The \code code for applying the described method is currently still proprietary, but it is planned to become publicly available in the near future, some time after the appearance of the present paper. It will become available at \url{https://github.com/mustang-project/}. The interested reader is welcome to contact the first author for further details.}

In summary, the method is aimed at the systematic and quantitative application of the formalism sketched in Section~\ref{sec:form2} to real data sets. This process requires five basic ingredients:
\begin{enumerate}
\item
selecting two maps of tracers that represent causally-related phases in a Lagrangian timeline as depicted in \autoref{fig:tschem};
\item
identifying emission peaks in this pair of maps;
\item
measuring the flux ratio of both maps around these peaks as a function of the spatial averaging scale;
\item
fitting a statistical model that describes the time-scale dependence of the resulting flux ratios (cf.~the tuning fork diagram of \autoref{fig:tuningfork});
\item
carrying out the full error propagation and calculating any derived quantities that follow from the time-scales.
\end{enumerate}
As we show in this paper, these steps enable the characterisation of the cloud-scale physics of star formation and feedback across a statistically representative galaxy sample.

In principle, the method can be applied to a pair of galaxy maps showing any tracer of interest, but physically meaningful results are only obtained if the tracer pair is related by a Lagrangian evolutionary step as in \autoref{fig:tschem}. Throughout this paper, we will mainly assume the example case in which `gas' (traced by e.g.~H{\sc i}, CO, HCN, HCO$^+$, or sub-mm dust continuum) turns into (young) `stars' (traced by e.g.~\halpha, FUV, or NUV). However, it is also possible to mask these tracers by any physical quantity for which one expects a monotonic change with time, in order to measure the time-scale on which this change takes place. For instance, if one assumes that molecular clouds evolve towards higher densities and excitation conditions during their collapse towards star formation, it is possible to set a maximum or minimum CO(3--2)$/$CO(1--0) ratio for which the CO(1--0) emission is shown, resulting in two CO maps of low- and high-excitation CO. A characteristic lifetime can be obtained for each of these, providing a time-scale for the evolution towards high-excitation conditions. Another example would be to mask by velocity dispersion in order to measure turbulent energy dissipation time-scales in clouds that evolve towards star formation. Similarly, ionized emission line ratios at optical wavelengths can be used to isolate individual stellar feedback mechanisms such as photoionising radiation, stellar winds, and supernovae, each of which take place on different time-scales, or physical quantities such as electron temperatures and densities \citep[e.g.][]{blair04,pellegrini10,mcleod15,mcleod16}. Masking maps of ionized emission lines based on these mechanisms or quantities can thus provide insight in the anatomy of cloud-scale stellar feedback. These are just a few examples -- there exist many more applications to a broad variety of astrophysical problems.

\begin{figure*}
\includegraphics[width=\hsize]{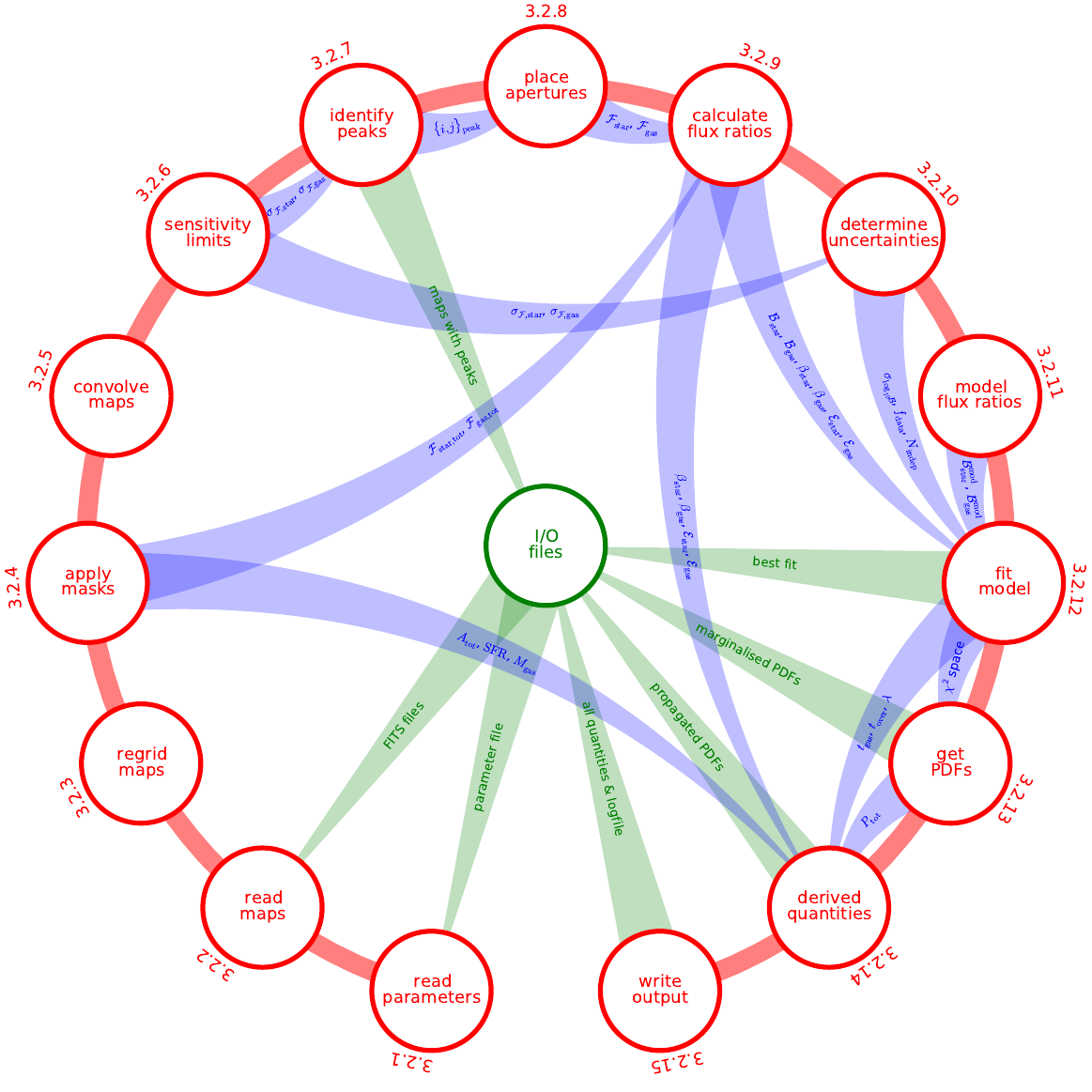}%
\vspace{-1mm}\caption{
\label{fig:schematic}
Schematic representation of the method that is presented and validated in this paper, specifically following the structure of the \code code. The red circles indicate key steps in the process, with section numbers referring to the subsections in Section~\ref{sec:detdes} that detail each step. The procedure runs from the bottom (`read parameters', Section~\ref{sec:stepreadpar}) in the clockwise direction along the red connections until it is back at the bottom (`write output', Section~\ref{sec:stepoutput}). Several of the steps interact with files on disk, by reading files or producing output. These steps are indicated with green wedges that connect to the green central `I/O files' circle. Temporary file output is not shown. The information content indicated in each wedge is contained by the circle at the wide end and flows towards the narrow end of the wedge. Likewise, the blue curved wedges indicate the flow of the highlighted variables or information between the several steps in the process. To avoid cluttering, these represent the most important (but non-exhaustive) subset of all information flows in the procedure.\vspace{-1mm}
}
\end{figure*}
The practical application of the above five steps requires several additional steps to be taken. \autoref{fig:schematic} shows the structure of the \code code, which has been developed for this purpose. Starting from the bottom of the schematic and proceeding in the clockwise direction, the method goes through the following steps (each also listing the subsection numbers in Section~\ref{sec:detdes} where that step is discussed in detail).
\begin{enumerate}
\item[(i) Section~\ref{sec:stepreadpar}:]
Read in the input parameter file that contains the settings for carrying out the analysis and use these to derive any additional parameters.
\item[(ii) Section~\ref{sec:stepreadmap}:]
Read in the FITS galaxy maps and determine the smallest possible aperture size given their spatial resolutions. One of these maps must show a phase from the timeline of \autoref{fig:tschem} with a known duration.
\item[(iii) Section~\ref{sec:stepregrid}:]
Regrid the galaxy maps by convolving them to the best common spatial resolution and changing the pixel grid to avoid extreme oversampling of a resolution element.
\item[(iv) Section~\ref{sec:stepmask}:]
Apply any masks or cuts in galactocentric radius that are specified in the parameter file to either restrict the analysis of the galaxy maps to these areas or to omit them from the analysis.
\item[(v) Section~\ref{sec:stepconvolve}:]
Convolve the galaxy maps to the range of spatial scales (i.e.~aperture sizes) specified in the parameter file using the chosen convolution kernel, with the goal of measuring the flux within a given radius around each position in the maps.
\item[(vi) Section~\ref{sec:stepsens}:]
Determine the sensitivity limit and flux zero point of each galaxy map by fitting a Gaussian to the low-flux end of the pixel flux probability distribution function (PDF).
\item[(vii) Section~\ref{sec:steppeaks}:]
Identify flux peaks in the two galaxy maps with \clfind,\footnote{There exist several modern, more accurate ways of defining clumps, overdensities, or structures and their properties in astrophysical data sets (e.g.~\citealt{rosolowsky06}, \citealt{henshaw16}, and {\sc Astrodendro}, see \url{http://www.astrodendro.org}). However, for the purpose of the presented method, we only need an algorithm to identify local maxima in two-dimensional maps. These peak coordinates should be measured reliably, but we do not use the clump properties derived by \clfind for any important steps in the analysis. This is a relatively simple task for which the results provided by \clfind are entirely adequate.} if desired using a different pair of maps than those used in the multi-scale flux integration, omit peaks with total fluxes below the sensitivity limit or those that should be excluded due to the area masking, and output images of the maps highlighting the positions of the included peaks.
\item[(viii) Section~\ref{sec:stepapertures}:]
Calculate the fluxes within the apertures of varying size focused on each peak, as well as the effective aperture size and areas accounting for masked pixels, and the distances between all possible peak pairs.
\item[(ix) Section~\ref{sec:stepfluxratios}:]
Calculate the flux ratio `bias' (i.e.~excess or deficit) relative to the galactic average for each set of peaks as in the tuning fork diagram of \autoref{fig:tuningfork}. For each aperture size and within apertures focused on each of the two sets of flux peaks, first calculate the total flux enclosed by the apertures in each of the two maps. This results in four total fluxes per aperture size: the flux in the first/second map around the peaks identified in the first/second map. For each aperture size and peak type, these fluxes are obtained by drawing Monte-Carlo realisations of peak subsets that have non-overlapping apertures, to ensure no pixels are counted twice, and by then taking the mean of the resulting total fluxes across all realisations. The flux ratio excess (i.e.~the top branch from \autoref{fig:tuningfork}) then follows by taking the peak sample from the first map and dividing the total flux around these peaks in the first map by that in the second map. The flux ratio deficit (i.e.~the bottom branch from \autoref{fig:tuningfork}) then follows by taking the peak sample from the second map and dividing the total flux around these peaks in the first map by that in the second map. These flux ratio biases are then normalised by the average flux ratio between both maps across the entire field.
\item[(x) Section~\ref{sec:steperrors}:]
Determine the uncertainties on the observed flux ratio bias measurements. The sources of uncertainty that should be considered are the sensitivity limits of the maps, which leads to flux uncertainties, and the variety of region masses, which manifests itself as a non-zero variance of the fluxes in individual apertures. Because the data points are flux ratios and thus originate from two fluxes, it is necessary to subtract the covariance between the numerator and the denominator, which reflects spatial trends in region properties. Finally, the uncertainty on each individual data point differs from the effective error bar when fitting a model to the data points, because the data points are not independent between different aperture sizes. Each data point on a branch in the tuning fork (\autoref{fig:tuningfork}) represents a different aperture size that is centred on the same set of peaks, implying that larger apertures (partially) contain the same flux as smaller apertures. We account for this by calculating the `independence fraction' of each data point relative to all other data points.
\item[(xi) Section~\ref{sec:stepmodel}:]
Derive a model that connects the flux ratio biases in the tuning fork diagram of \autoref{fig:tuningfork} to the evolutionary timeline of \autoref{fig:tschem}. Such a model was presented in \citetalias{kruijssen14}, which considered a random distribution of point-like regions that are situated at random positions on the evolutionary timeline and also allowed these regions to undergo flux evolution between the overlapping and `isolated' phases in \autoref{fig:tschem}. However, practical applications of this model also require it to deal with regions of finite densities, i.e.~regions characterized by extended emission rather than point particles. The present paper greatly improves on \citetalias{kruijssen14} by assuming that the regions follow a certain spatially-extended profile and then using the flux density contrast between the peaks and their immediate surroundings to determine the characteristic sizes of these regions in units of the region separation length.
\item[(xii) Section~\ref{sec:stepfit}:]
Carry out a reduced-$\chi^2$ fit of the model to the data points and their (independence-weighted) uncertainties to determine the quantities describing the time-evolution of independent regions. In the context of \autoref{fig:tschem}, we thus constrain the best-fitting values of the lifetime of the first phase $\tgas$ (assuming that the second phase provides the reference time-scale described in Section~\ref{sec:form2}), which in applications to molecular gas maps represents the molecular cloud lifetime, the duration of the overlap phase $\tover$, which in applications to molecular gas and SFR tracer maps represents the feedback time-scale, and the mean separation length between independent regions $\lambda$. Because we carry out a reduced-$\chi^2$ fit, the fitting process also returns the three-dimensional PDF of the above three model parameters. Additional quantities that are also returned by the fitting process (but are not specifically fitted for) are $\betastar$ and $\betagas$, which refer to the mean flux ratios of regions residing in the overlap phase relative to their `isolated' phases, i.e.~outside of the overlap phase (see Section~3.3 of \citetalias{kruijssen14}). These two parameters capture the (possibly complex) flux evolution of both tracers over the evolutionary timeline. This step also generates the output tuning fork diagram including the best-fitting model as a figure and as an ASCII table.
\item[(xiii) Section~\ref{sec:steppdfs}:]
Get the marginalised, one- and two-dimensional PDFs of each of the (pairs of) three free parameters in the model, i.e.~$\tgas$, $\tover$, and $\lambda$, and write the resulting figures and ASCII tables to disk. The error bars on the best-fitting values returned by the procedure refer to the 32nd percentile of the part of the one-dimensional PDF {\it below} the best-fitting value and the 68th percentile of the part of the PDF {\it above} the best-fitting value.
\item[(xiv) Section~\ref{sec:stepderived}:]
Calculate derived quantities using the fundamental free parameters $\tgas$, $\tover$, and $\lambda$. These quantities are the total star formation tracer lifetime including the overlap phase ($\tstar$), the total duration of the evolutionary timeline ($\tau$), the region radii ($\rstar$ and $\rgas$), the region size-to-separation ratios or filling factors ($\zetastar$ and $\zetagas$), the feedback outflow velocity ($\vfb$), the field-wide gas depletion time ($\tdepl$), the star formation efficiency per star formation event ($\sfe$), the star formation and mass removal rates per star formation event ($\mdotsf$ and $\mdotfb$), the instantaneous and time-averaged mass loading factors ($\etafb$ and $\etaavgfb$), and the feedback energy and momentum efficiencies ($\chifbe$ and $\chifbp$). The one-dimensional PDFs of each of these quantities are determined using Monte-Carlo error propagation, where we draw from the complete, three-dimensional PDF of $\tgas$, $\tover$, and $\lambda$ obtained from the fitting process as well as from the PDF of any other, independent quantities that are used to derive the above quantities. This typically leads to well-defined, nearly-noiseless cumulative PDFs. As in the previous step, each of the resulting PDFs are written to disk as figures and ASCII tables.
\item[(xv) Section~\ref{sec:stepoutput}:]
Write the output files as ASCII tables that summarise all constrained quantities and their error bars, as well as the logfile containing all of the command line analysis output. These are critical for evaluating and interpreting the results.
\end{enumerate}

The above steps can be carried out on modern personal computers or laptops within a reasonable computing time. On such systems, we have achieved typical runtimes of $1$--$20$ minutes per experiment, both for the experiments carried out in this paper and for the first applications of this method (\citealt{kruijssen18}; \citealt{haydon18}; Hygate et al.~in prep.; Schruba et al.~in prep.; Chevance et al.~in prep.; Ward et al. in prep.). This means that the method can readily be applied to observational and numerical data sets without requiring special-purpose hardware or supercomputing time.

\subsection{Detailed description of the method} \label{sec:detdes}
We now turn to a detailed description of the method summarised in Section~\ref{sec:qualdes} and \autoref{fig:schematic}. In this procedure, we extract the quantities describing the time evolution of \autoref{fig:tschem} from two galaxy maps that each show the emission from one phase in the timeline. For simplicity and clarity, we assume in this description that one of the two maps shows molecular gas traced by CO (referred to as `gas') and the other shows star formation traced by \halpha (referred to as `stars' or `stellar', even though it should technically be `young stars' or `massive stars'). Of these two maps, the stellar phase has a known duration and therefore acts as a `reference time-scale' that sets the absolute scale of the entire timeline of \autoref{fig:tschem} -- stellar population synthesis models show that the \halpha line is emitted over a $5~\myr$ time-scale \citep{haydon18}. Even though this particular example is used here, we expressly reiterate that the method is capable of handling any pair of emission maps that represent causally-related phases in a Lagrangian timeline as shown in \autoref{fig:tschem}. Other examples of tracers that could be used are atomic gas (i.e.~H{\sc I}), dense gas (e.g.~traced by HCN), 24~$\mu$m emission, ionized emission line tracers (e.g.~[O{\sc iii}] or [N{\sc ii}]), or UV emission, but these are by no means exhaustive.\footnote{Applications of the presented method to a wide variety of different tracers in the Large Magellanic Cloud will be presented by Ward et al.~in prep.}

\subsubsection{Input parameters} \label{sec:stepreadpar}
\begin{table*}
 \centering
  \begin{minipage}{\hsize}
  \caption{Flags to be set for the presented analysis}\label{tab:flags}\vspace{-1mm}
  \begin{tabular}{@{}l c l@{}}
  \hline
   Flag & Values (\textbf{default}) & Description \\
  \hline
  {\tt mask\_images} & \textbf{0}/1 & Mask images (\textbf{off}/on) \\
  {\tt generate\_plot} & 0/\textbf{1} & Generate output plots and save to PostScript files (off/\textbf{on}) \\
  {\tt derive\_phys} & 0/\textbf{1} & Calculate derived physical quantities as described in Section~\ref{sec:stepderived} (off/\textbf{on}) \\
  {\tt write\_output} & 0/\textbf{1} & Write results to output files (off/\textbf{on}) \\[1.5ex]
  {\tt use\_star2} & \textbf{0}/1 & Use a second map for identifying stellar peaks and use the default map for performing the flux calculation (\textbf{off}/on) \\
  {\tt use\_gas2} & \textbf{0}/1 & Use a second map for identifying gas peaks and use the default map for performing the flux calculation (\textbf{off}/on) \\[1.5ex]
  {\tt mstar\_ext} & \textbf{0}/1 & Mask areas in stellar map exterior to regions listed in a specified \ds region file (\textbf{off}/on) \\
  {\tt mstar\_int} & \textbf{0}/1 & Mask areas in stellar map interior to regions listed in a specified \ds region file (\textbf{off}/on) \\
  {\tt mgas\_ext} & \textbf{0}/1 & Mask areas in gas map exterior to regions listed in a specified \ds region file (\textbf{off}/on) \\
  {\tt mgas\_int} & \textbf{0}/1 & Mask areas in gas map interior to regions listed in a specified \ds region file (\textbf{off}/on) \\
  {\tt cut\_radius} & 0/\textbf{1} & Mask maps outside a specified radial interval within the galaxy (off/\textbf{on}) \\[1.5ex]
  \multirow{2}{*}{\tt set\_centre} & \hspace{-10pt}\ldelim\{{2}{10pt} 0 & Pixel coordinates of the galaxy centre are set to the central pixel of the map \\
   & \textbf{1} & Specify pixel coordinates of the galaxy centre (\textbf{default}) \\
  \multirow{2}{*}{\tt tophat$^\star$} & \hspace{-10pt}\ldelim\{{2}{10pt} 0 & Use Gaussian kernel to convolve maps to larger aperture sizes \\
   & \textbf{1} & Use tophat kernel to convolve maps to larger aperture sizes (\textbf{default}) \\
  \multirow{2}{*}{\tt loglevels$^\star$} & \hspace{-10pt}\ldelim\{{2}{10pt} 0 & Contour levels for peak identification are equally-spaced in linear space \\
   & \textbf{1} & Contour levels for peak identification are equally-spaced in logarithmic space (\textbf{default}) \\
  \multirow{2}{*}{\tt flux\_weight$^\star$} & \hspace{-10pt}\ldelim\{{2}{10pt} \textbf{0} & Peak positions correspond to brightest pixel in peak area (\textbf{default}) \\
   & 1 & Peak positions correspond to flux-weighted mean position of peak area \\
  {\tt calc\_ap\_area} & 0/\textbf{1} & Calculate the effective aperture area using the number of unmasked pixels within the target aperture area (off/\textbf{on}) \\
  \multirow{2}{*}{\tt tstar\_incl$^\star$} & \hspace{-10pt}\ldelim\{{2}{10pt} \textbf{0} & Reference time-scale $\tstarref$ does not include the overlap phase (i.e.~$\tstar=\tstarref+\tover$, \textbf{default}) \\
   & 1 & Reference time-scale $\tstarref$ includes the overlap phase (i.e.~$\tstar=\tstarref$) \\
  \multirow{3}{*}{\tt peak\_prof} & \hspace{-11pt}\ldelim\{{3}{11pt} 0 & Model independent regions as points \\
   & 1 & Model independent regions as constant-surface density discs \\
   & \textbf{2} & Model independent regions as two-dimensional Gaussians (\textbf{default}) \\
  \hline
\end{tabular}
$^\star$These flags are discussed specifically in Section~\ref{sec:stepreadpar}. For the other flags, the descriptions in this table are considered to be largely self-explanatory. Examples of setting some of these flags are given in the pertinent subsections below.\vspace{-1mm}
\end{minipage}
\end{table*}
In order to carry out the analysis depicted in \autoref{fig:schematic}, we must first specify a set of flags and input parameters. These are listed in Tables~\ref{tab:flags} and~\ref{tab:input}, respectively, and are listed in an input parameter file that is read by the \code code. Focusing first on \autoref{tab:flags}, the flags are divided into four main categories, separated by white space in the table. The first category concerns optional code modules or steps, the second enables the use of ancillary maps for carrying out the peak identification, the third covers the masking-related options, and the fourth lists the analysis options. \autoref{tab:flags} highlights the default choice of each of these flags, which will be used throughout the paper unless stated otherwise.

While the descriptions of the flags in \autoref{tab:flags} are mostly self-explanatory, we should comment in some detail on a subset for which the choice between the available options may not be obvious.
\begin{enumerate}
\item
The {\tt tophat} flag enables the use of either a two-dimensional Gaussian or tophat kernel to convolve the maps to larger aperture sizes. Technically, only the tophat kernel does this correctly -- for each pixel, it adds up all flux within the specified aperture area centred on that pixel. The Gaussian kernel adds up the surrounding flux too, but weighs it by a Gaussian profile. We do not use this kernel in the present paper, because it under-represents the distant pixels within the specified aperture size, but leave it as an available option for future work.
\item
The {\tt loglevels} flag specifies whether the contour levels that are used to identify peaks with \clfind are equally-spaced in linear or logarithmic space. In principle, there is no correct choice -- it depends on the maps to be analysed which option best identifies peaks of interest. Some trial-and-error is necessarily involved. However, given that the mass functions of molecular clouds and stellar clusters are characterized by power laws \citep[see e.g. the reviews by][]{portegieszwart10,dobbs14,kruijssen14c}, we prefer using logarithmically-spaced contour levels. As demonstrated below, this results in a satisfactory identification of peaks in the maps used in this paper.
\item
The {\tt flux\_weight} flag enables the peak position to correspond to the brightest pixel or the flux-weighted mean within the peak area. The latter option uses the area of the peak as defined by \clfind. Because there is no obvious physical definition of the peak edge (and hence its area), we prefer to use the brightest pixel.
\item
The {\tt tstar\_incl} flag indicates whether, in the context of the evolutionary timeline shown in \autoref{fig:tschem}, the specified `reference' time-scale $\tstarref$ (see Section~\ref{sec:form2}) includes the overlap phase (with duration $\tover$) or not. This depends entirely on the tracers used. In the present paper, we use pairs of maps generated from galaxy simulations. These map pairs can consist of two stellar maps covering star particles with two different, specified ranges of ages (used to test whether the method retrieves the correct evolutionary timeline, see Section~\ref{sec:starstar}), or of one such stellar map together with a gas map (used to correct the accuracy of the method under a variety of observational conditions, see Section~\ref{sec:gasstar}). In the cases where we use pairs of stellar maps, the reference time-scale includes the time overlap with the other stellar map. However, when using a stellar map and a gas map, the reference time-scale does not include this overlap phase, because new star particles may form as long as a region still contains gas. Similar considerations will hold in observational applications. When applying the presented method to measure a molecular cloud lifetime as traced by CO and using an ionized emission line like $\halpha$ or broadband UV to trace star formation, it is reasonable to assume that the corresponding reference time-scale does not include the overlap phase. $\halpha$ and UV emission trace unembedded, massive stars, which locally have blown out the residual gas while massive star formation may still be ongoing in other parts of the star-forming region \citep[e.g.][]{ginsburg16}. The emission lifetimes of SFR tracers such as $\halpha$ and FUV or NUV are set by stellar population synthesis models, which assume a coeval stellar population \citep[][also see \citealt{leroy12}]{haydon18}, and therefore do not include the time $\tover$ during which stars may still be forming --  the clock starts ticking when there is no molecular gas left.
\end{enumerate}

\begin{table*}
 \centering
  \begin{minipage}{\hsize}
  \caption{Input parameters of the presented analysis}\label{tab:input}\vspace{-1mm}
  \begin{tabular}{@{}l c l@{}}
  \hline
   Quantity [unit] & Default & Description \\
  \hline
  $D$ [\pc] & -- & Distance to galaxy \\
  $i$ [$^\circ$] & $0$ & Inclination angle \\
  $\phi$ [$^\circ$] & $0$ & Position angle \\
  $i_{\rm cen}$ [\pix] & -- & Index of $x$-axis coordinate of galaxy centre (only used if {\tt set\_centre}~$=1$, otherwise set to central pixel in map) \\
  $j_{\rm cen}$ [\pix] & -- & Index of $y$-axis coordinate of galaxy centre (only used if {\tt set\_centre}~$=1$, otherwise set to central pixel in map) \\
  $R_{\rm min}$ [\pc]$^\star$ & $0$ & Minimum inclination-corrected radius for analysis (only used if {\tt cut\_radius}~$=1$) \\
  $R_{\rm max}$ [\pc]$^\star$ & $10000$ & Maximum inclination-corrected radius for analysis (only used if {\tt cut\_radius}~$=1$) \\
  $\nsamp$$^\star$ & $10$ & Maximum number of pixels per unit FWHM of a map resolution element after regridding the maps \\
  $\theta_{\rm ast}$ [$^\circ$]$^\star$ & $10^{-6}$ & Allowed tolerance between the pixel coordinates and pixel dimensions in both maps \\
  $\nbins$$^\star$ & $20$ & Number of bins used during sensitivity limit calculation to sample the flux PDFs and fit Gaussians \\[1.5ex]
  $\lapmin$ [\pc]$^\star$ & $50$ & Minimum inclination-corrected aperture size (i.e.~diameter) to convolve the input maps to \\
  $\lapmax$ [\pc]$^\star$ & $6400$ & Maximum inclination-corrected aperture size (i.e.~diameter) to convolve the input maps to \\
  $\nap$$^\star$ & $8$ & Number of aperture sizes used to create logarithmically-spaced aperture size array in the range $[\lapmin,\lapmax]$ \\[1.5ex]
  $\npixmin$$^\star$ & $20$ & Minimum number of pixels for a valid peak (use $\npixmin=1$ to allow single points to be identified as peaks) \\
  $N_\sigma$ & $5$ & Flux multiple of the derived sensitivity limit needed for a peak to be included \\
  $\Delta\log_{10}\f_{\rm star}$$^\star$ & $2$ & Logarithmic range below flux maximum covered by flux contour levels for stellar peak identification \\
  $\delta\log_{10}\f_{\rm star}$$^\star$ & $0.5$ & Logarithmic interval between flux contour levels for stellar peak identification \\
  $\Delta\log_{10}\f_{\rm gas}$$^\star$ & $2$ & Logarithmic range below flux maximum covered by flux contour levels for gas peak identification \\
  $\delta\log_{10}\f_{\rm gas}$$^\star$ & $0.5$ & Logarithmic interval between flux contour levels for stellar peak identification \\
  $\nlinstar$$^\star$ & $11$ & Number of contours for stellar peak identification with range $\min{(\f_{\rm star})}$--$\max{(\f_{\rm star})}$ (only used if {\tt loglevels}~$=0$) \\
  $\nlingas$$^\star$ & $11$ & Number of contours for gas peak identification with range $\min{(\f_{\rm gas})}$--$\max{(\f_{\rm gas})}$ (only used if {\tt loglevels}~$=0$)\\[1.5ex]
  $\tstarref$ [\myr]$^\star$ & -- & Reference time-scale spanned by star formation tracer \\
  $\sigma_-(\tstarref)$ [\myr] & $0$ & Downwards uncertainty on reference time-scale \\
  $\sigma_+(\tstarref)$ [\myr] & $0$ & Upwards uncertainty on reference time-scale \\
  $t_{\rm gas,min}$ [\myr] & $0.1$ & Minimum value of $\tgas$ considered during fitting process \\
  $t_{\rm gas,max}$ [\myr] & $3000$ & Maximum value of $\tgas$ considered during fitting process \\
  $t_{\rm over,min}$ [\myr] & $0.01$ & Minimum value of $\tover$ considered during fitting process \\[1.5ex]
  $\nmcpeak$$^\star$ & $1000$ & Number of Monte-Carlo realisations of independent peak samples to be generated and averaged over \\
  $\ndepth$$^\star$ & $4$ & Maximum number of free parameter array refinement loops for obtaining best-fitting value \\
  $\ntry$$^\star$ & $101$ & Size of each free parameter array to obtain the best-fitting value \\
  $\nmcphys$$^\star$ & $10^6$ & Number of Monte-Carlo draws used for error propagation of derived physical quantities \\[1.5ex]
  $\log_{10}{X_{\rm star}}$$^\star$ & -- & Logarithm of conversion factor from map pixel value to an absolute SFR in $\msun~\yr^{-1}$ (only used if {\tt derive\_phys}~$=1$) \\
  $\sigma_{\rm rel}(X_{\rm star})$ & $0$ & Relative uncertainty (i.e.~$\sigma_x/x$) of $X_{\rm star}$ (only used if {\tt derive\_phys}~$=1$) \\
  $\log_{10}{X_{\rm gas}}$$^\star$ & -- & Logarithm of conversion factor from map pixel value to an absolute gas mass in $\msun$ (only used if {\tt derive\_phys}~$=1$) \\
  $\sigma_{\rm rel}(X_{\rm gas})$ & $0$ & Relative uncertainty (i.e.~$\sigma_x/x$) of $X_{\rm gas}$ (only used if {\tt derive\_phys}~$=1$) \\
  $\Psi_E$ [${\rm m}^2$~${\rm s}^{-3}$]$^\star$ & -- & Light-to-mass ratio of a desired feedback mechanism (only used if {\tt derive\_phys}~$=1$) \\
  $\psi_p$ [${\rm m}$~${\rm s}^{-2}$]$^\star$ & -- & Momentum output rate per unit mass of a desired feedback mechanism (only used if {\tt derive\_phys}~$=1$) \\
  \hline
\end{tabular}
$^\star$These parameters are discussed specifically in Section~\ref{sec:stepreadpar}. For the other parameters, the descriptions in this table are considered to be largely self-explanatory. Examples of setting some of these parameters are given in the pertinent subsections below.\vspace{-1mm}
\end{minipage}
\end{table*}
Turning to the description of the input parameters in \autoref{tab:input}, there are six main categories, again separated by white space in the table. The first category contains quantities describing the basic galaxy properties and manipulation of the maps, the second covers the definition of the aperture sizes to be used in the analysis, the third concerns the peak identification, the fourth describes the quantities needed to define the evolutionary timeline, the fifth sets the parameters for carrying out the fit of our model to the data, and the sixth lists conversion factors and constants needed to calculate the derived physical quantities discussed in Sections~\ref{sec:stepderived} and~\ref{sec:derivephys}.

As before, many of the input parameters in \autoref{tab:input} are self-explanatory, but we will discuss a subset of twelve items for which the choice of value is non-trivial.
\begin{enumerate}
\item
The parameters $R_{\rm min}$ and $R_{\rm max}$ can be chosen to limit the analysis to a certain galactocentric radial interval within a galaxy (if {\tt cut\_radius}~$=1$), which can be of great interest when studying any of the quantities constrained by our method, e.g.~the molecular cloud lifetime, star formation efficiency, or outflow velocity. However, care must be taken that the radial interval does not become too narrow, i.e.~when the area contains too few peaks (this statement is quantified in Section~\ref{sec:npeak}) or that it limits the maximum attainable effective aperture size such that the galactic average in \autoref{fig:tuningfork} cannot be reached. This requires some iteration after the first application of the analysis to a particular data set.
\item
The parameter $\nsamp$ indicates the maximum number of pixels used to sample the smallest aperture size [which is typically chosen to match the full width at half maximum (FWHM) of the native point spread function (PSF), see below] after regridding the input maps to the same pixel grid. For instance, for the default value of $\nsamp=10$ and a minimum aperture size of $\lap=50~\pc$, the maps will be placed on a grid with $\lap/\nsamp=5~\pc$ pixels. Classical Nyquist sampling corresponds to $\nsamp=2.5$, which yields maps with a factor of 16 fewer pixels than our default value and therefore results in improved performance on slow machines. However, we find that this sampling of the aperture size is too coarse to enable the peak identification process to identify all peaks in crowded regions. Setting $\nsamp=10$ enables accurate peak identification at an acceptable increase in computing time.
\item
The parameter $\theta_{\rm ast}$ defines the maximum allowed tolerance between the pixel coordinates and pixel dimensions of the maps. After the maps have been regridded to an optimal pixel scale (see Section~\ref{sec:stepregrid}), we verify in Section~\ref{sec:stepmask} that the pixel grids are the same to within $\theta_{\rm ast}$, based on the pixel coordinates and dimensions specified in the headers of the FITS files containing the galaxy maps. We adopt a default tolerance of $\theta_{\rm ast}\sim10^{-6}~{\rm deg}\sim3~{\rm mas}$, which corresponds to $\sim0.01~\pc$ in our simulated galaxy maps ($D=840~\kpc$), to $\sim1~\pc$ at a distance of $D\sim60~\mpc$, and to $\sim30~\pc$ at $z\sim3$. This default angle of $3~{\rm mas}$ is well below the spatial resolution used in this paper or the resolution generally achieved with modern observatories. This guarantees that the pixel grid cannot be a source of positioning errors. Of course, this assumes that the pixel coordinates specified in the FITS headers are correct. The astrometric precision of the maps should be verified manually before applying the presented method (see Section~\ref{sec:stepreadmap}).
\item
The parameter $\nbins$ sets the number of bins that is used to sample the pixel flux PDF and fit a Gaussian to determine the sensitivity limit of each map. The default number is $\nbins=20$, which is somewhat arbitrary but yields accurate results.
\item
The parameters $\lapmin$, $\lapmax$, and $\nap$ together define the array of aperture sizes (i.e.~diameters) to which the maps are convolved and at which the analysis is carried out. Specifically, the apertures are logarithmically spaced, such that:
\be
\label{eq:lap}
\lap(i)=\lapmin\left(\frac{\lapmax}{\lapmin}\right)^{\frac{i}{N_{\rm ap}-1}} ,
\ee
with $\{i\in\mathbb{N}~|~0\leq i\leq \nap-1\}$. The desired values of $\lapmin$, $\lapmax$, and $\nap$ strongly depend on the problem at hand. The minimum aperture size $\lapmin$ should be similar to the coarsest resolution across both maps, whereas $\lapmax$ should be large enough to enable the large-scale convergence of the curves in the tuning fork diagram of \autoref{fig:tuningfork} to the galactic average. The choice of maximum aperture size may therefore require some iteration. Likewise, there is no a priori rule for choosing the number of aperture sizes $\nap$. Most importantly, it should be large enough to sample the shape of the curves in the tuning fork diagram. The possible oversampling of these curves with a large number of aperture sizes (and hence observational data points) is harmless, because the statistical part of our analysis corrects for correlations between the data points (see Section~\ref{sec:steperrors}). However, using an unnecessarily large number of aperture sizes can greatly slow the analysis. Again, some iteration may be required to optimize this choice for the maps under consideration. For the tests performed in this paper, we find that the default values of $\lapmin$, $\lapmax$, and $\nap$ provide accurate results. In Section~\ref{sec:lapmin}, we vary $\lapmin$ and $\nap$ to carry out a resolution test and quantify the requirements for observational applications of the method.
\item
The parameter $\npixmin$ represents the minimum area of a peak identified by \clfind \citep[see Figure~2 of][]{williams94} to be considered in the remainder of the analysis. The default value of $\npixmin=20$ avoids unreliable detections without obstructing the identification of unresolved point sources. For illustration, the default pixel sampling rate $\nsamp=10$ means that the typical area within an FWHM is 75 pixels, implying that a minimum area of 20 pixels enables the identification of unresolved point sources. To apply the analysis to maps of point-like regions (see Section~\ref{sec:valid}), it is necessary to set $\npixmin=1$.
\item
The parameters $\Delta\log_{10}\f$ and $\delta\log_{10}\f$ (with subscripts `star' and `gas' referring to stars and gas, respectively) set the logarithmic range and separation of the contour levels used by \clfind to identify peaks (see Section~\ref{sec:steppeaks} for a summary of how this identification is performed, or \citealt{williams94} for a detailed discussion). Choosing the best values of these parameters requires the visual inspection of the maps produced by the analysis (see \autoref{fig:schematic}). If faint but relevant peaks are not identified, $\Delta\log_{10}\f$ should be increased, whereas it should be decreased if spurious peaks are identified. If adjacent but independent peaks are not distinguished, one should decrease $\delta\log_{10}\f$, whereas it should be increased if multiple peaks within a single independent region are each identified individually. Lacking a quantitative definition of `independent regions' other than the idealised description given in Section~\ref{sec:form1}, these choices are necessarily somewhat arbitrary. However, the details of the peak identification do not strongly affect the constrained quantities as long as an `obvious' set of physically relevant peaks is identified, because we carry out Monte-Carlo sampling of peaks to discard close pairs (see below and Section~\ref{sec:stepapertures}). The default values of $\Delta\log_{10}\f$ and $\delta\log_{10}\f$ provide a good starting point for applications of the method and are used for most of the experiments discussed in this work. The same approach applies to the choice of $\nlinstar$ and $\nlingas$. These quantities define the contour level spacing if linearly-spaced contours are chosen (i.e.~{\tt loglevels}~$=0$) and should be chosen by visual inspection of the maps produced by the analysis. In general, it is important to realise that the peak identification should have access to most of the tracer emission in the maps. If the peak positions of either tracer are biased to areas of the map that do not host most of the tracer emission, e.g.~due to differences in spatial structure and clumpiness, this negatively affects the accuracy of the method.
\item
The parameter $\tstarref$ represents the `reference time-scale' that is used to translate the derived {\it relative} evolutionary timeline to an {\it absolute} one (cf.~\autoref{fig:tschem}). Throughout this paper, we assume that $\tstarref$ refers to the SFR tracer map, because stellar population synthesis models like {\sc Starburst99} \citep{leitherer99} can provide an absolute `lifetime' for the emission from massive stars. Although the characteristic lifetimes of SFR tracers such as $\halpha$, FUV, and NUV can vary by more than an order of magnitude depending on the definition used \citep[Table 3]{leroy12}, they have been calibrated specifically for use in the presented method by \citet{haydon18}, who also considered the effect of sampling from the initial mass function (IMF) in low-SFR environments. These lifetimes are the best choice for observational applications of the method. Of course, once the lifetime of any other tracer (e.g.~CO) in the galaxy under consideration has been measured using our method, it becomes possible to use that lifetime as a reference time-scale for other measurements. In the specific context of our tests of the method using galaxy simulations in Section~\ref{sec:valid}, we use maps of the star particles in specific age bins, which grants us full control over the value of $\tstarref$.
\item
The parameter $\nmcpeak$ denotes the number of Monte-Carlo experiments used to draw independent peak samples from the parent sample at each aperture size. In other words, we generate $\nmcpeak$ different (but inevitably partially overlapping) subsets of peaks such that none of these peaks has any neighbours within $\lap$. As described in Section~\ref{sec:qualdes}, this is a necessary step to make sure that flux around peaks in crowded regions is not counted more than once. The total gas and stellar fluxes around each peak type (i.e.~gas peaks in the top branch of \autoref{fig:tuningfork}, as well as stellar peaks in the bottom branch) are then averaged over the $\nmcpeak$ experiments. Setting $\nmcpeak$ too low can result in run-to-run variations of the data points and thus the best-fitting quantities. It is therefore important to choose a sufficient number of Monte-Carlo samples, even if this goes at the expense of computing time (which scales as $\nmcpeak^2$ due to looping over all peak pairs). We find that at the default $\nmcpeak=1000$, the run-to-run variation is of the order 1--2 per cent, which is appreciably smaller than the uncertainties on the best-fitting quantities. This value requires a few minutes of runtime on modern laptops and is therefore used throughout the paper.
\item
The parameters $\ndepth$ and $\ntry$ govern the fitting process. This part of the analysis takes up a significant fraction of the computing time and can even dominate altogether. As discussed in Section~\ref{sec:qualdes}, we carry out the fit in a three-dimensional parameter space of $\tgas$, $\tover$, and $\lambda$. To achieve a reasonable computing time without sacrificing precision, we therefore iteratively refine the fitting grid by zooming in on the best-fitting point in parameter space. The parameter $\ndepth$ sets the maximum number of refinement steps (see Section~\ref{sec:stepfit} for details) and $\ntry$ sets the number of array elements in each of the three free parameters that is evaluated during each refinement step, resulting in a total number of $\ntry^3$ elements. This third-power scaling of the total number of elements means that it is very important to optimize the choice of $\ndepth$ and $\ntry$. For the default values of $\ndepth=4$ and $\ntry=101$, we obtain well-converged results within a reasonable runtime, to the extent that $\ndepth$ is typically not even reached because the condition for convergence is already satisfied during an earlier refinement step.
\item
The parameter $\nmcphys$ sets the number of Monte-Carlo draws used to perform the error propagation for the derived quantities in Section~\ref{sec:stepderived}. Again, this process can be quite time-intensive, implying that there is a trade-off between runtime and the smoothness of the PDFs of the derived physical quantities. For the default value of $\nmcphys=10^6$, the noise on the PDFs of the derived quantities is effectively unnoticeable, whereas it is entirely absent in the cumulative distribution functions (CDFs). To minimize the computing time, $\nmcphys$ can be lowered by up to two orders of magnitude, resulting in PDFs that have noticeable noise, but CDFs that are still smooth and well-defined. Throughout the paper, we adopt the default value listed in \autoref{tab:input}.
\item
The parameters $\log_{10}{X_{\rm star}}$ and $\log_{10}{X_{\rm gas}}$ represent the logarithm of the conversion factors from a pixel value to an absolute SFR (in $\msun~\yr^{-1}$) and an absolute gas mass (in $\msun$), respectively. Their values account for the conversion of the input maps' flux units to the physical units adopted in \code. As such, $\log_{10}{X_{\rm star}}$ and $\log_{10}{X_{\rm gas}}$ depend on the units of the maps under consideration as well as the distance of the galaxy, implying that these parameters need to be calculated prior to applying the method to a certain data set. The native input maps considered in the present paper show gas and stellar surface densities on a pixel scale of $\Delta x=14.25~\pc$ (see Section~\ref{sec:valid}), which means that the conversion factor to an SFR is given by
\be
\label{eq:xstar}
\log_{10}{X_{\rm star}}=\log_{10}\left[\left(\frac{\Delta x}{\pc}\right)^2\left(\frac{\tstar}{\yr}\right)^{-1}\right] ,
\ee
where the division by the stellar lifetime is necessary to convert a stellar surface density in a known age bin to a corresponding time-average SFR. For the experiments in this paper, assuming a value of $\tstar=1~\myr$ would imply $\log_{10}{X_{\rm star}}=-3.692$. Likewise, the conversion factor to a gas mass is given by
\be
\label{eq:xgas}
\log_{10}{X_{\rm gas}}=\log_{10}\left[\left(\frac{\Delta x}{\pc}\right)^2\right] ,
\ee
which for the experiments in this paper becomes $\log_{10}{X_{\rm gas}}=2.308$. These conversion factors are updated whenever the maps are regridded to a different pixel scale. Note that other data sets will require other expressions and values. The above examples are intended to illustrate the definition of these conversion factors. Also, we emphasize that the conversion factors are only used when calculating the derived physical quantities in Section~\ref{sec:derivephys}. The \klprinciple itself and the associated method to constrain the evolutionary timeline of \autoref{fig:tschem} both rely on flux {\it ratios} relative to the galactic average, implying that the conversion factors cancel.\footnote{This only holds if the conversion factors do not vary across the map. See Section~\ref{sec:stepreadmap} for a discussion of ways to apply the method to galaxies with non-uniform conversion factors.}
\item
Finally, the parameters $\Psi_E$ and $\psi_p$ are the light-to-mass ratio (or energy injection rate per unit mass) and momentum output rate per unit mass, respectively. They are only used when calculating the feedback energy and momentum efficiencies in Section~\ref{sec:derivephys}, which is done by comparing the effective output energy and momentum injection rates constrained by the method to the total rates implied by the young stellar populations in the star-forming regions. These parameters are needed to derive the total injection rates -- their values depend on the feedback mechanism under consideration and the time-scale over which the rates are averaged. For most feedback mechanisms, suitable numbers for these quantities can be obtained from stellar population synthesis models \citep[e.g.][]{leitherer99,agertz13}.
\end{enumerate}

All flags and input parameters are specified in an input file, which is read in by \code. Next to the elements listed in Tables~\ref{tab:flags} and~\ref{tab:input}, it also contains the full path of the maps and any \ds region files used in the analysis to mask the maps.\footnote{To create the \ds region files used for masking (see below), \ds is available at \url{http://ds9.si.edu}. In addition, \ds region files may be created manually, using the region file description that can be found at \url{http://ds9.si.edu/doc/ref/region.html}.} In addition, the input file includes a small number of quantities that are not listed in Tables~\ref{tab:flags} and~\ref{tab:input} because they are not relevant to the present paper, but are required to run the analysis. They are included in the \code documentation when it is made publicly available.

\subsubsection{Selecting and reading in galaxy maps} \label{sec:stepreadmap}
The next step in the analysis process is to read in the galaxy maps in FITS format to which the method is applied. In Section~\ref{sec:qualdes}, we explained that these maps can show any tracer pair of interest that follows a Lagrangian evolutionary connection as in \autoref{fig:tschem}. As stated previously, here we follow an example case in which `gas' turns into (young) `stars', but we reiterate that the presented method applies much more generally.

Irrespective of their physical nature, the pair of maps should satisfy a number of conditions.
\begin{enumerate}
\item
The maps should be two-dimensional and share the same dimensions and pixel grid. In the case of a spectral cube containing line emission data, this means one should use a moment-0 map. The pixel grid is defined using pixel coordinates $\{i,j\}$, with $\{i\in\mathbb{N}~|~0\leq i\leq N_{{\rm pix},x}-1\}$ and $\{j\in\mathbb{N}~|~0\leq j\leq N_{{\rm pix},y}-1\}$, where $\{N_{{\rm pix},x},N_{{\rm pix},y}\}$ represent the number of pixels in the $\{x,y\}$ directions. The pixel values represent the stellar and gas flux densities $\f_{{\rm star},ij}$ and $\f_{{\rm gas},ij}$, respectively.
\item
The astrometric precision of the pair of maps should be sufficient, such that positional uncertainties are considerably smaller than the resolution FWHM. The presented method correlates the spatial structure in pairs of tracer maps to derive the underlying evolutionary timeline. It therefore relies strongly on accurate position information. We have quantified this by carrying out experiments with a positional offset in one of the maps, which show that an acceptable astrometric precision is $\sim1/3$ of the FWHM. This is commonly achieved with modern observatories, which routinely achieve astrometric precision of the order of 10~per cent of the FWHM or better. For instance, ALMA achieves a precision of 0.05 times the FWHM, with a minimum of $3~{\rm mas}$.\footnote{See the ALMA Technical Handbook at \url{https://almascience.eso.org/documents-and-tools/}.} Maps for which the astrometric uncertainty exceeds $1/3$ of the FWHM cannot be used for measuring $\tover$, because such large positional offsets prohibit the statistical identification of physically co-spatial regions in both maps. If the astrometric uncertainties are not much smaller than the FWHM, the duration of the overlap phase $\tover$ (see \autoref{fig:tschem}) will be underestimated.
\item
One has to be relatively confident that each cloud or star-forming region that is visible in one tracer is also visible in the other tracer at some point in its lifecycle. The constrained lifetimes are a population average over all regions, meaning that if this condition is not satisfied, the lifetime includes a zero-duration contribution from those regions that never appear in one of the tracers. However, it does not necessarily pose a problem if regions never appear in either tracer, because then they are simply omitted from the population-average evolutionary timeline altogether. In such a case, the method returns the `visibility time-scale' and `visible separation length' of the tracers. The retrieved time-scales still match the `true' underlying lifetimes if the number of invisible regions is small or these regions are consistent with being a randomly drawn subset of the parent population. There are no general guidelines for dealing with region visibility, because its implications depends strongly on the question at hand and the tracers used. Specific examples are discussed in Sections~\ref{sec:expval} and~\ref{sec:limit}.
\item
Contamination should be minimized such that the emission from each tracer can be considered to be almost exclusively associated with the physical objects of interest. In the case of $\halpha$ emission, one would want to avoid emission from shock-heated galactic accretion or outflows that are not driven by stellar feedback. Likewise, 24$\mu{\rm m}$ emission may trace both young and evolved stars, which complicates its interpretation as a star formation tracer. In general, diffuse emission due to unresolved, low-mass objects or photon scattering at large distances from the emission sources should be removed from the maps. Ideally, this filtered emission is added back into the image by rescaling the pixel values. The main reason for filtering diffuse emission is that it does not belong to the population of peaks that is being studied. In extreme cases, the presence of a large flux reservoir without peaks in one of the tracers can result in a tracer {\it deficit} relative to the galactic average when focusing on the corresponding tracer peaks, because the diffuse flux contributes to the galactic average without contributing to the flux around the identified peaks (see Section~\ref{sec:diffuse} for a detailed discussion). This should be avoided. A new module in \code has the purpose of automatically filtering out diffuse emission \citep{hygate18}, but if this module is not used, the maps should be visually inspected and pre-processed if necessary.
\item
The emission maps should be as homogeneous as possible, avoiding major variations of extinction, excitation conditions (e.g.~temperature), and chemical abundances across the considered area. In principle, it is one of the method's main strengths that it is insensitive to galaxy-to-galaxy variations of the uncertain conversion factors between gas tracer flux and the gas mass that are well-known to hamper studies of extragalactic star formation \citep[e.g.][]{daddi10b}. However, if these quantities do vary strongly {\it within} a map (e.g. in the case of radial gradients in galaxies), this may pose a problem. In that case, one can still apply the method, but it should be acknowledged that a tracer lifetime will be obtained under the population-average conditions of the varying quantity. If this is undesirable (e.g.~due to a spatially-varying CO-to-H$_2$ conversion factor), it is recommended to either apply this conversion factor to the map beforehand, or to apply the method separately to radial bins in which any variation is within acceptable limits.
\item
The sensitivities or detection limits of the maps should enable a representative fraction of the tracer emission of interest to be recovered.
\item
The spatial resolution of the maps should enable the characteristic separation length between independent regions $\lambda$ to be resolved. This can only be verified after applying the method and is discussed in detail in Section~\ref{sec:lapmin}.
\item
The largest effective aperture size (after subtracting any masked pixels from the aperture) should enable the convergence towards the galactic average in \autoref{fig:tuningfork} to be reached. If this is not achieved, any masks or limits on the range of galactocentric radii should be relaxed to increase the largest effective aperture size. Should the galactic average still not be retrieved in any circular aperture (this is possible for irregular fields), the individual maps may be unsuitable for applying the method. The method may still be applied to such galaxies by combining the images of multiple galaxies at the same physical scale side-by-side in a single map, or combining the peak samples obtained for the individual galaxies, as long as these combinations are carried out prior to fitting the model. In this case, the method constrains the population-average evolutionary timeline of \autoref{fig:tschem} for the combined galaxy sample.
\item
If one wishes to recover absolute time-scales, then the lifetime of the tracer shown in one of the maps must be known, either directly from stellar population synthesis modelling \citep[e.g.~when using $\halpha$, FUV, or NUV, see][]{haydon18}, or by having been calibrated to such a reference time-scale in a separate application of the method. In other words, if the CO lifetime has been measured by applying the method to an $\halpha$ map and a CO map, then either of these can be used as a reference time-scale when measuring the lifetimes of H{\sc i} clouds across the same area.
\end{enumerate}
While it may seem that the above points set stringent limits on the applicability of the method, there currently exists a broad range of observational data sets that satisfy these conditions. These conditions are merely intended as guidelines to carefully consider when selecting the maps. In general, the \klprinciple should be thought of as a purely empirical method that as such will not necessarily yield a `wrong' answer if any of the above conditions are violated. However, the choice of maps and their properties together {\it define} the question that is answered by applying the method. Therefore, the interpretation of the results depends on to what extent these guidelines for selecting the maps are satisfied. In practice, some of the above guidelines cannot be evaluated in advance (e.g.~points vii and viii) and must be addressed a posteriori, after the analysis with \code is completed. The above list should be evaluated in addition to the guidelines for observational applications from Section~\ref{sec:guide}. We note that all of the above conditions are satisfied by the simulated maps that are used in this paper.

It is possible to select different maps for the peak identification and for the flux integration. An example where this may be of interest is when using CO(1--0) observations to trace molecular gas. These position-position-velocity data cubes can be quite noisy, meaning that a signal-masked moment-0 map showing only the emission above a minimum signal-to-noise level is often needed to select the real peaks in the map and avoid selecting noise peaks. However, this omits real flux from the map, because part of the emission below the signal-to-noise threshold is real. In this case, it may be desirable to use a different map that does include this emission when calculating the fluxes in the apertures focused on each peak. For instance, a moment-0 map that is generated without a signal-to-noise threshold, but instead uses a certain velocity mask around the velocity field of atomic gas (if available) may help to include and isolate all real CO emission even in the presence of noise. A flux calculation based on such a map would be accurate on scales larger than a few times the PSF, because the noise maxima and minima should then statistically cancel.

\begin{figure}
\includegraphics[width=\hsize]{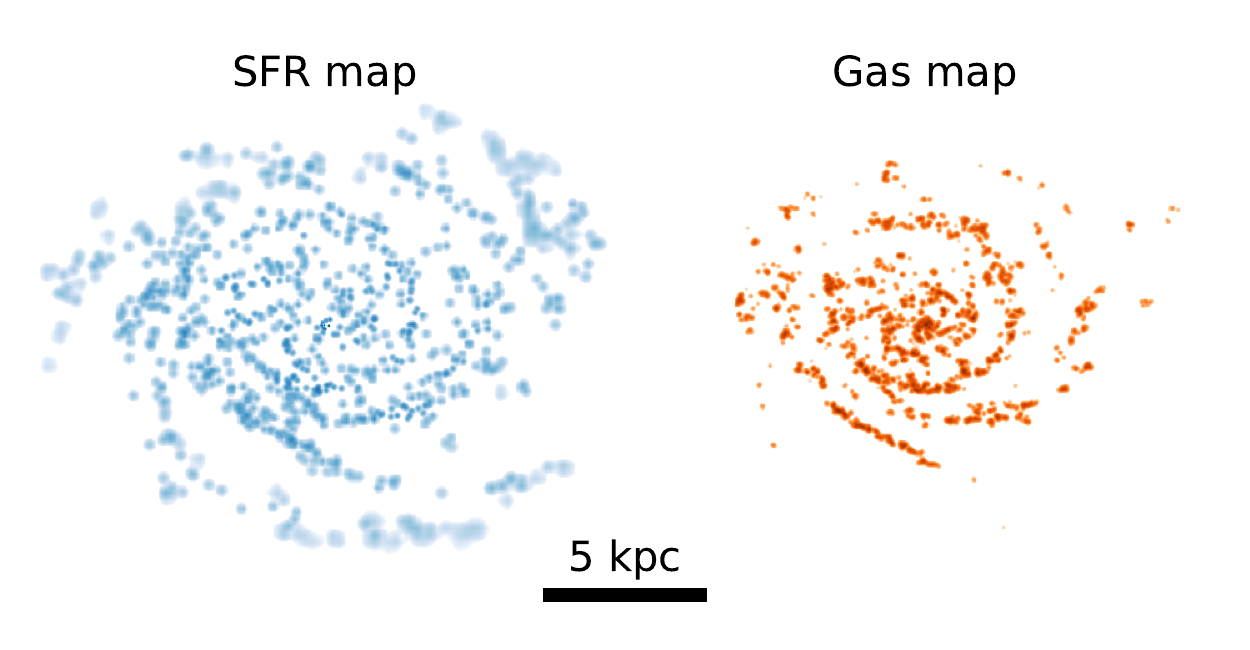}%
\vspace{-1mm}\caption{
\label{fig:stepmaps}
Example of two simulated galaxy maps (the high-resolution, extended-emission versions of experiment ID~37 in Section~\ref{sec:gasstar}) that are used for illustrating the method throughout this section. Shown are the SFR (left) and gas (right) maps on a logarithmic stretch over three orders of magnitude. The spatial scale is indicated by the scale bar at the bottom.\vspace{-1mm}
}
\end{figure}
Given a selected set of maps, \code reads in the FITS files. Throughout the description of the method, we will use a pair of example maps to illustrate the application of the method. These are taken from the high-resolution, extended emission simulated maps of experiment ID~37 in Section~\ref{sec:gasstar}. We describe later how these particular maps were generated -- at this point of the paper, they exclusively serve an illustrative purpose. The maps are presented in \autoref{fig:stepmaps} and represent the input for later illustrations of the analysis. When reading in the maps, \code immediately applies individual masks $\xi_{{\rm star},ij}$ and $\xi_{{\rm gas},ij}$ according to the \ds region files that are specified for each of them. These region files may be composites of any number of sub-regions and can be used to either include or exclude any pixels of which the pixel centres reside within the area enclosed by the region files. The values of excluded pixels are set to {\tt NaN} in each individually-masked map. The synchronisation of each of these masks across all maps is deferred to a later step, in which any cuts in galactocentric radius are also applied (see Section~\ref{sec:stepmask}).

After reading in the input file as described in Section~\ref{sec:stepreadpar} and subsequently reading in the maps, their FITS headers, and the \ds region files, the analysis is self-contained and does not require reading any further files. Before continuing the analysis, we compare the PSF sizes and determine the largest PSF size $l_{\rm PSF,max}$ across the set of maps. Assuming a Gaussian PSF, this size corresponds to an FWHM. Given the grid of aperture sizes $\lap$ defined by the input file and provided in equation~(\ref{eq:lap}), we determine the aperture size that is closest to the maximum PSF size in logarithmic space, i.e.~we minimize the quantity $|\log_{10}(\lap/l_{\rm PSF,max})|$, and set $\lapmin$ to the aperture size where this minimum is reached. This ensures that we only consider aperture sizes that are at least marginally resolved in both of the maps.

\subsubsection{Regridding the galaxy maps} \label{sec:stepregrid}
The runtime of the analysis increases with the area of the maps in units of square pixels. For this reason, we regrid the galaxy maps to a smaller number of pixels if the original pixel size $l_{\rm pix}$ allows doing so. As discussed in Section~\ref{sec:stepreadpar}, the target pixel size is $\lapmin/\nsamp$, which ensures that resolution elements are resolved by a sufficient number of pixels to reliably identify emission peaks. We therefore regrid the maps if the target pixel size $\lapmin/\nsamp>1.2l_{\rm pix}$, where the factor of 1.2 is added to skip the regridding if this yields less than a 30 per cent decrease of the map area in units of square pixels. In addition, the maps are convolved with a two-dimensional Gaussian PSF to a common resolution of $\lapmin$, provided that $\lapmin>1.05l_{\rm PSF}$. Again, the factor of $1.05$ is added to skip the convolution if only a 5 per cent change of resolution would be achieved. If the beam is non-circular, we use the effective beam width $l_{\rm PSF}=(l_{{\rm PSF},a}l_{{\rm PSF},b})^{1/2}$, where $l_{{\rm PSF},a}$ and $l_{{\rm PSF},b}$ indicate the beam width along the semi-minor and semi-major axes, respectively. The convolution of the maps then proceeds with a kernel FWHM of $(\lapmin^2-l_{\rm PSF}^2)^{1/2}$, which ensures that the final resolution equals $\lapmin$ irrespective of the original circular PSF size. The FITS header is updated with the new map dimensions $\{N_{{\rm pix},x},N_{{\rm pix},y}\}$, coordinates of the galactic centre $\{i_{\rm cen},j_{\rm cen}\}$ (which are taken to correspond to the central pixel of the image if ${\tt set\_centre}=0$), and pixel size $\theta_{\rm pix}$. Finally, the updated FITS file is written to disk for later use. The above procedure is applied to each of the maps under consideration.

After the regridding is completed, we update several of the quantities needed for the remainder of the analysis. 
\begin{enumerate}
\item
The original position of the galactic centre pixel $\{i_{\rm cen},j_{\rm cen}\}$ is converted to a fraction of the dimensions of the original maps in units of pixels $\{N_{{\rm pix},x},N_{{\rm pix},y}\}$. Interpolation of the pixel dimensions of the regridded maps then provides the position of the central pixel in the new maps and thus updates the original values of $\{i_{\rm cen},j_{\rm cen}\}$.
\item
The pixel size in $\pc$ is determined from the angular size $\theta_{\rm pix}$ in the FITS file header of the regridded map by writing
\be
\label{eq:pixtopc}
l_{\rm pix}=\frac{D\tan{\theta_{\rm pix}}}{\sqrt{\cos{i}}} ,
\ee
where the square root in the denominator accounts for the fact that a non-zero inclination only affects one of the pixel dimensions and thus does not enter linearly in the effective pixel size. If the distance or inclination angle are poorly constrained, we recommend carrying out the analysis for a reasonable range of distances or inclination angles to probe the impact of their uncertainties on the inferred quantities.
\item
We check that there are at least two elements in the aperture size array $\lap$ that exceed the pixel diagonal $\sqrt{2}l_{\rm pix}$ to enable a numerically (but not necessarily physically) meaningful application of the method. If this is not satisfied, the analysis is stopped to modify the grid of aperture sizes to increase $\lap$. This can happen if $\nsamp<\sqrt{2}$ and the dynamic range of $\lap$ is too small, or if the original pixel size was too large to enable regridding.
\item
If the smallest aperture size $\lapmin$ is smaller than the pixel diagonal, $\lapmin$ is changed to the first element of $\lap$ that does exceed the pixel diagonal. This can only happen if $\nsamp<\sqrt{2}$, or if the original pixel size was too large to enable regridding. Increasing $\lapmin$ ensures that all aperture sizes can be assigned accurate enclosed fluxes.
\item
Finally, we multiply the conversion factors $X_{\rm star}$ and $X_{\rm gas}$ that translate the pixel value to an absolute SFR in $\msun~\yr^{-1}$ and an absolute gas mass in $\msun$ by a factor of $(l_{\rm pix}/l_{\rm pix,orig})^2$ (with $l_{\rm pix,orig}$ the original pixel size) to account for the fact that the pixel size was changed by regridding the maps. This step is required because the regridding process treats the pixel values as surface densities and thus does not conserve the sum of all pixel values.
\end{enumerate}

\subsubsection{Applying the masks} \label{sec:stepmask}
Having regridded the maps on to an optimal pixel grid and after redefining several of the related quantities, the masks applied to each of the individual maps $\xi_{{\rm star},ij}$ and $\xi_{{\rm gas},ij}$ in Section~\ref{sec:stepreadmap} must be synchronised and applied uniformly to all maps. In addition, any desired cuts in galactocentric radius should be made here. Before proceeding, it should first be verified that the pixel grid of the maps is the same to within the specified tolerance angle $\theta_{\rm ast}$. We carry out the following checks.
\begin{enumerate}
\item
All maps must have identical dimensions $\{N_{{\rm pix},x},N_{{\rm pix},y}\}$.
\item
All maps must have identical equinoxes (e.g.~J2000).
\item
Any differences in pixel size $\theta_{\rm pix}$ between the maps must be within the specified tolerance, i.e.~$|\theta_{\rm pix,star}-\theta_{\rm pix,gas}|\leq\theta_{\rm ast}$.
\item
To definitively exclude any discrepancies between the maps, we carry out the brute-force test of checking whether all pixels in the map have the same right ascension $\alpha$ and declination $\delta$ within the specified tolerance, i.e. $|\alpha_{\rm star}-\alpha_{\rm gas}|\leq\theta_{\rm ast}$ and $|\delta_{\rm star}-\delta_{\rm gas}|\leq\theta_{\rm ast}$.
\end{enumerate}
If any of the above conditions is not satisfied, we abort the analysis. In such a case, the regridding should be redone and verified.

\begin{figure}
\includegraphics[width=\hsize]{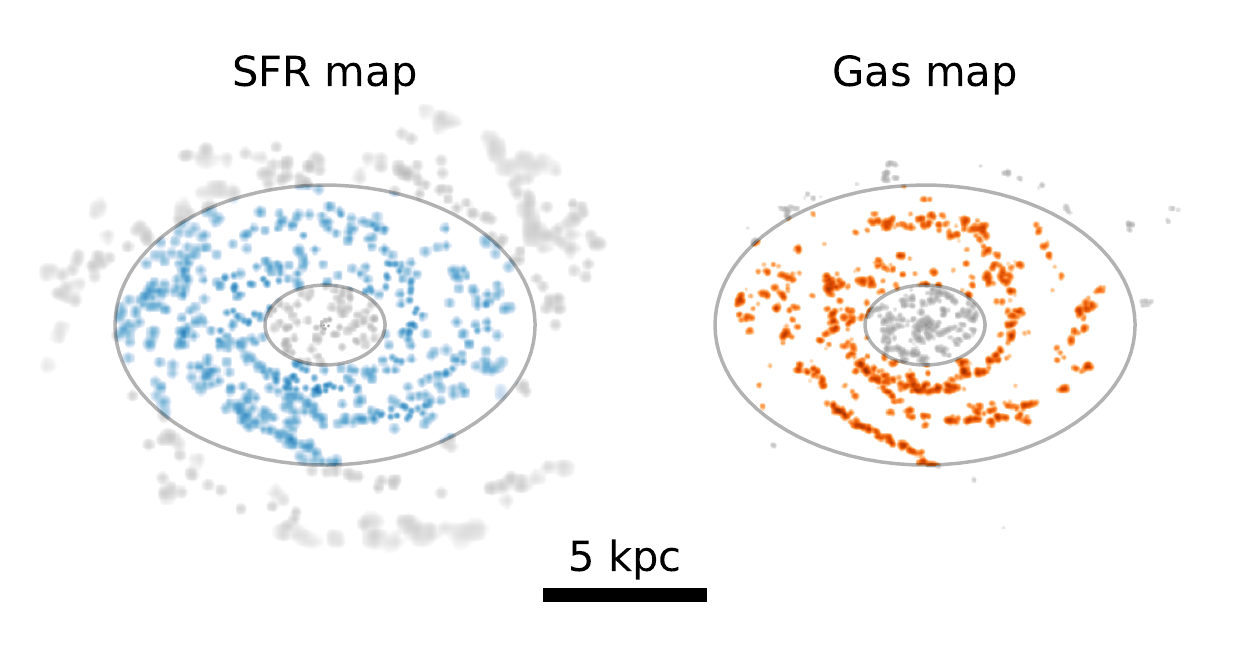}%
\vspace{-1mm}\caption{
\label{fig:stepmask}
Example of applying a galactocentric radius mask to the simulated galaxy maps from \autoref{fig:stepmaps}. Only radii $2<R/\kpc<7$ are included and the masked pixels are shown in greyscale. For the remainder of Section~\ref{sec:detdes}, any further illustrations of the method will be restricted to this radial interval.\vspace{-1mm}
}
\end{figure}

The individually-masked maps $\xi_{{\rm star},ij}$ and $\xi_{{\rm gas},ij}$ are then used to create a joint mask $\Xi_{ij}$ that is applied to all maps. If any pixel has a {\tt NaN} value in any of the maps, that pixel value is set to {\tt NaN} across all maps. We also create a mask array $\Xi_{ij}^0$ in which the masked pixels have a value of zero and the unmasked pixels have a value of unity. This array enables the straightforward calculation of effective aperture sizes in the later steps of the analysis, or the calculation of quantities that are integrated over the entire map. In addition, we perform any radial cuts specified in the input file, accounting for the position angle and inclination of the host galaxy. For each pixel, we calculate the $X_{\rm pix}$ and $Y_{\rm pix}$ distances to the galactic centre (which is taken to correspond to the central pixel of the image if ${\tt set\_centre}=0$) as
\be
\label{eq:xypix}
\begin{aligned}
X_{\rm pix}=D\tan(\theta_{\rm pix})&\left[\cos(\phi)(i_{\rm pix}-i_{\rm cen})-\right. \\
&\left.\sin(-\phi)(j_{\rm pix}-j_{\rm cen})\right] ,
\end{aligned}
\ee
and
\be
\begin{aligned}
Y_{\rm pix}=D\frac{\tan(\theta_{\rm pix})}{\cos(i)}&\left[\sin(-\phi)(i_{\rm pix}-i_{\rm cen})+\right. \\
& \left.\cos(\phi)(j_{\rm pix}-j_{\rm cen})\right] ,
\end{aligned}
\ee
for a position angle $\phi$. The galactocentric radius of each pixel is
\be
\label{eq:xpix}
R_{\rm pix}=\sqrt{X_{\rm pix}^2+Y_{\rm pix}^2} .
\ee
We remind the reader that the units of $i_{\rm pix}$, $i_{\rm cen}$, $j_{\rm pix}$, and $j_{\rm cen}$ are in units of pixels, whereas $X_{\rm pix}$ and $Y_{\rm pix}$ share the units of $D$ and represent distances in $\pc$. With the galactocentric radii of the pixels known, we set the values of the pixels across all maps to {\tt NaN} if they satisfy $R_{\rm pix}<R_{\rm min}$ or $R_{\rm pix}>R_{\rm max}$. \autoref{fig:stepmask} illustrates the application of a mask that selects galactocentric radii $2<R/\kpc<7$ for the example maps considered in this section.

After including any cuts in galactocentric radius in the total mask arrays $\Xi_{ij}$ and $\Xi_{ij}^0$, the mask is applied to all maps and the updated FITS files are written to disk for later use. A number of important quantities describing the maps are calculated as follows. The total area $A_{\rm tot}$ of the unmasked pixels in the map is
\be
\label{eq:totalarea}
A_{\rm tot}=l_{\rm pix}^2\sum_{i,j}\Xi_{ij}^0 .
\ee
The total stellar and gas flux in the map ($\f_{\rm star,tot}$ and $\f_{\rm gas,tot}$, respectively) are given by summation over all unmasked pixels:
\be
\label{eq:fstartot}
\f_{\rm star,tot}=l_{\rm pix}^2\sum_{i,j}\f_{{\rm star},ij}\Xi_{ij}^0 ,
\ee
and
\be
\label{eq:fgastot}
\f_{\rm gas,tot}=l_{\rm pix}^2\sum_{i,j}\f_{{\rm gas},ij}\Xi_{ij}^0 .
\ee
Likewise, we determine the total SFR and gas mass $M_{\rm gas}$ in the unmasked part of the map by summation over all unmasked pixels:
\be
\label{eq:sfrtot}
{\rm SFR}=X_{\rm star}\sum_{i,j}\f_{{\rm star},ij}\Xi_{ij}^0 ,
\ee
and
\be
\label{eq:mgastot}
M_{\rm gas}=X_{\rm gas}\sum_{i,j}\f_{{\rm gas},ij}\Xi_{ij}^0 .
\ee
The uncertainties on ${\rm SFR}$ and $M_{\rm gas}$ (including the statistical uncertainty) is captured in those of the conversion factors, i.e.
\be
\label{eq:sfrtoterr}
\sigma({\rm SFR})=\sigma_{\rm rel}(X_{\rm star}){\rm SFR} ,
\ee
and
\be
\label{eq:mgastoterr}
\sigma(M_{\rm gas})=\sigma_{\rm rel}(X_{\rm gas})M_{\rm gas} .
\ee
These quantities represent the most important macroscopic properties of the maps. Given that we will not use the original maps at any point of the analysis below, the term `map' from hereon refers to the unmasked part of a map, i.e.~where $\Xi_{ij}=1$. This excludes any masked pixels, unless specified otherwise.

\subsubsection{Convolution of the maps} \label{sec:stepconvolve}
The next step is to create the maps in which the emission peaks will be identified and to generate the convolved maps for determining the stellar and gas flux around the peak positions. For consistency with the minimum aperture size $\lapmin$ considered in the analysis, we do not identify the peaks in the original maps, but in maps that have been convolved to a resolution with an FWHM equal to $\lapmin$ (as in \autoref{fig:stepmask}). If no separate maps are specified for this purpose, the peaks will simply be identified in the regridded maps and no convolution is needed, because the maps used as input for the regridding were already convolved to a PSF size equal to $\lapmin$. However, if the peak identification does use separate maps, we convolve them with a two-dimensional Gaussian PSF with FWHM $(\lapmin^2-l_{\rm PSF}^2)^{1/2}$, analogous to convolution carried out prior to the regridding in Section~\ref{sec:stepreadmap}.

As described in Section~\ref{sec:qualdes}, the calculation of the flux around emission peaks is performed by convolution of the maps with a kernel of the desired size and shape. Various papers in the literature have used Gaussian kernels for this purpose, with a final FWHM equal to the size scale of interest \citep[e.g.][]{bigiel08,liu11,leroy13,leroy16}. While we have included the Gaussian kernel as an option in \code, we stress that this is not the optimal choice for the problem at hand, because it includes flux beyond the target radius and emphasizes emission close to the aperture centre. When aiming to obtain an unbiased measurement of the total flux within some radius of a given position in the map, this is achieved by convolution with a two-dimensional tophat kernel:
\be
\label{eq:tophat}
W(r,h)=\left\{\begin{array}{ll}
1 & \mbox{if } r \leq h \\
0 & \mbox{if } r > h .
\end{array} \right.
\ee
To define an aperture of size (i.e.~diameter) $\lap$, we set $h=\lap/2$. However, the numerical application of this kernel is carried out on the pixel grid, which means that the effective aperture radius in units of pixels becomes
\be
\label{eq:tophatpix}
\frac{h}{{\rm pix}}=\frac{\lap}{2\l_{\rm pix}}=\frac{\lap\sqrt{\cos{i}}}{D\tan{\theta_{\rm pix}}} .
\ee
This expression accounts for the fact that the effective pixel size increases with the inclination, which results in an aperture size in units of the number of pixels that decreases with the inclination.

\begin{figure}
\includegraphics[width=1.017\hsize]{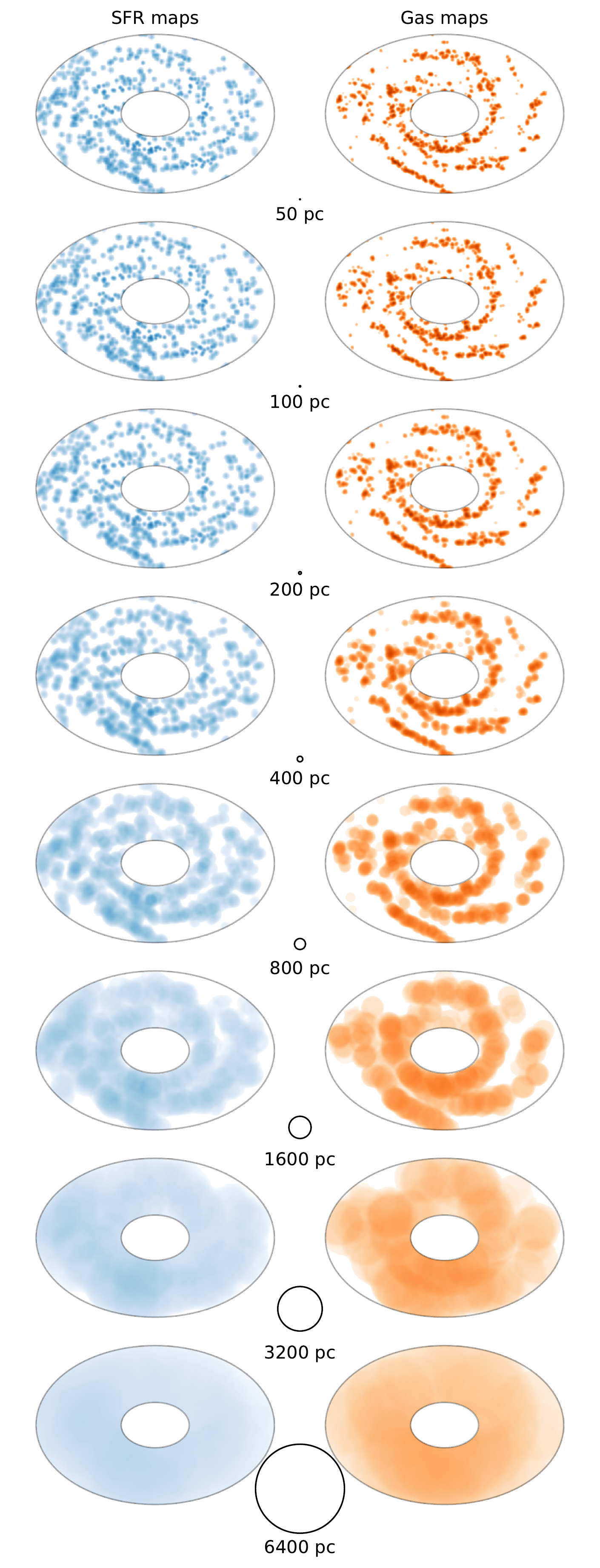}%
\vspace{-1mm}\caption{
\label{fig:stepconvolve}
Application of tophat convolution with different aperture sizes $\lap(i)$ to the pair of maps from \autoref{fig:stepmask}. The aperture sizes are shown in the middle of each row. The sequence of maps illustrates how each pixel value represents the mean flux surface density in a surrounding aperture.\vspace{-1mm}
}
\end{figure}
While the functional form of equation~(\ref{eq:tophat}) is very simple, its numerical application on a discretised pixel grid is more involved. If we define $W_{ij}$ as the kernel value of a pixel with coordinates $\{i,j\}$, then we trivially obtain $W_{ij}=1$ and $W_{ij}=0$ for any pixels that fall entirely within or outside the circular aperture boundary defined by $r=h$, respectively. However, the aperture boundary also intersects with a significant number of pixels, for which the kernel value should become a real number between zero and unity, i.e.~$\{W_{ij}\in\mathbb{R}~|~0\leq W_{ij}\leq1\}$. For each intersected pixel, we decompose the geometry of the intersection with the aperture into a combination of squares, rectangles, triangles, and/or circular segments, which can then be added or subtracted to analytically calculate the exact fraction of the pixel area $f_{\rm encl}$ that is enclosed by the aperture. The details of this calculation are provided in Appendix~\ref{sec:appcircseg}. By setting $W_{ij}=f_{\rm encl}$ for any pixels that are intersected by the aperture, we obtain a tophat kernel that achieves flux conservation at machine precision when convolved with the maps.

Having defined the convolution kernel, we obtain convolved stellar and gas maps for each aperture size, i.e.~$\f_{{\rm star},ij}(\lap)$ and $\f_{{\rm gas},ij}(\lap)$, as well as convolved mask arrays $\Xi_{ij}^0(\lap)$. Note that we do not subtract the PSF size from the kernel size as we did previously when regridding the maps and convolving them to a common Gaussian FWHM (see Section~\ref{sec:stepreadmap}), because the goal of the current convolution is to obtain the exact flux within the specified aperture radius. If the target aperture size $\lap\leq1.05l_{\rm PSF}$, \code returns a warning to notify the user that flux is being measured in an aperture smaller than the PSF, but proceeds none the less. The result of convolving the maps with tophat kernels of sizes $\lap(i)$ as defined in equation~(\ref{eq:lap}) is shown in \autoref{fig:stepconvolve} for the example pair of maps that is used throughout this section to illustrate the method. The sequence of maps nicely visualises how each pixel is assigned the mean flux surface density in a circular aperture centred on that position. While this does not noticeably affect the maps for aperture sizes $\lap\leq100~\pc$, it leads to the appearance of disc-like features on intermediate ($\lap=\{800,1600\}~\pc$) scales. Towards the largest aperture sizes, the maps become homogeneous and the `galactic average' introduced in \autoref{fig:tuningfork} is retrieved.

\subsubsection{Sensitivity limits} \label{sec:stepsens}
In order to enable a reliable peak identification and to quantify the uncertainties on the observed flux ratios, the sensitivity limits of the maps are determined next. Ideally, this should be done by fitting a one-dimensional Gaussian to the noise-dominated range of the pixel value PDF. However, it is non-trivial to estimate which range of pixel values are affected by noise without having a measurement of the sensitivity limit. For radio or sub-mm interferometric data, noise manifests itself as a normally-distributed, random component in the pixel value PDF that is symmetric around the background flux (which we here assume to be zero). For optical data, noise results from low photon counts and manifests itself as a Poisson distribution at pixel values above the background flux. These two cases require a different range of pixel values over which the sensitivity limit is characterized.

The range over which a Gaussian is fitted to the pixel value PDF is taken to be
\be
\label{eq:sensrange1}
\min{(\f_{ij})}\leq\f_{ij}\leq\f_{\rm ref} ,
\ee
where $\f_{ij}$ represents the array of all pixel values, following the notation of Section~\ref{sec:stepmask}, and we define
\be
\label{eq:sensrange2}
\f_{\rm ref}=\left\{\begin{array}{ll}
|\min{(\f_{ij})}| & \mbox{if } \min{(\f_{ij})}<0 \\
{\rm median}\,(\f_{ij}) & \mbox{if } \min{(\f_{ij})}\geq0 .
\end{array} \right.
\ee
The first of these two cases would typically apply to an interferometric image, whereas the second would typically apply to an optical image or simulated map, provided that the noise follows the pattern described above.

To carry out the fits and measure the sensitivity limits, the pixel value PDF of each image is binned over the above range, with a bin width chosen to enable an accurate Gaussian fit. We first determine the standard deviation of $\f_{ij}$ in the to-be-fitted range as
\be
\label{eq:sigmaf}
\sigma_\f=\sqrt{\langle\f_{ij}^2\rangle-\langle\f_{ij}\rangle^2}~\mbox{ for } \min{(\f_{ij})}\leq\f_{ij}\leq\f_{\rm ref} ,
\ee
and define the bin width as
\be
\label{eq:binwidth}
\delta_\f = \frac{\sigma_\f}{\nbins} .
\ee
After defining the bins, we generate the corresponding pixel value PDF and carry out a least-square fit of a one-dimensional Gaussian $p(\f)$ to the distribution. The sensitivity limits $\sigma_{\f,{\rm star}}$ and $\sigma_{\f,{\rm gas}}$ of the stellar and gas map, respectively, are then defined as the dispersion of the corresponding Gaussian fit. Likewise, the background flux levels $\f_{\rm back,star}$ and $\f_{\rm back,gas}$ of the stellar and gas map, respectively, are defined as the peak flux of each corresponding Gaussian fit.

While it is appealing to handle the sensitivity calculation automatically through the Gaussian fitting process described here, this represents a somewhat idealised situation that is not always achievable. For instance, the maps at hand may be signal masked or filtered to remove the diffuse emission (see Section~\ref{sec:diffuse}). In such cases, the noise properties have been modified and the sensitivity limit cannot be obtained by fitting a Gaussian to the pixel value PDF. The sensitivity limit should then be measured by hand using an unprocessed map and be provided to \code.

\subsubsection{Peak identification} \label{sec:steppeaks}
To identify emission peaks in each of the maps, we use the two-dimensional version of the publicly available \clfind algorithm \citep{williams94}. As mentioned in Section~\ref{sec:qualdes}, there exist more sophisticated identification methods, but the main goal in this step of the analysis is to obtain accurate peak coordinates, for which \clfind is entirely adequate. The peak identification is performed using the high-resolution, masked maps (see the top panels of \autoref{fig:stepconvolve}), unless separate maps have been provided for this purpose (see Section~\ref{sec:stepreadpar} and \autoref{tab:input}), in which case these separate maps will also have been masked (see Section~\ref{sec:stepmask}).

In brief, the \clfind algorithm loops over a predefined set of flux levels, starting at the highest level, and identifies closed contours for each of them. For the default choice of {\tt flux\_weight}~$=0$ (see \autoref{tab:flags}), the pixel with the highest flux value within the closed contour is taken to represent the peak position $\{i,j\}_{\rm peak}$.\footnote{If {\tt flux\_weight}~$=1$, then the peak position is taken to be the flux-weighted mean of all pixels associated with the peak. This option is available in \code (see \autoref{tab:flags}), but is not used in the present paper.} If this position has not previously been identified (i.e.~it is absent at higher flux levels), it is saved as a new peak. The lowest flux level in the set is only used to grow previously-defined peaks to their final sizes and no new peaks are assigned. The total flux of each peak $\f_{\rm peak}$ is taken to be the sum of all pixel values within its lowest contour. If the lowest contour encloses more than one peak, the pixels are assigned to individual peaks within the contour using a friend-of-friends algorithm. For further details, we refer the reader to \citet{williams94} and the \clfind website (see Section~\ref{sec:qualdes}).

To apply \clfind within \code to identify the emission peaks in the maps, we define the parameters listed in the third block of \autoref{tab:input}. First, we define the contours that are used by \clfind to perform the peak identification. For the default choice of using logarithmic spacing, we define the levels as
\be
\label{eq:levels}
\log_{10}\f_{\rm lev}(i)=\log_{10}\f_{\rm lev,max}-\frac{N_{\rm lev}-1-i}{N_{\rm lev}-1}\Delta\log_{10}\f ,
\ee
where $\{i\in\mathbb{N}~|~0\leq i\leq N_{\rm lev}-1\}$. In this expression, the maximum level is defined as
\be
\label{eq:maxlevel}
\log_{10}\f_{\rm lev,max}=\left\{\left\lfloor\frac{\log_{10}\max{(\f_{ij})}}{\delta\log_{10}\f}\right\rfloor-1\right\}\delta\log_{10}\f ,
\ee
which rounds down the maximum pixel value in the map to the nearest multiple of the level interval and subtracts another level interval. This latter step is made to ensure that the top level in which peaks are identified is sufficiently wide to encompass a significant fraction of the flux envelope around the brightest peaks and avoid fragmentation of these peaks. The number of levels follows as
\be
\label{eq:nlevel}
N_{\rm lev}=\frac{\Delta\log_{10}\f}{\delta\log_{10}\f}+1 .
\ee
These definitions imply that equation~(\ref{eq:levels}) covers a total logarithmic range equal to $\Delta\log_{10}\f$, with the highest level given by equation~(\ref{eq:maxlevel}) and using intervals between the levels of $\delta\log_{10}\f$. Finally, if linearly-spaced contours are preferred, the levels are defined as
\be
\label{eq:linlevels}
\f_{\rm lev}(i)=\min{(\f_{ij})}+\frac{i}{N_{\rm lin}-1}[\max{(\f_{ij})}-\min{(\f_{ij})}] ,
\ee
where $\{i\in\mathbb{N}~|~0\leq i\leq N_{\rm lin}-1\}$. Throughout the present paper, we will use the logarithmic spacing of equation~(\ref{eq:levels}), but the linear spacing of equation~(\ref{eq:linlevels}) is available as an option for future work.

\begin{figure}
\includegraphics[width=\hsize]{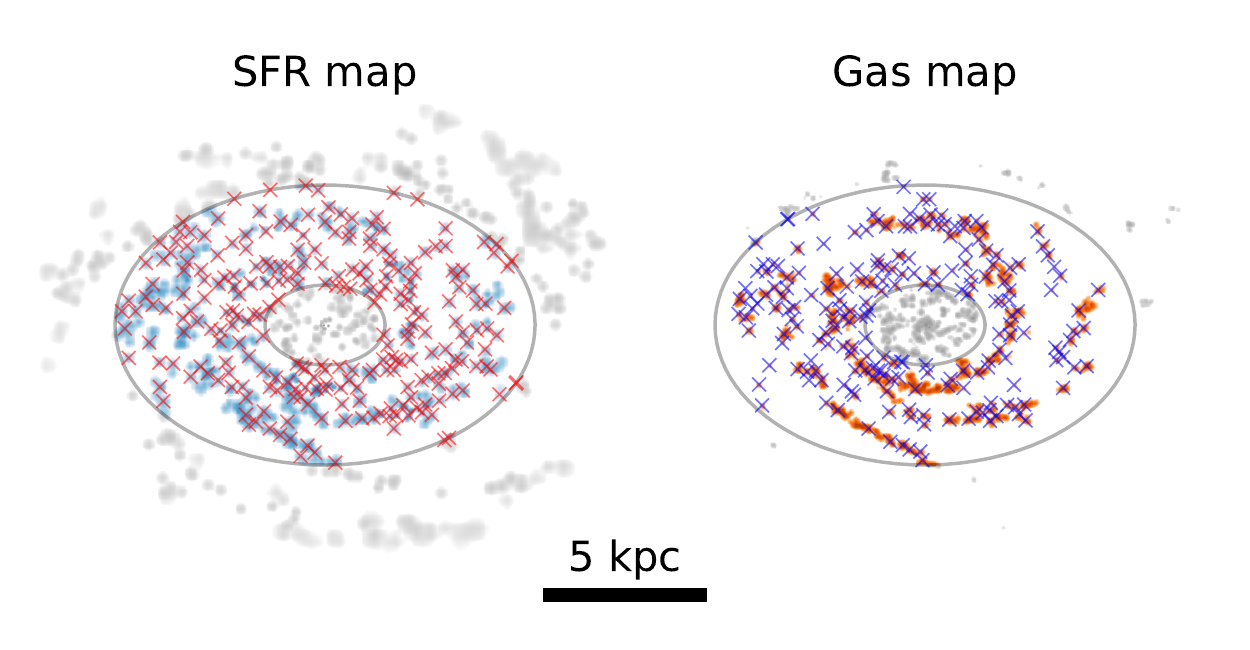}%
\vspace{-1mm}\caption{
\label{fig:steppeaks}
Outcome of the peak identification process for the example maps used throughout this section. The peak positions are indicated by red (left) and blue (right) crosses. Most of the peaks that one would visually identify are also identified by \clfind.\vspace{-1mm}
}
\end{figure}
For both maps, we run \clfind at the highest working resolution ($\lapmin$) with the specified flux levels. Any identified peaks are included if they satisfy the following two conditions.
\begin{enumerate}
\item
The total number of pixels assigned to the peak (i.e.~the peak area in pixels) should be at least $\npixmin$ (defined in \autoref{tab:input}).
\item
The total flux of the peak should exceed $N_\sigma\sigma_\f+\f_{\rm back}$, i.e.~it must be detected against the background flux with a total signal-to-noise ratio of at least $N_\sigma$ (defined in \autoref{tab:input}).
\end{enumerate}
Applying the above procedure to each of the maps results in two final samples of reliable stellar peaks and gas peaks, from which the peak positions $\{i,j\}_{\rm peak}$ will be used in the rest of the analysis. This provides a total of $N_{\rm peak}$ peaks, with $N_{\rm peak,star}$ stellar ones and $N_{\rm peak,gas}$ gas peaks. Finally, \code outputs images of the two maps on which the positions of the peaks are highlighted.

An outcome of the peak identification process is shown in \autoref{fig:steppeaks} for the standard example maps. The figure nicely illustrates that the default input parameters result in an accurate sample of peaks identified by \clfind. Applications of the method to other maps may require other values of these parameters, depending on the dynamic range of the pixel values, the relative degree of blending between peaks, the sensitivity limits of the maps, and the typical peak area in pixels. Because there is no quantitative rule for setting the input parameters that govern the peak identification, visual inspection of the output maps is required to verify if the outcome is satisfactory. It is important to reiterate that, if the identified peak sample is incomplete or biased to a certain part of the sample, the evolutionary timeline of \autoref{fig:tschem} that is constrained by the method is not necessarily incorrect, because the method is empirical in nature -- it applies to whatever subset of the peak sample is selected, as long as it is representative for the flux-emitting regions in the maps. The only way in which the outcome of the method can be truly incorrect, is if a significant fraction of the total flux in the map is inaccessible by the peak identification process. This occurs if a map contains a large reservoir of diffuse emission \citep{hygate18}, or if the depth of the peak identification ($\Delta\log_{10}\f$) is chosen to be too small to reach the peaks that contain most of the flux. As long as these situations are avoided, we find that the details of the peak identification process do not strongly affect the outcome of the method, implying that the derived timeline is robust.

\subsubsection{Aperture placement on peaks} \label{sec:stepapertures}
With the peak positions in hand, we calculate the total fluxes in circular apertures centred on these peaks, as well as the effective aperture areas and peak-to-peak distances. To quantify the flux within an aperture, we prefer to use the flux density in that aperture, which is equivalent to the pixel value at the peak position in the map convolved to the desired aperture size (see Section~\ref{sec:stepconvolve}), rather than the absolute flux within the aperture. To obtain an absolute flux rather than a flux density, the latter quantity should be multiplied by $\pi\lap^2/4$. However, the remainder of the presented analysis deals with flux {\it ratios} at each aperture size, implying that this geometric factor cancels. We therefore use $\f_{{\rm star},ij}(\lap,\{i,j\}_{\rm peak})$ and $\f_{{\rm gas},ij}(\lap,\{i,j\}_{\rm peak})$ to describe the flux within each aperture, where the terms in parentheses indicate that the flux is evaluated at each peak position $\{i,j\}_{\rm peak}$ in a map convolved to an aperture size $\lap$.

Even though the aperture size is formally defined by equation~(\ref{eq:lap}), each aperture focused on a particular peak has an effective aperture size $\lapeff\leq\lap$ because it may contain masked pixels. To account for this effect throughout the analysis, we calculate the included fraction of the intended aperture area in each case as
\be
\label{eq:aeff}
f_{A,ij}(\lap,\{i,j\}_{\rm peak})=\Xi_{ij}^0(\lap,\{i,j\}_{\rm peak}) .
\ee
This expression is so simple, because the pixel values in the convolved mask array $\Xi_{ij}^0(\lap)$ (in which the masked pixels have been set to zero) are equivalent to the unmasked fraction of the aperture area at each position. The effective aperture size then follows as
\be
\label{eq:lapeff}
\lapeff=\sqrt{\langle f_{A,ij}\rangle}\lap .
\ee
where $\langle f_{A,ij}\rangle$ takes the ensemble average over all peaks and is placed within the square-root, because the flux is proportional to the aperture area rather than its size. This quantity will be used in Section~\ref{sec:stepfluxratios} to place each measurement of the flux ratio excess or deficit at the correct effective aperture size in the tuning fork diagram of \autoref{fig:tuningfork}.

Finally, we calculate the distances between all peak pairs as
\be
\label{eq:distance}
d_{kl}=l_{\rm pix}|\{i,j\}_{{\rm peak},k}-\{i,j\}_{{\rm peak},l}| ,
\ee
where $\{k,l\in\mathbb{N}~|~0\leq k,l\leq N_{\rm peak}-1\}$ represent the indices used to select elements from the list of peaks. This distance is used in Section~\ref{sec:stepfluxratios} to draw Monte-Carlo realisations of peak samples with non-overlapping apertures and ensure that no pixels are counted twice in the flux ratio calculation.

\subsubsection{Gas-to-stellar flux ratio bias around peaks} \label{sec:stepfluxratios}
At this point in the analysis, all necessary flux levels have been determined. It is now possible to turn these fluxes into a gas-to-stellar flux ratio `bias' (i.e.~excess or deficit) relative to the galactic average as a function of the aperture size (see the cartoon of the tuning fork diagram in \autoref{fig:tuningfork}). This bias is defined as the ratio of two gas-to-stellar flux ratios, i.e.
\be
\label{eq:biasstar}
\bias_{\rm star}(\lap)=\frac{\rat_{\rm star}(\lap)}{\rat_{\rm tot}} ,
\ee
and
\be
\label{eq:biasgas}
\bias_{\rm gas}(\lap)=\frac{\rat_{\rm gas}(\lap)}{\rat_{\rm tot}} ,
\ee
for the bias towards stellar and gas peaks, which represent the bottom and top branches of the tuning fork diagram, respectively. The denominators represent the `galactic average' flux ratio, i.e.~the total gas-to-stellar flux ratio across the entire unmasked area of the maps, for which the fluxes have been calculated in Section~\ref{sec:stepmask}:
\be
\label{eq:rtot}
\rat_{\rm tot}=\frac{\f_{\rm gas,tot}}{\f_{\rm star,tot}} .
\ee
The numerators in equations~(\ref{eq:biasstar}) and~(\ref{eq:biasgas}) represent the gas-to-stellar flux ratio of apertures focused on stellar or gas peaks, respectively. In principle, such a ratio can be determined in two ways that have a different meaning. Firstly, one can obtain the gas-to-stellar flux ratio for each individual peak of a given type, after which the average ratio is taken across all peaks of that type. Secondly, one can calculate the total stellar flux and gas flux across all apertures focused on a given peak type and then take the ratio between these total aperture fluxes. The model of the \klprinciple follows the latter approach, because it is straightforward to predict analytically and is less sensitive to peak-to-peak variations or correlations between the local stellar and gas flux.

Before being able to provide the expressions for $\rat_{\rm star}(\lap)$ and $\rat_{\rm gas}(\lap)$ that are used in the method to represent the observed gas-to-stellar flux ratios, there is an important practical point to consider. Across each of the two peak samples, there will be a minimum distance between all peak pairs $d_{kl}$, as defined in equation~(\ref{eq:distance}). For aperture sizes $\lap\geq\min{(d_{kl})}$, this means that at least some apertures will overlap. If the total aperture flux is obtained by summation over all apertures irrespective of their (in)dependence, then pixels could be counted twice. In the best case [$\lapmax\sim\min{(d_{kl})}$], this would not significantly affect the observed gas-to-stellar flux ratio bias. However, for $\lapmax\gg\min{(d_{kl})}$, the observed flux ratio bias would count large numbers of pixels multiple times, particularly in regions of high peak density. This would lead to an inaccurate measure of the true flux ratio bias.

To address this problem, we draw Monte-Carlo realisations of independent peak samples for each peak type and aperture size, which will be averaged over to obtain the gas-to-stellar flux ratio. This is done by drawing candidate peaks in a random order. For each candidate peak, it is verified whether its distance to all previously drawn peaks is larger than the aperture size, i.e.:
\be
\label{eq:mindist}
\lap<\min{(\widehat{d_{kl}})} ,
\ee
where the hat indicates that this expression only considers the subset of previously drawn and accepted peaks. If equation~(\ref{eq:mindist}) is satisfied, then the peak candidate is accepted and added to the Monte-Carlo sample. If the condition is not satisfied, then the peak is erased from the candidate list. This process is repeated until all candidate peaks have been considered. The resulting realisations of the peak samples $\widehat{\{i,j\}}_{\rm peak,star}$ and $\widehat{\{i,j\}}_{\rm peak,gas}$ with their correspondingly-positioned apertures each only include independent apertures -- no pixel is included more than once in each of the samples. For each aperture size $\lap$, we follow the above procedure $\nmcpeak$ times to generate $\nmcpeak$ Monte-Carlo realisations of independent stellar and gas peak samples. At small apertures, these samples will hold close to the total number of peaks, because most peaks will have no neighbours closer than the minimum aperture size $\lapmin$. However, at large apertures (e.g.~$\lap=\lapmax$), the samples may contain only a handful of peaks, because the aperture sizes approach the diameter of the entire unmasked area in the map. The mean number of peaks considered at each aperture size is determined as
\be
\label{eq:nstarmc}
N_{\rm star,mc}(\lap)=\left\langle\widehat{N}_{\rm star}(\lap)\right\rangle_{\rm mc} ,
\ee
and
\be
\label{eq:ngasmc}
N_{\rm gas,mc}(\lap)=\left\langle\widehat{N}_{\rm gas}(\lap)\right\rangle_{\rm mc} ,
\ee
where the average $\langle\dots\rangle_{\rm mc}$ takes place over all Monte-Carlo realisations and the hat indicates that these expressions only consider the subset of accepted peaks.

\begin{figure}
\includegraphics[width=\hsize]{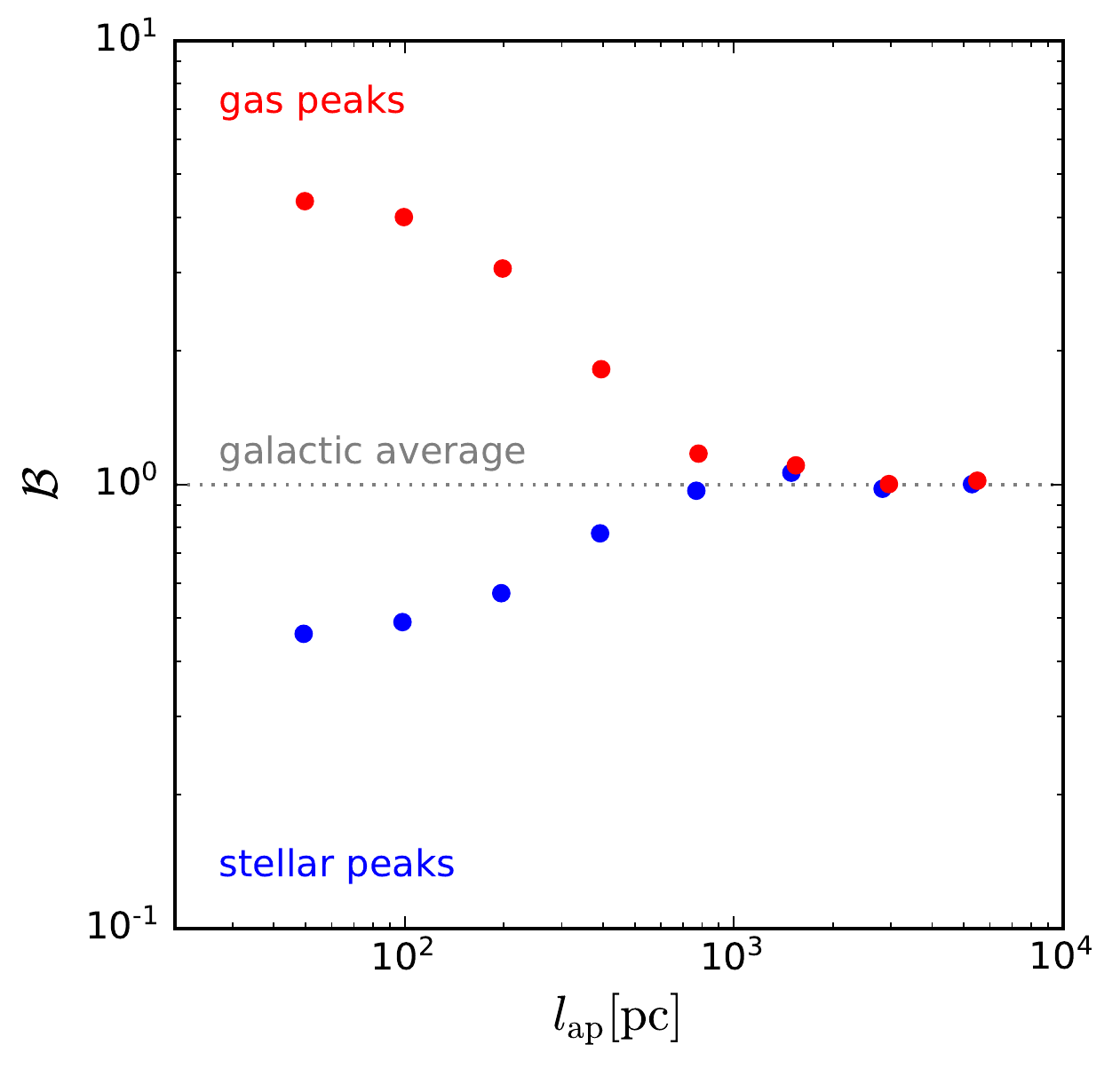}%
\vspace{-1mm}\caption{
\label{fig:stepfluxratios}
Example of the gas-to-stellar flux ratio bias as a function of the aperture size $\lap$ when focusing apertures on stellar peaks ($\bias_{\rm star}$, blue dots) or gas peaks ($\bias_{\rm gas}$, red dots) in the maps shown in \autoref{fig:steppeaks}. The dotted line indicates the `galactic average' gas-to-stellar flux ratio. Note the small differences in aperture size between the blue and red symbols towards the right. This reflects slightly different effective aperture areas due to masking, which affects both peak samples differently, because stellar and gas peaks are not co-spatial and thus apertures centred on the two different populations do not necessarily contain the same number of masked pixels (see the text for a detailed explanation). The characteristic `tuning fork' shape of this quantitative application of the method resembles the qualitative behaviour described in Section~\ref{sec:form2} and \autoref{fig:tuningfork}.\vspace{-1mm}
}
\end{figure}

With the Monte-Carlo realisation of the peak samples in hand, it is straightforward to calculate the gas-to-stellar flux ratios for apertures focused on stellar ($\rat_{\rm star}$) and gas ($\rat_{\rm gas}$) peaks as a function of the aperture size. We write
\be
\label{eq:rstar}
\rat_{\rm star}=\frac{\f_{\rm gas,star}}{\f_{\rm star,star}}=\frac{\left\langle\sum_{\widehat{\{i,j\}}_{\rm peak,star}}\f_{{\rm gas},ij}(\lap)\right\rangle_{\rm mc}}{\left\langle\sum_{\widehat{\{i,j\}}_{\rm peak,star}}\f_{{\rm star},ij}(\lap)\right\rangle_{\rm mc}} ,
\ee
and
\be
\label{eq:rgas}
\rat_{\rm gas}=\frac{\f_{\rm gas,gas}}{\f_{\rm star,gas}}=\frac{\left\langle\sum_{\widehat{\{i,j\}}_{\rm peak,gas}}\f_{{\rm gas},ij}(\lap)\right\rangle_{\rm mc}}{\left\langle\sum_{\widehat{\{i,j\}}_{\rm peak,gas}}\f_{{\rm star},ij}(\lap)\right\rangle_{\rm mc}} ,
\ee
where the average $\langle\dots\rangle_{\rm mc}$ again takes place over all Monte-Carlo realisations and the summation is performed only over the accepted, independent peaks in each realisation, as indicated by the hat in $\widehat{\{i,j\}}_{\rm peak}$. Note that these equations contain four different fluxes due to the fact that both flux types are measured around both peak types. This results in subscripts to the flux that contain two instances of `star' and `gas' (appearing after the first equality in these expressions), where we follow the convention that first word indicates the type of flux and the second word indicates the type of peak that the apertures are focused on. The gas-to-stellar flux ratios $\rat_{\rm star}$ and $\rat_{\rm gas}$ are substituted into equations~(\ref{eq:biasstar}) and~(\ref{eq:biasgas}) to obtain the flux ratio bias when focusing apertures on stellar and gas peaks, respectively. \autoref{fig:stepfluxratios} shows an example of the resulting gas-to-stellar flux ratio bias for the pair of maps used throughout this section, clearly highlighting how the qualitative behaviour of the tuning fork diagram in \autoref{fig:tuningfork} is reproduced in the quantitative application of the method.

Before continuing, we emphasize that these flux ratio biases are not affected by any uncertainties on the conversion factors from a stellar or gas flux to an SFR or gas mass, as long as these factors do not vary significantly across the maps. In that case, the conversion factors on scales of $\lap\la\lambda$ are similar to the map average ($\lap\sim\lapmax$), implying that the conversion factors cancel when taking the ratio of the two flux ratios in equations~(\ref{eq:biasstar}) and~(\ref{eq:biasgas}).

We also need to re-evaluate the effective aperture size of equation~(\ref{eq:lapeff}) after having generated the Monte-Carlo realisations, because the differing coverages of the masked pixels in each Monte-Carlo realisation of the peak sample lead to different effective aperture sizes around stellar and gas peaks. We calculate these as
\be
\label{eq:lapstar}
\lapstar=\left\langle\left[\frac{\sum_{\widehat{\{i,j\}}_{\rm peak,star}}f_{A,ij}(\lap)}{\widehat{N}_{\rm peak,star}}\right]^{1/2}\lap\right\rangle_{\rm mc} .
\ee
and
\be
\label{eq:lapgas}
\lapgas=\left\langle\left[\frac{\sum_{\widehat{\{i,j\}}_{\rm peak,gas}}f_{A,ij}(\lap)}{\widehat{N}_{\rm peak,gas}}\right]^{1/2}\lap\right\rangle_{\rm mc} .
\ee
for the aperture sizes for stellar and gas peaks, respectively. In these expressions, the term in square brackets averages the unmasked fraction of the aperture area over all peaks in a Monte-Carlo realisation, after which the square-root and multiplication with $\lap$ result in an effective aperture size for that realisation. The average $\langle\dots\rangle_{\rm mc}$ then provides the final effective aperture sizes by averaging over all realisations. Note that these revised aperture sizes do not affect the gas-to-stellar flux ratios of equations~(\ref{eq:rstar}) and~(\ref{eq:rgas}), because the aperture sizes in the numerator and denominator of each of these expressions are identical. The change only affects the horizontal positions of the data points in \autoref{fig:stepfluxratios}.

In addition to determining the gas-to-stellar flux ratio bias as a function of size scale around emission peaks, we also use the Monte-Carlo sampling to obtain the necessary constraints on the flux evolution of independent regions. As explained in \citetalias{kruijssen14}, regions may undergo any form of flux evolution during the evolutionary timeline of \autoref{fig:tschem}. The \klprinciple discretises that timeline into three phases and allows one to measure their relative durations. Because this method is based on flux ratio measurements, it effectively uses the mean flux level integrated over each of these phases. If one is not interested in the overlap phase, but only in determining the ratio between the gas and stellar lifetimes, then the measurement is therefore entirely independent of the flux evolution of each tracer. Only when concerned with the duration of the overlap phase it is necessary to account for time-averaged relative flux level changes of a tracer between the different phases. To capture this evolution, we determine the quantities $\betastar$ and $\betagas$, which refer to the mean flux ratios of regions residing in the overlap phase relative to their `isolated' phases, i.e.~outside of the overlap phase. For the purpose of our method, these two parameters fully capture the (possibly complex) flux evolution of both tracers over the evolutionary timeline. Assuming that there exists a way of determining which peaks reside in the overlap and isolated phases, $\betastar$ and $\betagas$ can be defined as
\be
\label{eq:betastar}
\betastar=\left\langle\frac{\widehat{N}_{\rm peak,star,iso}\sum_{\widehat{\{i,j\}}_{\rm peak,star,over}}\f_{{\rm star},ij}(\lapmin)}{\widehat{N}_{\rm peak,star,over}\sum_{\widehat{\{i,j\}}_{\rm peak,star,iso}}\f_{{\rm star},ij}(\lapmin)}\right\rangle_{\rm mc} ,
\ee
and
\be
\label{eq:betagas}
\betagas=\left\langle\frac{\widehat{N}_{\rm peak,gas,iso}\sum_{\widehat{\{i,j\}}_{\rm peak,gas,over}}\f_{{\rm gas},ij}(\lapmin)}{\widehat{N}_{\rm peak,gas,over}\sum_{\widehat{\{i,j\}}_{\rm peak,gas,iso}}\f_{{\rm gas},ij}(\lapmin)}\right\rangle_{\rm mc} ,
\ee
where $\widehat{N}_{\rm peak,star,iso}$, $\widehat{N}_{\rm peak,gas,iso}$, $\widehat{N}_{\rm peak,star,over}$, and $\widehat{N}_{\rm peak,gas,over}$ represent the number of peaks in a Monte-Carlo realisation that reside in the isolated stellar, isolated gas, stellar overlap, and gas overlap phases, respectively, and the subscripts to $\widehat{\{i,j\}}$ refer to the peak coordinates of these same subsets of peaks. Equations~(\ref{eq:betastar}) and~(\ref{eq:betagas}) determine the mean flux level within an aperture of size $\lapmin$ for isolated peaks and for peaks residing in the overlap phase, take the ratio between both flux levels, and average the resulting ratio over all Monte-Carlo realisations. They thus represent the `overlap-to-isolated flux ratios' of independent regions.

The key question is now how one determines whether a peak resides in the isolated or overlap phase. Without further knowledge regarding the evolutionary timeline of \autoref{fig:tschem}, there is no objective way of making that decision. One solution could be to define a critical flux contrast between stellar and gas flux to consider a peak in the isolated or overlap phase. However, this would not only be arbitrary, but most importantly it would also be degenerate with the idea that the duration of the overlap phase can be constrained by measuring flux ratios. Fortunately, there exists an objective way of distinguishing between isolated and overlap-phase peaks if one assumes that the identified peaks homogeneously sample the complete evolutionary timeline and are thus representative for the underlying flux evolution. This is a strong assumption -- so strong in fact, that in its most extreme interpretation it would enable one to derive the evolutionary time-scales of \autoref{fig:tschem} by simply counting the identified peaks and using the relative counts as a proxy for the relative time-scales. However, it would be a grave mistake to do so, because it results in a timeline that is entirely dependent on the peak identification process, which (as discussed in Section~\ref{sec:steppeaks}) relies on by-eye verification and is therefore much too subjective to draw such an immediate quantitative conclusion.

Here, we propose a much weaker version of the above assumption. If the peaks reasonably homogeneously sample the evolutionary timeline of converting one tracer into the next, which happens in the discussed example by converting gas into stars and expelling the residual gas by feedback, then a peak sample that is sorted by decreasing gas-to-stellar flux ratio reflects a trend of increasing evolutionary age. The sorted peak sample can then be roughly divided into peaks residing in the overlap and isolated phases as a function of the evolutionary time-scale ratios $f_{\rm star,over}=\tover/\tstar$ and $f_{\rm gas,over}=\tover/\tgas$, by requiring that the fraction of peaks residing in a certain phase must be the same as the fraction of the tracer lifetime covered by that phase. Because these time-scales are constrained during the fitting process (see Section~\ref{sec:stepfit}), such a dependence would imply functional dependences of $\betastar(f_{\rm star,over})$ and $\betagas(f_{\rm gas,over})$, allowing both quantities to be derived as a byproduct of the fitting process. This version of assuming that the identified peaks are representative for the evolutionary timeline is much weaker than in the earlier example of directly using peak counts to constrain the evolutionary timeline, because (as we show in the example below) $\betastar$ and $\betagas$ only vary weakly with $f_{\rm star,over}$ and $f_{\rm gas,over}$, respectively -- typically by a few tens of per cent over factors of several in the underlying time-scales.

\begin{figure}
\includegraphics[width=\hsize]{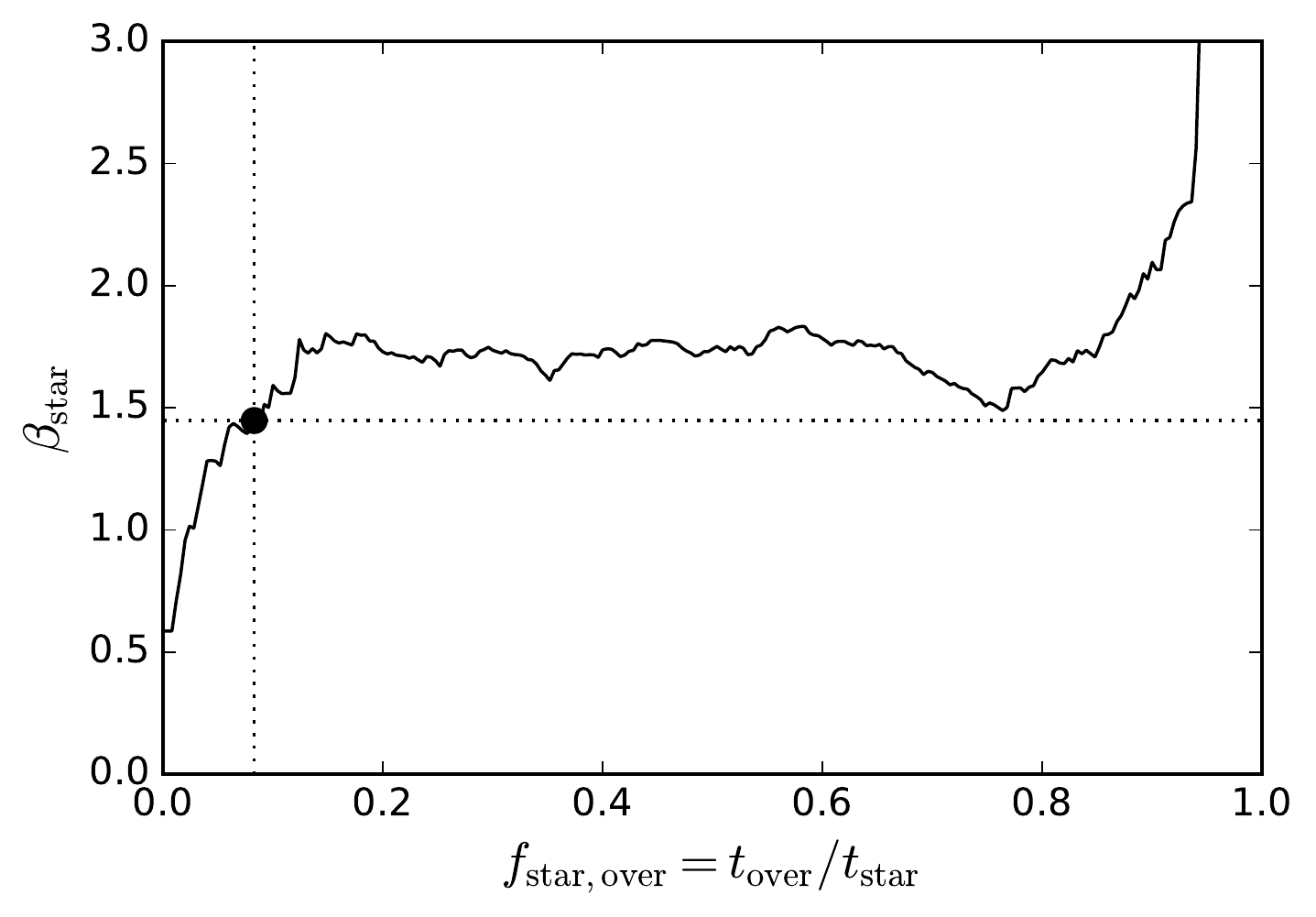}%
\vspace{-1mm}\caption{
\label{fig:stepbeta}
Dependence of the overlap-to-isolated stellar flux ratio $\betastar$ on the relative duration of the overlap phase $f_{\rm star,over}=\tover/\tstar$ for the example maps that are used throughout this section. The flux ratio $\betastar$ captures the (possibly complex) flux evolution of the SFR tracer in the presented method. The best fitting value of $\tover$ and $\tstar$ (see Section~\ref{sec:stepfit}) and the corresponding best-fitting value of $\betastar$ are shown by the black dot and the dotted lines. In this example, $\betastar$ is well-constrained, because it varies only weakly over most of the range of $f_{\rm star,over}$.\vspace{-1mm}
}
\end{figure}
Quantitatively, the above method for deriving $\betastar(f_{\rm star,over})$ and $\betagas(f_{\rm gas,over})$ is applied as follows. First, the arrays for $f_{\rm star,over}$ and $f_{\rm gas,over}$ are discretised by writing
\be
\label{eq:foverstar}
f_{\rm star,over}(k)=\frac{k}{\widehat{N}_{\rm peak,star}-1} ,
\ee
with $\{k\in\mathbb{N}~|~0\leq k\leq \widehat{N}_{\rm peak,star}-1\}$, and
\be
\label{eq:fovergas}
f_{\rm gas,over}(l)=\frac{l}{\widehat{N}_{\rm peak,gas}-1} ,
\ee
with $\{l\in\mathbb{N}~|~0\leq l\leq \widehat{N}_{\rm peak,gas}-1\}$. For each Monte-Carlo realisation and for each $f_{\rm star,over}(k)$:
\begin{enumerate}
\item
we assign the $\widehat{N}_{\rm peak,star,iso}=k+1$ stellar peaks that have the lowest $\f_{{\rm gas},ij}(\lapmin)/\f_{{\rm star},ij}(\lapmin)$ ratio to the isolated stellar phase, which thus defines $\widehat{\{i,j\}}_{\rm peak,star,iso}$;
\item
we assign the $\widehat{N}_{\rm peak,star,over}=\widehat{N}_{\rm peak,star}-k-1$ stellar peaks that have the highest $\f_{{\rm gas},ij}(\lapmin)/\f_{{\rm star},ij}(\lapmin)$ ratio to the overlap phase, which thus defines $\widehat{\{i,j\}}_{\rm peak,star,over}$.
\end{enumerate}
Vice versa, for each Monte-Carlo realisation and for each $f_{\rm gas,over}(l)$:
\begin{enumerate}
\item
we assign the $\widehat{N}_{\rm peak,gas,iso}=l+1$ gas peaks that have the lowest $\f_{{\rm star},ij}(\lapmin)/\f_{{\rm gas},ij}(\lapmin)$ ratio to the isolated gas phase, which thus defines $\widehat{\{i,j\}}_{\rm peak,gas,iso}$;
\item
we assign the $\widehat{N}_{\rm peak,gas,over}=\widehat{N}_{\rm peak,gas}-l-1$ gas peaks that have the highest $\f_{{\rm star},ij}(\lapmin)/\f_{{\rm gas},ij}(\lapmin)$ ratio to the overlap phase, which thus defines $\widehat{\{i,j\}}_{\rm peak,gas,over}$.
\end{enumerate}
This way, we can evaluate equations~(\ref{eq:betastar}) and~(\ref{eq:betagas}) and obtain $\betastar(f_{\rm star,over})$ and $\betagas(f_{\rm gas,over})$ for use in the fitting process of Section~\ref{sec:stepfit}. An example of $\betastar(f_{\rm star,over})$ is given in \autoref{fig:stepbeta}, which illustrates the earlier statement that $\beta$ typically varies weakly with $f_{\rm over}$. In addition to $\betastar(f_{\rm star,over})$, \autoref{fig:stepbeta} also shows the best-fitting value to $\betastar$ that results from the fitting process in Section~\ref{sec:stepmodel} below. When fitting the observed gas-to-stellar flux ratio bias as a function of the aperture size, we obtain the time-scales $\tgas$ and $\tover$ for a given $\tstar$. This immediately provides $f_{\rm star,over}$ and $f_{\rm gas,over}$, implying that $\betastar$ and $\betagas$ are byproducts of the fitting process. We demonstrate in Section~\ref{sec:validbeta} that these overlap-to-isolated flux ratios give an accurate representation of the underlying flux evolution.

Finally, we determine a set of quantities making use of the flux distribution in the maps that are used at later points in the analysis. The first of these is the flux density contrast of emission peaks relative to the average flux density on a size scale $\lap$ for each tracer. This contrast will be used to determine the region sizes in units of the mean separation length $\lambda$ (i.e.~their filling factor) when fitting the model in Sections~\ref{sec:stepmodel} and~\ref{sec:stepfit}. Because the region separation length is unknown at this stage, we calculate the flux density contrast relative to all size scales $\lap$ and interpolate this later to the desired size scale for measuring the background flux density. Analogously to equations~(\ref{eq:rstar}) and~(\ref{eq:rgas}), this contrast can be expressed as
\be
\label{eq:excstar}
\begin{aligned}
\exc_{\rm star}(\lap)=&\frac{\left\langle\sum_{\widehat{\{i,j\}}_{\rm peak,star}}\f_{{\rm star},ij}(\lapmin)\right\rangle_{\rm mc}}{\left\langle\sum_{\widehat{\{i,j\}}_{\rm peak,star}}\f_{{\rm star},ij}(\lap)\right\rangle_{\rm mc}} \\
&\times\frac{N_{\rm star,mc}(\lap)}{N_{\rm star,mc}(\lapmin)} ,
\end{aligned}
\ee
and
\be
\label{eq:excgas}
\begin{aligned}
\exc_{\rm gas}(\lap)=&\frac{\left\langle\sum_{\widehat{\{i,j\}}_{\rm peak,gas}}\f_{{\rm gas},ij}(\lapmin)\right\rangle_{\rm mc}}{\left\langle\sum_{\widehat{\{i,j\}}_{\rm peak,gas}}\f_{{\rm gas},ij}(\lap)\right\rangle_{\rm mc}} \\
&\times\frac{N_{\rm gas,mc}(\lap)}{N_{\rm gas,mc}(\lapmin)} .
\end{aligned}
\ee
These expressions effectively yield the mean flux density (averaged over all Monte-Carlo realisations and all peaks) at the smallest aperture size in units of the mean flux density at larger aperture sizes. As stated above, the resulting flux density contrasts of single peaks relative to the background population will be used when fitting the model in Sections~\ref{sec:stepmodel} and~\ref{sec:stepfit}.

Secondly, we calculate a quantity that is similar to the flux density contrasts of equations~(\ref{eq:excstar}) and~(\ref{eq:excgas}) above, but instead measures the flux density contrast between the flux on some size scale $\lap$ and the entire map. This `global' flux density contrast can be used in observational applications of the method to account for large-scale morphological features of the target system (or the spatial clustering of independent regions) when converting map-averaged flux densities to the mean flux density per independent region, which in turn can be useful for deriving a variety of physical quantities (see e.g.~Section~\ref{sec:expval}). As before, this requires us to calculate the flux density contrast as a function of size scale $\lap$ and interpolate this to the mean separation length $\lambda$ once it has been obtained from the fitting process in Section~\ref{sec:stepfit}. The contrast is expressed as
\be
\label{eq:excstarglob}
\exc_{\rm star,glob}(\lap)=\frac{\left\langle\sum_{\widehat{\{i,j\}}_{\rm peak,star}}\f_{{\rm star},ij}(\lap)\right\rangle_{\rm mc}}{N_{\rm star,mc}(\lap)}\frac{A_{\rm tot}}{\f_{\rm star,tot}} ,
\ee
and
\be
\label{eq:excgasglob}
\exc_{\rm gas,glob}(\lap)=\frac{\left\langle\sum_{\widehat{\{i,j\}}_{\rm peak,gas}}\f_{{\rm gas},ij}(\lap)\right\rangle_{\rm mc}}{N_{\rm gas,mc}(\lap)}\frac{A_{\rm tot}}{\f_{\rm gas,tot}} ,
\ee
These expressions effectively yield the mean flux density (averaged over all Monte-Carlo realisations and all peaks) at an aperture size $\lap$ in units of the mean flux density of the entire map. As stated above, an example of using the resulting flux density contrasts is provided in Section~\ref{sec:expval}.

\subsubsection{Determination of uncertainties} \label{sec:steperrors}
Section~\ref{sec:stepfluxratios} presents a derivation of the gas-to-stellar flux ratio bias when focusing on stellar and gas peaks, but these are meaningless without an estimate of the uncertainties (cf.~\autoref{fig:stepfluxratios}). The data points also require error bars in order to constrain our best-fitting statistical model (see Section~\ref{sec:stepmodel}) to the data in Section~\ref{sec:stepfit}. We now describe how the uncertainties are quantified. This includes a discussion of how we account for the covariance between the data points in the tuning fork diagram -- after all, apertures of different sizes are focused on the same peaks, implying that they are at least partially sharing the same information.

We start by identifying the main sources of uncertainty in the data points of equations~(\ref{eq:rstar}) and~(\ref{eq:rgas}), which are shown in \autoref{fig:stepfluxratios}.
\begin{enumerate}
\item
The stellar and gas fluxes used in each of the data points are uncertain due to the finite signal-to-noise level of the data. To include this uncertainty, we use the measurements of the sensitivity limits $\sigma_{\f,{\rm star}}$ and $\sigma_{\f,{\rm gas}}$ in Section~\ref{sec:stepsens}.
\item
The peak flux levels exhibit an intrinsic variance because they do not all represent the same SFR or mass. This variance arises because the region mass function is not a delta function, but also because the individual regions may undergo some flux evolution, as discussed in Section~\ref{sec:stepfluxratios}. It leads to an uncertainty on the derived gas-to-stellar flux ratio biases, because we have used a finite number of peaks to determine these biases. For an infinite number of regions, this uncertainty would approach zero.
\item
Finally, we should account for the covariance between the numerator and denominator of the gas-to-stellar flux ratios in equations~(\ref{eq:rstar}) and~(\ref{eq:rgas}). It is possible that bright peaks from both populations are in close proximity to one another, either because of the evolutionary connection between both peak populations (e.g.~a bright stellar region emerging from a bright gas region), or because there may exist a systematic environmental variation of peak brightnesses in the considered maps. These effects are captured by subtracting the covariance between the stellar and gas flux from the total uncertainty.
\end{enumerate}

Mathematically, we account for the above sources of uncertainty as follows. We first assume that the uncertainties on the aperture flux ratios are much larger than those on the galactic average flux ratios. This allows us to set the relative uncertainties of the flux ratio biases equal to the relative uncertainties of the gas-to-stellar flux ratios around stellar and gas peaks, i.e.
\be
\label{eq:bstarerror}
\frac{\sigma_{\bias,{\rm star}}(\lapstar)}{\bias_{\rm star}(\lapstar)}=\frac{\sigma_{\rat,{\rm star}}(\lapstar)}{\rat_{\rm star}(\lapstar)} ,
\ee
and
\be
\label{eq:bgaserror}
\frac{\sigma_{\bias,{\rm gas}}(\lapgas)}{\bias_{\rm gas}(\lapgas)}=\frac{\sigma_{\rat,{\rm gas}}(\lapgas)}{\rat_{\rm gas}(\lapgas)} .
\ee
The relative uncertainties of the flux ratios are then defined as
\be
\label{eq:rstarerror}
\begin{aligned}
\frac{\sigma_{\rat,{\rm star}}(\lapstar)}{\rat_{\rm star}(\lapstar)}=&\left(\sigma_{\rm sens,star}^2+\sigma_{\rm star,star}^2+\sigma_{\rm gas,star}^2\right.\\
& \left.-2\sigma_{\rm star,star}\sigma_{\rm gas,star}\rho_{\rm star}\right)^{1/2} ,
\end{aligned}
\ee
and
\be
\label{eq:rgaserror}
\begin{aligned}
\frac{\sigma_{\rat,{\rm gas}}(\lapgas)}{\rat_{\rm gas}(\lapgas)}=&\left(\sigma_{\rm sens,gas}^2+\sigma_{\rm star,gas}^2+\sigma_{\rm gas,gas}^2\right.\\
& \left.-2\sigma_{\rm star,gas}\sigma_{\rm gas,gas}\rho_{\rm gas}\right)^{1/2} ,
\end{aligned}
\ee
respectively, where each of the uncertainties in parentheses represent relative uncertainties and the negative terms represent the covariance between the stellar and gas flux in the apertures. Each of these terms is defined below. As before, in the subscripts containing a comma, the word after the comma indicates the peak type that the apertures are focused on. The first term represents the uncertainty due to the finite signal-to-noise level (or non-zero sensitivity), the second term represents the uncertainty on the stellar flux, the third term represents the uncertainty on the gas flux, and the fourth term represents the covariance of the second and third terms. Finally, all terms have a dependence on the aperture size, but these dependences are omitted in the notation for brevity.

For the uncertainties due to the finite signal-to-noise level, we write
\be
\label{eq:sensstarerror}
\begin{aligned}
\sigma_{\rm sens,star}^2=&\left[\left(\frac{\sigma_{\f,{\rm star}}}{\f_{\rm star,star}}\right)^2+\left(\frac{\sigma_{\f,{\rm gas}}}{\f_{\rm gas,star}}\right)^2\right] \\
& \times\left(\frac{\lapstar}{\lapmin}\right)^{-2}N_{\rm star,mc} ,
\end{aligned}
\ee
and
\be
\label{eq:sensgaserror}
\begin{aligned}
\sigma_{\rm sens,gas}^2=&\left[\left(\frac{\sigma_{\f,{\rm star}}}{\f_{\rm star,gas}}\right)^2+\left(\frac{\sigma_{\f,{\rm gas}}}{\f_{\rm gas,gas}}\right)^2\right] \\
& \times\left(\frac{\lapgas}{\lapmin}\right)^{-2}N_{\rm gas,mc} ,
\end{aligned}
\ee
for apertures focused on stellar and gas peaks, respectively. As in equations~(\ref{eq:rstar}) and~(\ref{eq:rgas}), the double subscripts of the total fluxes follow the order of indicating the flux type before the peak type. The relative uncertainty on the total flux due to noise in these expressions follows normal statistics. Firstly, as mentioned in Sections~\ref{sec:stepregrid} and~\ref{sec:stepconvolve}, the pixel flux values represent surface densities, implying that their uncertainty due to noise should decrease with the square-root of the number of pixels in the aperture $N_{\rm pix,ap}$ and thus $\sigma^2\propto N_{\rm pix,ap}^{-1}\propto\lap^{-2}$. Secondly, the flux ratios of equations~(\ref{eq:rstar}) and~(\ref{eq:rgas}) include the summation over all peaks in the Monte-Carlo realisation, which means that the uncertainty $\sigma$ on the numerator and denominator should scale with $N_{\rm gas,mc}^{1/2}$ and $N_{\rm star,mc}^{1/2}$, respectively. Both of these dependences are included in equations~(\ref{eq:sensstarerror}) and~(\ref{eq:sensgaserror}). The division by the total flux in these equations converts the absolute uncertainties to relative ones.

The uncertainties due to the intrinsic variance of the peak flux (e.g.~by flux evolution or a region mass spectrum) are also straightforward to define. Specifically, we take the standard deviation of the aperture fluxes for each of the four flux measurements (i.e.~each of the two flux types around each of the two peak types) and convert this to a relative uncertainty on the sum of all flux measurements by writing
\be
\label{eq:starstarerror}
\sigma_{\rm star,star}^2=\frac{\langle\f_{{\rm star},ij}^2\rangle_{\rm peak,star}-\langle\f_{{\rm star},ij}\rangle_{\rm peak,star}^2}{\f_{\rm star,star}^2}N_{\rm star,mc} ,
\ee
and
\be
\label{eq:gasstarerror}
\sigma_{\rm gas,star}^2=\frac{\langle\f_{{\rm gas},ij}^2\rangle_{\rm peak,star}-\langle\f_{{\rm gas},ij}\rangle_{\rm peak,star}^2}{\f_{\rm gas,star}^2}N_{\rm star,mc} ,
\ee
for apertures focused on stellar peaks (i.e.~$\rat_{\rm star}$). Likewise, for apertures focused on gas peaks (i.e.~$\rat_{\rm gas}$) we define
\be
\label{eq:stargaserror}
\sigma_{\rm star,gas}^2=\frac{\langle\f_{{\rm star},ij}^2\rangle_{\rm peak,gas}-\langle\f_{{\rm star},ij}\rangle_{\rm peak,gas}^2}{\f_{\rm star,gas}^2}N_{\rm gas,mc} ,
\ee
and
\be
\label{eq:gasgaserror}
\sigma_{\rm gas,gas}^2=\frac{\langle\f_{{\rm gas},ij}^2\rangle_{\rm peak,gas}-\langle\f_{{\rm gas},ij}\rangle_{\rm peak,gas}^2}{\f_{\rm gas,gas}^2}N_{\rm gas,mc} .
\ee
In these expressions, the numerator represents the variance (i.e.~the standard deviation squared) of the flux in a single aperture across the population of stellar (subscript `peak,star') or gas (subscript `peak,gas') peaks, the denominator represents the total flux that is used in the flux ratios of equations~(\ref{eq:rstar}) and~(\ref{eq:rgas}), and the factor $N$ at the end converts the variance of the flux in a single aperture to that of the total flux.

\begin{figure}
\includegraphics[width=\hsize]{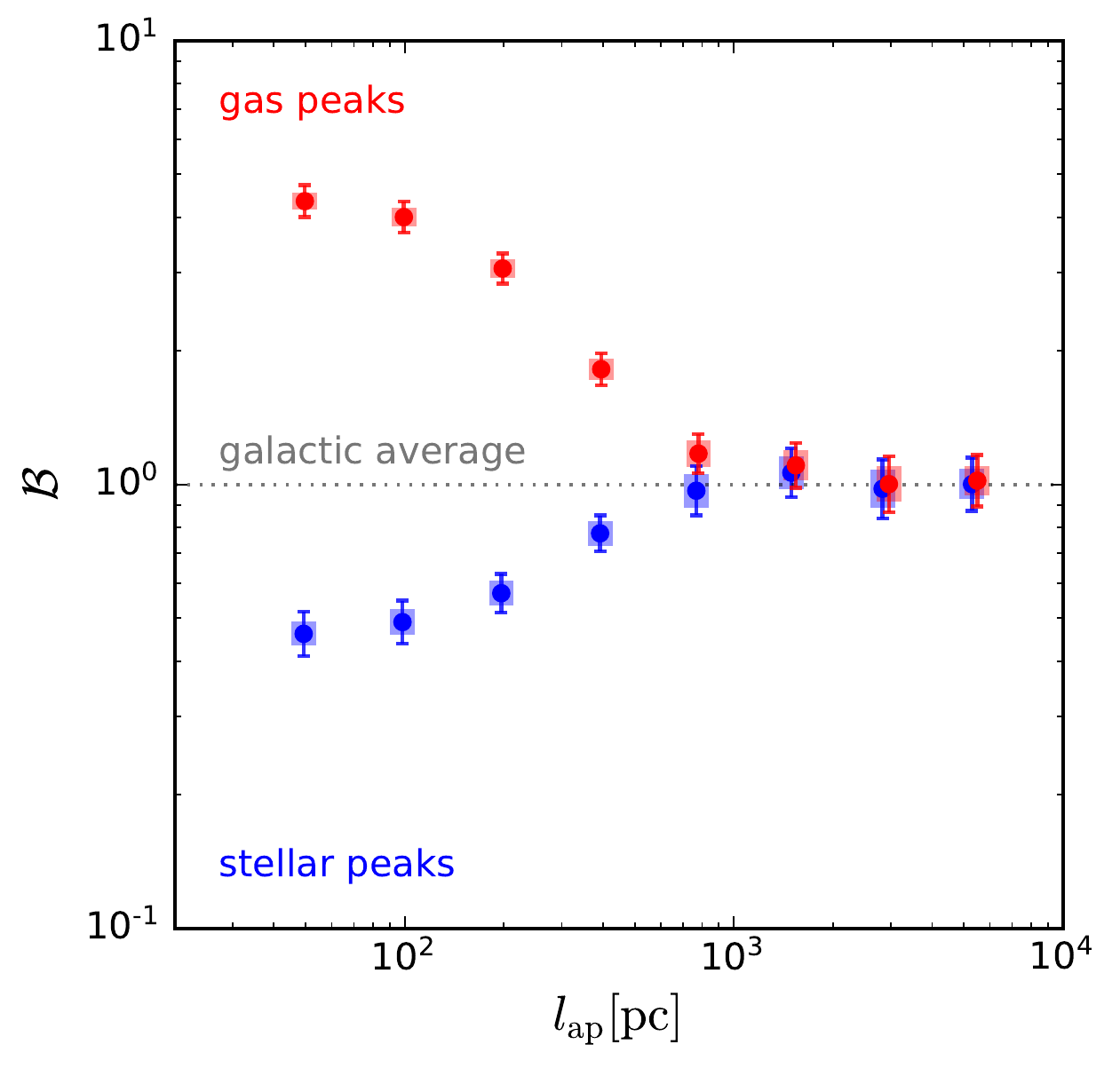}%
\vspace{-1mm}\caption{
\label{fig:steperrors}
Illustration of the uncertainties on the example tuning fork diagram of \autoref{fig:stepfluxratios}. The error bars with hats indicate the uncertainty on each individual data point, without accounting for the covariance between the data points. By contrast, the red (top) and blue (bottom) shaded areas behind each of the data points indicates the effective uncertainty range that accounts for the correlation between the data points (see the text). These `effective' uncertainties are used when obtaining the best-fitting model in Section~\ref{sec:stepfit} and provide the correct point of reference when visually assessing the quality of a fit.\vspace{-1mm}
}
\end{figure}

Having defined all relative uncertainty terms that contribute to equations~(\ref{eq:rstarerror}) and~(\ref{eq:rgaserror}), we should subtract the relative covariance between the stellar and gas fluxes in the apertures, because $\rat_{\rm star}$ and $\rat_{\rm gas}$ represent the ratios of these fluxes. For apertures focused on stellar and gas peaks, the relative covariances are 
\be
\label{eq:starcov}
\begin{aligned}
{\rm cov}(\f)_{\rm peak,star}=&\langle\f_{{\rm star},ij}\f_{{\rm gas},ij}\rangle_{\rm peak,star} \\
&-\langle\f_{{\rm star},ij}\rangle_{\rm peak,star}\langle\f_{{\rm gas},ij}\rangle_{\rm peak,star} ,
\end{aligned}
\ee
and
\be
\label{eq:gascov}
\begin{aligned}
{\rm cov}(\f)_{\rm peak,gas}=&\langle\f_{{\rm star},ij}\f_{{\rm gas},ij}\rangle_{\rm peak,gas} \\
&-\langle\f_{{\rm star},ij}\rangle_{\rm peak,gas}\langle\f_{{\rm gas},ij}\rangle_{\rm peak,gas} ,
\end{aligned}
\ee
respectively. These covariances define the correlation coefficients between the stellar and gas fluxes in the apertures in equations~(\ref{eq:rstarerror}) and~(\ref{eq:rgaserror}) as
\be
\label{eq:starcorr}
\rho_{\rm star}=\frac{{\rm cov}(\f)_{\rm peak,star}}{\sigma_{\rm star,star}\sigma_{\rm gas,star}} ,
\ee
for apertures focused on stellar peaks and
\be
\label{eq:gascorr}
\rho_{\rm gas}=\frac{{\rm cov}(\f)_{\rm peak,gas}}{\sigma_{\rm star,gas}\sigma_{\rm gas,gas}} ,
\ee
for apertures focused on gas peaks, respectively. Together, equations~(\ref{eq:rstarerror})--(\ref{eq:gascorr}) define the uncertainty on each individual data point in equations~(\ref{eq:biasstar}) and~(\ref{eq:biasgas}). These are converted to uncertainties on the logarithmic flux ratio bias by writing
\be
\label{eq:logbiaserror}
\sigma_{\log_{10}{\bias}}=\frac{1}{\ln{10}}\frac{\sigma_\bias}{\bias} .
\ee
The result of the presented uncertainty calculation is illustrated by the error bars in \autoref{fig:steperrors} for the example maps used throughout this section.

To determine the variances and covariances in equations~(\ref{eq:starstarerror})--(\ref{eq:gascov}), we calculate the flux averages across all peaks, instead of evaluating the variance for each Monte-Carlo realisation of non-overlapping aperture samples and then taking the mean variance, or calculating the variance of the total flux across all Monte-Carlo realisations. As a result, equations~(\ref{eq:starstarerror})--(\ref{eq:gascov}) may seem lacking in two regards.
\begin{enumerate}
\item
Firstly, these expressions do not account for the fact that overlapping apertures are omitted in each Monte-Carlo realisation. However, this is not necessarily incorrect, because pixels contained by multiple apertures are more likely to be drawn. Determining the variance for each realisation would overestimate the variance due to the range of environments covered by the maps. In the extreme example of large aperture sizes, the flux variance in each Monte-Carlo realisation may be based on a small number of peaks in widely different parts of a galaxy, leading to a large aperture-to-aperture flux variance, even if the total flux may be similar in each Monte-Carlo realisation. Such a scenario arises naturally when the same environments are included in each realisation. Overestimating the variance in this way would be highly undesirable.
\item
Alternatively, one could take the variance of the total fluxes across all Monte-Carlo realisations. However, this would underestimate the variance -- if none of the peaks is overlapping, then the total flux variance between the Monte-Carlo realisations will be zero, because all peaks are included in each realisation. This would give the erroneous impression that the variance is constrained to infinite precision, which gravely underestimates the variance at small aperture sizes. However, the non-zero variance only reflects the point that the total number of peaks in the maps is finite. This finite peak sample itself should be seen as one possible realisation of an underlying physical distribution of peak fluxes, which implies a non-zero variance of the total flux even if all peaks are included.
\end{enumerate}
We find that taking the variance of all aperture fluxes, irrespective of whether an aperture is included or not, provides an accurate middle ground between the above two extremes. It accounts for the fact that pixels contained by multiple apertures are more likely to be drawn in each Monte-Carlo realisation, while also properly accounting for the fact that the peak population itself represents one possible realisation of the underlying physical flux distribution function. The uncertainties due to the intrinsic variance of the peak flux are therefore defined by averaging over all peaks in equations~(\ref{eq:starstarerror})--(\ref{eq:gascov}).

The error bars derived above are corrected for the correlation between the numerator and denominator in the gas-to-stellar flux ratios of equations~(\ref{eq:rstar}) and~(\ref{eq:rgas}), but they do not account for the fact that the data points in the tuning fork diagram are not independent. This dependence arises because the apertures of different sizes are focused on the same peaks, implying that some fraction of the flux contained by apertures of a certain size is also present in apertures of other sizes. We correct for this dependence between the data points during the fitting process in Section~\ref{sec:stepfit} by calculating their normalised `independence fraction', which represents the fraction of the flux in a data point that is independent from the flux in all other data points, and using this quantity to weigh the contribution of each data point to the goodness-of-fit statistic. The sum of these independence fractions corresponds to the total number of independent data points that the model from Section~\ref{sec:stepmodel} is fitted to. The net effect of accounting for the correlation between data points is to decrease the number of degrees of freedom, leading to a larger difference between the model and the data. This is equivalent to decreasing the error bars on the data points, which is illustrated by the red and blue shaded areas behind each data point in \autoref{fig:steperrors}. These shaded areas represent `effective' uncertainties that should be used for visually evaluating the quality of a fit. In the text below, we derive these effective uncertainties.

First, we determine for each aperture size which fraction of the flux is independent from the flux in the other apertures. For the stellar and gas flux around stellar peaks, we define the fraction of the flux at an aperture size $\lap(j)$ that is also contained by an aperture of (smaller) size $\lap(i)$, with $\{i,j\in\mathbb{N}|i\leq j\}$, to be
\be
\label{eq:nstarstar}
N_{\rm star,star}=\frac{\f_{\rm star,star}[\lap(i)]}{\f_{\rm star,star}[\lap(j)]}\frac{N_{\rm star,mc}[\lap(j)]}{N_{\rm star,mc}[\lap(i)]}\left[\frac{\lap(i)}{\lap(j)}\right]^2 ,
\ee
and
\be
\label{eq:ngasstar}
N_{\rm gas,star}=\frac{\f_{\rm gas,star}[\lap(i)]}{\f_{\rm gas,star}[\lap(j)]}\frac{N_{\rm star,mc}[\lap(j)]}{N_{\rm star,mc}[\lap(i)]}\left[\frac{\lap(i)}{\lap(j)}\right]^2 ,
\ee
respectively. Similarly, for the stellar and gas flux around gas peaks, we define the fraction of the flux at an aperture size $\lap(j)$ that is also contained by an aperture of (smaller) size $\lap(i)$, with $\{i,j\in\mathbb{N}|i\leq j\}$, to be
\be
\label{eq:nstargas}
N_{\rm star,gas}=\frac{\f_{\rm star,gas}[\lap(i)]}{\f_{\rm star,gas}[\lap(j)]}\frac{N_{\rm gas,mc}[\lap(j)]}{N_{\rm gas,mc}[\lap(i)]}\left[\frac{\lap(i)}{\lap(j)}\right]^2 ,
\ee
and
\be
\label{eq:ngasgas}
N_{\rm gas,gas}=\frac{\f_{\rm gas,gas}[\lap(i)]}{\f_{\rm gas,gas}[\lap(j)]}\frac{N_{\rm gas,mc}[\lap(j)]}{N_{\rm gas,mc}[\lap(i)]}\left[\frac{\lap(i)}{\lap(j)}\right]^2 ,
\ee
respectively. As before, we have followed the convention that the first word of a subscript indicates the type of flux and the second word indicates the peak type that the apertures are focused on. We symmetrise these four arrays by setting
\be
\label{eq:ntrans}
N(j,i|i\leq j)=N(i,j|i\leq j) ,
\ee
for each of them. As a result, array elements with $i>j$ indicate the fraction of the flux at an aperture size $\lap(i)$ that is also contained by an aperture of (smaller) size $\lap(j)$. For all four arrays, setting $i=j$ gives $N=1$, indicating the trivial result that an aperture contains all of the flux within that aperture.

Equations~(\ref{eq:nstarstar})--(\ref{eq:ngasgas}) define the fraction of the flux in a larger aperture that is `reused' from a smaller aperture, for each aperture size pair. The above definitions imply that by performing a summation over one of the dimensions in these `reuse' fraction arrays, one obtains the non-integer number of data points that are constituted by the flux contained within an aperture of size $\lap(i)$, i.e.
\be
\label{eq:ndata}
N_{\rm data}(i)=\sum_j N(i,j) ,
\ee
which returns a number $1<N_{\rm data}[\lap(i)]\leq\nap$. We then define the `independence fraction' of the flux at an aperture size $\lap(i)$ as the inverse of this number
\be
\label{eq:fdata}
f_{\rm data}(i)=\frac{1}{N_{\rm data}}=\frac{1}{\sum_j N(i,j)} ,
\ee
which takes values of $0<f_{\rm data}<1$. Because the flux ratio biases of equations~(\ref{eq:biasstar}) and~(\ref{eq:biasgas}) that are shown in \autoref{fig:steperrors} represent flux ratios, we define the independence fraction of each data point as the mean of the fluxes in the numerator and the denominator, resulting in
\be
\label{eq:fdatastar}
f_{\rm data,star}(i)=\frac{1}{2\sum_j N_{\rm star,star}(i,j)}+\frac{1}{2\sum_j N_{\rm gas,star}(i,j)} ,
\ee
and
\be
\label{eq:fdatagas}
f_{\rm data,gas}(i)=\frac{1}{2\sum_j N_{\rm star,gas}(i,j)}+\frac{1}{2\sum_j N_{\rm gas,gas}(i,j)} ,
\ee
respectively. The total number of independent data points in each of the arms of the tuning fork diagram then follows as
\be
\label{eq:nindepstar}
N_{\rm indep,star}=\sum_i f_{\rm data,star}(i) ,
\ee
and
\be
\label{eq:nindepgas}
N_{\rm indep,gas}=\sum_i f_{\rm data,gas}(i) ,
\ee
for the bottom arm and the top arm of the diagram, focusing on stellar peaks and gas peaks, respectively. The total number of independent data points is therefore smaller than twice the number of aperture sizes, i.e.
\be
\label{eq:nindep}
N_{\rm indep}=N_{\rm indep,star}+N_{\rm indep,gas}<2\nap .
\ee

As stated above, the correlation between data points is accounted for during the fitting process of Section~\ref{sec:stepfit} by weighing the contribution of each data point to the goodness-of-fit statistic by its normalised independence fraction. We also decrease the number of degrees of freedom by using the total number of independent data points $N_{\rm indep}$ rather than the total number of data points $2\nap$. The net effect of these changes is equivalent to decreasing the uncertainties $\sigma_\bias$ to becoming `effective uncertainties' $\sigma_\bias'$. The weights and degrees of freedom are discussed in detail in Section~\ref{sec:stepfit}, but here we provide these effective uncertainties. Relative to the error bars defined in equations~(\ref{eq:bstarerror}) and~(\ref{eq:bgaserror}), we obtain correction factors of
\be
\label{eq:sigmastareff}
\frac{\sigma_{\log_{10}\bias_{\rm star}}'(i)}{\sigma_{\log_{10}\bias_{\rm star}}(i)}=\sqrt{\frac{N_{\rm indep}}{2f_{\rm data,star}(i)\nap}\frac{N_{\rm indep}-N_{\rm fit}}{2\nap-N_{\rm fit}}} ,
\ee
and
\be
\label{eq:sigmagaseff}
\frac{\sigma_{\log_{10}\bias_{\rm gas}}'(i)}{\sigma_{\log_{10}\bias_{\rm gas}}(i)}=\sqrt{\frac{N_{\rm indep}}{2f_{\rm data,gas}(i)\nap}\frac{N_{\rm indep}-N_{\rm fit}}{2\nap-N_{\rm fit}}} ,
\ee
for the gas-to-stellar flux ratio bias when focusing on stellar peaks and gas peaks, respectively. The square-roots in these expressions contain two ratios. The first term reflects the inverse of the weight that is assigned to the data points when evaluating the goodness-of-fit statistic. The second term represents the decrease of the number of degrees of freedom by only counting the effective number of independent data points, where $N_{\rm fit}$ refers to the number of free parameters (this is $N_{\rm fit}=3$ for the problem considered here, see Section~\ref{sec:stepfit} for details). The above two expressions provide the effective uncertainties shown by the shaded areas in \autoref{fig:steperrors}.

\subsubsection{Model for interpreting the gas-to-stellar flux ratio bias} \label{sec:stepmodel}
We now turn to the derivation of the theoretical model for the gas-to-stellar flux ratio bias that will be fitted to the observed tuning fork diagram of \autoref{fig:steperrors} in Section~\ref{sec:stepfit} below. The presentation in the original \klprinciple paper already included a theoretical model for the flux ratio bias (see their Appendix~C), but the model presented here eliminates an important assumption of the previous version. Previously, the model assumed that all independent regions in the map resemble point sources, implying that the surface densities approach infinity in the limit of $\lap\downarrow0$. This then leads to a flux ratio bias that diverges at small aperture sizes if $\tover=0$ \citepalias[see Figure~4 of][]{kruijssen14}. In the present paper, we describe the regions by spatial profiles with finite central densities, which is obviously appropriate for real-Universe regions. This implies that a non-diverging flux ratio bias at small aperture sizes does not necessarily require $\tover>0$. This change is critical for the accuracy of observational applications of the method. After presenting the derivation of the new model, we will briefly discuss the effect on the predicted flux ratio bias of changing the model parameters. These include the to-be-fitted, free parameters $\tgas$, $\tover$, and $\lambda$, as well as input parameters of the method ($\tstar$ and {\tt peak\_prof}) and parameters that have been extracted from the maps in the previous sections ($\betastar$, $\betagas$, $\exc_{\rm star}$, and $\exc_{\rm gas}$).

In deriving the new model, we will start by taking the result from \citetalias{kruijssen14} and then modify it to introduce the peak surface density profiles. In the original model, the gas-to-stellar flux ratio bias is obtained by counting regions within the aperture. The flux from the peak type (i.e.~stars or gas) that the apertures are focused on is constituted by the sum of the central peak and the statistically expected background flux from other peaks. By contrast, the flux from the other tracer is constituted by the sum of the statistically expected background flux from other peaks and the central peak if it spends any time in the overlap phase. This latter contribution depends on the time-scales involved -- for stellar peaks, the probability that the central peak also contains gas flux is $f_{\rm star,over}=\tover/\tstar$, whereas the probability that gas peaks also contain stellar flux is $f_{\rm gas,over}=\tover/\tgas$. This contribution should also be corrected for any evolution of the peak flux, which may result in fluxes during the overlap phase that differ from those in the isolated phase (as captured by the parameters $\betastar$ and $\betagas$, see Section~\ref{sec:stepfluxratios}).

The expressions for the gas-to-stellar flux ratio bias are derived in Appendix~C of \citetalias{kruijssen14} and we refer the interested reader to that earlier work for the details of the derivation. The gas-to-stellar flux ratio bias when focusing apertures on stellar peaks (corresponding to the bottom branch in the tuning fork diagram) is
\be
\label{eq:biasstarmodold}
\bias_{\rm star}^{\rm KL14}=
\frac{\left[1+\betagas^{-1}\left(\frac{\tgas}{\tover}-1\right)\right]^{-1}+\frac{\tstar}{\mbox{$\tau$}}\left(\frac{\lap}{\mbox{$\lambda$}}\right)^2}
       {1+\frac{\tstar}{\mbox{$\tau$}}\left(\frac{\lap}{\mbox{$\lambda$}}\right)^2} .
\ee
When focusing on gas peaks (top branch in the tuning fork diagram), the bias is given by
\be
\label{eq:biasgasmodold}
\bias_{\rm gas}^{\rm KL14}=
\frac{1+\frac{\tgas}{\mbox{$\tau$}}\left(\frac{\lap}{\mbox{$\lambda$}}\right)^2}
       {\left[1+\betastar^{-1}\left(\frac{\tstar}{\tover}-1\right)\right]^{-1}+\frac{\tgas}{\mbox{$\tau$}}\left(\frac{\lap}{\mbox{$\lambda$}}\right)^2} ,
\ee
Each of the variables in these expressions have already been introduced, but for clarity we briefly repeat them. The time-scale $\tau=\tstar+\tgas-\tover$ is the total duration of the evolutionary timeline in \autoref{fig:tschem}, $\tstar$ is the duration of the stellar phase, $\tgas$ is the duration of the gas phase, $\tover$ is the duration of the overlap phase, $\lambda$ is the mean separation length between independent regions, $\lap$ is the aperture size, and $\betastar$ and $\betagas$ represent the flux ratios between the overlap phase and the isolated phase of stellar peaks and gas peaks, respectively. In both of the above two expressions, the terms containing the square of the aperture size represent the background flux from other peaks, the `1' represents the peak that the aperture is focused on, and the terms containing $\tover$ represent the contribution from that peak to the other tracer if it resides in the overlap phase. A change relative to \citetalias{kruijssen14} that is not immediately visible in these expressions is that $\betastar$ and $\betagas$ are now treated as functions of $f_{\rm star,over}=\tover/\tstar$ and $f_{\rm gas,over}=\tover/\tgas$. In the exploration of the model behaviour below, we will treat these as independent parameters, but when fitting the model to the observed tuning fork diagram in Section~\ref{sec:stepfit}, $\betastar$ and $\betagas$ are not independent variables.

As a consequence of fully counting the peak that is being focused on (as well as fully counting any contribution from this peak to the other tracer if it resides in the overlap phase), irrespective of the aperture size, these formulations implicitly assume that the peak profiles are delta functions, i.e.~that their radii are infinitesimally small. As a result, gas-to-stellar flux ratios of equations~(\ref{eq:biasstarmodold}) and~(\ref{eq:biasgasmodold}) diverge when $\tover\downarrow0$. This can be seen directly by taking the limits of $\lap\downarrow0$, which yields
\be
\label{eq:biasstarlimitold}
\lim_{\lap\downarrow0}\bias_{\rm star}^{\rm KL14}=\left[1+\betagas^{-1}\left(\frac{\tgas}{\tover}-1\right)\right]^{-1} ,
\ee
and
\be
\label{eq:biasgaslimitold}
\lim_{\lap\downarrow0}\bias_{\rm gas}^{\rm KL14}=1+\betastar^{-1}\left(\frac{\tstar}{\tover}-1\right) .
\ee
If $\tover\downarrow0$, these expressions approach zero and infinity, respectively. This result led us to conclude in \citetalias{kruijssen14} that the flattening of the tuning fork diagram at small aperture sizes provides a way of measuring $\tover$. While this is still true, we pointed out above that the real-Universe situation is more complex, because the flattening of the tuning fork diagram is not unique to a non-zero duration of the overlap phase. Real molecular clouds or star-forming regions are not infinitesimally small and their finite central densities provide another way of avoiding divergence.

We account for the peak surface density profiles by introducing an extra factor in equations~(\ref{eq:biasstarmodold}) and~(\ref{eq:biasgasmodold}) that takes values between 0 and 1 and represents the fraction of the flux from the central peak that is contained within the aperture. This factor is only added to the terms that represent the stellar and gas flux contribution from the central peak. For the randomly-distributed, background population of spatially-extended peaks in and around the aperture, the flux contributions from peaks with centres outside of the aperture cancel on geometric grounds with the flux loss from peaks with centres within the aperture. As a result, the background flux in the aperture is exclusively set by the number surface density of the peaks and is insensitive to their individual surface density profiles.\footnote{We thank Sophie Kruijssen for pointing out this logic, which has led to an important simplification of the model.} This allows us to formulate the updated version of the \klprinciple model as
\be
\label{eq:biasstarmod}
\bias_{\rm star}^{\rm mod}=
\frac{f_{\rm gas}\left[1+\betagas^{-1}\left(\frac{\tgas}{\tover}-1\right)\right]^{-1}+\frac{\tstar}{\mbox{$\tau$}}\left(\frac{\lap}{\mbox{$\lambda$}}\right)^2}
       {f_{\rm star}+\frac{\tstar}{\mbox{$\tau$}}\left(\frac{\lap}{\mbox{$\lambda$}}\right)^2} ,
\ee
and
\be
\label{eq:biasgasmod}
\bias_{\rm gas}^{\rm mod}=
\frac{f_{\rm gas}+\frac{\tgas}{\mbox{$\tau$}}\left(\frac{\lap}{\mbox{$\lambda$}}\right)^2}
       {f_{\rm star}\left[1+\betastar^{-1}\left(\frac{\tstar}{\tover}-1\right)\right]^{-1}+\frac{\tgas}{\mbox{$\tau$}}\left(\frac{\lap}{\mbox{$\lambda$}}\right)^2} ,
\ee
where $f_{\rm star}$ and $f_{\rm gas}$ represent the fraction of the central peak flux that is contained by the aperture, with the subscripts referring to stellar peaks and gas peaks respectively. These expressions are used throughout the rest of this paper.

Equations~(\ref{eq:biasstarmod}) and~(\ref{eq:biasgasmod}) do not yet show how $f_{\rm star}$ and $f_{\rm gas}$ depend on the aperture size. We will specify this below for a small number of useful peak surface density profiles. However, in defining the above expressions, we have already made an important assumption. The enclosed flux fraction of the central peak is assumed to differ between both tracers, because a factor of $f_{\rm gas}$ appears in the numerator and a factor of $f_{\rm star}$ appears in the denominator of equations~(\ref{eq:biasstarmod}) and~(\ref{eq:biasgasmod}). This means that we allow the flux from both tracers to follow different profiles, which provides some useful flexibility and is consistent with observations \citep[e.g.][]{walker15,walker16}. However, we also assume that there is no difference in profile shape between peaks of a given tracer that reside in the isolated and overlap phases, because the same factors of $f_{\rm star}$ and $f_{\rm gas}$ are used in equations~(\ref{eq:biasstarmod}) and~(\ref{eq:biasgasmod}). This may introduce a minor inconsistency in cases where there is considerable time-evolution of the peak size. In practical applications, however, this typically happens on size scales below the resolution limit, because we convolve the maps to a common resolution $\lapmin$ (see Section~\ref{sec:stepregrid}). It is demonstrated in Section~\ref{sec:valid} that the method yields accurate measurements of $\tover$, to such an extent that any effect of this inconsistency must fall within acceptable limits. Indeed, we have verified that the results do not change (i.e.~they stay well within the obtained uncertainties) when dividing the peak sample in peaks residing in the isolated and overlap phases before determining the flux density profiles. It is therefore undesirable to add another layer of complexity that could address this, in particular because we have no a priori constraints on how the peak profiles change with evolutionary stage.

We consider three different types of surface density profiles to model the peak emission, all of which are available in \code. However, only the third of these profiles will be used throughout the rest of the paper, for reasons outlined below. Each of these profiles is normalised to have a total flux fraction of unity (i.e.~$\lim_{\lap\rightarrow\infty}f(\lap)=1$), so that at infinitely large aperture sizes the entire peak resides within the aperture. The profiles are as follows.
\begin{enumerate}
\item
A two-dimensional delta function that is centred on the peak position, resulting in an enclosed flux fraction as a function of aperture size that is given by
\be
\label{eq:prof1}
f_\delta(\lap)=1 .
\ee
This corresponds to the case derived in \citetalias{kruijssen14}, where we assumed that the peaks correspond to points.
\item
A two-dimensional, constant surface density disc that is centred on the peak position, resulting in an enclosed flux fraction as a function of aperture size that is given by
\be
\label{eq:prof2}
f_{\rm disc}(\lap)=\left\{\begin{array}{ll}
(\lap/2r_{\rm peak})^2 & \mbox{if } \lap < 2r_{\rm peak} \\
1 & \mbox{if } \lap \geq 2r_{\rm peak} .
\end{array} \right.
\ee
In this expression, $r_{\rm peak}$ represents the radius of the constant surface density disc. We discuss below how this radius is obtained from the galaxy maps.
\item
A two-dimensional Gaussian centred on the peak position, resulting in an enclosed flux fraction as a function of aperture size that is given by
\be
\label{eq:prof3}
f_{\rm Gauss}(\lap)=1-\exp{\left[-\frac{1}{2}\left(\frac{\lap}{2r_{\rm peak}}\right)^2\right]} .
\ee
In this expression, $r_{\rm peak}$ represents the dispersion of the two-dimensional Gaussian. We discuss below how this scale radius is obtained from the galaxy maps.
\end{enumerate}

The second and third of these profiles require the characteristic peak radius $r_{\rm peak}$ to be known. However, it is straightforward to obtain this from the galaxy maps once a peak profile has been assumed to approximate the flux density profiles of the peaks in the maps. In Section~\ref{sec:stepfluxratios}, we measured the flux density contrast of emission peaks to the average flux density on a size scale $\lap$ in equations~(\ref{eq:excstar}) and~(\ref{eq:excgas}). This flux ratio contrast can be related to the ratio between the central surface density of peaks and the background surface density of the peak population. In turn, this allows us to express the peak radius $r_{\rm peak}$ in units of the region separation length $\lambda$. For the stellar and gas flux, the background number densities of independent regions are
\be
\label{eq:surfbackstar}
\Sigma_{\rm back,star}=\frac{\tstar}{\tau}\left(\frac{\pi\lambda^2}{4}\right)^{-1} ,
\ee
and
\be
\label{eq:surfbackgas}
\Sigma_{\rm back,gas}=\frac{\tgas}{\tau}\left(\frac{\pi\lambda^2}{4}\right)^{-1} ,
\ee
respectively. In these expressions, the term in parentheses uses the mean separation length between independent regions to represent the area per region, whereas the multiplications by  $\tstar/\tau$ and $\tgas/\tau$ reduce the resulting total number densities to only refer to the number densities of stellar peaks and gas peaks, respectively.

For a constant surface density disc normalised to unity, the central surface density is
\be
\label{eq:surfdisc}
\Sigma_{\rm disc}=\frac{1}{\pi r_{\rm peak}^2} ,
\ee
whereas for a two-dimensional Gaussian normalised to unity, the central surface density is
\be
\label{eq:surfgauss}
\Sigma_{\rm Gauss}=\frac{1}{2\pi r_{\rm peak}^2} .
\ee
With the expected background and central flux densities in hand, it is now possible to relate these to the flux density contrasts $\exc_{\rm star}$ and $\exc_{\rm gas}$ as a function of aperture size obtained in equations~(\ref{eq:excstar}) and~(\ref{eq:excgas}) above (see Section~\ref{sec:stepfluxratios}). This will provide the peak radius $r_{\rm peak}$ in units of the region separation length $\lambda$.

Because $\exc_{\rm star}$ and $\exc_{\rm gas}$ represent flux density contrasts measured relative to a variable aperture size, there is some freedom in choosing the size scale on which the flux density contrast is evaluated. We find that it is desirable to define $\exc_{\rm star}$ and $\exc_{\rm gas}$ on a size scale $\lambda$, because on larger scales the considered maps may contain voids or other morphological features that change the surface number density of regions from the density in the direct vicinity of the peaks implied by the best-fitting $\lambda$. By setting the size scale to $\lambda$, the background surface number density of neighbouring peaks best corresponds to the measured region separation length. In practice, this means that when fitting the observed tuning fork diagram in Section~\ref{sec:stepfit} below, $\exc_{\rm star}$ and $\exc_{\rm gas}$ are both varied with $\lambda$ and are constrained as a byproduct of the fitting process.

On a size scale $\lambda$, the measured flux ratio contrasts defined in equations~(\ref{eq:excstar}) and~(\ref{eq:excgas}) can be expressed in terms of the above central and background number densities as 
\be
\label{eq:excstarsurf}
\exc_{\rm star}\equiv\exc_{\rm star}(\lambda)=\frac{\Sigma_{\rm peak}+\Sigma_{\rm back,star}}{(\pi\lambda^2/4)^{-1}+\Sigma_{\rm back,star}} ,
\ee
and
\be
\label{eq:excgassurf}
\exc_{\rm gas}\equiv\exc_{\rm gas}(\lambda)=\frac{\Sigma_{\rm peak}+\Sigma_{\rm back,gas}}{(\pi\lambda^2/4)^{-1}+\Sigma_{\rm back,gas}} ,
\ee
where $\Sigma_{\rm peak}$ refers to $\Sigma_{\rm disc}$ or $\Sigma_{\rm Gauss}$ depending on the adopted profile. In these two expressions, the numerator represents the flux density on a size scale $\lapmin$, consisting of the central surface density of the peak (first term) and the background population (second term), and the denominator represents the flux density on a size scale $\lambda$, consisting of the entire central peak (first term) and the background population (second term).

Rearranging the above expressions allows the characteristic peak radii to be expressed in terms of quantities that are either measured directly from the maps ($\exc_{\rm star}$ and $\exc_{\rm gas}$), or represent free parameters during the fitting process in Section~\ref{sec:stepfit} ($\tau$, $\tgas$, and $\lambda$). If we represent the peaks with constant surface density discs, we obtain for the radii of stellar peaks
\be
\label{eq:rstardisc}
r_{\rm star}=\frac{\lambda}{2}\sqrt{\frac{\tau/\tstar}{\exc_{\rm star}(1+\tau/\tstar)-1}} ,
\ee
and for the radii of gas peaks
\be
\label{eq:rgasdisc}
r_{\rm gas}=\frac{\lambda}{2}\sqrt{\frac{\tau/\tgas}{\exc_{\rm gas}(1+\tau/\tgas)-1}} .
\ee
Likewise, if we represent the peaks with two-dimensional Gaussian profiles, we obtain for the radii of stellar peaks
\be
\label{eq:rstargauss}
r_{\rm star}=\frac{\lambda}{2}\sqrt{\frac{\tau/\tstar}{2\exc_{\rm star}(1+\tau/\tstar)-2}} ,
\ee
and for the radii of gas peaks
\be
\label{eq:rgasgauss}
r_{\rm gas}=\frac{\lambda}{2}\sqrt{\frac{\tau/\tgas}{2\exc_{\rm gas}(1+\tau/\tgas)-2}} .
\ee
Equations~(\ref{eq:rstardisc})--(\ref{eq:rgasgauss}) are substituted into the enclosed flux fractions as a function of aperture size from equations~(\ref{eq:prof2}) and~(\ref{eq:prof3}). Substitution of the result into equations~(\ref{eq:biasstarmod}) and~(\ref{eq:biasgasmod}) yields the predicted gas-to-stellar flux ratio bias when focusing apertures on stellar peaks and gas peaks, respectively, in terms of measurable quantities and free parameters of the fitting process in Section~\ref{sec:stepfit}.

When applying the method to real-Universe observations or to the simulated galaxy maps in Section~\ref{sec:valid}, the Gaussian peak surface density profile $f_{\rm Gauss}$ is generally the best choice to make. Above all, it is versatile -- in the limit of a very large flux ratio excess $\exc$, the peak radii approach zero and hence the possible range of profiles includes a reasonable approximation of the delta function profile $f_\delta$ to represent point particles. This feature is also shared by the constant surface density disc profile $f_{\rm disc}$, but the Gaussian profile is more suitable for two other reasons. Firstly, the regridding of the maps in Section~\ref{sec:stepregrid} is accompanied by a convolution of the maps with a Gaussian PSF to a common resolution. This means that unresolved peaks have Gaussian profiles by definition and resolved peaks at least have a Gaussian component. Secondly, real-Universe molecular clouds and star-forming regions generally follow centrally concentrated morphologies, rather than having constant surface densities \citep[e.g.][]{walker15,walker16}. Gaussian profiles provide a better representation of this behaviour. Of course, it is straightforward to select any other desired analytical profile and derive expressions equivalent to equations~(\ref{eq:rstardisc})--(\ref{eq:rgasgauss}) for that particular choice. This may yield differences in the best-fitting quantities, but for the above reasons we expect these to be minor. In the remainder of this paper, we use the Gaussian profile to represent the peaks in the map.

Substitution of the stellar and gas peak radii into the Gaussian profile of equation~(\ref{eq:prof3}) provides the enclosed flux fraction of stellar and gas peaks as a function of the aperture size, i.e.
\be
\label{eq:gaussstar}
f_{\rm star}(\lap)=1-\exp{\left[\frac{1-\exc_{\rm star}\left(1+\tau/\tstar\right)}{\tau/\tstar}\left(\frac{\lap}{\lambda}\right)^2\right]} ,
\ee
and
\be
\label{eq:gaussgas}
f_{\rm gas}(\lap)=1-\exp{\left[\frac{1-\exc_{\rm gas}\left(1+\tau/\tgas\right)}{\tau/\tgas}\left(\frac{\lap}{\lambda}\right)^2\right]} ,
\ee
respectively. We will use these two expressions in combination with the gas-to-stellar flux ratios of equations~(\ref{eq:biasstarmod}) and~(\ref{eq:biasgasmod}) to fit the observed tuning fork diagram that was obtained in Section~\ref{sec:steperrors}.

To avoid overly long expressions, we do not explicitly provide the (trivial) substitution of the peak density profiles into equations~(\ref{eq:biasstarmod}) and~(\ref{eq:biasgasmod}), but it is useful to consider the limit of $\lap\downarrow0$ to highlight the difference in behaviour relative to the original model of \citetalias{kruijssen14}. For small aperture sizes, we obtain
\be
\label{eq:biasstarlimit}
\lim_{\lap\downarrow0}\bias_{\rm star}^{\rm mod}=\frac{1}{\exc_{\rm star}\left(1+\frac{\mbox{$\tau$}}{\tstar}\right)}+\frac{1-\exc_{\rm star}^{-1}\left(1+\frac{\mbox{$\tau$}}{\tstar}\right)^{-1}}{1+\betagas^{-1}\left(\frac{\tgas}{\tover}-1\right)} ,
\ee
for the gas-to-stellar flux ratio bias when focusing on stellar peaks and
\be
\label{eq:biasgaslimit}
\lim_{\lap\downarrow0}\bias_{\rm gas}^{\rm mod}=\left[\frac{1}{\exc_{\rm gas}\left(1+\frac{\mbox{$\tau$}}{\tgas}\right)}+\frac{1-\exc_{\rm gas}^{-1}\left(1+\frac{\mbox{$\tau$}}{\tgas}\right)^{-1}}{1+\betastar^{-1}\left(\frac{\tstar}{\tover}-1\right)}\right]^{-1} ,
\ee
when focusing on gas peaks. It is straightforward to verify that we obtain the old limits of equations~(\ref{eq:biasstarlimitold}) and~(\ref{eq:biasgaslimitold}) when we consider point particles and thus let $\exc_{\rm star}\rightarrow\infty$ and $\exc_{\rm gas}\rightarrow\infty$. However, the change relative to the old limits is best illustrated by considering the case of $\tover\downarrow0$, which yields
\be
\label{eq:biasstarlimit2}
\lim_{\tover\downarrow0}\left(\lim_{\lap\downarrow0}\bias_{\rm star}^{\rm mod}\right)=\frac{1}{\exc_{\rm star}\left(1+\frac{\mbox{$\tau$}}{\tstar}\right)} ,
\ee
and
\be
\label{eq:biasgaslimit2}
\lim_{\tover\downarrow0}\left(\lim_{\lap\downarrow0}\bias_{\rm gas}^{\rm mod}\right)=\exc_{\rm gas}\left(1+\frac{\mbox{$\tau$}}{\tgas}\right) ,
\ee
for the bottom and top branches of the tuning fork diagram, respectively. These limits show immediately that setting $\exc_{\rm star}\rightarrow\infty$ and $\exc_{\rm gas}\rightarrow\infty$ to represent point-like regions results in a gas-to-stellar flux ratio bias that diverges to $0$ or $\infty$ for stellar and gas peaks, respectively, as in \citetalias{kruijssen14}. However, they also quantify that for peak surface density profiles with a finite central density, the gas-to-stellar flux ratio bias does not diverge to $0$ or $\infty$ towards small aperture sizes as in \citetalias{kruijssen14}, but instead takes values proportional to the (inverse) flux density contrast, i.e.~$\exc^{-1}$ or $\exc$. The two factors in these limits represent the contribution from the peak flux relative to a size scale $\lambda$ (the factor $\exc$) and from the background peak population within $\lambda$ relative to the galactic average (the factor $1+\tau/t$). The first of these factors follows from the definition of $\exc$ and the second is easily verified by evaluating equations~(\ref{eq:biasstarmod}) and~(\ref{eq:biasgasmod}) at $\lap=\lambda$. By combining these two factors, we obtain the central flux ratio bias of a peak relative to the galactic average. The qualitative behaviour of these limits is also as expected when setting $\tover=0$. Firstly, a low filling factor (high $\exc$) implies a large difference between the central peak and the background population within $\lambda$, leading to a strong bias. Secondly, a rarity of tracer peaks within an aperture of size $\lambda$ (i.e.~when $\tstar$ or $\tgas$ is much smaller than $\tau$) implies a large difference between the flux density on a size scale $\lambda$ and the galactic average, also leading to a strong bias.

The immediate implication of equations~(\ref{eq:biasstarlimit2}) and~(\ref{eq:biasgaslimit2}) is that the flattening of the tuning fork diagram towards small aperture sizes is still a measure of the duration of the overlap phase $\tover$, but there is a degeneracy with the central flux ratio excess of peaks relative to the background. However, this degeneracy is lifted by directly obtaining the peak flux ratio excess from the maps, as discussed in Section~\ref{sec:stepfluxratios}. In practice, a finite flux ratio excess provides a minimum amount of flattening. Any additional flattening indicates a non-zero duration of the overlap phase and thus still provides a way of directly measuring $\tover$.

\begin{table}
 \centering
 \begin{minipage}{62mm}
  \caption{Model parameter values explored in \autoref{fig:stepmodel}}\label{tab:model}\vspace{-1mm}
  \begin{tabular}{c c c c c c c c}
   \hline
   Quantity & low & middle & high \\ 
   \hline
   \multicolumn{4}{c}{free parameter}\\
   $\tgas~[\myr]$ & $5$ & $\mathbf{10}$ & $20$ \\ 
   $\tover~[\myr]$ & $0$ & $\mathbf{2}$ & $4$ \\ 
   $\lambda~[\pc]$ & $100$ & $\mathbf{300}$ & $1000$ \\[1.5ex]
   \multicolumn{4}{c}{function of free parameters}\\
   $\tstar~[\myr]$ & $2.5$ & $\mathbf{5}$ & $10$ \\ 
   $\betastar$ & $0.5$ & $\mathbf{1}$ & $2$ \\ 
   $\betagas$ & $0.5$ & $\mathbf{1}$ & $2$ \\
   $\exc_{\rm star}$ & $2.5$ & $\mathbf{5}$ & $10$ \\ 
   $\exc_{\rm gas}$ & $2.5$ & $\mathbf{5}$ & $10$ \\[1.5ex]
   \multicolumn{4}{c}{fixed parameter}\\
   {\tt peak\_prof} & $\mathrm{points}$ & $\mathrm{discs}$ & $\mathbf{Gaussians}$ \\ 
   \hline
  \end{tabular} \\ 
  The boldface numbers indicate the fiducial values used in each of the panels in  \autoref{fig:stepmodel}.\vspace{-1mm}
 \end{minipage}
\end{table}
We now turn to a brief discussion of how the gas-to-stellar flux ratio biases as a function of aperture size vary with each of the quantities involved in our model. \autoref{tab:model} lists the nine quantities that together determine the flux ratio biases of equations~(\ref{eq:biasstarmod}) and~(\ref{eq:biasgasmod}), as well as three typical values for each of these quantities that we will consider during a quantitative illustration of the model below. The first three of these ($\tgas$, $\tover$, and $\lambda$) represent the three free parameters in the model and are obtained by fitting the model in Section~\ref{sec:stepfit} to the observed gas-to-stellar flux ratio bias of Section~\ref{sec:steperrors} and \autoref{fig:steperrors}. The second set of five quantities ($\tstar$, $\betastar$, $\betagas$, $\exc_{\rm star}$, and $\exc_{\rm gas}$) depend on the three free parameters and are therefore constrained during the fitting process, but are not free parameters themselves. The default definition of $\tstar=\tstarref+\tover$ implies a dependence on the duration of the overlap phase, but if the flag {\tt tstar\_incl}~$=1$ (see \autoref{tab:flags}), then $\tstar=\tstarref$ is fixed in the input file and becomes independent of the free parameters. In this example, we specify $\tstar$, implying that $\tstarref$ changes when $\tover$ changes, which effectively corresponds to {\tt tstar\_incl}~$=1$. By contrast, $\betastar(f_{\rm star,over})$ and $\betagas(f_{\rm gas,over})$ always depend on the free parameters, because they depend on the ratios $\tover/\tstar$ and $\tover/\tgas$, respectively, as described in Section~\ref{sec:stepfluxratios} and illustrated in \autoref{fig:stepbeta}. Likewise, $\exc_{\rm star}(\lambda)$ and $\exc_{\rm gas}(\lambda)$ both depend on the mean separation length $\lambda$ (see Section~\ref{sec:stepfluxratios}). While these five quantities will vary as a function of $\tgas$, $\tover$, and $\lambda$ in practical applications of the method, we consider them as free parameters in the example given here, with the goal of isolating their effects. The final quantity in \autoref{tab:model} ({\tt peak\_prof}) represents a fixed parameter that is set in the input parameter file. This quantity therefore influences the shape of the modelled tuning fork diagram, but does not vary during the fitting process of Section~\ref{sec:stepfit}.

\begin{figure*}
\includegraphics[width=\hsize]{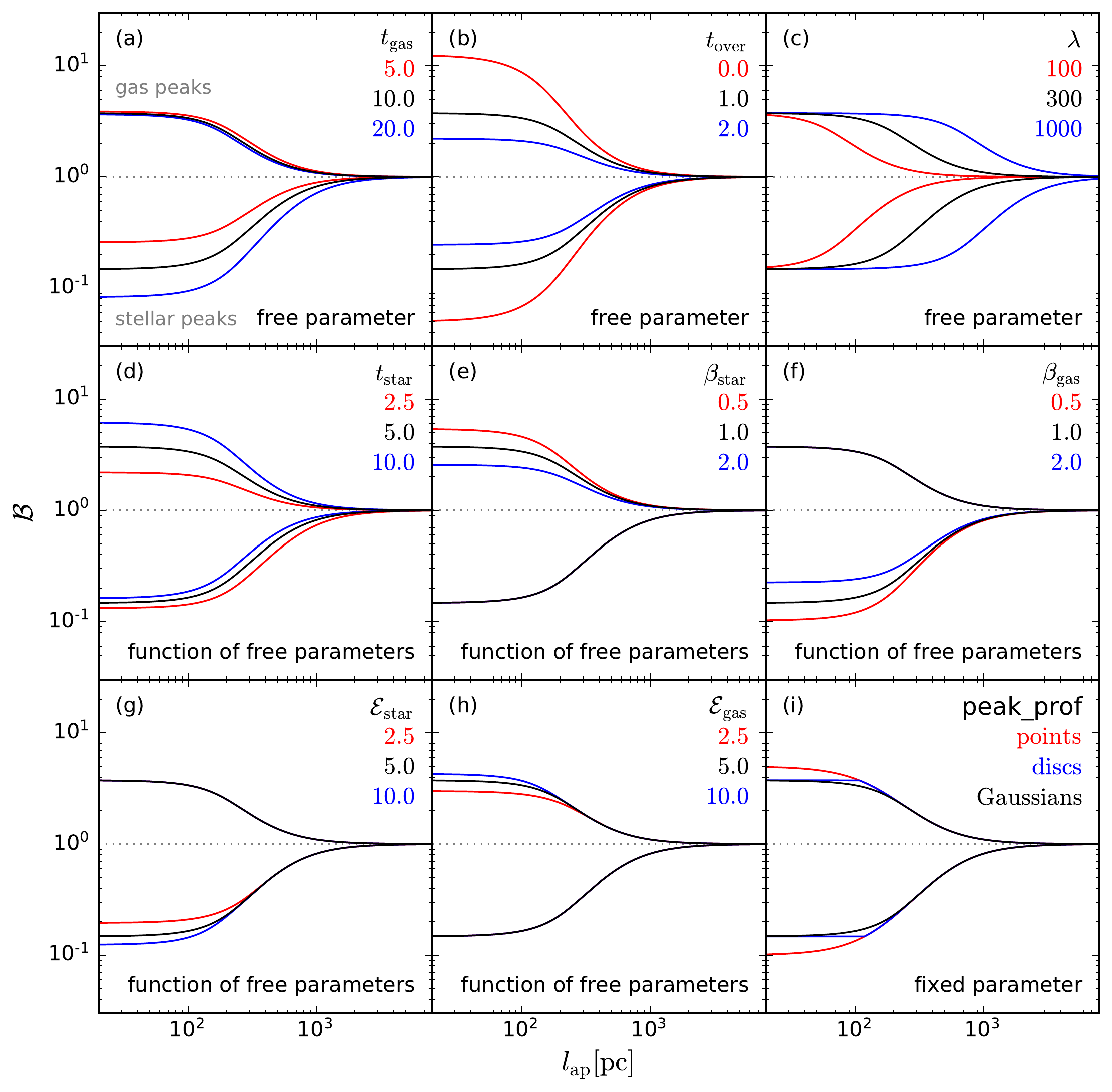}%
\vspace{-1mm}\caption{
\label{fig:stepmodel}
Predicted gas-to-stellar flux ratio bias as a function of the aperture size when focusing apertures on stellar peaks (bottom branches) or gas peaks (top branches), as indicated by the grey labels in panel (a). In each panel, the effect of varying one model quantity is shown relative to the fiducial parameter set (see \autoref{tab:model}). The values in the top right corner of each panel show the values used, with the fiducial choice shown in black. In the bottom right corner of each panel, we list the nature of each quantity. The quantities varied in panels (a)--(c) are all {\it free parameters} in the model, meaning that they are fitted for in Section~\ref{sec:stepfit}. The quantities varied in panels (d)--(h) are all a function of the three free parameters in the top row, implying that while they will be constrained during the fitting process, they are not free parameters as such. Finally, the quantity {\tt peak\_prof} that is varied in panel (i) is fixed prior to starting the fitting process. The figure shows that each of the free parameters has a unique effect on the shapes of the curves, implying that they are non-degenerate and can therefore be measured by fitting our model to observed tuning fork diagrams.\vspace{-1mm}
}
\end{figure*}
\autoref{fig:stepmodel} systematically shows how each of the nine quantities from \autoref{tab:model} affect the shape of the modelled gas-to-stellar flux ratio bias as a function of the aperture size. The main result of the parameter study in the figure is that each of the nine quantities in general, but especially the three free parameter in particular, have a unique effect on the shapes of the curves. This means that the gas-to-stellar flux ratio bias is a non-degenerate function of these quantities, allowing the free parameters to be measured by fitting the model to observed tuning fork diagrams. The basic behaviour is the same as in the schematic of \autoref{fig:tuningfork}. When focusing apertures on stellar (gas) peaks, the gas-to-stellar flux ratio is smaller (larger) than the galactic average. This gas-to-stellar flux ratio bias increases towards decreasing aperture sizes, because the central peak contributes a larger fraction to the total flux ratio in the aperture when a smaller population of other peaks is included. However, the bias does not grow indefinitely towards small apertures. As discussed in the above derivation, this arises for two reasons. Firstly, if the stellar and gas phases overlap in time, then there is a non-zero probability that a peak resides in the overlap phase and emission from both tracers is present. This limits the gas-to-stellar flux ratio bias when preferentially selecting one type of tracer. Secondly, peaks have finite central surface densities, implying that for the smallest aperture sizes, the gas-to-stellar flux ratio bias reflects the overdensity of the peaks relative to the background population of other peaks, as demonstrated in equations~(\ref{eq:biasstarlimit2}) and~(\ref{eq:biasgaslimit2}).

Starting with the top row of \autoref{fig:stepmodel}, we see in panel~(a) that increasing $\tgas$ decreases the bottom branch of the tuning fork diagram, i.e.~it leads to a larger gas-to-stellar flux deficit when focusing on stellar peaks. This happens because for larger $\tgas$, the overlap phase (which sets the bias at small aperture sizes) takes up a smaller fraction of the total duration of the gas phase, implying that stellar peaks residing in the overlap phase contribute a smaller fraction of the galactic gas flux. Due to the relatively smaller `damping' effect of the gas flux from stellar peaks in the overlap phase the bottom branch flattens less if $\tgas$ is larger. By contrast, changing the duration of the gas phase does not affect the relative contribution of the stellar flux from gas peaks in the overlap phase to the galactic stellar flux, because this is mainly set by the ratio $\tover/\tstar$ in equation~(\ref{eq:biasgasmod}). The flattening of the top branch is therefore only weakly affected by changing $\tgas$. There is a small decrease of the aperture size below which the gas-to-stellar flux ratio bias sets in when increasing $\tgas$. This is easily understood -- gas peaks become more common as $\tgas$ increases, causing their number density to increase too. Therefore, a smaller aperture size is required to isolate the contribution from the gas peak that the aperture is focused on.

The second free parameter is the duration of the overlap phase $\tover$. Panel~(b) shows that the effect of increasing $\tover$ is that the probability that the central peak in the aperture also contains emission from the other tracer increases. This limits the gas-to-stellar flux ratio bias relative to the galactic average that can be achieved at the smallest aperture sizes and therefore leads to a stronger flattening of the model tuning fork diagram towards larger $\tover$. The figure again illustrates the key difference relative to \citetalias{kruijssen14}, namely that the branches of the tuning fork do not diverge towards small aperture sizes when $\tover=0$, due to the finite central surface densities of the two-dimensional Gaussians used here. Most importantly, the behaviour due to $\tover$ fundamentally differs from the influence of $\tgas$ on the gas-to-stellar flux ratio bias. While the latter affects the vertical position of one of the branches and only weakly affects the horizontal position of the other branch, the former symmetrically sets the vertical positions of both branches.

The third free parameter is the mean separation length between independent regions $\lambda$. Its effect on the shape of the tuning fork is easy to understand. Panel~(c) shows that increasing the separation length means that the entire model tuning fork diagram shifts to larger aperture sizes, because recovering the galactic average by encompassing a sufficient number of regions in the aperture requires a larger aperture area. Crucially, the effect of the separation length is orthogonal to that of $\tgas$ and $\tover$, as it controls the overall horizontal position of the diagram.

We briefly summarise the effects of the three free parameters during the fitting process of our model.
\begin{enumerate}
\item
The ratio between $\tgas$ and $\tstar$ sets the vertical asymmetry of the model tuning fork diagram.
\item
The duration of the overlap phase $\tover$ sets the flattening of the tuning fork towards small aperture sizes.
\item
The mean separation length of independent regions $\lambda$ sets the horizontal position of the model tuning fork.
\end{enumerate}

The quantities in panels~(d)--(h) of \autoref{fig:stepmodel} depend on the free parameters $\tgas$, $\tover$, and $\lambda$ from the first row. When changing either of these free parameters during the fitting process, their effect on the tuning fork diagram will therefore be accompanied by the change seen in one or more of panels~(d)--(h). The duration of the stellar phase $\tstar=\tstarref+\tover$ is varied in panel~(d) and mirrors the behaviour of changing $\tgas$ in panel~(a). In addition, because $\tstar$ increases with $\tover$, the increased flattening of the curves towards large values of $\tover$ is accompanied by a slight increase of the gas-to-stellar flux ratio excess in the top branch and an even smaller inwards shift of the bottom branch. We note that the effect of changing $\tstar$ on the bottom branch in panel~(d) is larger than that of changing $\tgas$ on the top branch in panel~(a). This is caused by the fact that we have set $\tgas>\tstar$ in this example. As a result of this dissimilarity, the flattening of the top branch in panel~(a) is affected more strongly by the high probability that young stellar peaks reside in the overlap phase and are therefore included when focusing on gas peaks, which does not depend on $\tgas$, whereas the flattening of the bottom branch in panel~(d) is affected more strongly by the flux density contrast of the central peak relative to its immediate surroundings, which does depend on $\tstar$. This difference can be quantitatively verified by inspection of equations~(\ref{eq:biasstarlimit}) and~(\ref{eq:biasgaslimit}).

Panel~(e) shows the influence of changing the overlap-to-isolated flux ratio of stellar emission. Unsurprisingly, this only affects the top branch, where the incidence of stellar emission from gas peaks in the overlap phase flattens the tuning fork. This flattening becomes stronger if the stellar emission from peaks in the overlap phase is brighter than from those in the isolated phase (i.e.~$\betastar>1$) and weakens if the stellar emission from peaks in the overlap phase is fainter (i.e.~$\betagas<1$). The same behaviour is seen for the bottom branch in Panel~(f), which shows the influence of changing the overlap-to-isolated flux ratio of gas emission. Because $\betastar$ and $\betagas$ depend on $\tover/\tstar$ and $\tover/\tgas$, respectively, a change of the model tuning fork diagram due to any of the three time-scales can be accompanied by a change as seen in panels~(e) or~(f). This could hamper the accuracy of the method if the corresponding changes in $\betastar$ and $\betagas$ would systematically cancel the impact of changing the above time-scales. However, there is no obvious reason why the relations between $\betastar$ and $\betagas$ should scale with these time-scale ratios in a particular direction, or even monotonically. The applications to simulated maps in Section~\ref{sec:valid}, as well as early applications of the method to observational data (e.g.~\citealt{kruijssen18}; Hygate et al.~in prep.; Chevance et al.~in prep.) show no strong systematic patterns in these relations.

If the relation between $\betastar(f_{\rm star,over})$ and $\betagas(f_{\rm gas,over})$ would be monotonic, we would still not expect any significant change of the tuning fork diagram due to the propagation of time-scale changes through $\betastar$ and $\betagas$, because $\betastar$ and $\betagas$ typically only change by a small amount across the range of $\tover/\tstar$ and $\tover/\tgas$ (see e.g.~\autoref{fig:stepbeta}, where $\betastar=1.3$--$1.7$ for $\tover/\tstar=0.05$--$0.85$ and compare this dynamic range to \autoref{fig:stepmodel}). Even if this would occur in individual cases, it is important to note that the effects of $\betastar$ and $\betagas$ differ fundamentally from the free parameters in panels~(a)--(c). The greatest similarity is between $\tgas$ and $\betagas$, both of which affect the vertical position of the bottom branch. In that case, a systematic decrease of $\betagas$ with $\tover/\tgas$ would mean that the widening of the tuning fork by increasing $\tgas$ would be compensated by the accompanying decrease of $\tover/\tgas$, the resulting increase of $\betagas$, and the corresponding vertical compression of the tuning fork. In such a situation, $\tgas$ would be ill-constrained, but this is unlikely to occur for two reasons. Firstly, in the vast majority of the experiments considered in this paper, we find that $\betastar$ and $\betagas$ weakly increase with $\tover/\tstar$ and $\tover/\tgas$, respectively (see e.g.~\autoref{fig:stepbeta}). This trend is opposite to what would be needed for changes in the tuning fork diagram due to the evolutionary timeline to be compensated by changes in $\betastar$ and $\betagas$. Secondly, even in this rather specific case, the detailed change of the tuning fork due to varying $\tgas$ and $\betagas$ differs. The former affects the vertical position of the bottom branch over the entire range of aperture sizes, whereas the latter only affects the part where the overlap phase dominates the flux ratio bias in the aperture (i.e.~where $\lap<\lambda$). In summary, we therefore do not expect any systematic variation of $\betastar$ or $\betagas$ with $\tover/\tstar$ or $\tover/\tgas$ to negatively influence the accuracy with which the free parameters can be constrained.

Panels~(g) and~(h) demonstrate the influence of changing the flux ratio contrast $\exc_{\rm star}$ and $\exc_{\rm gas}$ between the central peak and the flux measured on a size scale $\lambda$. The effect of changing these quantities on the model tuning fork diagram is similar to that of $\betastar$ and $\betagas$, with the key difference that $\exc_{\rm star}$ and $\exc_{\rm gas}$ are functions of $\lambda$ rather than $\tover/\tstar$ or $\tover/\tgas$. We see that increasing the flux ratio contrast of single peaks relative to the background leads to a larger gas-to-stellar flux ratio excess (when focusing on gas peaks) or deficit (when focusing on stellar peaks) at the smallest aperture sizes. This is easily understood, because on these size scales, the central surface density of the peak sets the flux ratio bias. Because $\exc_{\rm star}$ and $\exc_{\rm gas}$ depend on $\lambda$, a change of $\lambda$ can be accompanied by a change as seen in panels~(g) or~(h). As before, this could hamper the accuracy of the method if the corresponding changes in $\exc_{\rm star}$ and $\exc_{\rm gas}$ would systematically cancel the impact of changing the separation length $\lambda$. Fortunately, \autoref{fig:stepmodel} clearly shows that the effects of changing $\lambda$ and $\exc$ are orthogonal -- the former causes a horizontal shift of the model tuning fork, while the latter vertically shifts a single branch. Therefore, we do not expect any systematic variation of $\exc_{\rm star}$ or $\exc_{\rm gas}$ with $\lambda$ to negatively influence the accuracy with which the free parameters can be constrained.

The final quantity in panel~(i) of \autoref{fig:stepmodel} is fixed and therefore does not change during the fitting process. However, it is still useful to understand its impact of the gas-to-stellar flux ratio bias as a function of aperture size. Panel~(i) shows how the shape of the surface density profile chosen to represent the peaks affects the resulting model tuning fork diagram. Because this only concerns the central peak in each aperture, we expect this choice to affect the shape only for apertures $\lap<\lambda$. The delta function profile, reflecting point-like regions, corresponds to the model from \citetalias{kruijssen14}. If we instead represent the regions as constant surface density discs with radius $r_{\rm peak}$, the tuning fork abruptly flattens once $\lap<2r_{\rm peak}$, because there the enclosed surface density no longer increases with decreasing aperture size. Adopting a two-dimensional Gaussian profile results in the same central gas-to-stellar flux ratio bias as for the disc profile, but the additional flattening sets in at larger aperture sizes, because the extended nature of the Gaussian profile leads to flux loss from the aperture already for $\lap>2r_{\rm peak}$. While the effective change from the point particle representation to the two-dimensional Gaussian may seem minor, it is critical for accurately retrieving the duration of the overlap phase. As we will discuss in Section~\ref{sec:stepderived}, this change is therefore instrumental for deriving any feedback-related physical quantities.

Thanks to its purely analytical form, the derived model for the gas-to-stellar flux ratio bias is easy to interpret and readily applicable to the type of observational measurement made in Sections~\ref{sec:stepreadpar}--\ref{sec:steperrors}. By fitting the model to observed `tuning fork diagrams', the six quantities in the top two rows of \autoref{fig:stepmodel} can be obtained, which provide a direct measurement of the evolutionary timeline of independent star-forming regions. With this timeline in hand, it is possible to derive a broad range of other quantities describing the cloud-scale physics of star formation and feedback.

\subsubsection{Fitting the model to the data} \label{sec:stepfit}
Having obtained the observed gas-to-stellar flux ratio bias and its uncertainties as a function of the aperture size in Section~\ref{sec:steperrors} and having derived a model to describe its behaviour in Section~\ref{sec:stepmodel}, we now describe how the model is fitted to the data. The best fit is obtained by minimising the goodness-of-fit statistic $\chired$ over $\{\tgas,\tover,\lambda\}$ space. We then marginalise the resulting three-dimensional PDF to obtain the one-dimensional PDF of each free parameter (see Section~\ref{sec:steppdfs}). At the end of the fitting process, a figure of the best-fitting tuning fork diagram is written to disk, together with ASCII tables containing the data points, their error bars, and the best-fitting model.

Before performing the fit, we need to define the arrays describing the values of the free parameters that are evaluated. Because the parameter space under consideration is three-dimensional, increasing the number of elements $\ntry$ in these arrays significantly slows down the fitting procedure. For that reason, we attain the desired precision of the fit not by using a high value of $\ntry$, but by iteratively `zooming in' on the best-fitting part of parameter space and refining the free parameter arrays. We define $\ndepth$ as the maximum number of zooms (including the first, unrefined fit), with a default value of $\ndepth=4$ (see \autoref{tab:input}). Note that most fits do not reach this number of iterations, but finish earlier to avoid omitting too much of parameter space (see below). When using the default number of $\ntry=101$ array elements per free parameter (and thus $\ntry^3$ elements in total), this approach allows the fit to reach a precision that falls comfortably within the error bars on each free parameter.

The free parameter arrays are defined as follows. For the duration of the gas phase we use
\be
\label{eq:tgasarr}
\tgas(i)=\tgasmin\left(\frac{\tgasmax}{\tgasmin}\right)^{\frac{i+1/2}{\ntry}} ,
\ee
where $\{i\in\mathbb{N}~|~0\leq i\leq \ntry-1\}$ and the range of $\tgas$ is specified in the input parameter file (see \autoref{tab:input}), with default limits of $\tgasmin=0.1~\myr$ and $\tgasmax=5000~\myr$. Note that we do not allow $\tgasmax$ to exceed the gas depletion time $\tdepl=M_{\rm gas}/{\rm SFR}$, which would be unphysical. If $\tdepl<\tgasmax$, we set $\tgasmax=\tdepl$. Likewise, for the duration of the overlap phase we define
\be
\label{eq:toverarr}
\tover(j)=\tovermin\left(\frac{\tovermax}{\tovermin}\right)^{\frac{j+1/2}{\ntry}} ,
\ee
where $\{j\in\mathbb{N}~|~0\leq j\leq \ntry-1\}$ and the default value of $\tovermin=0.01~\myr$ (see \autoref{tab:input}) effectively corresponds to $\tover=0$, but is chosen to avoid divergence in the model expressions for the gas-to-stellar flux ratio bias (see Section~\ref{sec:stepmodel}). The upper limit of the range of $\tover$ follows from the fact that the duration of the overlap phase cannot exceed the lifetime of the stellar or gas tracer. We therefore set $\tovermax=\tgasmax$ in the default case of {\tt tstar\_incl}~$=0$ (i.e.~$\tstar$ does not include the duration of the overlap phase) and $\tovermax=\min{(\tstar,\tgasmax)}$ if {\tt tstar\_incl}~$=1$. Finally, for the mean separation length we use
\be
\label{eq:lambdaarr}
\lambda(k)=\lambdamin\left(\frac{\lambdamax}{\lambdamin}\right)^{\frac{k+1/2}{\ntry}} ,
\ee
where $\{k\in\mathbb{N}~|~0\leq k\leq \ntry-1\}$ and we define the limits as $\lambdamin=0.3\min{(\lapstar,\lapgas)}$ and $\lambdamax=3\max{(\lapstar,\lapgas)}$. For all three free parameters, the defined limits effectively enclose (even more than) the physically reasonable part of parameter space.

By defining the three arrays in equations~(\ref{eq:tgasarr})--(\ref{eq:lambdaarr}) as we have, the lowest and highest values in the array do not correspond to the specified limits. This is caused by the addition of $1/2$ in each of the exponents. However, this offset of half a logarithmic separation step enables the straightforward integration of the PDF using these same arrays. In logarithmic space, the integration steps are constant and the array elements are situated in the centre of each step. The integration steps are defined as
\be
\label{eq:dtgas}
\frac{{\rm d}\tgas(i)}{\tgas(i)}=\left[\left(\frac{\tgasmax}{\tgasmin}\right)^\frac{1}{2\ntry}-\left(\frac{\tgasmax}{\tgasmin}\right)^{-\frac{1}{2\ntry}}\right] ,
\ee
for the duration of the gas phase,
\be
\label{eq:dtover}
\frac{{\rm d}\tover(j)}{\tover(j)}=\left[\left(\frac{\tovermax}{\tovermin}\right)^\frac{1}{2\ntry}-\left(\frac{\tovermax}{\tovermin}\right)^{-\frac{1}{2\ntry}}\right] ,
\ee
for the duration of the overlap phase, and
\be
\label{eq:dlambda}
\frac{{\rm d}\lambda(k)}{\lambda(k)}=\left[\left(\frac{\lambdamax}{\lambdamin}\right)^\frac{1}{2\ntry}-\left(\frac{\lambdamax}{\lambdamin}\right)^{-\frac{1}{2\ntry}}\right] ,
\ee
for the mean separation length.

Across all $\ntry^3$ elements covered by $\{\tgas(i),\tover(j),\lambda(k)\}$, we use equations~(\ref{eq:biasstarmod}) and~(\ref{eq:biasgasmod}) to calculate the predicted gas-to-stellar flux ratio bias when focusing on stellar peaks ($\bias_{\rm star}^{\rm mod}$) or gas peaks ($\bias_{\rm gas}^{\rm mod}$). If {\tt tstar\_incl}~$=0$, the duration of the stellar phase is defined as $\tstar=\tstarref+\tover(j)$ and we thus update its value for each $j$. Likewise, the values of $\betastar$ and $\betagas$ depend on $f_{\rm star,over}=\tover(j)/\tstar$ and $f_{\rm gas,over}=\tover(j)/\tgas(i)$, respectively. The dependence of both quantities on the time-scales that the fitting process explores is quantified in Section~\ref{sec:stepfluxratios}. For each combination of $\tgas(i)$ and $\tover(j)$, we linearly interpolate the grids of $f_{\rm star,over}$ and $f_{\rm gas,over}$ to obtain the appropriate values of $\betastar$ and $\betagas$. In the same way, the values of $\exc_{\rm star}$ and $\exc_{\rm gas}$ depend on $\lambda(k)$. These dependences are quantified in equations~(\ref{eq:excstar}) and~(\ref{eq:excgas}). We interpolate $\exc(\lambda)$ in double-logarithmic space to obtain the values of $\exc_{\rm star}$ and $\exc_{\rm gas}$ for each explored value of $\lambda(k)$. The above process provides us with the predicted tuning fork diagram for each point in parameter space, i.e.~$\bias_{\rm star}^{\rm mod}(i,j,k,\lap)$ and $\bias_{\rm gas}^{\rm mod}(i,j,k,\lap)$.

For each point in parameter space, the goodness-of-fit statistic $\chired$ is calculated by evaluating the difference between the observations and the model prediction in logarithmic space. We define this difference as
\be
\label{eq:diffstar}
\Delta_{\rm star}[\lapstar(l)]=\left\{\frac{\log_{10}[\bias_{\rm star}(\lapstar)/\bias_{\rm star}^{\rm mod}(\lapstar)]}{\sigma_{\log_{10}\bias,{\rm star}}(\lapstar)}\right\}^2 ,
\ee
for the gas-to-stellar flux ratio bias when focusing on stellar peaks (bottom branch in the tuning fork diagram) and
\be
\label{eq:diffgas}
\Delta_{\rm gas}[\lapgas(l)]=\left\{\frac{\log_{10}[\bias_{\rm gas}(\lapgas)/\bias_{\rm gas}^{\rm mod}(\lapgas)]}{\sigma_{\log_{10}\bias,{\rm gas}}(\lapgas)}\right\}^2 ,
\ee
for the gas-to-stellar flux ratio bias when focusing on gas peaks (top branch in the tuning fork diagram). In both expressions, we have $\{l\in\mathbb{N}~|~0\leq l\leq \nap-1\}$, analogously to equation~(\ref{eq:lap}). The $\chi^2$ statistic of the point in parameter space then follows as
\be
\label{eq:chi}
\chi^2=\sum_{l=0}^{N_{\rm ap}-1}\Delta_{\rm star}[\lapstar(l)]w_{\rm star}(l) + \Delta_{\rm gas}[\lapgas(l)]w_{\rm gas}(l) ,
\ee
where $w_{\rm star}(l)$ and $w_{\rm gas}(l)$ represent the weights of the data points on the bottom and top branches of the tuning fork diagram, respectively. These weights were briefly mentioned in Section~\ref{sec:steperrors} and here we define them formally. As discussed in Section~\ref{sec:steperrors}, none of the data points in the tuning fork diagram is truly independent, because they reflect the flux ratios in apertures of different sizes that are focused on the same emission peaks. We defined the `independence fractions' $f_{\rm data,star}(l)$ and $f_{\rm data,gas}(l)$ of the data points in equations~(\ref{eq:fdatastar}) and~(\ref{eq:fdatagas}). These quantities represent the fraction of each data point that is truly independent from all other data points. When calculating the goodness-of-fit statistic, we should weigh the difference between model and observation for each data point by its independence fraction, so that highly independent data points contribute more strongly than correlated data points. This ensures that each independent measurement has the same total contribution to the goodness-of-fit statistic. The weights are defined by normalising the independence fractions to their mean value, i.e.
\be
\label{eq:wstar}
w_{\rm star}(l)=f_{\rm data,star}\left(\frac{N_{\rm indep}}{2\nap}\right)^{-1} ,
\ee
and
\be
\label{eq:wgas}
w_{\rm gas}(l)=f_{\rm data,gas}\left(\frac{N_{\rm indep}}{2\nap}\right)^{-1} .
\ee
In these weights, $N_{\rm indep}$ represents the total number of independent data points as defined in equation~(\ref{eq:nindep}), which is the sum of the independence fractions across both branches of the tuning fork diagram. The quantity $N_{\rm indep}/2\nap$ thus represents the mean independence fraction across all data points. By normalising the independence fractions to their mean value, we guarantee that the mean weight is unity. That way, any change of $\chi^2$ due to weighing the data points in equation~(\ref{eq:chi}) only reflects a change of their relative contributions, rather than a change of the absolute number of data points. This absolute number is decreased by independence fractions smaller than unity, but that effect is accounted for by decreasing the number of degrees of freedom below, not by modifying the weights of the data points.

The $\chi^2$ statistic is converted to the goodness-of-fit statistic $\chired$ by division by the number of degrees of freedom $N_{\rm deg}$, i.e.
\be
\label{eq:chired}
\chired(i,j,k)=\frac{\chi^2(i,j,k)}{N_{\rm deg}} ,
\ee
where the indices $(i,j,k)$ have been added to emphasize that the described process is repeated for each point in parameter space. The number of degrees of freedom is defined as
\be
\label{eq:ndeg}
N_{\rm deg}=N_{\rm indep}-N_{\rm fit} .
\ee
In this expression, the number of free parameters is $N_{\rm fit}=3$. If none of the data points in the tuning fork diagram would have been correlated, then $N_{\rm indep}=2\nap$ by definition. However, as discussed at length, the fact that the data points are not independent implies that $N_{\rm indep}<2\nap$. This leads to a decrease of the number of degrees of freedom, which in turn causes the goodness-of-fit statistic $\chired$ to increase.

The best-fitting solution across the covered parameter space corresponds to the element where $\chired$ is minimal, i.e.
\be
\label{eq:chimin}
\chimin\equiv\min{[\chired(i,j,k)]} ,
\ee
with best-fitting values of $\{\tgas(i_{\rm best}),\tover(j_{\rm best}),\lambda(k_{\rm best})\}$. However, this best fit may be improved by refining the free parameter arrays in the space around the best fit. A refinement step consists of first updating the minimum and maximum parameter values in equations~(\ref{eq:tgasarr})--(\ref{eq:lambdaarr}), defining the refined free parameter arrays, and then repeating the fitting process. This way, we carry out a maximum of $\ndepth$ fitting loops. The criterion for refinement makes use of the one-dimensional PDFs of the free parameters, which are described in the error analysis of Section~\ref{sec:steppdfs} below. We therefore defer the quantitative description of the refinement process to that discussion.

\begin{figure}
\includegraphics[width=\hsize]{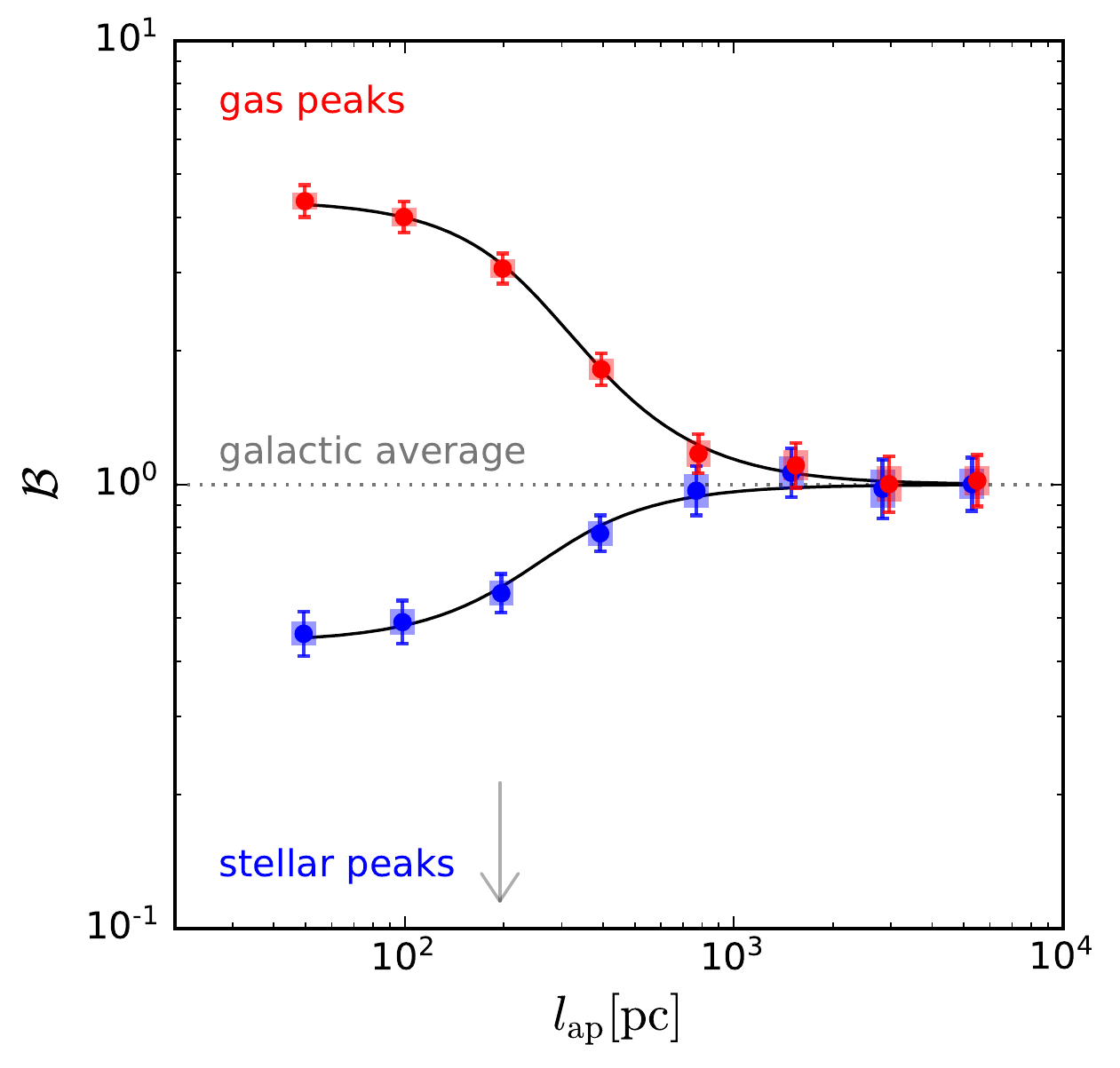}%
\vspace{-1mm}\caption{
\label{fig:stepfit}
Best-fitting model to the example tuning fork diagram of \autoref{fig:steperrors}. When visually assessing the quality of the fit, the reader should use the shaded areas as the effective error bars, because they account for the correlation between the data points and the resulting decrease of the number of degrees of freedom (see the text in Sections~\ref{sec:steperrors} and~\ref{sec:stepfit}). By fitting the model of Section~\ref{sec:stepmodel}, we obtain the best-fitting values and their uncertainties of the lifetimes of gas peaks $\tgas$, the coexistence time-scale of gas and stellar peaks $\tover$, and the characteristic peak separation length $\lambda$. The aperture size where $\lap=\lambda$ is highlighted by the grey arrow.\vspace{-1mm}
}
\end{figure}
After the last refinement step is completed, the final best-fitting solution has been obtained. \code outputs a figure of the tuning fork diagram, figures of $\betastar$ and $\betagas$ as a function of the time fraction spent in the overlap phase (i.e.~\autoref{fig:stepbeta}), as well as ASCII tables that contain the data points, their error bars, and the best-fitting model. \autoref{fig:stepfit} shows the best-fitting solution for the example maps used throughout this section. The figure demonstrates that the method provides an accurate representation of the `observed' gas-to-stellar flux ratio bias, which in this case is reflected in the goodness-of-fit statistic of $\chimin=0.20$. The best-fitting quantities are $\tgas=3.06~\myr$, $\tover=0.90~\myr$, and $\lambda=194~\pc$, whereas the additional quantities that are also constrained during the fitting process are $\tstar=10.90~\myr$, $\betastar=1.45$ (as was already shown in \autoref{fig:stepbeta}), $\betagas=0.44$, $\exc_{\rm star}=1.34$, and $\exc_{\rm gas}=1.49$. We emphasize that these numbers have little-to-no physical meaning, because we applied the method to a simulated data set in this example. In Section~\ref{sec:steppdfs}, we use these numbers to show that the relative uncertainties on the best-fitting quantities are small, implying that the fitting process provides precise measurements of these quantities.

\subsubsection{Error analysis and parameter space refinement} \label{sec:steppdfs}
After having performed a first fit using the unrefined free parameter arrays of equations~(\ref{eq:tgasarr})--(\ref{eq:lambdaarr}), equation~(\ref{eq:chired}) provides the goodness-of-fit statistic for each element in parameter space. It is straightforward to convert this statistic to a three-dimensional PDF of the free parameters by writing
\be
\label{eq:pdf3d}
P_{\rm tot}(\tgas,\tover,\lambda)\equiv\frac{{\rm d}^3p}{{\rm d}\tgas{\rm d}\tover{\rm d}\lambda}=I_\chi^{-1}{\rm e}^{-\chired/2} ,
\ee
with a normalisation given by
\be
\label{eq:pdf3dnorm}
\begin{aligned}
I_\chi=&\iiint {\rm e}^{-\chired/2}{\rm d}\tgas{\rm d}\tover{\rm d}\lambda \\
=&\sum_{i,j,k}{\rm e}^{-\chired(i,j,k)/2}{\rm d}\tgas(i){\rm d}\tover(j){\rm d}\lambda(k) ,
\end{aligned}
\ee
where the second equality accounts for the discretisation of parameter space into the free parameter arrays of equations~(\ref{eq:tgasarr})--(\ref{eq:lambdaarr}). The one-dimensional, differential PDFs of the individual free parameters then follow trivially, for the duration of the gas phase:
\be
\label{eq:pdftgas}
P_{\rm 1D}(\tgas)\equiv\frac{{\rm d}p}{{\rm d}\tgas}=\iint P_{\rm tot}(\tgas,\tover,\lambda){\rm d}\tover{\rm d}\lambda ,
\ee
for the duration of the overlap phase:
\be
\label{eq:pdftover}
P_{\rm 1D}(\tover)\equiv\frac{{\rm d}p}{{\rm d}\tover}=\iint P_{\rm tot}(\tgas,\tover,\lambda){\rm d}\tgas{\rm d}\lambda ,
\ee
and for the region separation length:
\be
\label{eq:pdflambda}
P_{\rm 1D}(\lambda)\equiv\frac{{\rm d}p}{{\rm d}\lambda}=\iint P_{\rm tot}(\tgas,\tover,\lambda){\rm d}\tgas{\rm d}\tover .
\ee
These are converted to their cumulative forms by defining
\be
\label{eq:pdfcum}
P_{\rm 1D}(X\leq x)\equiv\int_{x_{\rm min}}^{x}P_{\rm 1D}(X){\rm d}X ,
\ee
for each of the PDFs in equations~(\ref{eq:pdftgas})--(\ref{eq:pdflambda}). The numerical implementations in \code of the six (differential and cumulative) PDFs described by equations~(\ref{eq:pdftgas})--(\ref{eq:pdfcum}) are discretised analogously to the second equality of equation~(\ref{eq:pdf3dnorm}). Finally, we also calculate the three different two-dimensional PDFs by separately integrating out each single variable in equation~(\ref{eq:pdf3d}). These two-dimensional distributions can be used to visualise the degree of correlation and thus degeneracy between the three free parameters.

To compute the uncertainties on the best-fitting values of the free parameters, we use the one-dimensional cumulative PDFs. For most of the applications discussed in Section~\ref{sec:valid} and the early observational applications of the presented method, the differential PDFs are asymmetric and non-Gaussian. One way of appropriately representing this asymmetry would be to define the uncertainties by using the 1$\sigma$ percentiles in a Gaussian, i.e.~16 and 84 per cent. However, this can result in situations where the best-fitting value is found outside of the 16--84 percentile range, leading to negative uncertainties. To avoid such a (rare) situation, we choose to define the uncertainties used in this work differently. Given a best-fitting value $x_{\rm best}$, we divide the variable range in two around $x_{\rm best}$ and define the uncertainties analogously to the Gaussian case as the 32$^{\rm nd}$ percentile of the part of the PDF below $x_{\rm best}$ and the 68$^{\rm th}$ percentile of the part of the PDF above $x_{\rm best}$. If the PDF is Gaussian, this corresponds to the formal definition of the 1$\sigma$ uncertainties, but for asymmetric PDFs it deviates.

Mathematically, the above procedure results in the following steps. First, we define the percentiles of the 1$\sigma$ uncertainties in a Gaussian PDF as usual, i.e.
\be
\label{eq:percgauss1}
p_{\rm min,Gauss}=1-\frac{1}{2}\left[1+{\rm erf}\left(\frac{1}{\sqrt{2}}\right)\right]\approx 0.16 ,
\ee
for the downward uncertainty, and
\be
\label{eq:percgauss2}
p_{\rm max,Gauss}=\frac{1}{2}\left[1+{\rm erf}\left(\frac{1}{\sqrt{2}}\right)\right]\approx 0.84 ,
\ee
for the upward uncertainty. We can normalise these to the range on either side of the best-fitting value in a Gaussian PDF (i.e.~$p_{\rm best,Gauss}=0.5$) and thus determine an equivalent fraction of that range corresponding to the above percentiles by writing
\be
\label{eq:fracgauss1}
f_{\rm min,Gauss}=\frac{p_{\rm min,Gauss}}{p_{\rm best,Gauss}}\approx 0.32 ,
\ee
for the downward uncertainty, and
\be
\label{eq:fracgauss2}
f_{\rm max,Gauss}=\frac{p_{\rm max,Gauss}-p_{\rm best,Gauss}}{1-p_{\rm best,Gauss}}\approx 0.68 ,
\ee
for the upward uncertainty. The corresponding percentiles around the best-fitting value $x_{\rm best}$ then follow by multiplication with the PDF ranges on either side of $x_{\rm best}$, i.e.
\be
\label{eq:perc1}
\begin{aligned}
p_{\rm min}&=f_{\rm min,Gauss}P_{\rm 1D}(x\leq x_{\rm best}) \\
&\approx 0.32P_{\rm 1D}(x\leq x_{\rm best}) ,
\end{aligned}
\ee
for the downward uncertainty, and
\be
\label{eq:perc2}
\begin{aligned}
p_{\rm max}&=P_{\rm 1D}(x\leq x_{\rm best})+f_{\rm max,Gauss}[1-P_{\rm 1D}(x\leq x_{\rm best})] \\
&\approx 0.32P_{\rm 1D}(x\leq x_{\rm best})+0.68 ,
\end{aligned}
\ee
for the upward uncertainty. For the Gaussian value of $P_{\rm 1D}(x\leq x_{\rm best})=0.5$, these reduce to equations~(\ref{eq:percgauss1}) and~(\ref{eq:percgauss2}). The parameter values $x_-$ and $x_+$ at the percentiles of equations~(\ref{eq:perc1}) and~(\ref{eq:perc2}) are obtained by numerically solving
\be
\label{eq:xmin}
\int_{x_{\rm min}}^{x_-}P_{\rm 1D}(x){\rm d}x=p_{\rm min} ,
\ee
for $x_-$ and
\be
\label{eq:xmax}
\int_{x_{\rm min}}^{x_+}P_{\rm 1D}(x){\rm d}x=p_{\rm max} ,
\ee
for $x_+$. This results in downward and upward error bars on the best-fitting values of
\be
\label{eq:sigmin}
\sigma_-(x)=x_{\rm best}-x_- ,
\ee
and
\be
\label{eq:sigmax}
\sigma_+(x)=x_+-x_{\rm best} ,
\ee
respectively.

The uncertainties obtained through the above procedure are used after each iteration of the fitting process to determine whether it requires another iteration and, if so, within which value range the free parameter arrays of equations~(\ref{eq:tgasarr})--(\ref{eq:lambdaarr}) should be re(de)fined. At the end of a complete fitting step with these arrays, we determine the part of the three-dimensional parameter space that corresponds to the 3$\sigma$ confidence ellipsoid and encloses 99.73~per~cent of the total probability, with the intention of restricting the fitting range to this volume during the next fitting step. The surface corresponding to this ellipsoid has a goodness-of-fit parameter value of
\be
\label{eq:chisurf}
\chired=\chimin+\frac{\Delta\chi^2}{N_{\rm deg}} ,
\ee
where $\Delta\chi^2$ depends on the number of free parameters. The problem under consideration has $N_{\rm fit}=3$, which for the 3$\sigma$ ellipsoid results in $\Delta\chi^2=14.2$ \citep[chapter~15.6]{press92}. For each of the free parameter arrays of equations~(\ref{eq:tgasarr})--(\ref{eq:lambdaarr}), we then determine at each element what the minimum value of $\chired$ is across the full range of both other free parameters. The refined array for the next fitting step is restricted to the range of elements where $\chired$ is smaller than the reference value of equation~(\ref{eq:chisurf}), plus one element on either side of this range. This removes any free parameter values for which no elements in the complete, three-dimensional $\chired$ array fall within the 3$\sigma$ ellipsoid.

\begin{figure*}
\includegraphics[width=\hsize]{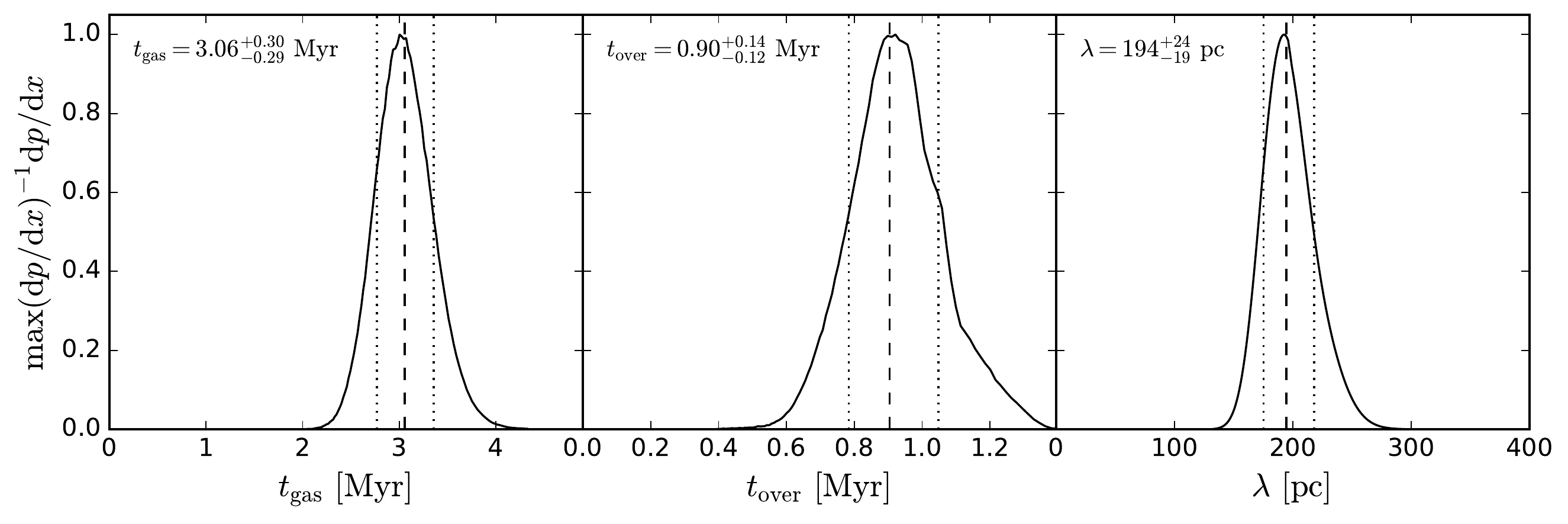}%
\vspace{-1mm}\caption{
\label{fig:steppdfs}
One-dimensional PDFs of $\tgas$ (left), $\tover$ (middle), and $\lambda$ (right), which are the free parameters in the fit of the tuning fork diagram in \autoref{fig:stepfit}. The PDFs are normalised to their maximum probability density. Vertical dashed lines indicate the best-fitting values (also indicated in the top left of each panel), whereas the vertical dotted lines represent the 1$\sigma$ uncertainty range as defined in the text. These three PDFs are representative of the PDFs that are typically obtained when applying the method and illustrate that they are close to symmetric and Gaussian.\vspace{-1mm}
}
\end{figure*}
The above steps update the values of $\tgasmin$, $\tgasmax$, $\tovermin$, $\tovermax$, $\lambdamin$, and $\lambdamax$ that are used in equations~(\ref{eq:tgasarr})--(\ref{eq:lambdaarr}) for the next fitting step, with the added condition that if $\tovermin>\tgasmin$, we set $\tovermin=\tgasmin$ to avoid the unphysical situation that the minimum duration of the overlap phase exceeds the duration of the gas phase for some part of the range of $\tgas$. Even if this enables a next fitting step in principle, we carry out four additional checks of the new free parameter limits, three of which can prevent entering the next fitting step. These checks are as follows.
\begin{enumerate}
\item
If the best-fitting value resides within four array elements of the edge of a new parameter space range, which can happen in the rare situation that $\chired$ increases very steeply from $\chimin$ to $\chimin+\Delta\chi^2/N_{\rm deg}$, the free parameter range at that edge is expanded. For lower limits, we expand the updated minimum through multiplication by $x_{\rm min}/x_{\rm max}$, whereas for upper limits, the updated maximum is expanded through multiplication by $x_{\rm max}/x_{\rm min}$. In either case, the updated minimum (maximum) is not allowed to fall below (exceed) the value set at the beginning of the fitting process, as defined in Section~\ref{sec:stepfit} for equations~(\ref{eq:tgasarr})--(\ref{eq:lambdaarr}). This check does not prevent the next fitting step, but it ensures that the best-fitting solution does not end up at the edge of parameter space.
\item
If both the downward and upward 1$\sigma$ ranges of all free parameters reach within four array elements of the edge of the current, unrefined parameter space, the refinement is aborted and the fitting process is concluded. This check is added as a safety feature against removing relevant parts of parameter space.
\item
If the absolute logarithmic difference between all six of the old and new parameter space limits is smaller than a certain tolerance value $\delta_{\rm lim}$, another refinement step is deemed unnecessary, because the added precision is negligible. We use a value of $\delta_{\rm lim}=0.01$, which implies that if the change of all parameter space limits is smaller than 2.3~per~cent, the refinement is aborted and the fitting process is concluded.
\item
Finally, we cap the number of fitting steps at $\ndepth$, corresponding to a maximum of $\ndepth-1$ refinements. For the default value of $\ndepth=4$ (see \autoref{tab:input}, this cap is often not reached. Instead, the refinement is aborted because the absolute logarithmic difference between the old and new limits falls below the specified tolerance (as discussed in the previous check).
\end{enumerate}

The fitting process continues through the cycle of obtaining a fit and refining the free parameter arrays until any of the three exit conditions specified above are met. Once it is completed, we obtain the final best-fitting values and their uncertainties. For the example galaxy maps used throughout this section, these are $\tgas=3.06_{-0.29}^{+0.30}~\myr$, $\tover=0.90_{-0.12}^{+0.14}~\myr$, and $\lambda=194_{-19}^{+23}~\pc$. As stated before in Section~\ref{sec:stepfit}, due to the nature of the experiment these values have no particular physical meaning, but they highlight that the quantities are very well-constrained, with relative errors of 16~per~cent or smaller in all three cases. Finally, \code writes the differential one-dimensional PDFs to disk in ASCII format and also outputs figures of all nine free parameter PDFs calculated here, i.e.~the three differential one-dimensional PDFs, the three cumulative one-dimensional PDFs, and the three two-dimensional PDFs of each free parameter pair. The best-fitting values and their uncertainties are also indicated in these plots.

For the familiar example maps, the one-dimensional differential PDFs of $\tgas$, $\tover$, and $\lambda$ are shown in \autoref{fig:steppdfs}. These PDFs are well-behaved in that they are quite symmetric and even fairly close to Gaussian, as evidenced by the fact that the 1$\sigma$ lines intersect with the PDFs near the expected $\exp{(-1/2)}\approx0.61$ times the peak value. The kinks in the $\tover$ PDF result from the discrete nature of the relation between $\betastar$, $\betagas$, and $\tover$ (see \autoref{fig:stepbeta}), which is unavoidable given that any galaxy map hosts a finite number of emission peaks. We note that these PDFs are representative for the applications of our method in Section~\ref{sec:valid}, as well as the ongoing observational applications. Having said that, it is possible to retrieve asymmetric PDFs if the tuning fork diagram provides less strong constraints than in this example (e.g.~when one of the branches remains nearly flat over the full range of aperture sizes, which is indicative of a long duration of the overlap phase). For this reason, we emphasize that future applications of the method should generally present the PDFs rather than the best-fitting values and their uncertainties, especially when the downward and upward uncertainties on one of the best-fitting values are dissimilar.

\subsubsection{Calculation of derived quantities and error propagation} \label{sec:stepderived}
As the final quantitative step of the method, we use the constraints on the evolutionary timeline of independent star-forming regions and their characteristic separation lengths to derive additional physical quantities describing the star formation and feedback process. Unsurprisingly, having access to the evolutionary time-scales provides a powerful avenue for constraining these physics. By carrying out Monte-Carlo sampling from the complete, three-dimensional PDF of $\tgas$, $\tover$, and $\lambda$ from Section~\ref{sec:steppdfs}, we self-consistently obtain the one-dimensional PDFs of all derived quantities and fully account for the covariance between the three free parameters. This error propagation process also provides us with the complete covariance matrix between all constrained quantities. In what follows, we first discuss the expressions used to obtain each of the derived quantities, before treating the Monte-Carlo error propagation.

We first describe a set of quantities that represent {\it byproducts} of the fitting process, in that they play a role in the equations describing the model in Section~\ref{sec:stepmodel}, but do not act as free parameters in the fitting process, because they are functions of $\tgas$ and $\tover$. In Section~\ref{sec:stepreadpar}, we described that the input file contains the stellar `reference time-scale' used to convert the relative timeline constrained by the method to an absolute timeline. However, we also introduced the flag {\tt tstar\_incl}~$=\{0,1\}$, which encodes whether the reference time-scale excludes or includes the overlap time-scale, respectively. A physical example of the first case would be \halpha emission, because star-forming regions with a non-zero age spread may continue forming stars while they are already visible in \halpha, implying that only when star formation has ceased and any residual star-forming gas is expelled, the final massive star has formed and the stellar evolutionary `clock' of the reference time-scale starts ticking. By contrast, in the numerical experiments carried out in Section~\ref{sec:starstar}, we use star particles within a certain age range to provide the stellar map and reference time-scale, in which case the reference time-scale is equivalent to this age range and includes the duration of the overlap phase. Mathematically, these cases correspond to
\be
\label{eq:tstar1}
\tstar=\tstarref+\tover ,
\ee
if {\tt tstar\_incl}~$=0$ and $\tstarref$ does not include $\tover$, and
\be
\label{eq:tstar2}
\tstar=\tstarref ,
\ee
if {\tt tstar\_incl}~$=1$ and $\tstarref$ does include $\tover$. Throughout this paper, we will indicate which definition of $\tstarref$ we are using for each experiment. Similarly to $\tstar$, the total duration of the evolutionary timeline shown in \autoref{fig:tschem} follows by combining the time-scales
\be
\label{eq:tau}
\tau=\tstar+\tgas-\tover ,
\ee
where the subtraction of $\tover$ is needed to avoid counting it twice, because the duration of the overlap phase is included both in $\tstar$ and $\tgas$.

The next quantities that are obtained as byproducts of the fitting process are $\betastar$ and $\betagas$. They are defined by equations~(\ref{eq:betastar}) and~(\ref{eq:betagas}), where they are given as a function of $f_{\rm star,over}=\tover/\tstar$ and $f_{\rm gas,over}=\tover/\tgas$, respectively (also see \autoref{fig:stepbeta}). \code stores the numerical relations $\betastar(f_{\rm star,over})$ and $\betagas(f_{\rm gas,over})$ from Section~\ref{sec:stepfluxratios} and uses these after constraining the best fit to the tuning fork diagram to determine the values of $\betastar$ and $\betagas$. The PDFs of these quantities are determined below using the same relations in the Monte-Carlo error propagation process.

The third set of quantities that are constrained as byproducts of the fitting process are $\exc_{\rm star}$, $\exc_{\rm gas}$, $\exc_{\rm star,glob}$, and $\exc_{\rm gas,glob}$, which are defined in equations~(\ref{eq:excstar})--(\ref{eq:excgasglob}) as a function of a size scale $\lap$ that is set to $\lap=\lambda$ in Section~\ref{sec:stepmodel}. As with $\betastar$ and $\betagas$, \code stores the relations between these quantities and $\lambda$ and uses these after obtaining the best-fitting value of $\lambda$ to determine the corresponding values of $\exc_{\rm star}$, $\exc_{\rm gas}$, $\exc_{\rm star,glob}$, and $\exc_{\rm gas,glob}$. These relations are also used for determining the PDFs of these quantities by Monte-Carlo error propagation.

We now turn to a number of {\it derived} quantities that are themselves not used directly in the fitting process, but follow immediately from the quantities determined so far. Firstly, the radii of the stellar and gas emission peaks, $\rstar$ and $\rgas$, are defined by equations~(\ref{eq:rstardisc})--(\ref{eq:rgasgauss}), depending on whether it is assumed that the peak profiles follow constant surface density discs or two-dimensional Gaussians. Throughout this paper, we adopt the latter case, as this is more representative of typical emission peak profiles, both in observations and in our simulated maps (see Sections~\ref{sec:stepmodel} and~\ref{sec:valid}). From these radii, we immediately derive the relative filling factors of the peaks, i.e.~their diameters in units of the mean separation length of independent regions. For the stellar peaks, we write this quantity as
\be
\label{eq:zetastar}
\zeta_{\rm star}=\frac{2\rstar}{\lambda} ,
\ee
while for the gas peaks this becomes
\be
\label{eq:zetagas}
\zeta_{\rm gas}=\frac{2\rgas}{\lambda} .
\ee
In Section~\ref{sec:valid}, we find that a reliable measurement of the duration of the overlap phase $\tover$ and, hence, the characterisation of feedback-related quantities requires $\zetastar$ and $\zetagas$ to be lower than a certain maximum value (also see Appendix~\ref{sec:appblending}). These quantities therefore represent an important benchmark in assessing the accuracy of the method's results.

Finally, by combining the mean separation length and the duration of the overlap phase, we obtain the characteristic velocity $\vfb$ by which the residual gas is removed from independent star-forming regions by feedback:
\be
\label{eq:vfb}
\vfb=\frac{\lambda}{2\tover} .
\ee
The physical meaning of this velocity depends on the feedback mechanism responsible for removing the gas. If the gas is kinetically removed, the feedback velocity simply represents the outflow velocity averaged over the distance from the feedback source to half the region separation length $\lambda/2$. However, if the gas tracer is rendered invisible by a change in phase or emissivity (e.g.~through photoevaporation, ionization, excitation conditions, or a chemical reaction), then the feedback velocity corresponds to the velocity of the phase transition or emissivity front. In that case, the relevant length scale is not the region separation length, but the gas region radius $\rgas$. The same applies if the gas becomes undetectable by dilution once the shell escapes the cloud boundary. These situations may arise in regions of low gas surface density, where molecular clouds are embedded in an atomic-dominated medium \citep[e.g.][]{blitz06,krumholz09b}. The feedback velocity is then defined as
\be
\label{eq:vfbr}
\vfbr=\frac{\rgas}{\tover} .
\ee
Both feedback velocities are calculated by \code. Most generally, $\vfb$ and $\vfbr$ bracket the range of possible feedback velocities, but distinguishing between these is possible through a variety of metrics. In the context of the presented method, the duration of the overlap phase $\tover$ may rule out the importance of certain feedback mechanisms, e.g.~when the gas is removed on a time-scale shorter than $3~\myr$, it is certain that core-collapse SNe are not responsible \citep[e.g.][]{ekstrom12}. In addition, we specify the feedback energy and momentum coupling efficiencies in equations~(\ref{eq:chie}), (\ref{eq:chier}), (\ref{eq:chip}) and~(\ref{eq:chipr}) below, which indicate whether a given feedback mechanism provides the energy or momentum required for the removal of the gas over length scales $\lambda/2$ or $\rgas$. Independent constraints from ancillary observations can also help. For instance, it is possible to directly measure the expansion velocities of shell-like features around individual star-forming regions. Comparing these to $\vfb$ as defined above provides a conclusive way of identifying the relevant length scale -- if the typical expansion velocity is similar to $\vfbr$ rather than $\vfb$, this means that the gas tracer vanishes at a radius $\rgas$ rather than $\lambda/2$, due to dilution or a change in phase or emissivity.

The final set of derived physical quantities is referred to as {\it composite} quantities, because they depend on additional quantities that are not directly constrained by the method itself. In particular, by combining the obtained time-scales and size scales with the absolute SFR and gas mass (or their surface densities), we obtain several of the key quantities describing cloud-scale star formation and feedback. This is an important implication of the presented method, albeit not necessarily surprising. Recall the example given in equation~(\ref{eq:sfrelation}), which showed that the degeneracy between the star formation time-scale and the star formation efficiency can be lifted by directly measuring one of these. This is exactly what we have done in the method presented above.

Firstly, we define the SFR surface density as
\be
\label{eq:sigmasfr}
\Sigma_{\rm SFR}=\frac{{\rm SFR}}{A_{\rm tot}} ,
\ee
which uses the total area and the SFR defined in equations~(\ref{eq:totalarea}) and~(\ref{eq:sfrtot}). The uncertainty on $\Sigma_{\rm SFR}$ is assumed to be dominated by the uncertainty of the conversion factor $X_{\rm star}$ and we therefore set the relative uncertainty $\sigma_{\rm rel}(\Sigma_{\rm SFR})=\sigma_{\rm rel}(X_{\rm star})$. Likewise, the gas surface density is defined as
\be
\label{eq:sigmagas}
\Sigma_{\rm gas}=\frac{M_{\rm gas}}{A_{\rm tot}} ,
\ee
which uses the total area and the gas mass defined in equations~(\ref{eq:totalarea}) and~(\ref{eq:mgastot}). Again, we assume that the uncertainty on the conversion factor $X_{\rm gas}$ dominates and therefore set the relative uncertainty $\sigma_{\rm rel}(M_{\rm gas})=\sigma_{\rm rel}(X_{\rm gas})$. The ratio of the gas mass and the SFR (surface densities) corresponds to the gas depletion time
\be
\label{eq:tdepl}
\tdepl=\frac{\Sigma_{\rm gas}}{\Sigma_{\rm SFR}} ,
\ee
which represents the time it takes to exhaust the presently available gas reservoir at the present SFR.

As shown by equation~(\ref{eq:sfrelation}), the ratio between the gas mass and the SFR (surface densities) provides the ratio between the star formation efficiency and the star formation time-scale. Given that our method provides the duration of the gas phase (over which star formation can proceed), we obtain the mean star formation efficiency per unit cloud lifetime by dividing the surface density of the formed stars $\tgas\Sigma_{\rm SFR}$ by the gas surface density:
\be
\label{eq:sfe}
\sfe=\frac{\tgas\Sigma_{\rm SFR}}{\Sigma_{\rm gas}}=\frac{\tgas}{\tdepl} ,
\ee
where the second equality shows that this is equivalent to the ratio between the duration of the gas phase (or cloud lifetime) and the gas depletion time. By constraining the evolutionary timeline of cloud-scale star formation, we can now determine whether star formation is rapid and inefficient or slow and efficient.

The expressions derived so far allow us to formulate a set of important dimensionless quantities describing cloud-scale feedback. These quantities are similar to those used in galaxy formation simulations, but here they describe feedback on a size scale $\lambda$, whereas in simulations they are often considered on a larger spatial scale. Firstly, a comparison of the instantaneous SFR and outflow rates per region of equations~(\ref{eq:mdotsf}) and~(\ref{eq:mdotfb}) provide the instantaneous mass loading factor, i.e.~the outflow rate during the overlap phase in units of the SFR during the gas phase:
\be
\label{eq:etainst}
\etafb=\frac{\mdotfb}{\mdotsf}=\frac{(1-\sfe)\tgas}{\sfe\tover}=\frac{\tdepl-\tgas}{\tover} .
\ee
This expression does not reflect the total mass budget, because it considers the instantaneous {\it rates} of star formation and mass outflow, both of which take place on different time-scales ($\tgas$ and $\tover$). By contrast, the time-averaged mass loading factor does reflect the total mass budget, i.e.~it balances the total mass ejected by feedback $(1-\sfe)\Sigma_{\rm gas}$ by the total stellar mass formed $\tgas\Sigma_{\rm SFR}$ and thus becomes a simple ratio of either star formation efficiencies or time-scales:
\be
\label{eq:etaavg}
\etaavgfb=\frac{(1-\sfe)\Sigma_{\rm gas}}{\tgas\Sigma_{\rm SFR}}=\frac{1-\sfe}{\sfe}=\frac{\tdepl}{\tgas}-1 .
\ee
While the mass loading factor of equation~(\ref{eq:etainst}) may be appropriate for comparison to observational estimates of the mass loading factor based on observed mass outflow rates and SFRs, the definition in equation~(\ref{eq:etaavg}) is appropriate for comparison to theoretical models, which often consider the total ejected mass in terms of the total stellar mass \citep[e.g.][]{krumholz17}.

Likewise, it is possible to compare the total energy and momentum imparted by feedback on the ISM to the total injected energy and momentum, resulting in the feedback energy and momentum coupling efficiencies $\chifbe$ and $\chifbp$. The total imparted kinetic energy per unit area is given by
\be
\label{eq:efb}
E_{\rm fb}=\frac{1}{2}(1-\sfe)\Sigma_{\rm gas}v^2 ,
\ee
where $v$ is $\vfb$ or $\vfbr$. The total injected energy per unit area is
\be
\label{eq:etot}
E_{\rm tot}=\sfe\Sigma_{\rm gas}\Psi_E\tover ,
\ee
with $\Psi_E$ the energy output rate per unit mass (i.e.~the light-to-mass ratio). The first half of this expression ($\sfe\Sigma_{\rm gas}$) represents the total mass surface density of new-born stars, whereas the second half ($\Psi_E\tover$) represents the total energy output per unit mass. With these two energy budgets in hand, it is straightforward to define the feedback energy efficiency as the ratio between $E_{\rm fb}$ and $E_{\rm tot}$:
\be
\label{eq:chie}
\chifbe=\frac{E_{\rm fb}}{E_{\rm tot}}=\frac{(1-\sfe)\vfb^2}{2\sfe\Psi_E\tover} ,
\ee
for expansion out to $\lambda/2$ and
\be
\label{eq:chier}
\chifber=\frac{E_{\rm fb}}{E_{\rm tot}}=\frac{(1-\sfe)\vfbr^2}{2\sfe\Psi_E\tover} ,
\ee
if the gas vanishes already at $\rgas$ by dilution or a change in phase or emissivity. Clearly, the feedback energy efficiency depends on the light-to-mass ratio $\Psi_E$, which should be specified. However, the appropriate value of $\Psi_E$ depends on the problem under consideration. With the output from stellar population synthesis models, it is possible to define this parameter for individual feedback mechanisms or any combination thereof. Typical values range from $\Psi_E=10^{-4}$--$10^{0}~{\rm m}^2~{\rm s}^{-3}$, which covers core-collapse SNe, stellar winds, and radiative feedback \citep[e.g.][]{agertz13}. A feedback energy efficiency in excess of unity ($\chifbe>1$) indicates that the feedback mechanisms accounted for by the adopted value of $\Psi_E$ are insufficient to drive the observed feedback energy imparted on the ISM, whereas $\chifbe<1$ provides constraints on the inefficiency of feedback energy coupling to the ISM due to e.g.~the porosity of the ISM or radiative losses.

Analogously to the total imparted kinetic energy per unit area, we define the total imparted momentum per unit area as
\be
\label{eq:pfb}
p_{\rm fb}=(1-\sfe)\Sigma_{\rm gas}v ,
\ee
where $v$ again refers to $\vfb$ or $\vfbr$. The total injected momentum per unit area is
\be
\label{eq:ptot}
p_{\rm tot}=\sfe\Sigma_{\rm gas}\psi_p\tover ,
\ee
with $\psi_p$ the momentum output rate per unit mass, which has units of acceleration. As in equation~(\ref{eq:etot}), the first half of this expression ($\sfe\Sigma_{\rm gas}$) represents the total mass surface density of new-born stars, whereas the second half ($\psi_p\tover$) represents the total momentum output per unit mass. With these two momentum budgets in hand, it is straightforward to define the feedback momentum efficiency as the ratio between $p_{\rm fb}$ and $p_{\rm tot}$:
\be
\label{eq:chip}
\chifbp=\frac{p_{\rm fb}}{p_{\rm tot}}=\frac{(1-\sfe)\vfb}{\sfe\psi_p\tover} .
\ee
for expansion out to $\lambda/2$ and
\be
\label{eq:chipr}
\chifbpr=\frac{p_{\rm fb}}{p_{\rm tot}}=\frac{(1-\sfe)\vfbr}{\sfe\psi_p\tover} .
\ee
if the gas vanishes already at $\rgas$ by dilution or a change in phase or emissivity. Evaluating these expressions requires the momentum output rate per unit mass $\psi_p$ to be specified. As before, the appropriate value can be chosen to represent individual feedback mechanisms or to reflect their combined effect. For stellar winds, typical values are $\psi_p=10^{-11}$--$10^{-10}~{\rm m}~{\rm s}^{-2}$ \citep[e.g.][]{agertz13}. Analogously to the feedback energy efficiency, a feedback momentum efficiency in excess of unity ($\chifbp>1$) indicates that the feedback mechanisms accounted for by the adopted value of $\psi_p$ are insufficient to drive the observed feedback momentum imparted on the ISM, whereas $\chifbp<1$ provides constraints on the inefficiency of the feedback momentum coupling to the ISM due to e.g.~the porosity of the ISM or opposing forces (such as the ambient gas pressure).

In general, a comparison of the measured $\chifbe$ and $\chifbp$ helps determine whether outflow is energy-driven or momentum-driven. If $\Psi_E$ and $\psi_p$ are chosen such that they contain all possible contributions to the injected energy and momentum and either $\chifbe$ or $\chifbp$ exceeds unity, this indicates that the {\it other} driving process dominates, because the observed outflow cannot be generated when $\chi_{\rm fb}>1$. When both $\chifbe\ga1$ and $\chifbp\ga1$, this suggests that the gas is rendered invisible at a radius $\rgas$ rather than $\lambda/2$, necessitating the use of $\chifber$ and $\chifbpr$ instead. If all four coupling efficiencies exceed unity, this either points towards a problem in the input parameters chosen for the analysis, leading to inaccurate measurements of $\tover$, $\vfb$, $\vfbr$, or $\sfe$, or indicates that the (models motivating the) choice of $\Psi_E$ and $\psi_p$ are incorrect or incomplete. The former situation would require the user to revisit the input and correct any mistakes, but the latter case would count as a falsification of the tested feedback description and would hint at the importance of additional feedback mechanisms.

Finally, the obtained characteristic time-scales and size scales governing cloud-scale star formation and feedback can be used to derive absolute properties of the average cloud or star-forming region in the considered maps. This can be done by multiplying the mean surface densities across the maps with the region area $\pi(\lambda/2)^2$. However, this requires the strong assumption that the number densities of regions on a size scale $\lambda$ are similar to that across the entire map, i.e.~that the regions are randomly distributed in space. While a similar assumption is made in deriving the model of Section~\ref{sec:stepmodel}, it is much stronger here. In the general context of the model, we assume a random distribution of independent regions (see Section~\ref{sec:morphology}) on size scales up to a few times the mean separation length $\lambda$. On larger size scales, the flux ratio biases of the tuning fork diagram converge to the galactic average, irrespectively of (possibly strong) deviations from a random spatial distribution. However, it is a much stronger assumption to obtain a region mass by multiplying the map-averaged surface density with the region separation length, because this conversion will be affected by large-scale morphological features such as spiral arms, rings, and bars. Due to the associated uncertainty of calculating the absolute properties of independent regions, we intend to consider these quantities in more detail in future work and only provide two examples here.

We determine the typical absolute SFRs and feedback outflow rates of the independent star-forming regions described by our model. The mean separation length $\lambda$ provides the characteristic size scale within which star formation proceeds, implying a region-averaged SFR during the gas phase of
\be
\label{eq:mdotsf}
\mdotsf=\pi\left(\frac{\lambda}{2}\right)^2\Sigma_{\rm SFR}=\pi\left(\frac{\lambda}{2}\right)^2\frac{\sfe\Sigma_{\rm gas}}{\tgas} .
\ee
The first expression given here is trivial -- it simply multiplies the area-averaged SFR surface density by the area per region to get the SFR per region. The second expression obtains the SFR surface density in terms of gas tracer properties to enable a straightforward comparison to equation~(\ref{eq:mdotfb}) below. Analogously to equation~(\ref{eq:mdotsf}), we can derive the region-averaged mass outflow (or removal) rate due to feedback during the overlap phase:
\be
\label{eq:mdotfb}
\mdotfb=\pi\left(\frac{\lambda}{2}\right)^2\frac{(1-\sfe)\Sigma_{\rm gas}}{\tover} ,
\ee
where the numerator denotes the gas surface density of the residual gas and the denominator represents the feedback time-scale (i.e.~the duration of the overlap phase). We reiterate that both equations~(\ref{eq:mdotsf}) and~(\ref{eq:mdotfb}) assume a homogeneous distribution of independent regions, without large-scale morphological features or a strong spatial clustering of the regions, such that the average surface densities $\Sigma_{\rm SFR}$ and $\Sigma_{\rm gas}$ yield meaningful SFRs or masses when multiplying by $\pi(\lambda/2)^2$. When applying the method to galaxies with prominent morphological features, the risk exists that the global surface density underestimates the surface density in the direct vicinity of star-forming regions or gas clouds. In such a situation, the above expressions for the absolute SFR and mass outflow rate are accurate when a part of the galaxy with relatively homogeneous structure is selected by masking. Alternatively, it is possible to correct for morphology using the flux density contrasts $\exc_{\rm star,glob}$ and $\exc_{\rm gas,glob}$ from equations~(\ref{eq:excstarglob}) and~(\ref{eq:excgasglob}). We do not provide a detailed derivation here, but defer it to a future application of the method for which a morphological correction is needed.

After having determined all of the quantities described in this section for the best-fitting model of the tuning fork diagram, we obtain their PDFs through Monte-Carlo error propagation. To fully account for any covariance between the three free parameters $\tgas$, $\tover$, and $\lambda$, it is necessary to draw from their complete three-dimensional PDF of equation~(\ref{eq:pdf3d}). Drawing numbers from a discretised PDF is trivial in one dimension, because one simply interpolates the cumulative one-dimensional PDF, but this is not possible when dealing with a three-dimensional PDF. We therefore draw each Monte-Carlo realisation of the three free parameters through three correlated, one-dimensional representations of the relevant parts of the three-dimensional PDF.

First, we draw a value $\tgasmc$ from its marginalised, one-dimensional cumulative PDF by numerically solving
\be
\label{eq:drawtgas}
q_l = P_{\rm 1D}(\tgas\leq\tgasmc) ,
\ee
for $\tgasmc$, where $q_l$ represents a random number $\{q_l\in\mathbb{R}|0\leq q_l\leq1\}$. In practice, this requires interpolating the discretised form of $P_{\rm 1D}(\tgas<\tgasmc)$.\footnote{When drawing the Monte-Carlo realisations from the three-dimensional PDF, the interpolations are carried out on the logarithmic grids of equations~(\ref{eq:tgasarr})--(\ref{eq:lambdaarr}), but they are linear in the random number and the cumulative probability.} To obtain $\tovermc$, we evaluate the marginalised, two-dimensional PDF of $\tgas$ and $\tover$ at $\tgasmc$, resulting in the one-dimensional PDF of $\tover$ given the value of $\tgasmc$. This one-dimensional PDF is obtained by interpolation between the two discrete one-dimensional PDFs:
\be
\label{eq:drawpdf2}
\begin{aligned}
&P_{\rm 2D}[\tover\leq\tovermc|\tgas(i)], \\
&P_{\rm 2D}[\tover\leq\tovermc|\tgas(i+1)] ,
\end{aligned}
\ee
where $i$ denotes the index such that $\tgas(i)\leq\tgasmc\leq\tgas(i+1)$. We then draw $\tovermc$ by numerically solving:
\be
\label{eq:drawtover}
q_{l+1} = P_{\rm 2D}(\tover\leq\tovermc|\tgasmc) ,
\ee
for $\tovermc$, where the increment of the subscript in $q_{l+1}$ indicates that we use a new random number. As before, the drawn value $\tovermc$ is obtained by interpolating the discretised form of $P_{\rm 2D}(\tover<\tovermc|\tgasmc)$. To obtain $\lambdamc$, we evaluate the complete three-dimensional PDF of $\tgas$, $\tover$, and $\lambda$ at $\{\tgasmc,\tovermc\}$, resulting in the one-dimensional PDF of $\lambda$ given the values of $\tgasmc$ and $\tovermc$. This one-dimensional PDF is obtained by interpolation between the four discrete one-dimensional PDFs:
\be
\label{eq:drawpdf3}
\begin{aligned}
&P_{\rm 3D}[\lambda\leq\lambdamc|\tgas(i),\tover(j)], \\
&P_{\rm 3D}[\lambda\leq\lambdamc|\tgas(i),\tover(j+1)], \\
&P_{\rm 3D}[\lambda\leq\lambdamc|\tgas(i+1),\tover(j)], \\
&P_{\rm 3D}[\lambda\leq\lambdamc|\tgas(i+1),\tover(j+1)],
\end{aligned}
\ee
where $i$ is defined as before and $j$ denotes the index such that $\tover(j)\leq\tovermc\leq\tover(j+1)$. We then draw $\lambdamc$ by numerically solving:
\be
\label{eq:drawlambda}
q_{l+2} = P_{\rm 3D}(\lambda\leq\lambdamc|\tgasmc,\tovermc) ,
\ee
for $\lambdamc$, where we have again used a new random number. As before, the drawn value $\lambdamc$ is obtained by interpolating the discretised form of $P_{\rm 3D}(\lambda\leq\lambdamc|\tgasmc,\tovermc)$.

Through the above procedure, we obtain a Monte-Carlo realisation of a fit to the observed tuning fork diagram, i.e.~$\{\tgasmc,\tovermc,\lambdamc\}$, by drawing from the three-dimensional PDF. In addition, we generate Monte-Carlo realisations of the duration of the reference time-scale $\tstarref$ and the SFR and gas surface densities $\Sigma_{\rm SFR}$ and $\Sigma_{\rm gas}$. The uncertainties on each of these quantities are defined either directly in the input file (see \autoref{tab:input}) or by analytical error propagation, as detailed in the discussion of equations~(\ref{eq:sigmasfr}) and~(\ref{eq:sigmagas}). We assume that the underlying PDFs are Gaussian,\footnote{We allow asymmetric error bars on $\tstarref$ in \autoref{tab:input}. If the upward and downward uncertainties differ, we use the mean uncertainty and still generate the uncertainties from a symmetric Gaussian.} implying that the Monte-Carlo realisations of these three quantities are obtained by drawing Gaussian random numbers. This is repeated a total of $\nmcphys$ times and thus requires a total of $N_{\rm rnd}=3\nmcphys$ random numbers $\{q_l\in\mathbb{R}|0\leq q_l\leq1\}$ and the same number of Gaussian random numbers $\{q_l'\in\mathbb{R}\}$, where we adopt a default value of $\nmcphys=10^6$ for a total of $N_{\rm rnd}=6\times10^6$ random numbers. We find that $10^6$ Monte-Carlo realisations yield PDFs of the derived quantities (obtained below) that are essentially noise-free and well-defined.

\begin{figure}
\includegraphics[width=\hsize]{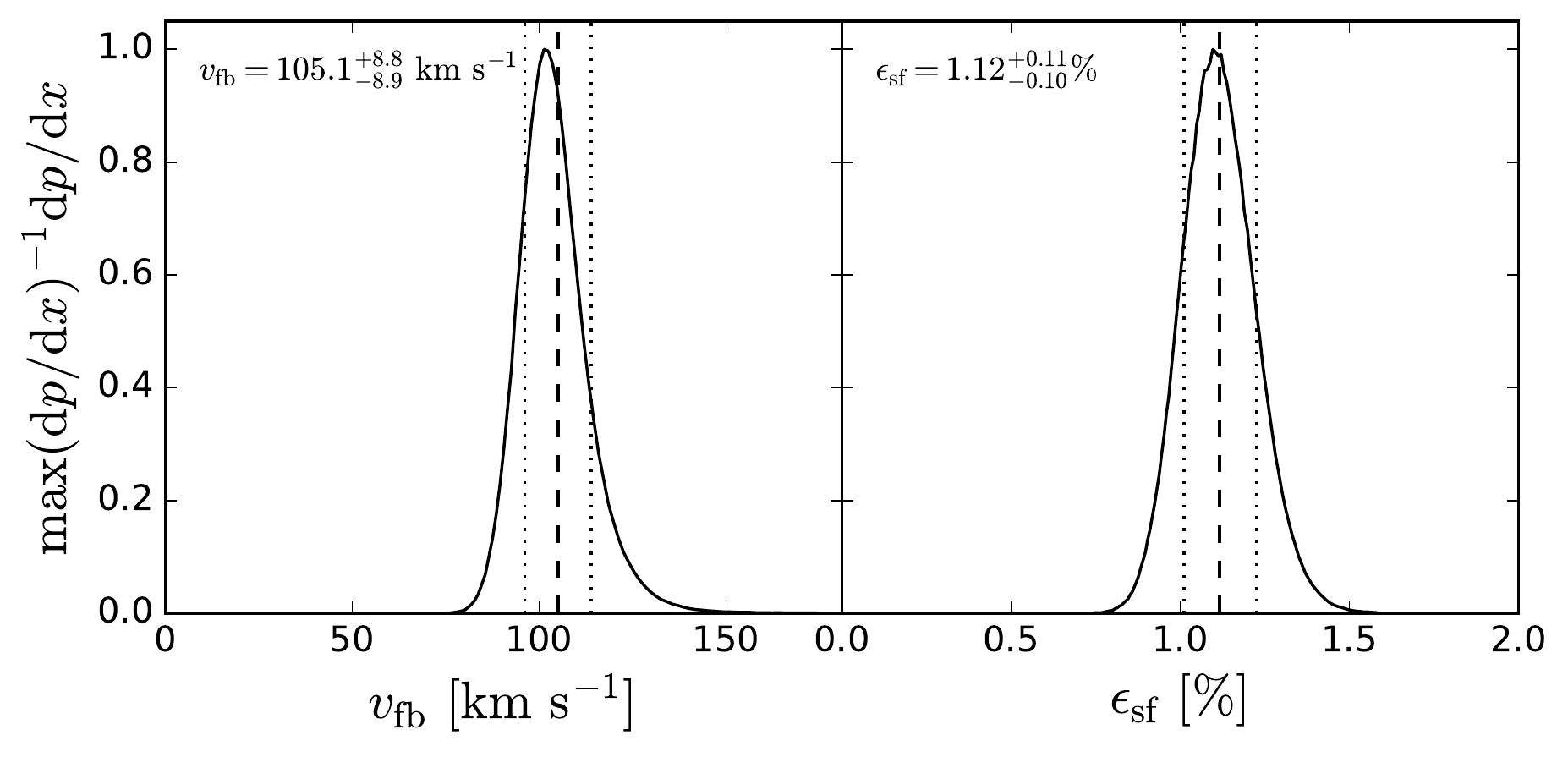}%
\vspace{-1mm}\caption{
\label{fig:stepderived}
One-dimensional PDFs of the feedback outflow velocity $\vfb$ (left) and the star formation efficiency per star formation event $\sfe$ (right), providing two examples of the quantities that can be derived from the fit in \autoref{fig:stepfit}. The PDFs are normalised to their maximum probability density. Vertical dashed lines indicate the best-fitting values (also indicated in the top left of each panel), whereas the vertical dotted lines represent the 1$\sigma$ uncertainty range as defined in the text. These two PDFs are representative for the PDFs of the derived quantities and illustrate that they are well-defined by the Monte-Carlo error propagation procedure described in the text.\vspace{-1mm}
}
\end{figure}

For each Monte-Carlo realisation, we calculate the quantities described in this section. The total ensemble of Monte-Carlo realisations then provides the marginalised, one-dimensional PDF for each of these quantities. We obtain the upward and downward uncertainties from these PDFs in the same way as for the three free parameters $\tgas$, $\tover$, and $\lambda$, as described in equations~(\ref{eq:percgauss1})--(\ref{eq:sigmax}) of Section~\ref{sec:steppdfs}. In addition, we obtain the covariance and correlation matrices for all quantity pairs $\{X_i,X_j\}$ as usual, i.e.
\be
\label{eq:cov}
{\rm cov}(X_i,X_j)=\langle X_iX_j\rangle-\langle X_i\rangle\langle X_j\rangle ,
\ee
and
\be
\label{eq:corr}
\rho_{ij}=\frac{{\rm cov}(X_i,X_j)}{\sigma_i\sigma_j} ,
\ee
for the covariance and correlation matrix, respectively, where $\langle\dots\rangle$ takes the mean over all Monte-Carlo realisations. In equation~(\ref{eq:corr}), the quantities $\sigma_i$ and $\sigma_j$ represent the standard deviations of the Monte-Carlo samples of $X_i$ and $X_j$. The covariance and correlation matrices are written to disk in ASCII format. \code writes each differential one-dimensional PDF to disk in ASCII format too and also outputs figures of all PDFs calculated here, i.e.~the differential and cumulative one-dimensional PDFs for each derived quantity. The best-fitting values and their uncertainties are indicated in these plots. For illustration, \autoref{fig:stepderived} shows the differential PDFs of the feedback velocity $\vfb$ and the star formation efficiency per star formation event $\sfe$ for the example maps used throughout Section~\ref{sec:method}. The PDFs are well-defined and show almost no statistical noise. In addition, the uncertainty ranges on the quantities are small, with relative errors less than 10 per cent, showing that even the derived quantities can be determined to high precision.

\begin{table*}
 \centering
  \begin{minipage}{\hsize}
  \caption{Quantities constrained by the presented analysis}\label{tab:output}\vspace{-1mm}
  \begin{tabular}{@{}l c c l@{}}
  \hline
   Quantity [unit] & Class$^a$ & Equation & Description \\
  \hline
  $\chi^2_{\rm red,min}$ & {\it fundamental} & \ref{eq:chired} & Goodness-of-fit statistic \\
  $\tgas$ [\myr] & {\it fundamental} & \ref{eq:pdf3d} & Best-fitting gas tracer lifetime (e.g.~the cloud lifetime) \\
  $\tover$ [\myr] & {\it fundamental} & \ref{eq:pdf3d} & Best-fitting overlap lifetime (e.g.~the feedback time-scale) \\
  $\lambda$ [\pc] & {\it fundamental} & \ref{eq:pdf3d} & Best-fitting mean separation length of independent regions (e.g.~the fragmentation length) \\[1.5ex]
  $\tstar$ [\myr] & {\it byproduct} & \ref{eq:tstar1}, \ref{eq:tstar2} & Star formation tracer lifetime\\
  $\tau$ [\myr] & {\it byproduct} & \ref{eq:tau} & Total region lifetime\\
  $\betastar$ & {\it byproduct} & \ref{eq:betastar} & Ratio between the mean flux of overlapping and isolated star formation tracer peaks \\
  $\betagas$ & {\it byproduct} & \ref{eq:betagas} & Ratio between the mean flux of overlapping and isolated gas tracer peaks \\
  $\exc_{\rm star}$ & {\it byproduct} & \ref{eq:excstar} & Ratio between the central peak flux density and that on a size scale $\lambda$ for star formation tracer peaks \\
  $\exc_{\rm gas}$ & {\it byproduct} & \ref{eq:excgas} & Ratio between the central peak flux density and that on a size scale $\lambda$ for gas tracer peaks \\
  $\exc_{\rm star,glob}$ & {\it byproduct} & \ref{eq:excstarglob} & Ratio between the flux density on a size scale $\lambda$ and the map average for star formation tracer peaks \\
  $\exc_{\rm gas,glob}$ & {\it byproduct} & \ref{eq:excgasglob} & Ratio between the flux density on a size scale $\lambda$ and the map average for gas tracer peaks \\[1.5ex]
  $\rstar$ [\pc] & {\it derived} & \ref{eq:rstardisc}, \ref{eq:rstargauss} & Disc radius or Gaussian dispersion radius of star formation tracer peaks \\
  $\rgas$ [\pc] & {\it derived} & \ref{eq:rgasdisc}, \ref{eq:rgasgauss} & Disc radius or Gaussian dispersion radius of gas tracer peaks \\
  $\zetastar$ & {\it derived} & \ref{eq:zetastar} & Star formation tracer peak concentration parameter\\
  $\zetagas$ & {\it derived} & \ref{eq:zetagas} & Gas tracer peak concentration parameter\\
  $\vfb$ [\kms] & {\it derived} & \ref{eq:vfb} & Feedback-driven expansion velocity of ejecta if gas vanishes at $\lambda/2$ \\
  $\vfbr$ [\kms] & {\it derived} & \ref{eq:vfbr} & Feedback-driven expansion velocity of ejecta if gas vanishes at $\rgas$ \\[1.5ex]
  $\Sigma_{\rm SFR}$ [\msun~\yr$^{-1}$~\pc$^{-2}$] & {\it composite} & \ref{eq:sigmasfr} & SFR surface density \\
  $\Sigma_{\rm gas}$ [\msun~\pc$^{-2}$] & {\it composite} & \ref{eq:sigmagas} & Gas surface density \\
  $\tdepl$ [\gyr] & {\it composite} & \ref{eq:tdepl} & Gas depletion time \\
  $\sfe$ & {\it composite} & \ref{eq:sfe} & Star formation efficiency per star formation event \\
  $\etafb$ & {\it composite} & \ref{eq:etainst} & Instantaneous mass loading factor \\
  $\etaavgfb$ & {\it composite} & \ref{eq:etaavg} & Time-integrated mass loading factor \\
  $\chifbe$ & {\it composite} & \ref{eq:chie} & Feedback energy efficiency using $\vfb$ \\
  $\chifber$ & {\it composite} & \ref{eq:chier} & Feedback energy efficiency using $\vfbr$ \\
  $\chifbp$ & {\it composite} & \ref{eq:chip} & Feedback momentum efficiency using $\vfb$ \\
  $\chifbpr$ & {\it composite} & \ref{eq:chipr} & Feedback momentum efficiency using $\vfbr$ \\
  $\mdotsf$ [\msun~\yr$^{-1}$] & {\it composite} & \ref{eq:mdotsf} & SFR per star formation event during the gas phase \\
  $\mdotfb$ [\msun~\yr$^{-1}$] & {\it composite} & \ref{eq:mdotfb} & Feedback mass removal rate per star formation event during the overlap phase \\
  \hline
\end{tabular}\\
$^a$ The listed classes of quantities are {\it fundamental} (obtained directly from the fitting process), {\it byproducts} (of the fitting process), {\it derived} (from fundamental quantities), or {\it composite} (obtained using additional constants such as conversion factors).\vspace{-1mm}
\end{minipage}
\end{table*}

\subsubsection{Model output} \label{sec:stepoutput}
At this point in the application of the method, all quantities of interest have been calculated and the information needed for further analysis should be written to disk. \autoref{tab:output} lists all constrained quantities for which PDFs and uncertainties are calculated and for which ASCII tables and figures have been produced. In the final step of the process, \code generates an ASCII table that contains the best-fitting quantities from \autoref{tab:output} and their uncertainties, as well as the absolute SFR and gas mass $M_{\rm gas}$, the number of stellar and gas peaks $N_{\rm peak,star}$ and $N_{\rm peak,gas}$, and the minimum aperture size used $\lapmin$. In addition, \code outputs a single table row containing all of these quantities, which is straightforward to copy into a master table of multiple runs. Finally, a log file is written to disk, which contains all terminal output produced during steps described in this section.

The text files generated during this final step add to the output files written to disk at earlier steps, as summarised in \autoref{fig:schematic}. Upon completing the application of the method to a pair of galaxy maps, we have access to:
\begin{enumerate}
\item
The masked maps showing the identified peak positions.
\item
Figures and ASCII tables of the observed and best-fitting tuning fork diagram.
\item
ASCII tables of $\betastar(f_{\rm star,over})$, $\betagas(f_{\rm gas,over})$, $\exc_{\rm star}(\lambda)$, and $\exc_{\rm gas}(\lambda)$.
\item
Figures of the marginalised, two-dimensional PDFs of the three free parameters.
\item
Figures and ASCII tables of the marginalised, one-dimensional PDFs of all constrained quantities (see \autoref{tab:output}).
\item
The correlation and covariance matrices of all constrained quantities.
\item
The output files listing the best-fitting values and uncertainties of all constrained quantities.
\item
The log file of the terminal window.
\end{enumerate}

In future work, we aim to extend the method further. In particular, we have included the option of closing the loop in \autoref{fig:schematic} and iteratively applying the method, such that the output of one analysis run can be used to modify the input maps and rerun the analysis with the updated maps \citep{hygate18}. This approach can be used to filter out diffuse emission from the maps that does not belong to independent regions as defined by the application of the method, or plausibly to account for distinct morphological features in the galaxy maps such as strong spiral arms or rings.

\section{Validation using disc galaxy simulations} \label{sec:valid}
To test the method described in Section~\ref{sec:method}, we require pairs of galaxy maps to which the method can be applied. For this purpose, we carry out hydrodynamical simulations of isolated disc galaxies and create a large number of different maps from these simulations, showing their gas and stellar content in a variety of ways. Because these are simulated rather than observed systems, we can perform a variety of controlled experiments to assess the accuracy of the method. These simulations are carried out using commonly-used, sub-grid models for star formation \citep[e.g.][]{katz92} and feedback (see \citealt{hu14} for a detailed description of the code used). Even though such models result in galaxies with macroscopic properties that provide a reasonable match to the observed galaxy population \citep[e.g.][]{vogelsberger14,schaye15}, the star formation and feedback prescriptions are certainly inadequate when concerned with the cloud-scale processes discussed here \citep[e.g.][]{hopkins14}. However, this is not a concern in the context of the problem at hand. The goal of this work is not to accurately model cloud-scale star formation and feedback, but to determine whether the presented method accurately retrieves the quantities describing what happens in the simulated maps, irrespective of whether the underlying physics are correct. If the method successfully extracts these quantities, then it can be applied to observations of real galaxies to motivate improved models for star formation and feedback in galaxy simulations. As we show in this section, the method indeed provides accurate measurements of these quantities and is suitable for application to a large variety of observed galaxies.

\subsection{Disc galaxy models} \label{sec:models}
We briefly describe the setup and properties of the numerical disc galaxy models used. In order, we summarise the hydrodynamical method used, the cooling, star formation, and feedback model, the initial conditions, and the resulting properties of the isolated disc galaxy models.

\subsubsection{Hydrodynamical method}
The galaxy simulations presented here have been performed with the smoothed particle hydrodynamics \citep[SPH,][]{monaghan92} code {\sc P-Gadget-3} \citep{springel05c}, using the {\sc SPHGal} implementation of the hydrodynamics solver by \citet{hu14}. SPH is a Galilean invariant, Lagrangian method for hydrodynamical simulations that adopts a particle representation of the fluid. The implementation by \citet{hu14} contains a number of important improvements relative to classical SPH flavours and alleviates many of the numerical issues that affected SPH models in recent years (see \citealt{agertz07} and \citealt{hu14} for discussions). In brief, these improvements are as follows.
\begin{enumerate}
\item
We use a Wendland $C^4$ smoothing kernel \citep{dehnen12} to enable the use of $N_{\rm ngb}=200$ neighbouring particles without suffering from the pairing instability \citep{price12}. Using this high number of neighbours reduces the noise originating from the `$E_0$ error' in the SPH momentum equation \citep{read10} and improves the convergence rate.
\item
The classical density-entropy formulation of SPH is unable to accurately model contact discontinuities or fluid mixing. To remedy this, we use the pressure-entropy formulation of the SPH equations of motion \citep[e.g.][]{ritchie01,saitoh13,hopkins13c}. This formulation smoothes over the pressure rather than the density and therefore avoids spurious pressure jumps, which fundamentally improves the numerical approximation of the force field.
\item
We include a strong artificial viscosity limiter as defined in \citet{hu14} and progressing from the weak limiter from \citet{cullen10}. This updated limiter keeps the fluid from being too viscous in a differentially rotating disc, while retaining the ability to capture shocks properly.
\item
We adopt an artificial thermal energy conduction term to dampen the entropy jumps at hydrodynamic shocks, which smoothes the resulting pressure and force fields \citep[see e.g.][]{read12}. This term only operates across shocks (not shearing flows) and is highly pertinent in pressure-entropy SPH, because pressure jumps can exceed density jumps by orders of magnitude.
\item
Finally, the time steps of particles in the vicinity of strong shocks are limited to ensure that they are similar to within a factor of a few \citep{saitoh09}. In addition, particles become active whenever they are subject to a feedback energy injection \citep{durier12}.
\end{enumerate}
Together, these improvements are chosen to optimise the accuracy of the hydrodynamical modelling in differentially-rotating disc galaxies \citep{hu14}, as considered in this paper. The simulations include the evolutionary cycling between gas and star particles, as well as a live dark matter halo.

\subsubsection{Cooling, star formation, and feedback}
The thermal evolution of the gas particles proceeds according to the metal enrichment and cooling scheme from \citet{aumer13}, which traces the 11 elements H, He, C, N, O, Ne, Mg, Si, S, Ca, and Fe. Once a star particle forms (see below), it inherits the abundances from its parent gas particle. The chemical enrichment includes ejecta from SNe types Ia and II, as well as AGB stars. After metals are deposited into the ISM, they are advected with the flow and undergo metal diffusion as in \citet{aumer13} to account for turbulent mixing. The resulting cooling rate of each particle is obtained for each element by assuming the gas is optically thin and subject to the UV/X-ray background \citep{wiersma09}. We adopt a cooling floor of $T=10~{\rm K}$, which is not reached in practice (see Section~\ref{sec:discprops}).

Gas particles are eligible for star formation once they have temperatures $T<1.2\times10^4~{\rm K}$ and hydrogen particle densities $n_{\rm H}>0.5~\cmc$ (assuming atomic gas). Star-forming gas particles are converted into stars stochastically, according to the prescription from \citet{katz92}:
\be
\label{eq:rhosf}
\rho_{\rm SFR}=\epsilon\frac{\rhog}{\tdyn} ,
\ee
where $\rho_{\rm SFR}$ is the SFR volume density, $\rhog$ is the gas particle volume density, $\epsilon=0.02$ is the star formation efficiency per dynamical time, and
\be
\label{eq:tdyn}
\tdyn=\frac{1}{\sqrt{4\pi G\rhog}} ,
\ee
is the dynamical time. In practice, this expression implies that, during a timestep ${\rm d}t$, the probability that a gas particle is converted into a star particle is given by
\be
\label{eq:psf}
p_{\rm sf}=1-\exp{\left(-\epsilon{\rm d}t/\tdyn\right)} .
\ee
The effective timescale of the exponential is thus $\tdyn/\epsilon$, which at the minimum density for star formation corresponds to 1.6~Gyr. In practice, clouds consist of a large number of particles (recall that $N_{\rm ngb}=200$), implying that they begin forming stars on a correspondingly shorter time-scale.

The deposition of mass, (radial) momentum, and (thermal) energy into the ISM is driven by SN explosions, which adds these to the 10 gas particles closest to the star particle according to the kernel weighting. Type II SNe take place after $3~\myr$, whereas Type Ia SNe occur continuously between $0.1$--$1~\gyr$. SN ejecta correspond to $\sim19$ per cent of the particle mass and carry $10^{51}~{\rm erg}$ of energy, with a velocity of $3000~\kms$. The momentum is added to the gas particles according to an inelastic collision and the remaining energy added to the thermal energy budget of these particles. See \citet{scannapieco06} and \citet{hu14} for further details.
\begin{figure*}
\includegraphics[width=\hsize]{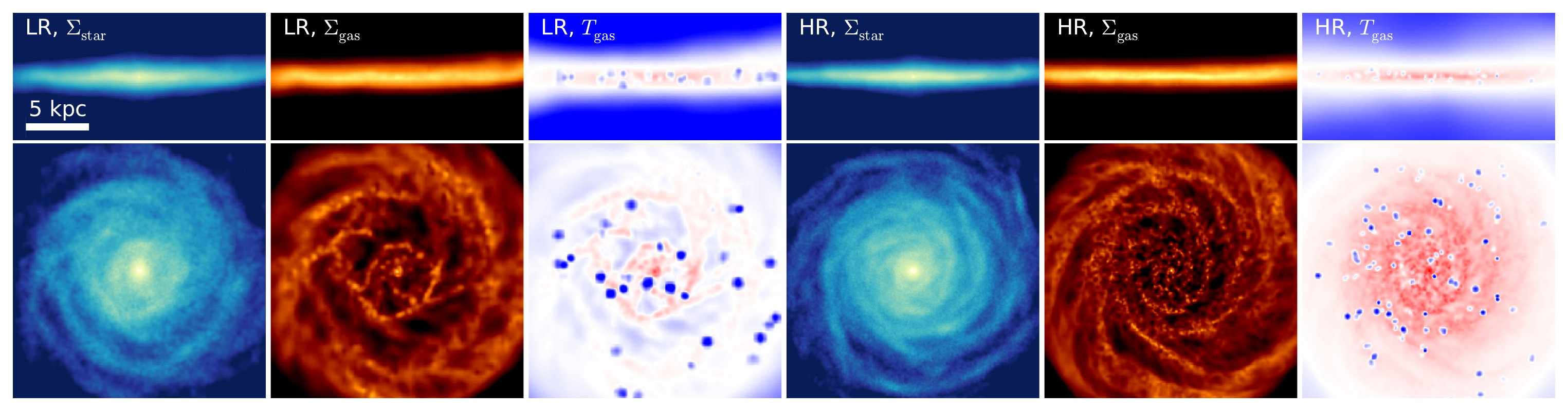}%
\vspace{-1mm}\caption{
\label{fig:maps}
Maps of the simulated galaxies, showing (from left to right) the stellar surface density, gas surface density, and density-weighted temperature along the line of sight, for the low-resolution simulation (LR, left three columns) and the high-resolution simulation (HR, right three columns). The top row shows the edge-on views of the galaxies, whereas the bottom row shows the face-on views. The colour scale is logarithmic in all panels, with stellar surface densities covering $\log_{10}(\Sigma/\msun~\pc^{-2})=0.5$--$3.5$ in the edge-on maps and $\log_{10}(\Sigma/\msun~\pc^{-2})=0.5$--$2.5$ in the face-on maps. The upper limits on the gas surface densities are 0.5~dex lower. For both the edge-on and face-on temperature maps, \{red, white, blue\} corresponds to $\log_{10}(T/{\rm K})=\{3, 4, 5\}$. All panels are $20~\kpc$ on a side and the top-left panel includes a scale bar for reference.\vspace{-1mm}
}
\end{figure*}

\subsubsection{Initial conditions}
\begin{table}
 \centering
  \begin{minipage}{77mm}
  \caption{Particle resolution of the simulated galaxies}\label{tab:ics}\vspace{-1mm}
  \begin{tabular}{@{}l c c c c @{}}
  \hline
   Component & $N_{\rm LR}$ & $N_{\rm HR}$ & $m_{\rm LR}~[\msun]$ & $m_{\rm HR}~[\msun]$ \\
  \hline
  Halo & $400000$ & $1000000$ & $2.3\times10^6$ & $9.0\times10^5$ \\
  Stellar disc & $462000$ & $2310000$ & $1.4\times10^4$ & $2.7\times10^3$ \\
  Gas disc & $308000$ & $1540000$ & $1.4\times10^4$ & $2.7\times10^3$ \\
  Bulge & $20000$ & $100000$ & $1.4\times10^4$ & $2.7\times10^3$ \\
  \hline
\end{tabular}\vspace{-1mm}
\end{minipage}
\end{table}
We generate the initial conditions of two isolated disc galaxies according to the method by \citet{springel05b}. The galaxies consist of a live dark matter halo, a mixed gas-stellar disc, and a stellar bulge. The halo is taken to follow a \citet{hernquist90} profile with an equivalent \citet{navarro97} concentration parameter of $c=12$, a virial mass of $M_{\rm 200}=9.0\times10^{11}~\msun$, and a virial radius of $r_{200}=2.2\times10^2~\kpc$. We set the halo spin parameter to $\lambda=J|E|^{1/2}/GM_{200}^{5/2}=0.030$, where $J$ is the halo angular momentum and $E$ is the halo energy. The disc has a total mass of $M_{\rm d}=1.05\times10^{10}~\msun$ and an angular momentum fraction set to be equal to the disc mass fraction. The stellar disc constitutes 60 per cent of the total disc mass ($M_{\rm d,s}=6.3\times10^9~\msun$) and follows an exponential profile with a scale radius of $r_{\rm d,s}=3.1~\kpc$ and a scale height of $h_{\rm d}=0.14~\kpc$. The gas disc constitutes 40 per cent of the total disc mass ($M_{\rm d,g}=4.2\times10^9~\msun$) and follows a composite radial profile, consisting of an exponential profile with a scale radius of $r_{\rm d,g}=6.3~\kpc$ (holding 5 per cent of the gas mass) and a constant surface density H{\sc i} disc that extends out to $r_{\rm H{\sc I}}=12.5~\kpc$ (holding 95 per cent of the gas mass). Finally, the galaxies host a \citet{hernquist90} bulge with mass $M_{\rm b}=2.7\times10^8~\msun$ and a scale length of $r_{\rm b}=0.23~\kpc$. Over the course of the simulation, $h_{\rm d}$ and $r_{\rm b}$ increase somewhat, by up to 40 per cent. The above parameters have been chosen to be somewhat representative of near-$L^\star$, flocculent spiral galaxies in the local Universe.

To sample the three-dimensional mass distribution of the galaxy models, we use four families of particles with numbers and masses listed in \autoref{tab:ics}. The dependence on numerical resolution is assessed by considering two galaxy models with different resolutions. The low-resolution (subscript `LR') model consists of a total of $1.19\times10^6$ particles, whereas the high-resolution (subscript `HR') model consists of a total of $4.95\times10^6$ particles. Gravitational forces between the particles are calculated using a softening length for the halo particles of $h_{\rm halo}=100~\pc$ and for the baryonic particles in the \{low, high\}-resolution simulations of $h_{\rm LR,bary}=20~\pc$ and $h_{\rm HR,bary}=10~\pc$, respectively. Note that the smoothing lengths enclosing $N_{\rm ngb}=200$ neighbours used for determining the local fluid properties are considerably larger than the above softening lengths, with typical map-averaged values of $h_{\rm smooth,LR}\sim250~\pc$ and $h_{\rm smooth,HR}\sim100~\pc$ (see Section~\ref{sec:expder}). Finally, the particle masses listed in \autoref{tab:ics} are not sufficient to resolve the cold gas, because the mass contained in an SPH kernel is $N_{\rm ngb}m_{\rm HR,g}\sim5\times10^5~\msun$. However, they are sufficient to marginally resolve the Jeans mass at the onset of cloud formation, which suffices for the main goals of this paper (see Section~\ref{sec:discprops}).
\begin{figure*}
\includegraphics[width=\hsize]{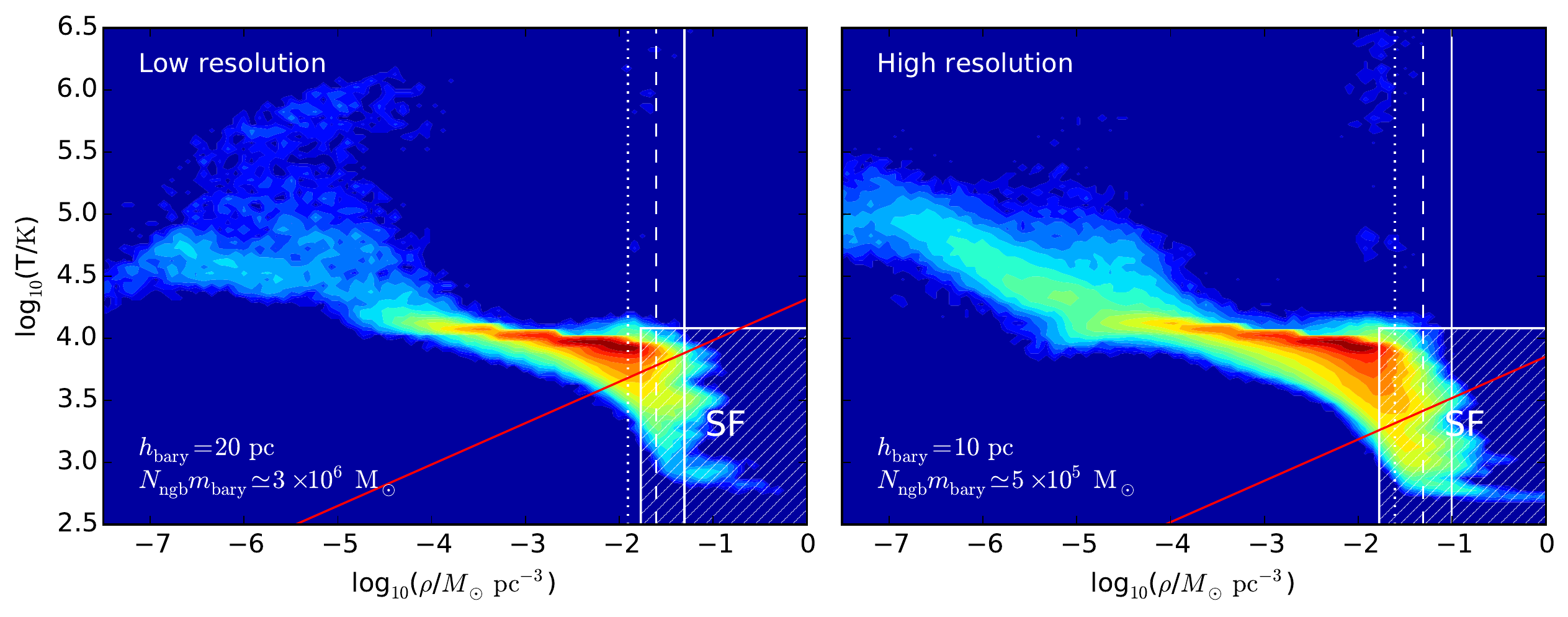}%
\vspace{-1mm}\caption{
\label{fig:phase}
Phase diagram of the gas particles in the low-resolution (left) and high-resolution (right) simulations. The part of phase space where gas particles are eligible for star formation is represented by the hatched region, indicated with `SF'. The vertical dotted, dashed, and solid lines indicate the critical volume densities above which the simulated gas maps are generated in Section~\ref{sec:gasstar}. Converting densities from $\msun~\pc^{-3}$ to $\cmc$ or $m_{\rm H}~\cmc$ requires the addition of $1.25~\dex$ (assuming a mean molecular weight $\mu=2.3$) or $1.61~\dex$, respectively. These phase diagrams show that the simulations resolve cooling down to temperatures of several $100~{\rm K}$ and densities of $\sim10~\cmc$, covering the onset of molecular cloud condensation. The Jeans mass is resolved with at least 200 particles for fluid elements above the red diagonal line, which applies to all gas at the onset of the collapse towards star formation, in the top left corner of the hatched region. At both resolutions, some feedback-heated and shocked gas in star-forming regions is present at high densities and temperatures.\vspace{-1mm}
}
\end{figure*}

\subsubsection{Simulated disc galaxy properties} \label{sec:discprops}
We run the disc galaxy simulations for a total duration of $t_{\rm run}=2.2~\gyr$ before generating the maps on which the method of Section~\ref{sec:method} is to be tested. For reference, we show the stellar and gas particle maps at $t=t_{\rm run}$ in \autoref{fig:maps}. The time covered by the simulations exceeds the dynamical time at the virial radius and corresponds to over a dozen dynamical times within the galaxy itself (at $R<10~\kpc$). The galaxies achieve equilibrium on the order of a few disc dynamical times, well before the snapshots at $t_{\rm run}$ that we will consider throughout the rest of this paper. Indeed, \autoref{fig:maps} shows that the simulated galaxies are stable and do not exhibit any obvious transient morphological features. This is confirmed by visual inspection of all preceding snapshots.

To assess the behaviour of the ISM, \autoref{fig:phase} shows the phase diagram of the gas discs at $t=t_{\rm run}$, demonstrating that the ISM in the simulations is stable too. Star formation takes place in the regime where cooling from the atomic phase has set in and the Jeans mass is sufficiently well-resolved to identify where the cooling track sets in. Some hot and dense feedback ejecta can be seen along the vertical white lines in the top right of each panel. The increased resolution of the high-resolution simulation enables more efficient gas cooling than in the low-resolution simulation due to the higher densities. In Section~\ref{sec:gasstar}, we will be testing the method of this paper by using gas maps of these galaxies above minimum volume densities indicated by the vertical white lines. \autoref{fig:phase} shows that these volume density thresholds trace star-forming gas in all but one (LR, dotted line) case. The total SFR in the simulations averaged over the $10$--$300~\myr$ preceding these snapshots is $0.2$--$0.3~\msun~\yr^{-1}$.

\subsection{Age-binned stellar maps} \label{sec:starstar}
As a first test of the method presented in Section~\ref{sec:method}, we exclusively use maps of the star particles in the simulations. The major advantage of using star particles is that their ages are known. By only displaying the star particles in specified age ranges, the duration of the evolutionary phase displayed in each map is known, allowing the accuracy of the method to be directly quantified. In this subsection, we use these `age-binned' stellar maps to assess how well the method retrieves the `cloud lifetime' $\tgas$, the `feedback time-scale' $\tover$, and the overlap-to-isolated flux ratios $\betastar$ and $\betagas$. Even though we only use stellar maps in this subsection and all phases of the evolutionary timeline therefore represent stellar phases, we keep referring to the first phase as `gas' (this includes its associated quantities such as $\tgas$ and $\betagas$) to remain consistent with the rest of the paper.

\subsubsection{Procedure for creating the maps} \label{sec:starproc}
To generate age-binned maps of the star particles in the galaxy models, we first need to define the age bins for which the particles are displayed. The duration of the stellar phase $\tstar$ represents the width of the age bin for which the star particles are shown in the stellar map. This time-scale is taken to be known and is used during the fitting process to determine the other time-scales. The duration of the `gas' phase $\tgas$ represents the width of the age bin for which the star particles are shown in the `gas' map. The duration of the `overlap' phase $\tover$ represents the age range for which star particles are included in both maps. We add a fourth time-scale $\toff$ that represents the offset of the age bins from an age of zero ($t=0$). For each experiment, these four time-scales define the adopted age bins. The stellar phase includes the star particles in the age bin spanning $t=\langle \toff+\tgas-\tover, \toff+\tgas+\tstar-\tover\rangle$, whereas the `gas' phase includes the star particles in the age bin spanning $t=\langle \toff, \toff+\tgas\rangle$. The above definitions are consistent with the evolutionary timeline in \autoref{fig:tschem} and imply that the youngest star particles are shown in the `gas' map, whereas the oldest star particles are shown in the stellar map. As will be discussed in Section~\ref{sec:starexp}, we have set $\toff=0$ for all but two experiments.

In addition to the above time-scales, we specify the overlap-to-isolated flux ratios $\betastar$ and $\betagas$. To do so, we first identify the star particles in the age range $t=\langle \toff+\tgas-\tover, \toff+\tgas\rangle$, which corresponds to the overlap phase. When creating the stellar map, we then multiply the masses of these star particles by $\betastar$, while for the `gas' map, we multiply their masses by $\betagas$. Note that out of the six quantities discussed here, only $\tstar$ is used as an input parameter in the method of Section~\ref{sec:method}. The method is not informed of the values of the other five quantities. Instead, we are using the experiments carried out here to assess how well the best-fitting values of these quantities match their input values.

Having defined the age bins for which the star particles are shown in each map, we generate the surface density maps of the particles in these bins. These maps are square with a diameter of $20~\kpc$ and $1403$~pixels, corresponding to a pixel scale of $14.25~\pc$. Two types of maps are made for each galaxy model. The `point particle' surface density maps show the distribution of particles as-is, for each pixel adding up the enclosed mass of the particles (modulo a factor $\beta$ if it contains particles in the overlap phase) and dividing by the pixel area. At the adopted numerical resolution of the simulations and the pixel scale of $14.25~\pc$, the typical number of particles per pixel with ages in the selected age bins is zero or unity. By contrast, the `extended emission' surface density maps account for the extended morphology expected for real star-forming regions. In the case of gas particles, this extended emission is set by the smoothing kernel and smoothing length, but normally there is no need to smoothen the star particles (only their gravitational force is softened). For consistency with the gas maps discussed in Section~\ref{sec:gasproc} below, we generate the extended emission surface density maps of the star particles by calculating their smoothing length following the same procedure as for the gas particles, i.e.~by using the 200 nearest neighbours to define the smoothing length and adopting a Wendland $C^4$ smoothing kernel \citep{dehnen12} to distribute the mass around each particle position. This choice implies coarser resolution in low-density regions and is made for consistency with P-Gadget, in which the same kernel is used to smoothen the gas particles in the simulations (as discussed in Section~\ref{sec:models}).

\begin{figure}
\includegraphics[width=\hsize]{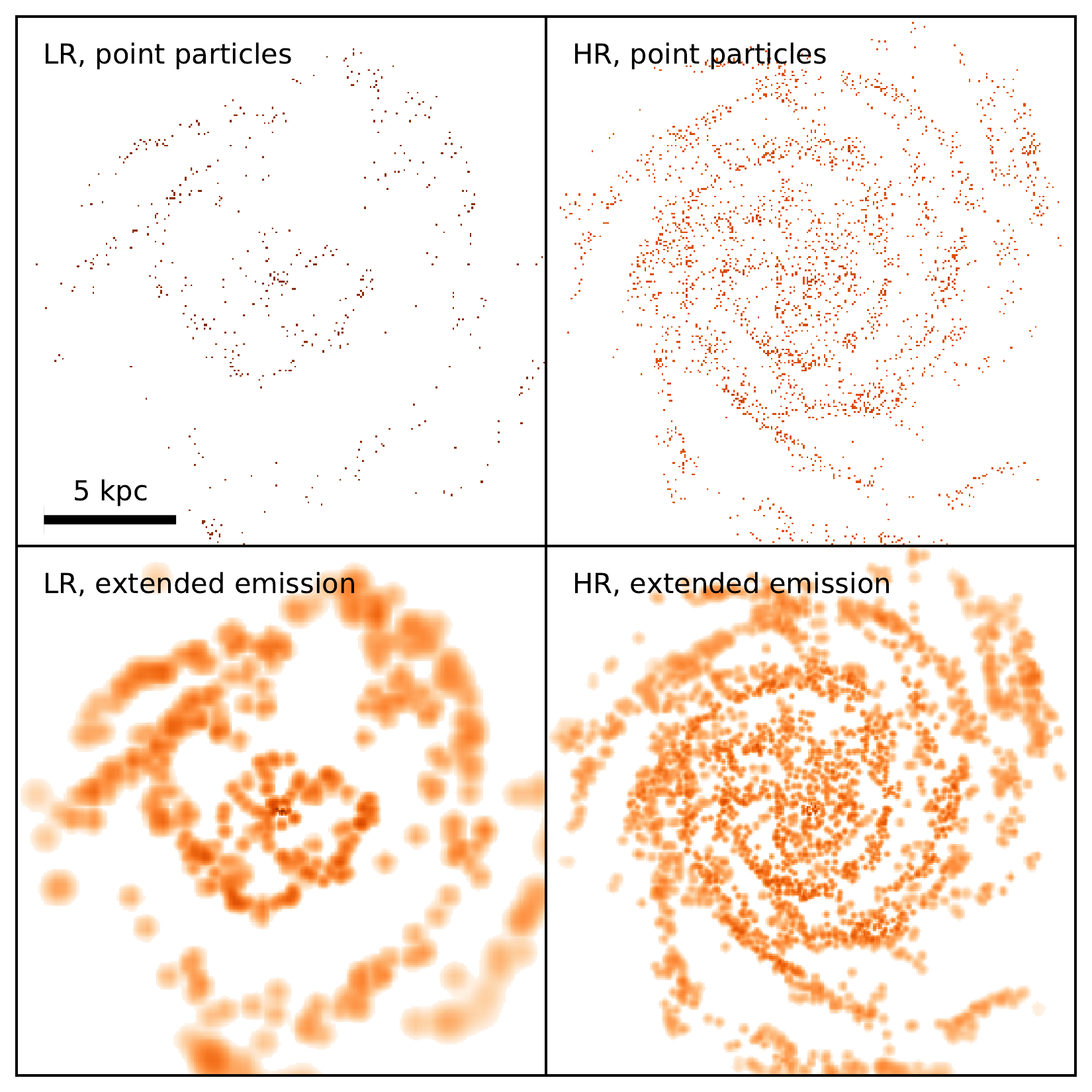}%
\vspace{-1mm}\caption{
\label{fig:starmaps}
The four different types of `gas' maps used in Section~\ref{sec:starstar}. From left to right, top to bottom, these are maps of the low-resolution (LR) point particle distribution, the high-resolution (HR) point particle distribution, the LR extended emission (generated by using a smoothing kernel, see the text), and the HR extended emission. In these examples, the maps show the distribution of star particles in the age range $t=\langle0, 30\rangle~\myr$. In Section~\ref{sec:starexp}, we present the 22 different sets of age ranges (and thus maps) for which our method is tested, for a total of $4\times22=88$ experiments.\vspace{-1mm}
}
\end{figure}

The maps resulting from the above procedure are written to disk in the FITS file format and include a FITS header that contains their key properties, consistent with common conventions for observational data delivery. For the particular case of $\tgas=30~\myr$ and $\toff=0~\myr$, \autoref{fig:starmaps} shows four examples of a typical `gas' map that these experiments are carried out with, illustrating the similarities and differences between the two types of maps (`point particles' and `extended emission') for both disc galaxy models (`low resolution' and `high resolution'). Because we are using age-binned stellar maps both to generate the stellar maps and the `gas' maps, the stellar maps have the same general properties as the `gas' maps shown in \autoref{fig:starmaps}. In contrast to the pairs of age-binned stellar maps discussed here, Section~\ref{sec:gasstar} below concerns a second set of experiments that each combine a stellar map with an actual gas map of the gas particles in the disc galaxy models.
\begin{table}
 \centering
 \begin{minipage}{78mm}
  \caption{Experiments carried out using age-binned stellar maps}\label{tab:starruns_in}\vspace{-1mm}
  \begin{tabular}{c c c c c c c c}
   \hline
   ID & $\tstar$ & $\tgas$ & $\tover$ & $\toff$ & $\betastar$ & $\betagas$ & Symbol\\ 
   \hline
   $1$ & $3$ & $3$ & $0$ & $0$ & $1.0$ & $1.0$ & $\color{black}\blacksquare$ \\ 
   $2$ & $10$ & $10$ & $0$ & $0$ & $1.0$ & $1.0$ & $\color{black}\blacksquare$ \\ 
   $3$ & $30$ & $30$ & $0$ & $0$ & $1.0$ & $1.0$ & $\color{black}\blacksquare$ \\ 
   $4$ & $100$ & $100$ & $0$ & $0$ & $1.0$ & $1.0$ & $\color{black}\blacksquare$ \\[1.5ex] 
   $5$ & $1$ & $3$ & $0$ & $0$ & $1.0$ & $1.0$ & $\color{blue}\rotatebox[origin=c]{90}{$\blacktriangle$}$ \\ 
   $6$ & $3$ & $1$ & $0$ & $0$ & $1.0$ & $1.0$ & $\color{blue}\rotatebox[origin=c]{270}{$\blacktriangle$}$ \\ 
   $7$ & $10$ & $30$ & $0$ & $0$ & $1.0$ & $1.0$ & $\color{blue}\rotatebox[origin=c]{90}{$\blacktriangle$}$ \\ 
   $8$ & $30$ & $10$ & $0$ & $0$ & $1.0$ & $1.0$ & $\color{blue}\rotatebox[origin=c]{270}{$\blacktriangle$}$ \\ 
   $9$ & $1$ & $10$ & $0$ & $0$ & $1.0$ & $1.0$ & $\color{blue}\blacktriangledown$ \\ 
   $10$ & $10$ & $1$ & $0$ & $0$ & $1.0$ & $1.0$ & $\color{blue}\blacktriangle$ \\ 
   $11$ & $3$ & $30$ & $0$ & $0$ & $1.0$ & $1.0$ & $\color{blue}\blacktriangledown$ \\ 
   $12$ & $30$ & $3$ & $0$ & $0$ & $1.0$ & $1.0$ & $\color{blue}\blacktriangle$ \\[1.5ex] 
   $13$ & $10$ & $10$ & $2$ & $0$ & $1.0$ & $1.0$ & $\color{red}\blacksquare$ \\ 
   $14$ & $10$ & $10$ & $4$ & $0$ & $1.0$ & $1.0$ & $\color{red}\blacksquare$ \\ 
   $15$ & $10$ & $10$ & $8$ & $0$ & $1.0$ & $1.0$ & $\color{red}\blacksquare$ \\[1.5ex] 
   $16$ & $10$ & $30$ & $4$ & $0$ & $1.0$ & $1.0$ & $\color{green}\rotatebox[origin=c]{90}{$\blacktriangle$}$ \\ 
   $17$ & $30$ & $10$ & $4$ & $0$ & $1.0$ & $1.0$ & $\color{green}\rotatebox[origin=c]{270}{$\blacktriangle$}$ \\[1.5ex] 
   $18$ & $10$ & $10$ & $4$ & $0$ & $0.5$ & $1.0$ & $\color{cyan}\blacksquare$ \\ 
   $19$ & $10$ & $10$ & $4$ & $0$ & $1.0$ & $0.5$ & $\color{cyan}\blacksquare$ \\ 
   $20$ & $10$ & $10$ & $4$ & $0$ & $0.5$ & $0.5$ & $\color{cyan}\blacksquare$ \\[1.5ex] 
   $21$ & $10$ & $30$ & $4$ & $500$ & $1.0$ & $1.0$ & $\color{magenta}\rotatebox[origin=c]{90}{$\blacktriangle$}$ \\ 
   $22$ & $30$ & $3$ & $0$ & $500$ & $1.0$ & $1.0$ & $\color{magenta}\blacktriangle$ \\ 
   \hline
  \end{tabular} \\ 
  All time-scales are in units of Myr.\vspace{-1mm}
 \end{minipage}
\end{table}

\subsubsection{Description of the experiments} \label{sec:starexp}
\autoref{tab:starruns_in} summarises the experiments carried out using the age-binned stellar maps, specifying the values of $\tstar$, $\tgas$, $\tover$, $\betastar$, and $\betagas$ for each experiment. It also lists the symbols used to indicate each experiment in the figures presented in Section~\ref{sec:validtgas} and \ref{sec:validtover}. The experiments are divided into six main categories, reflected by the differently coloured symbols in the final column.
\begin{enumerate}
\item
The first set of experiments (black symbols) is aimed at investigating the role of galactic dynamics. This is the simplest possible experiment, with equal stellar and `gas' time-scales ($\tstar=\tgas$), no overlap ($\tover=0~\myr$), no age offset ($\toff=0~\myr$), and no flux difference between particles in the isolated and overlap phases ($\betastar=1$ and $\betagas=1$). By spanning $1.5~\dex$ in total duration (ranging from just a few $\myr$ in the first experiment to more than an orbital time in the fourth experiment), we can assess whether the gradual dispersion of groups of young star particles affects the accuracy of our method.
\item
The second set of experiments (blue symbols) adds a first level of additional complexity by considering various degrees of asymmetry between $\tstar$ and $\tgas$. In the first four experiments, both time-scales differ only by a factor of three, whereas the final four experiments have time-scales differing by an order of magnitude. These experiments are intended to test how well the method retrieves the $\tgas$ if it is very dissimilar from the reference time-scale $\tstar$.
\item
The third set of experiments (red symbols) reverts to equal stellar and `gas' time-scales, but adds a non-zero overlap ($\tover>0~\myr$). These experiments allow us to determine how well our method retrieves $\tover$ for various durations of the overlap phase.
\item
The fourth set of experiments (green symbols) combines the second and third sets, but considering dissimilar $\tstar$ and $\tgas$ and including an overlap phase. This setup reminisces the situation one would expect to occur in nature. We consider two cases, one where $\tgas>\tstar$ and one where $\tgas<\tstar$.
\item
The fifth set of experiments (cyan symbols) is an extension of the third set, this time varying the overlap-to-isolated flux ratios $\betastar$ and $\betagas$. We consider two cases where only one of these quantities deviates from unity (meaning that either $\betastar=0.5$ or $\betagas=0.5$, which corresponds to the particles being half as bright when they are in the overlap phase than when they are in the isolated phase) and also include an experiment where both $\betastar=0.5$ and $\betagas=0.5$. These experiments are included to evaluate how well our method retrieves $\tover$ when there is any flux evolution of independent regions during the covered timeline, as well as to see how well we retrieve the parameters $\betastar$ and $\betagas$ that capture this flux evolution.
\item
The sixth set of experiments (magenta symbols) considers a large age offset of $\toff=500~\myr$ from the star particle formation time. One experiment represents a typical set of time-scales (cf.~the fourth set) and one reflects the extreme contrast between the stellar and `gas' time-scale (cf.~the second set). These experiments consider cases where any initial structure has entirely been erased by galactic dynamics and is aimed at testing how well the method works for smoothed distributions of particles or regions.
\end{enumerate}

The above sets of experiments are carried out using the `point particle' or the `extended emission' surface density maps (see Section~\ref{sec:starproc}), both for the low and high-resolution simulations. We thus have a total of 88 experiments using age-binned stellar maps. Some of the parameters used in these experiments deviate from the default values in Section~\ref{sec:method}. Firstly, we are dealing with maps showing star particles with specified age ranges, for which the reference time-scale includes the time for which these age ranges overlap. We therefore set {\tt tstar\_incl}~$=1$, indicating that $\tstarref$ includes $\tover$ and thus $\tstar=\tstarref$. Secondly, we set $\npixmin=1$ for the point particle experiments, to enable single pixels to be identified as peaks. Thirdly, for the low-resolution runs we set $\lapmin=100~\pc$ to remove sub-resolution apertures, as well as $\Delta\log_{10}\f=2.5$ and $\delta\log_{10}\f=0.25$ for the identification of both the stellar peaks and gas peaks, because the low-resolution maps used in this work require a finer and deeper contour level spacing to identify the relevant emission peaks.

\subsubsection{Accuracy of the `cloud lifetime' $\tgas$} \label{sec:validtgas}

\begin{figure*}
\includegraphics[width=\hsize]{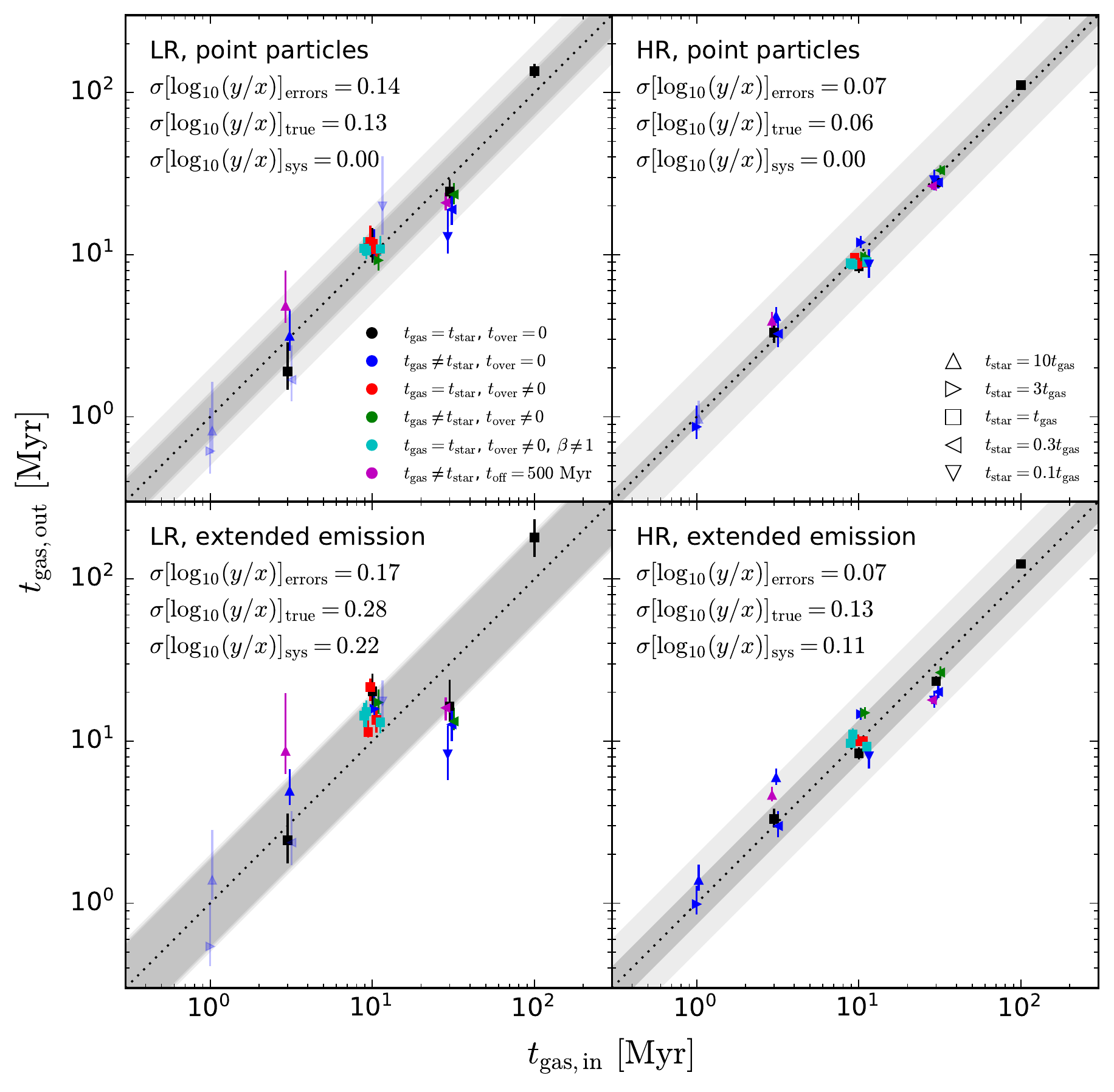}%
\vspace{-1mm}\caption{
\label{fig:tgas}
Accuracy of the `gas' phase lifetimes ($\tgas$) retrieved by our new method, for the low-resolution (LR, left panels) and high-resolution (HR, right panels) simulations, using point particles (top panels) and extended emission (bottom panels). Shown is the best-fitting value $\tgasout$ as a function of the input value $\tgasin$ specified for each experiment (see \autoref{tab:starruns_in}). The data points indicate the results for each experiment (see \autoref{tab:starruns_in} and the legends in the top panels for the meanings of the symbols), with error bars indicating the 16th and 84th percentiles of the $\tgasout$ PDF. Note that only five values of $\tgasin=\{1, 3, 10, 30, 100\}~\myr$ are considered in these experiments. The small horizontal offsets of the data points relative to these exact values have been added to improve the legibility of the figure. Transparent symbols do not pass the guidelines for the reliable application of our method presented in Section~\ref{sec:guide} below and should therefore be omitted from further analysis. The dotted line indicates the one-to-one agreement and the grey areas illustrate different amounts of scatter. The lightest shade represents a factor of two, the middle shade represents the standard deviation of $\log_{10}(\tgasout/\tgasin)$ (i.e.~the scatter around the diagonal) for all data points (including the unreliable, transparent ones), and the darkest shade indicates the standard deviation of $\log_{10}(\tgasout/\tgasin)$ exclusively for the opaque (reliable) data points. Three different measures of the scatter around the one-to-one relation are given in the top left corner of each panel. The first (subscript `errors') is the scatter around the one-to-one relation expected for the error bars of the opaque data points (this is effectively the mean error bar). The second (subscript `true') is the actual scatter around the one-to-one relation for the opaque data points (corresponding to the darkest shade of grey). The third (subscript `sys') is the difference between the first and second values (defined as $\sigma_{\rm sys}^2=\sigma_{\rm true}^2-\sigma_{\rm errors}^2$, with the additional requirement that $\sigma_{\rm sys}^2\geq0$), which is the part of the scatter that cannot be accounted for by the error bars. As such, this represents the `systematic' inaccuracy of the method. In all cases, the intrinsic accuracy of the method is $\sigma_{\rm sys}\la0.2~\dex$, whereas at the high resolutions reached for local-Universe galaxies with telescopes like ALMA, the accuracy is even higher with $\sigma_{\rm sys}\la0.1~\dex$.\vspace{-1mm}
}
\end{figure*}

The first test carried out using the $4\times22=88$ experiments presented in Section~\ref{sec:starexp} is to determine how well the duration of the `gas' phase ($\tgas$) is retrieved by our new method. The quantity $\tgas$ can have a variety of physical meanings depending on the (gas) tracers used. For the specific, real-Universe example of tracing (molecular) gas with CO, this quantity refers to the time spent by a region in a CO-bright state preceding a star formation event, commonly referred to as the molecular cloud lifetime. As discussed in Section~\ref{sec:method}, our method provides the complete PDF for each of the quantities constrained in the fitting process, including $\tgas$. \autoref{fig:tgas} shows the best-fitting value of $\tgas$ (referred to as $\tgasout$) as a function of the input value of $\tgas$ from \autoref{tab:starruns_in} (referred to as $\tgasin$) for all 88 experiments.

\paragraph{Point particles}
Because the method assumes that independent regions can be easily identified and the evolutionary timeline is well-sampled, the best agreement between input and output values is expected for the high-resolution point particle experiments (top-right panel). A comparison between the scatter of the data points around the one-to-one relation ($\sigma_{\rm true}$) in the four panels of \autoref{fig:tgas} shows that this is indeed the case. The logarithmic scatter in the top-right panel is just $0.06~\dex$, corresponding to just $\sim14$~per~cent. The fact that this scatter is similar in magnitude to the mean error bar size of $0.07~\dex$ means that the observed scatter is entirely accounted for by the uncertainties on the best-fitting values $\tgasout$. In other words, the method does not have any fundamental systematic biases in determining $\tgas$ for the high-resolution point particle maps.

Even though this subset of experiments is expected to return the best results, the close agreement is non-trivial. For instance, the method assumes that independent regions are randomly distributed in space up to a spatial scale of a few times the mean separation length $\lambda$, which corresponds to $\sim1~\kpc$ in these experiments. As is evident from \autoref{fig:starmaps}, the distribution of particles in the simulation is substructured. It is encouraging that the deviation from a random distribution seen here does not strongly influence the accuracy of the method. In Section~\ref{sec:fails2}, we will discuss more extreme situations in which it may be important to account for substructure in the spatial distribution of independent regions. The method also assumes that the SFR has been constant to within the error bars on the best-fitting value of $\tgas$ -- any excess star formation during one phase relative to the other phase will proportionally bias the measured time-scales. Again, the good agreement observed here shows that variations of the SFR in the simulated isolated disc galaxy with an M33-like mass are sufficiently small in principle to avoid biases in the measured value of the `gas' phase lifetime.

The agreement between input and output is less pronounced in the low-resolution point particle experiments (top-left panel). With a smaller number of independent regions, the evolutionary timeline of \autoref{fig:tschem} is more sparsely sampled, implying that the Poisson noise of the star formation history increases and the measured time-scales are more prone to being biased. The number of independent regions scales inversely with the particle mass, which is a factor of five larger than in the high-resolution simulation. We thus expect the Poisson noise to be a factor of $\sqrt{5}\approx2.2$ larger than in the high-resolution simulation, which is indeed the difference between the scatter observed in the top-left and top-right panels ($0.13/0.06\approx2.2$). This `true' scatter around the one-to-one relation is consistent with the scatter expected for the size of the error bars ($\sigma_{\rm errors}$), indicating that the error bars on the measurements are accurate.

The scatter around the one-to-one relation does not need to be random. Because these experiments all use the same simulation snapshot, deviations from a constant SFR may affect all experiments covering a similar age range in the same way. Especially for the low numbers of independent regions considered in the top-left panel of \autoref{fig:tgas}, this can lead to a systematic deviation of the measured time-scales, which explains at least part of the discrepant data points at $\tgasin=30~\myr$. If we consider the example of experiment 11, the number ratio between the particles in the stellar and `gas' phases is $0.08~\dex$ (20~per~cent) higher than expected for the relative durations of these phases. This causes the retrieved value of $\tgas$ to be $0.08~\dex$ lower than it should be and represents yet another example of Poisson noise, illustrating that the sampling statistics of having smaller numbers of regions contribute to many of the differences between the top-left and top-right panels. In addition, the spatial structure of these maps deviates more strongly from a random distribution than for the high-resolution experiments (see \autoref{fig:starmaps}). As we will discuss in Section~\ref{sec:fails2}, this substructure may also contribute somewhat to the deviation from the one-to-one relation.

\paragraph{Extended emission}
The bottom two panels of \autoref{fig:tgas} show the results of the experiments based on maps of the extended emission, i.e.~for a non-zero smoothing length (see \autoref{fig:starmaps}). Using extended emission maps rather than point particle maps again increases the scatter around the one-to-one relation, as would be expected. After all, compared to the point particle maps it is less trivial to identify independent regions as peaks when their emission is smoothed over a certain area. This smoothing also adds an emission background around each region that may overlap in space with neighbouring regions and may be mistaken for a time overlap (see Section~\ref{sec:validtover}).

In the method of Section~\ref{sec:method}, we deal with the above issues by representing the independent regions as two-dimensional Gaussians. In our formalism, each Gaussian stands for one region, implying that its two-dimensional integral must be equal to unity. This property is used to relate the surface density contrast between peaks and the average across the map to the width (i.e.~the Gaussian dispersion) of each peak. Mathematically, this is the correct approach and it works very well in practice when peaks are marginally resolved in observed galaxy maps, because the observational point spread function typically resembles a Gaussian (see Section~\ref{sec:derivephys}). However, it is possible (e.g.~due to substructure, blending, highly-resolved independent regions, or a different point spread or smoothing function\footnote{We note that the shape of the Wendland $C^4$ smoothing kernel used here to generate the maps is very similar to a Gaussian.}) that the assumption of Gaussian surface density profiles is invalid. In this case, the measured time-scales may be a bad representation of the true evolutionary timeline.

The bottom panels of \autoref{fig:tgas} demonstrate that any deviations resulting from the above assumptions is modest. The excess of the scatter beyond the mean error bar size (which is a measure for the inaccuracy of the method) indeed increases, but it remains within reasonable limits. The experiments using high-resolution extended emission maps have a spatial resolution that is comfortably obtained with ALMA. For these experiments, the excess scatter around the one-to-one relation is $0.11~\dex$ (which is less than $30$~per~cent). While this residual scatter is non-zero, it is an encouraging accuracy given the discussed assumptions. It is also sufficient for astrophysical purposes. Current discussions in the literature regarding the molecular cloud lifetime discuss order-of-magnitude differences \citep[e.g.][]{dobbs14,jeffreson18}, which our method can plausibly distinguish at the $>8\sigma$ level.

\paragraph{General conclusions}
We see that across all panels of \autoref{fig:tgas}, the largest outliers (ignoring the transparent symbols) are upward or downward-pointing triangles. These symbols represent the experiments in which $\tstar$ and $\tgas$ differ by a factor of 10. Throughout the experiments carried out for this study (also see Section~\ref{sec:tstar}), we have noticed that the most accurate results are obtained when $\tstar$ and $\tgas$ are similar. This occurs because the `tuning fork' diagram of \autoref{fig:tuningfork} becomes asymmetric for dissimilar values of $\tstar$ and $\tgas$, which moves one of the two branches in the diagram closer to the $\bias=1$ line, where the shape of the diagram becomes less sensitive to the underlying time-scales. This can limit the accuracy of the method.

To optimize the reliability of the results, the tracer for which the time-scale is known (i.e.~the `reference' time-scale, which in practice will often refer to the star formation tracer lifetime $\tstar$) should be chosen to be as similar as possible to the measured time-scale (which in practice will often refer to the cloud lifetime $\tgas$). This may require iteration or running multiple experiments with different reference maps (e.g.~using $\halpha$, FUV, and NUV, see \citealt{haydon18}) to converge on the best tracer to use. \autoref{fig:tgas} shows that the inaccuracy of the upward and downward-pointing triangles is not unacceptable -- the deviations from the one-to-one relation are generally less than a factor of two and there are several such symbols that lie right on top of the one-to-one relation. However, larger deviations would begin to impede astrophysical applications of the method. For practical applications, we therefore recommend an a posteriori evaluation to discard results for which the difference between $\tstar$ and $\tgas$ exceeds an order of magnitude. By contrast, time-scale measurements can be considered reliable when $|\log_{10}(\tstar/\tgas)|\leq1$.

\subsubsection{Accuracy of the `feedback time-scale' $\tover$} \label{sec:validtover}
\begin{figure*}
\includegraphics[width=\hsize]{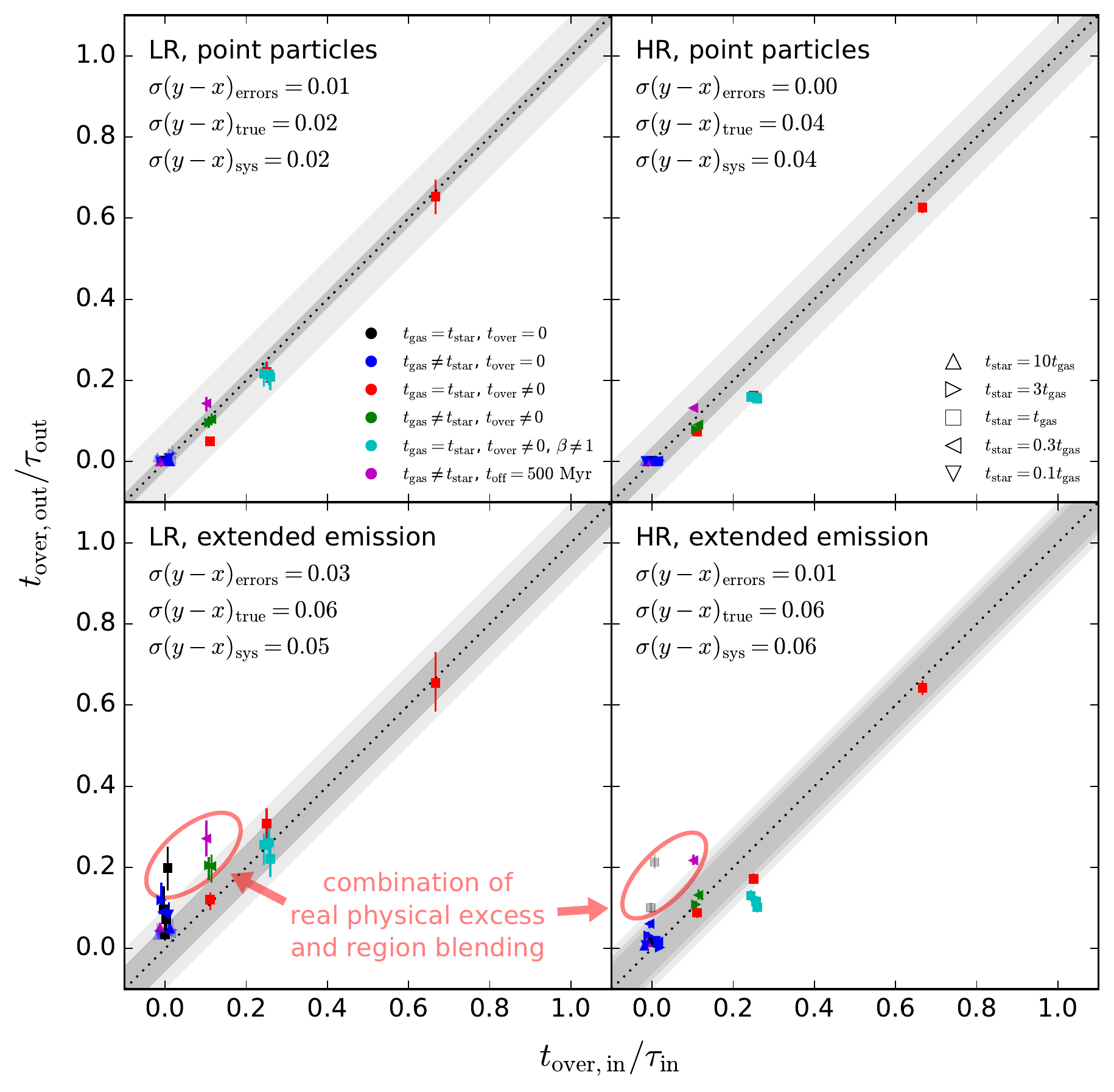}%
\vspace{-1mm}\caption{
\label{fig:tover}
Accuracy of the `overlap' phase lifetimes ($\tover$) retrieved by our new method, normalised to the total duration of the timeline ($\tau$), for the low-resolution (LR, left panels) and high-resolution (HR, right panels) simulations, using point particles (top panels) and extended emission (bottom panels). Shown is the best-fitting value $\toverout/\tauout$ as a function of the input value $\toverin/\tauin$ specified for each experiment (see \autoref{tab:starruns_in}). The data points indicate the results for each experiment (see \autoref{tab:starruns_in} and the legends in the top panels for the meanings of the symbols), with error bars indicating the 16th and 84th percentiles of the $\toverout/\tauout$ PDF. Note that only four values of $\toverin/\tauin=\{0, 0.11, 0.25, 0.67\}$ are considered in these experiments. The small horizontal offsets of the data points relative to these exact values have been added to improve the legibility of the figure. Transparent symbols do not pass the guidelines for the reliable application of our method presented in Section~\ref{sec:guide} below and should therefore be omitted from further analysis. The red ellipses and annotation in the bottom panels indicate real outliers that are not due to any inaccuracy of the method (part of this $\tover/\tau$ excess is caused by the fact that the extended emission maps contain peaks consisting of multiple particles that have a non-zero age spread, see the text). The dotted line indicates the one-to-one agreement and the grey areas illustrate different amounts of scatter. The lightest shade represents a scatter of $0.1$ (i.e.~$10$~per~cent of the timeline duration $\tau$), the middle shade represents the standard deviation of $\toverout/\tauout-\tgasin/\tauin$ (i.e.~the scatter around the diagonal) for all data points (including the unreliable, transparent ones), and the darkest shade indicates the standard deviation of $\toverout/\tauout-\tgasin/\tauin$ exclusively for the opaque (reliable) data points. Three different measures of the scatter around the one-to-one relation are given in the top left of each panel. The first (subscript `errors') is the scatter around the one-to-one relation expected for the error bars of the opaque data points (this is effectively the mean error bar). The second (subscript `true') is the actual scatter around the one-to-one relation for the opaque data points (corresponding to the darkest shade of grey). The third (subscript `sys') is the difference between the first and second values (defined as $\sigma_{\rm sys}^2=\sigma_{\rm true}^2-\sigma_{\rm errors}^2$, with the additional requirement that $\sigma_{\rm sys}^2\geq0$), which is the part of the scatter that cannot be accounted for by the error bars. As such, this represents the `systematic' inaccuracy of the method. The intrinsic accuracy of the method is $\sigma_{\rm sys}\la0.05$ (see the text), implying that $\tover$ can be determined to about 5~per~cent of $\tau$. Any $\tover<0.05\tau$ is therefore consistent with $\tover=0$.\vspace{-1mm}
}
\end{figure*}

The second test carried out using the $88$ experiments from Section~\ref{sec:starexp} is to determine how well the duration of the `overlap' phase ($\tover$) is retrieved by our method. Like $\tgas$ before, $\tover$ can have a variety of physical meanings depending on the gas and star formation tracers used. For the specific, real-Universe example of tracing (molecular) gas with CO and star formation with an ionization tracer like $\halpha$, this quantity refers to the time it takes the CO emission from a molecular cloud to disappear once the first ionising photons have started to emerge from the embedded star-forming region. We refer to this duration as the feedback time-scale. Analogously to Section~\ref{sec:validtgas}, we obtain the complete PDF of $\tover$ and show the best-fitting value of $\tover$ (referred to as $\toverout$) as a function of the input value of $\tover$ from \autoref{tab:starruns_in} (referred to as $\toverin$) for all 88 experiments in \autoref{fig:tover}. On both axes, we normalise $\tover$ to the total duration of the timeline $\tau$, thus comparing the input fraction of a region's lifetime during which it is visible in both maps to the retrieved fraction.

\paragraph{Point particles}
As in Section~\ref{sec:validtgas}, we expect the point particle experiments to yield the best results, owing to the straightforward identification of independent regions. The top panels of \autoref{fig:tover} indeed demonstrate good agreement, with standard deviations around the one-to-one relation ($\sigma_{\rm true}$) of $2$--$4$~per~cent in $\tover/\tau$. In both cases, this scatter is larger than expected based on the errors, indicating that there is a systematic uncertainty associated with the method. This uncertainty does not take the form of a systematic bias, but represents an increase of the statistical uncertainty. As previously, there are two possible sources of the added scatter. Firstly, it is possible that the substructure in the spatial distribution of regions somehow influences $\tover$ more strongly than $\tgas$ (on which the influence for the amount of substructure in these simulations is negligible, see Section~\ref{sec:validtgas}). Secondly, deviations from a constant SFR could lead to biases in the measurements of $\tover$. This should affect $\tover$ more strongly than $\tgas$, because it covers a shorter duration and thus is more susceptible to Poisson noise.

We test the hypothesis that variations in the SFR are responsible for the main outliers in the top panels of \autoref{fig:tover}. In the left-hand panel, the main outlier is experiment 13 (the red square at $\{\toverin/\tauin, \toverout/\tauout\}=\{0.11,0.05\}$), whereas in the right-hand panel, the main outliers are experiments 14, 18, 19, and 20 (the cluster of squares at $\{\toverin/\tauin, \toverout/\tauout\}=\{0.25,0.16\}$). The latter four experiments cover the same age bins and only differ in terms of the overlap-to-isolated flux ratios $\betastar$ and $\betagas$ (see Section~\ref{sec:validbeta}). The fact that these four points show an almost identical discrepancy compared to the input values while sharing the same timeline coverage strongly suggests that SFR variations are the cause of their deviation. We verify this by comparing the fraction $f_M(\tover)$ of the total stellar mass in each map formed during $\tover$ to the corresponding time fractions $\toverin/\tauin$ and $\toverout/\tauout$. If SFR variations cause the observed offsets, then $f_M(\tover)$ should match the measured value of $\toverout/\tauout$ rather than the stellar mass fraction expected for a constant SFR $\toverin/\tauin$.

For experiment 13 in the top-left panel of \autoref{fig:tover}, we find that $f_M(\tover)=0.05$, whereas for the cluster of points around experiment 14 in the top-right panel, we find that $f_M(\tover)=0.16$. Within the displayed precision, both of these numbers are identical to the retrieved values of $\toverout/\tauout$ ($0.05$ and $0.16$, respectively). Therefore, we conclude that the systematic uncertainty of $\sigma_{\rm sys}=2$--$4$~per~cent identified in the top panels of \autoref{fig:tover} is caused by deviations from a constant SFR during the covered part of the timeline. This is a real effect that will also affect observational applications of our method. Assuming that the simulations considered in this paper provide a reasonable representation of SFR variations in disc galaxies, we recommend a minimum uncertainty of $5$~per~cent in $\tover/\tau$, even if the formal error is smaller. In practice, this means that $\tover/\tau<0.05$ is consistent with no overlap and $\tover/\tau>0.95$ is consistent with completely overlapping phases.

\paragraph{Extended emission}
Contrary to the good performance of the method in retrieving $\tover/\tau$ when applying it to the point particle maps in the top panels of \autoref{fig:tover}, the bottom panels of \autoref{fig:tover} show a group of experiments with measured values of $\tover$ that are significantly higher than the input values (highlighted by the light red ellipses). This excess of $\tover$ only occurs in some (but not all) of the experiments with low $\toverin/\tauin$. While this deviation from the one-to-one relation may look like a major inaccuracy of the method at first sight, it is largely physical in nature.

When determining how well $\tover$ is retrieved from the extended emission maps discussed in \autoref{fig:starmaps}, it is important to realise that only the point particle maps may exhibit one-to-one agreement between input and output. In these maps, it is possible to control the value of $\tover$ by specifying the age ranges for which the particles are included in the map, because individual star particles have a single formation time. This does not hold for the extended emission maps, because their independent regions consist of groups of star particles that have a non-zero, intrinsic age spread -- the star formation history of each progenitor cloud is not a delta function. As a result, the descendant groups of star particles may be present in both maps even in the experiments for which we specify $\toverin=0$. Therefore, any value of $\tover$ retrieved from the extended emission maps should correspond to the sum of the number specified in \autoref{tab:starruns_in} and the age spread of each region. This explains why mostly experiments with low $\toverin/\tauin$ show the $\tover$ excess. In these experiments, the intrinsic age spread of the regions in the simulation is more likely to dominate over the imposed age spread $\toverin$.

It is important to realise that $\tover$ can only contain an intrinsic age spread for regions that consist of a sufficient number of star particles to sample this age spread. In practice, the mean number of particles per region depends on the total duration of each phase. A very rough estimate of this number can be made by dividing the number of regions in the point particle experiments by the number of regions in the extended emission experiments (cf.~the tables in Appendix~\ref{sec:appstarstar}). A minimum of two particles per region is required for both tracers to sample any age spread at all. This is only consistently achieved in experiments 3 and 4 (see \autoref{tab:starruns_in}), where the duration of both phases exceeds $30~\myr$ and the mean number of star particles per region is 4--15 for each phase. The other experiments have less than two particles per region on average, implying that the age spread is only resolved in a subset of the regions in the map. This dependence on the numerical sampling implies that the excess of $\tover$ depends on the total duration of the timeline, and is only fully included in experiments 3 and 4. For phase durations shorter than $\sim30~\myr$, the $\tover$ excess is smaller than the true age spread, because the likelihood of sampling the timeline with a sufficient number of particles to cover the full overlap phase is smaller than unity.

We reiterate that the above increase of the intrinsic age spread with the total duration of the timeline is entirely numerical in nature and results from the quantisation of the mass reservoir into particles of $1.4\times10^4~\msun$ (low resolution) and $2.7\times10^3~\msun$ (high resolution). In real star-forming regions, this type of quantisation could potentially take place in ionization-based star formation tracers such as \halpha, because such emission is generated by individual massive stars. However, such a quantisation will have a much weaker effect than in our simulations, because roughly one O-star is formed per $100~\msun$, assuming a normal IMF \citep[e.g.][]{chabrier03}. A single star particle in our high-resolution simulation would host 27 O-stars, i.e.~the sampling rate of a real star-forming region is a factor of $\sim30$ higher, meaning that an age spread like the one we see in the simulations would be well-sampled already for $\tstar\sim1~\myr$ rather than the limit of $\tstar\sim30~\myr$ that we see in the experiments. The commonly-used star formation tracers \halpha, FUV, and NUV all cover time-scales well in excess of $1~\myr$ \citep{haydon18}. Therefore, we conclude that the method is capable of accurately retrieving the overlap (or feedback) time-scale from observed (molecular) gas and star formation tracer maps.

The bottom panels of \autoref{fig:tover} contain two additional transparent data points (indicative of not satisfying the guidelines for the reliable application of our method presented in Section~\ref{sec:guide} below) relative to those in \autoref{fig:tgas}. These are the largest outliers in the bottom-right panel and correspond to experiments 3 and 4, which have large-to-extreme $\tover$ excesses of $\toverout=4.87^{+0.58}_{-0.62}~\myr$ and $\toverout=39.27^{+3.13}_{-2.74}~\myr$, respectively. Visual inspection of the maps reveals extreme blending of the regions, caused by the long timelines covered by these experiments (see \autoref{tab:starruns_in}) and the resulting high spatial density of the regions. The large discrepancy between the input and output value of $\tover$ in these experiments highlights that blending can negatively affect the accuracy of measuring $\tover$ with our method.

Interestingly, blending does not influence the accuracy of the $\tgas$ measurements in Section~\ref{sec:validtgas}, because it affects both phases of the timeline in \autoref{fig:tschem} in proportion to their number densities in the maps and therefore in proportion to their durations, resulting in the same {\it relative} change of either time-scale. Fundamentally, our method measures the relative time-scale difference between both phases, implying that the measurement of $\tgas$ is not affected by blending. However, this does not hold when measuring $\tover$, because that quantity is obtained by measuring the correlation between both phases in excess of the statistical correlation expected for a random distribution in the plane. It is easy to picture the influence of blending on this measurement with an example. If one would convolve both the stellar and `gas' maps with a point spread function that matches the size of the simulated galaxy, both phases would be fully correlated in the convolved maps and the ratio between both maps would equal the galactic average everywhere. Naturally, this should result in a duration of the overlap phase equal to the total duration of the timeline, irrespective of what the true duration of the overlap phase would have been in the unconvolved maps. This extreme example shows that blending adds to the true time overlap between the phases.

In order to minimize the effects of blending on the measurement of $\tover$ with our method, two potential effects should be accounted for. Firstly, blending can inhibit the identification of independent regions, because the blending decreases the contrast between the flux density minimum in between adjacent peaks and the peaks themselves. For a logarithmic spacing $\delta\log_{10}{\f}$ between the flux levels used to identify peaks (see Section~\ref{sec:method}), there exists a minimum distance between the peaks (in units of the peak radius) such that they can still be identified. The mathematical derivation of this distance is provided in Appendix~\ref{sec:appblending}, but here we give the ratios between the separation lengths and the peak radius needed to identify the peaks in our experiments. For the \{low, high\} resolution simulations, we use a logarithmic spacing of $\delta\f=\{0.25, 0.5\}$, which requires that $\lambda/r>\{3.14, 3.84\}$ or, when expressed in terms of the region concentration parameter $\zeta\equiv2r/\lambda$, requires that $\zeta<\{0.64, 0.52\}$.

Secondly, the blending of adjacent peaks also causes the emission of each individual region to be increased by some factor. This leads to an overestimation of the duration of the overlap phase, because blending combines the emission from regions with independent (and therefore most likely different) ages into a single region. We quantify this effect in Appendix~\ref{sec:appblending}, where we consider two adjacent, two-dimensional Gaussian peaks and calculate as a function of $\lambda/r$ which fraction of the emission on one side of the equidistance line is constituted by contamination from the neighbour. This fraction steeply decreases as a function of distance and should be lower than the desired accuracy of the $\tover$ measurement. The top panels of \autoref{fig:tover} demonstrate that the intrinsic accuracy of the method is $\sim0.05$ in $\tover/\tau$ due to deviations from a constant SFR. Therefore, it is desirable that much less than $5$~per~cent of the emission in any region comes from neighbouring emission peaks. The condition that peaks are separated by $\sim1.7$ times their FWHM (corresponding to $\lambda/r>4$ or $\zeta\equiv2r/\lambda<0.5$) results in a contamination percentage of $\sim5$~per~cent. This satisfies the above requirements and we will use this condition to determine if blending may have influenced the measurements of $\tover$. If regions are marginally resolved, this requirement translates to having a resolution (FWHM) of $\lambda/1.7$ or better. This is discussed further in Section~\ref{sec:lapmin}, where we carry out resolution tests designed to explore these issues further.

The above upper limits on $\zeta$ (and hence blending) successfully identify experiments 3 and 4 in the bottom-right panel of \autoref{fig:tover} as an inaccurate measurement of the duration of the overlap phase. It also successfully identifies several of the experiments using the real gas maps of the simulations in Section~\ref{sec:gasstar} below, which upon visual inspection are all clearly affected by blending as well. However, it fails to identify experiments 4 and 21 in the bottom-left panel of \autoref{fig:tover}, which also have large $\tover$ excesses due to blending. We expect that this is caused by a large radial variation of $\zeta$ within the maps, which implies that the mean $\zeta$ is not representative for a non-negligible fraction of the regions, especially towards the inner galaxy. Therefore, we recommend to always carry out a visual inspection of the maps to verify if they show any region blending. This is particularly useful because the measured value of $\zeta$ is an average over the entire map. If there is substantial spatial variation of $\zeta$ (e.g.~as a function of galactocentric radius), then it is possible that blending can only be identified by visual inspection, after which parts of the maps may be masked to obtain a more homogeneous and representative measurement of $\zeta$. We have expanded the method with a new way of filtering out emission on size scales exceeding the region separation length $\lambda$ \citep{hygate18}, which will help to avoid blending without human intervention.

\paragraph{General conclusions}
In summary, we see that the method is capable of measuring the fraction of the timeline in which both phases overlap ($\tover/\tau$) with an accuracy of about $0.05$ or $5$~per~cent. A higher accuracy could be obtained in principle, but fluctuations of the SFR imply that our assumption of a constant SFR is violated, thus limiting the attainable accuracy. We propose that the above accuracy is appropriate for disc galaxies, but will naturally be worse in systems with strongly varying SFRs, i.e.~starburst systems like dwarf galaxies or mergers. Should the star formation history of the observed field be known over the entire duration of the timeline $\tau$ (e.g.~by stellar population synthesis fitting), then it is possible to correct the time-scales retrieved by our method for variations in the SFR over the fraction of the timeline covered by the stellar phase.

The method's accuracy may be decreased if no measures are taken to avoid the blending of regions in the tracer maps used. For this reason, we recommend that $\max{(\zetastar, \zetagas)}<\zetacrit$ (to allow the regions to be identified by the clump finding algorithm; see Appendix~\ref{sec:appblending} for the derivation of $\zetacrit$ as a function of $\delta\log_{10}{\f}$) and $\max{(\zetastar, \zetagas)}<0.5$ (to avoid the contamination of peaks by nearby regions). In case of a strong spatial variation of $\zeta$ across either of the maps, we recommend to carry out a visual inspection to verify the degree of blending and to mask the maps if necessary, or to filter out diffuse emission on size scales larger than $\lambda$ \citep{hygate18}. If any of these conditions is not satisfied and region blending does affect the analysis, then the measured value of $\tover$ is not necessarily incorrect, but it reflects the duration for which both tracers coexist at the blending scale. This composite value of $\tover$ may then exceed the `Lagrangian' duration of the overlap phase when following an individual region through its evolution.

As discussed at the beginning of this section, we expect that real-Universe applications of the method's ability to constrain the duration of the overlap phase will help provide insight in (molecular cloud destruction by) stellar feedback. Of course, the physical meaning of $\tover$ depends on the adopted tracers. In the example of tracing (molecular) gas with CO and star formation with $\halpha$, $\tover$ corresponds to the feedback time-scale, i.e.~the time over which excess CO emission disappears once ionising photons start to escape from a region.

In order to distinguish between various feedback mechanisms, it is essential that the accuracy of the $\tover$ measurement allows resolving the time-scale differences expected between these mechanisms. For instance, ruling out SNe in favour of early (e.g.~photoionization) feedback would require an accuracy of the order $3~\myr$. In view of the expected $\sim5$~per~cent accuracy in $\tover/\tau$, this example would necessitate the use of gas and star formation tracers that together cover $\tau<60~\myr$. This can be ensured by adopting tracers with short lifetimes, e.g.~$\halpha$ (which contributes $\tstar\sim5~\myr$, \citealt{haydon18}) to trace star formation. Contrary to some star formation tracers, the lifetime of the molecular gas tracer is not known a priori. If the molecular gas tracer [e.g.~$^{12}$CO(1--0) or $^{12}$CO(2--1)] turns out to be longer-lived than $\tgas\sim50~\myr$, one can probe feedback-driven gas dispersal by switching to a higher-density gas tracer such as $^{13}$CO(1--0) or HCN \citep[e.g.][]{leroy17}, which are expected to have much shorter lifetimes. In addition to using the time-scale associated with ionising photons to determine which feedback mechanism is responsible for molecular cloud destruction, it should be possible to distinguish between feedback mechanisms in even more detail if additional tracers can be identified that are associated with single mechanisms (see Section~\ref{sec:outlook}). Combining the above considerations, our method should then allow measurements of the feedback time-scale in galaxies with an accuracy better than $1~\myr$, which is sufficient to distinguish between several of the possible feedback mechanisms. The above distinctions between feedback mechanisms share the benefit of not depending on feedback models. However, it will also be possible to directly compare the cloud destruction time-scales predicted for individual feedback mechanisms (or combinations thereof) to measurements of $\tover$ obtained with our method.

\subsubsection{Accuracy of the overlap-to-isolated flux ratio $\beta$} \label{sec:validbeta}
Next to the gas phase lifetime $\tgas$ and the duration of the overlap phase $\tover$, our method also measures the mean flux ratio between regions in the overlap phase to those not in the overlap phase (which we refer to as being `isolated' in time rather than spatially). This overlap-to-isolated flux ratio $\beta$ is calculated for both tracers, implying that we have measurements of $\betastar$ and $\betagas$. As discussed in Section~\ref{sec:method} and \citetalias{kruijssen14}, knowledge of this flux ratio is sufficient to account for the effect of the flux evolution history of either tracer on the retrieved time-scales. In \autoref{tab:starruns_in}, all experiments with a non-zero value of $\tover$ result in a meaningful measurement of $\betastar$ and $\betagas$. For most experiments, we have set both quantities to unity, but experiments 18--20 are chosen to explore the effects of varying $\beta$ on the best-fitting quantities.

\autoref{tab:starruns_beta} shows the mean obtained values of $\betastar$ and $\betagas$ for different subsets of experiments. We divide up the experiments in the four types of \autoref{fig:starmaps}, i.e.~low-resolution point particles, high-resolution point particles, low-resolution extended emission, and high-resolution extended emission. For each of these, the mean output values of $\betastar$ and $\betagas$ are calculated for the experiments where the input values are set to unity (columns indicated by $\langle\beta\rangle_{\rm 1.0}$) and to $0.5$ (columns indicated by $\langle\beta\rangle_{\rm 0.5}$). We only consider experiments for which $\toverin>0$. As expected, the values of $\betastar$ and $\betagas$ in the point particle experiments are in excellent agreement with the input values. This shows that the method is capable of capturing the mean flux evolution of independent regions by correctly retrieving the overlap-to-isolated flux ratio, even if the error bars may be underestimated by up to a factor of 2.\footnote{We obtain this factor by considering that across eight numbers, only two or three should deviate by more than $1\sigma$, whereas currently this applies to half of the measurements. Increasing the error bars by a factor of 2 would result in two numbers deviating by more than $1\sigma$.}. For the extended emission experiments with input value $\beta=0.5$, the agreement is generally poor. However, as in Section~\ref{sec:validtover}, this is a real effect rather than a flaw in the method, because the regions in the extended emission maps consist of collections of particles. This means that it is no longer possible to control which particles reside in the overlap phase, implying that the overlap-to-isolated flux ratio of these composite regions should be in the range $\beta=0.5$--$1.0$. \autoref{tab:starruns_beta} shows that this is indeed the case to within the error bars. While this broad range clearly shows that the extended emission experiments provide no particularly strong test of the method's accuracy in retrieving the overlap-to-isolated flux ratios $\betastar$ and $\betagas$, it is encouraging that the output values are consistent with the expected range.
\begin{table}
 \centering
 \begin{minipage}{\hsize}
  \caption{Mean best-fitting values of $\beta$ using age-binned stellar maps}\label{tab:starruns_beta}\vspace{-1mm}
  \begin{tabular}{l c c c c}
   \hline
   Runs & $\langle\betastar\rangle_{\rm 1.0}$ & $\langle\betastar\rangle_{\rm 0.5}$ & $\langle\betagas\rangle_{\rm 1.0}$ & $\langle\betagas\rangle_{\rm 0.5}$\\ 
   \hline
   LR points & $0.99_{-0.02}^{+0.03}$ & $0.52_{-0.02}^{+0.04}$ & $0.94_{-0.03}^{+0.04}$ & $0.48_{-0.00}^{+0.02}$ \\ 
   HR points & $0.98_{-0.00}^{+0.01}$ & $0.50_{-0.00}^{+0.01}$ & $0.93_{-0.01}^{+0.01}$ & $0.47_{-0.00}^{+0.01}$ \\ 
   LR extended & $0.95_{-0.11}^{+0.26}$ & $0.55_{-0.05}^{+0.09}$ & $1.12_{-0.10}^{+0.39}$ & $0.87_{-0.02}^{+0.03}$ \\ 
   HR extended & $1.02_{-0.03}^{+0.03}$ & $0.74_{-0.04}^{+0.03}$ & $0.91_{-0.02}^{+0.03}$ & $0.62_{-0.04}^{+0.05}$ \\ 
   \hline
  \end{tabular} \vspace{-1mm}
 \end{minipage}
\end{table}

\subsection{Gas and stellar maps} \label{sec:gasstar}
Our second set of tests is based on maps of both the gas and the star particles in the simulations. Having established the accuracy of the method with the controlled experiments of Section~\ref{sec:starstar}, we can now turn to a more realistic set of experiments. By using the actual distributions of gas particles in the simulations and combining these with the `age-binned' stellar maps as before, we can investigate dependences that may affect practical applications of the method to observations. Specifically, we consider the effects of
\begin{enumerate}
\item
diffuse emission in the maps, i.e.~emission emerging from scales $>\lambda$;
\item
the (stellar) reference time-scale, reflecting the star formation tracer used \citep[e.g.][]{leroy12,haydon18};
\item
the volume density threshold above which the gas is visible, reflecting the gas tracer used \citep[e.g.][]{shirley15,leroy17};
\item
the spatial resolution, reflecting the observational setup and the target distance or redshift;
\item
the galaxy inclination angle;
\item
the number of independent regions in the maps, represented by the identified number of peaks.
\end{enumerate}

\subsubsection{Procedure for creating the maps} \label{sec:gasproc}
For the experiments discussed here, the age-binned maps of the star particles are generated for different age ranges in exactly the same way as described in Section~\ref{sec:starproc}. To generate maps of the gas particles in the galaxy models, we first need to define the minimum volume densities above which the particles are displayed. In the phase diagrams of \autoref{fig:phase}, vertical white lines indicate volume densities of $\rho=\{0.5,1,2\}~\cmc$ (low resolution) and $\rho=\{1,2,4\}~\cmc$ (high resolution). As shown by the hatched region, these critical densities provide a reasonable match to the minimum density for particles to become star-forming ($\rho=0.7~\cmc$) and together span roughly an order of magnitude, across which different fractions of the star-forming ISM are traced. Therefore, we use the three volume densities indicated in each panel of \autoref{fig:phase} to generate the gas particle maps, where the central values of $\rhominlr=1~\cmc$ (low resolution) and $\rhominhr=2~\cmc$ (high resolution) act as the default choices.

In addition to the above reference time-scales $\tstar$ and minimum gas volume densities $\rhomin$, we specify two other quantities that are investigated here. Firstly, the spatial resolution of the maps is specified by changing $\lapmin$ in Table~\ref{tab:input} and setting $\nap$ such that the other aperture sizes remain unchanged. For instance, to change the spatial resolution to 200~pc instead of the default 50~pc, we set $\lapmin=200~\pc$ and $\nap=6$ (instead of the default $\nap=8$) to retain the same aperture size spacing as before. Secondly, we vary the inclination angle of the galaxy maps. This is achieved by rotating the simulation output around the $x$-axis by an angle in the range $i=[0,90]^\circ$ prior to generating the maps. The remaining quantities of interest (diffuse emission fraction and number of independent regions) are not specified, but vary sufficiently across the set of experiments to draw meaningful conclusions about their influence on the results. We note that out of the quantities used to generate the maps, only $\tstar$ and $i$ are used as input parameters in the method of Section~\ref{sec:method}. The method is not informed of the values of the other quantities. Having said that, the choice of stellar and gas tracer (each associated with their own time-scale or minimum density, respectively) needs to be considered when setting the input parameters $X_{\rm star}$ and $X_{\rm gas}$ for converting between pixel values and SFR and gas mass.

Having defined the subsets of particles that are shown in each map, we generate the surface density maps of these particles as described in Section~\ref{sec:starproc}. The maps are square with a projected diameter of $20~\kpc$ and $1403$~pixels, corresponding to a projected pixel scale of $14.25~\pc$ before correcting for inclination.\footnote{A non-zero inclination results in non-circular resolution elements that are elongated in the direction of the inclination by a factor of $(\cos{i})^{-1}$, implying that the effective resolution is larger by a factor of $(\cos{i})^{-1/2}$ [see equation~(\ref{eq:pixtopc})].} As before, two types of maps are made for each galaxy model. The `point particle' surface density maps show the distribution of particles as-is, for each pixel adding up the enclosed mass of the particles and dividing by the pixel area. At the adopted numerical resolution of the simulations and the pixel scale of $14.25~\pc$, the typical number of gas particles per pixel is zero or exceeds unity -- contrary to the stellar particles, which each generally occupy their own pixel (see \autoref{fig:starmaps}). This difference arises because high-density gas particles cluster more strongly than the young star particles, which are stochastically spawned from the high-density gas particles. As a result, the point particle maps display a morphology that is broadly similar to the `extended emission' surface density maps, which account for the extended morphology expected for real star-forming regions. This extended emission is obtained by using a smoothing kernel with a smoothing length chosen to be consistent with the treatment of the gas particles in the simulations (see Section~\ref{sec:models}), i.e.~by using the 200 nearest neighbours to define the smoothing length and adopting a Wendland $C^4$ smoothing kernel \citep{dehnen12} to distribute the mass around each particle position. As before, this results in coarser resolution in low-density regions.

\begin{figure}
\includegraphics[width=\hsize]{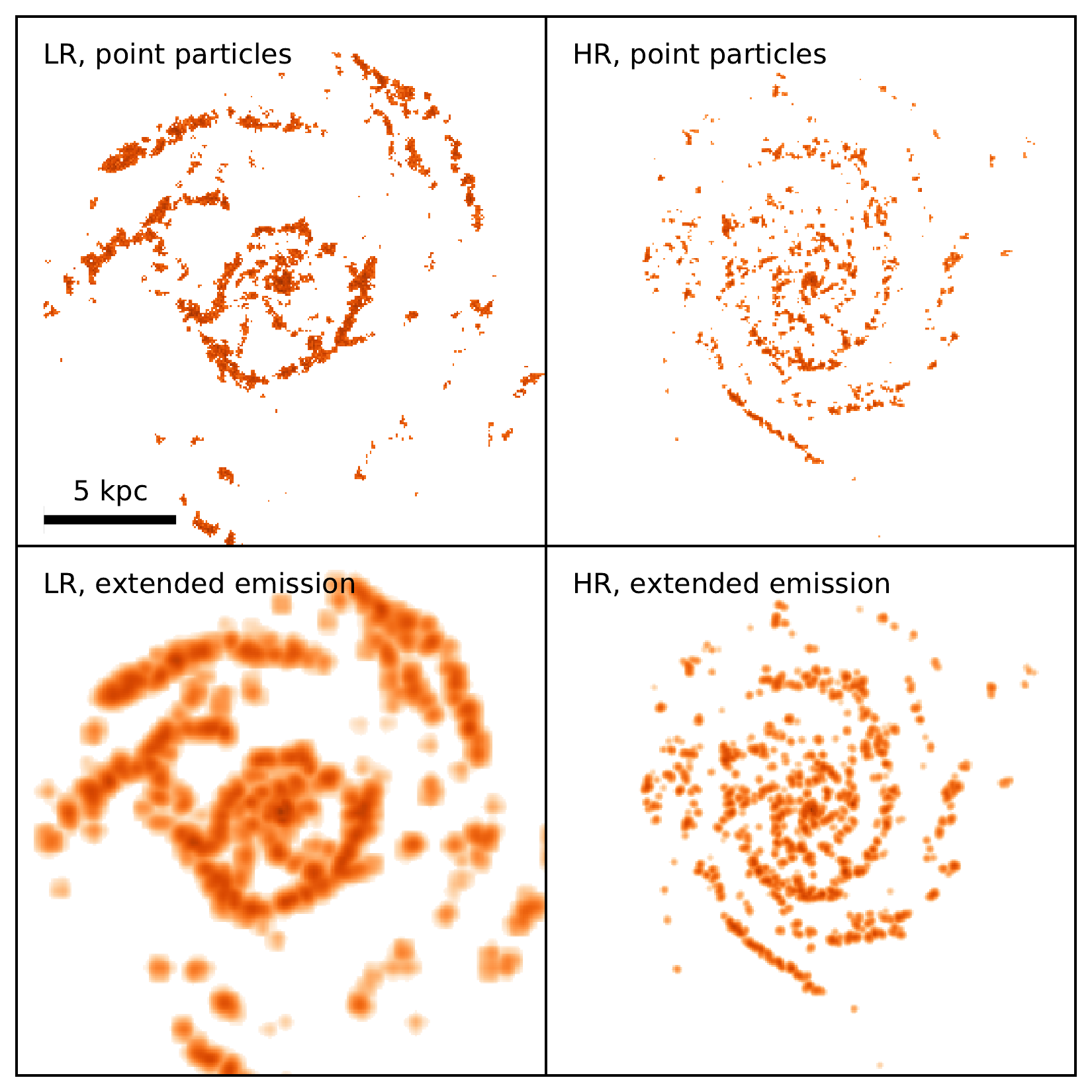}%
\vspace{-1mm}\caption{
\label{fig:gasmaps}
The four different types of `gas' maps used in Section~\ref{sec:gasstar}. From left to right, top to bottom, these are maps of the low-resolution (LR) point particle distribution, the high-resolution (HR) point particle distribution, the LR extended emission (generated by using a smoothing kernel, see the text), and the HR extended emission. In these examples, the \{LR, HR\} maps show the distribution of gas particles with densities in excess of $\rhomin=\{1,2\}~\cmc$, respectively. In Section~\ref{sec:gasexp}, we present 50 different sets of map pairs for which our method is tested, for a total of $4\times50=200$ experiments.\vspace{-1mm}
}
\end{figure}
The maps resulting from the above procedure are written to disk in the FITS file format and include a FITS header that contains their key properties, consistent with common conventions for observational data delivery. The header contains the resolution of the map to determine the minimum aperture size used (see Section~\ref{sec:stepregrid}), which is taken to be 50~per cent of the minimum smoothing length across the displayed subset of particles. This translation from the smoothing length to the resolution (or FWHM) is a reasonable approximation, because the Wendland $C^4$ kernel adopted here has an FWHM of 0.56 times the smoothing length. \autoref{fig:gasmaps} shows four examples of a typical gas map that these experiments are carried out with, illustrating the similarities and differences between the two types of maps (`point particles' and `extended emission') for both disc galaxy models (`low resolution' and `high resolution'), adopting $\rhomin=\{1,2\}~\cmc$ (for \{low, high\} resolution) and $i=0^\circ$. As expected due to the increased clustering of high-density gas particles relative to young stellar particles, the point particle and extended emission maps differ considerably less than seen in the stellar maps from Section~\ref{sec:starstar}. In the following experiments, the stellar maps have the same general properties as before (see \autoref{fig:starmaps}).

\subsubsection{Description of the experiments} \label{sec:gasexp}
\autoref{tab:gasruns_in} summarises the experiments carried out using the combination of age-binned stellar maps and gas maps above a certain critical volume density, specifying the values of $\tstar$, $i$, $\rho_{\rm min,ref}$, $\lapminlr$, and $\lapminhr$ for each experiment. Because the high-resolution simulation reaches higher gas volume densities than the low-resolution simulation, we set $\{\rhominlr,\rhominhr\}=\{1,2\}\rho_{\rm min,ref}$ for the \{low, high\}-resolution experiments. The experiments are divided into four main categories.
\begin{enumerate}
\item
The first set of experiments is aimed at investigating the role of the adopted SFR tracer through varying the reference time-scale. This is an important experiment, because the duration of the young stellar phase cannot be freely chosen in practical applications of the method -- it depends on the availability of the observational data. It is therefore necessary to identify any biases of the derived timeline as a function of $\tstar$ or $\tstar/\tgas$. By spanning $2.5~\dex$ in the reference time-scale (ranging from just $1~\myr$ in the first experiment to more than an orbital revolution in the eleventh experiment), we can assess whether the choice of reference time-scale affects the accuracy of our method.
\item
The second set of experiments investigates the choice of gas tracer by considering various minimum volume densities for the gas to be visible. For both the low and high-resolution simulations, the densities cover a factor of 4. As stated above, the minimum densities for the high-resolution simulations are taken to be twice the reference values listed here. As a result, the low and high-resolution experiments together span an order of magnitude in minimum density. For each minimum density, we consider a set of three different reference time-scales $\tstar$. These experiments are intended to show if applying the method to different gas tracers results in different time-scales, so that a multi-tracer chronology of cloud condensation and collapse can be obtained from observations.
\item
The third set of experiments probes the dependence on the observational setup and target distance or redshift by varying the spatial resolution through the minimum aperture size $\lapmin$. We consider spatial resolutions in the range $\lapmin=25$--$800~\pc$, each at three different reference time-scales $\tstar$, with the goal of determining the critical value of the size scale ratio $\lapmin/\lambda$ above which our method fails. This will limit the cosmic distances out to which the method can be applied with current and upcoming observational facilities and, hence, how representatively the galaxy population can be covered.
\item
The fourth set of experiments varies the galaxy inclination angle $i$ to identify any corresponding limits on the applicability of our method. As before, we consider three different reference time-scales for each inclination angle. This set of experiments is expected to show some degeneracy with the third set (spatial resolution), because increasing the inclination results in an increase of the effective resolution scale by a factor of $(\cos{i})^{-1/2}$.
\end{enumerate}
As outlined in Section~\ref{sec:gasproc}, we additionally investigate how the accuracy of the results changes as a function of the number of independent regions hosted in each of the maps and make a first estimate of how diffuse emission affects the results \citep[this latter dependence will be quantified in more detail by][]{hygate18}. Both of these are relevant tests for applications of our method to observations. The diffuse fraction cannot always be controlled a priori \citep[e.g.][]{pety13,calduprimo16} and obtaining an understanding of its effects is thus important. By contrast, if we can identify a minimum number of independent regions needed for reliable applications of the method, this can be an important driver for choosing the spatial coverage in observational campaigns.
\begin{table}
 \centering
 \begin{minipage}{\hsize}
  \caption{Experiments carried out using gas maps}\label{tab:gasruns_in}\vspace{-1mm}
  \begin{tabular}{c c c c c c}
   \hline
   ID & $t_{\rm star}$ & $i$ & $\rho_{\rm min,ref}$ & $\lapminlr$ & $\lapminhr$\\ 
      & $[\myr]$ & $[^\circ]$ & $[\mh~\cmc]$ & $[\pc]$ & $[\pc]$\\ 
   \hline
   $1$ & $1$ & $0.0$ & $1$ & $100$ & $50$ \\ 
   $2$ & $1.75$ & $0.0$ & $1$ & $100$ & $50$ \\ 
   $3$ & $3$ & $0.0$ & $1$ & $100$ & $50$ \\ 
   $4$ & $6$ & $0.0$ & $1$ & $100$ & $50$ \\ 
   $5$ & $10$ & $0.0$ & $1$ & $100$ & $50$ \\ 
   $6$ & $17.5$ & $0.0$ & $1$ & $100$ & $50$ \\ 
   $7$ & $30$ & $0.0$ & $1$ & $100$ & $50$ \\ 
   $8$ & $60$ & $0.0$ & $1$ & $100$ & $50$ \\ 
   $9$ & $100$ & $0.0$ & $1$ & $100$ & $50$ \\ 
   $10$ & $175$ & $0.0$ & $1$ & $100$ & $50$ \\ 
   $11$ & $300$ & $0.0$ & $1$ & $100$ & $50$ \\[1.5ex] 
   $12$ & $3$ & $0.0$ & $0.5$ & $100$ & $50$ \\ 
   $13$ & $10$ & $0.0$ & $0.5$ & $100$ & $50$ \\ 
   $14$ & $30$ & $0.0$ & $0.5$ & $100$ & $50$ \\ 
   $15$ & $3$ & $0.0$ & $2$ & $100$ & $50$ \\ 
   $16$ & $10$ & $0.0$ & $2$ & $100$ & $50$ \\ 
   $17$ & $30$ & $0.0$ & $2$ & $100$ & $50$ \\[1.5ex] 
   $18$ & $3$ & $0.0$ & $1$ & $25$ & $25$ \\ 
   $19$ & $10$ & $0.0$ & $1$ & $25$ & $25$ \\ 
   $20$ & $30$ & $0.0$ & $1$ & $25$ & $25$ \\ 
   $21$ & $3$ & $0.0$ & $1$ & $50$ & $100$ \\ 
   $22$ & $10$ & $0.0$ & $1$ & $50$ & $100$ \\ 
   $23$ & $30$ & $0.0$ & $1$ & $50$ & $100$ \\ 
   $24$ & $3$ & $0.0$ & $1$ & $200$ & $200$ \\ 
   $25$ & $10$ & $0.0$ & $1$ & $200$ & $200$ \\ 
   $26$ & $30$ & $0.0$ & $1$ & $200$ & $200$ \\ 
   $27$ & $3$ & $0.0$ & $1$ & $400$ & $400$ \\ 
   $28$ & $10$ & $0.0$ & $1$ & $400$ & $400$ \\ 
   $29$ & $30$ & $0.0$ & $1$ & $400$ & $400$ \\ 
   $30$ & $3$ & $0.0$ & $1$ & $800$ & $800$ \\ 
   $31$ & $10$ & $0.0$ & $1$ & $800$ & $800$ \\ 
   $32$ & $30$ & $0.0$ & $1$ & $800$ & $800$ \\[1.5ex] 
   $33$ & $3$ & $33.6$ & $1$ & $100$ & $50$ \\ 
   $34$ & $10$ & $33.6$ & $1$ & $100$ & $50$ \\ 
   $35$ & $30$ & $33.6$ & $1$ & $100$ & $50$ \\ 
   $36$ & $3$ & $48.2$ & $1$ & $100$ & $50$ \\ 
   $37$ & $10$ & $48.2$ & $1$ & $100$ & $50$ \\ 
   $38$ & $30$ & $48.2$ & $1$ & $100$ & $50$ \\ 
   $39$ & $3$ & $60.0$ & $1$ & $100$ & $50$ \\ 
   $40$ & $10$ & $60.0$ & $1$ & $100$ & $50$ \\ 
   $41$ & $30$ & $60.0$ & $1$ & $100$ & $50$ \\ 
   $42$ & $3$ & $70.5$ & $1$ & $100$ & $50$ \\ 
   $43$ & $10$ & $70.5$ & $1$ & $100$ & $50$ \\ 
   $44$ & $30$ & $70.5$ & $1$ & $100$ & $50$ \\ 
   $45$ & $3$ & $80.4$ & $1$ & $100$ & $50$ \\ 
   $46$ & $10$ & $80.4$ & $1$ & $100$ & $50$ \\ 
   $47$ & $30$ & $80.4$ & $1$ & $100$ & $50$ \\ 
   $48$ & $3$ & $87.1$ & $1$ & $100$ & $50$ \\ 
   $49$ & $10$ & $87.1$ & $1$ & $100$ & $50$ \\ 
   $50$ & $30$ & $87.1$ & $1$ & $100$ & $50$ \\ 
   \hline
  \end{tabular}\vspace{-1mm}
 \end{minipage}
\end{table}

The above sets of experiments are carried out using the `point particle' or the `extended emission' surface density maps (see Section~\ref{sec:gasproc}), both for the low and high-resolution simulations. We thus have a total of 200 experiments using pairs of age-binned stellar maps and gas maps. Some of the parameters used in these experiments deviate from the default values in Section~\ref{sec:method}. Firstly, the inclinations and spatial resolutions (i.e.~minimum aperture sizes) of the maps vary on a case-by-case basis, as shown by \autoref{tab:gasruns_in}. We therefore adjust $i$, $\lapmin$, and $\nap$ accordingly for each experiment. Secondly, we adjust the range of galactocentric radii to avoid large empty areas in the maps and set $R_{\rm max}=10~\kpc$ ($7~\kpc$) for the low-resolution (high-resolution) simulations. Thirdly, we set $\npixmin=1$ for the point particle experiments, to enable single pixels to be identified as peaks. Finally, for the low-resolution runs we choose a default resolution of $\lapminlr=100~\pc$ to remove sub-resolution apertures, as well as $\Delta\log_{10}\f=2.5$ and $\delta\log_{10}\f=0.25$ for the identification of both the stellar peaks and gas peaks, because the low-resolution maps used in this work require a finer and deeper contour level spacing to identify the relevant emission peaks. Contrary to Section~\ref{sec:starexp}, these experiments use the default value of {\tt tstar\_incl}~$=0$ (see \autoref{tab:flags}), implying that $\tstarref$ represents the duration of the isolated stellar phase. This choice is motivated by the fact that a gas concentration may form star particles with a certain age spread, implying that the total lifetime of the group of star particles spawned can exceed the duration of the stellar age bin used for generating the map.

In the remainder of this section, we will evaluate how the free parameters constrained by the method (i.e.~$\tgas$, $\tover$, and $\lambda$) are affected by changing the parameters listed in \autoref{tab:gasruns_in}, as well as by the number of independent regions. Each of these effects is discussed in a dedicated subsection and makes use of $2\times2\times50$ experiments (cf.~\autoref{tab:gasruns_in}).

\subsubsection{Expected values of the free parameters} \label{sec:expval}
Before discussing the measurements of the `cloud lifetime' $\tgas$, the `feedback time-scale' $\tover$, and the region separation length $\lambda$ in each of the experiments discussed above, we derive which values we expect to obtain based on the baryonic physics captured in the simulations. This is helpful for assessing the accuracy of the method, even if this is not quite straightforward, because the experiments discussed in this section are less controlled than those in Section~\ref{sec:starstar}. There, the duration of the entire evolutionary timeline was predefined, whereas the current set of experiments only specifies the duration of the (isolated) stellar phase $\tstarref$. The values of $\tgas$ and $\tover$ are emergent properties of the simulations. We remind the reader that the goal is not to reproduce the star formation and feedback properties of galaxies in the real Universe, but to retrieve what happens in the simulations. In terms of the sub-grid recipes for star formation and feedback, the simulations differ markedly from observational results. As such, the expected values derived here are not necessarily relevant in the context of the observational literature, but only serve as a point of reference when interpreting the experiments summarised in Section~\ref{sec:gasexp}.

The expected duration of the gas phase $\tgasexp$ is obtained by assuming that feedback from star formation destroys the dense gas that is displayed in the gas maps. Fundamentally, star formation in the simulations is possible for all gas particles above the critical density for star formation\footnote{This critical density corresponds to hydrogen particle density of $n_{\rm H}=0.5~\cmc$ and a total mass density of $\rhocrit=0.68~\mh~\cmc$ for an assumed hydrogen mass fraction of $0.76$.} $\rhocrit$ and is represented as a Poisson process that takes place on the dynamical time defined in equation~(\ref{eq:tdyn}), with a star formation efficiency $\epsilon=0.02$. These numbers result in an effective star formation time-scale of $\tdyn/\epsilon$, which is about $1.6~\gyr$ at $\rhog=\rhocrit$. This would be the expected gas lifetime if the gas maps would show all gas above the critical density, i.e.~$\rhomin=\rhocrit$. However, \autoref{tab:gasruns_in} shows that most maps have higher minimum gas densities, implying not only shorter dynamical times, but also that some star formation is taking place below the gas visibility threshold. To phrase this in terms of the Lagrangian timeline of \autoref{fig:tschem}, the gaseous progenitors of some of the young star particles never reached the gas densities high enough to have been visible in the gas map. The effect of this invisibility in the context of the formalism used here is that the expected gas lifetime is decreased in proportion to the fraction of young star particles for which the progenitors reached densities $\rhog\geq\rhomin$. Statistically, this fraction can be calculated as the probability of a gas particle to have been unaffected by star formation prior to reaching these densities.

It is important to emphasise that the conversion of gas into stars below the density threshold of the simulated gas map is unique to these simulations (as well as many other simulations in the literature) and does not apply to observed gas maps. In the real Universe, the collapse of gas clouds continues to extremely high densities, before the densest gas pockets reach their eventual end state as stars. Gas tracers with increasing critical densities therefore remain the progenitors of these stars and do not represent a phase that can simply be skipped. Indeed, the correlation between gas and star formation becomes tighter with increasing critical density \citep[e.g.][]{usero15}, up to values ($n>10^5~\cmc$, see e.g.~\citealt{leroy17}) that are three orders of magnitude higher than the critical density of a classical molecular gas tracer like CO(1-0). Of course, it may be that the gas tracer is not detected (e.g.~at low metallicity), but that is a matter of sensitivity and can be alleviated with a longer observing time. It does not correspond to a fundamental absence of the gas tracer under consideration, contrary to the stochastic star formation model used in the simulations.

In Appendix~\ref{sec:apppgas}, we derive an expression for the probability of gas particles to be unaffected by star formation until reaching a density $\rhog$, under the assumption that the gas density increases with time according to the gravitational free-fall of a homogeneous sphere. This assumption is appropriate for simulations that resolve the disc scale height, which is achieved for our high-resolution simulation, but not for the low-resolution simulation. We account for this by introducing a correction factor below. Note that in the statement `unaffected by star formation', we include any of the $N_{\rm ngb}$ particles within a particle's smoothing kernel. The final expression obtained this way is
\be
\label{eq:pgas}
p_{\rm gas}(\rhog)=\left[\frac{1+\sqrt{1-(\rhocrit/\rhog)^{1/3}}}{1-\sqrt{1-(\rhocrit/\rhog)^{1/3}}}\right]^{-\sqrt{3/2}\epsilon N_{\rm ngb}} ,
\ee
which is defined for $\rhog\geq\rhocrit$ and behaves as expected: for $\rhog=\rhocrit$, we obtain $p_{\rm gas}=1$, whereas in the limit $\rhog\rightarrow\infty$, we find $p_{\rm gas}=0$.

We can now combine the gas particle `survival fraction' with the time-scale for star formation to obtain an expression for the expected lifetime of gas with densities $\rhog\geq\rhomin$
\be
\label{eq:tgasexp}
\tgasexp(\rhomin)=\left[\frac{\tdyn(\rhomin)}{\epsilon}+t_{\rm fb}(\rhomin)\right]\frac{p_{\rm gas}(\rhomin)}{f_t} ,
\ee
where we include the time expected for feedback to remove the gas once star particles have formed ($t_{\rm fb}$, as defined below). The expression also includes a correction factor $f_t\geq1$, which applies to simulations that do not resolve the disc scale height and thus have insufficient resolution to estimate the mid-plane free-fall time using a kernel average. For the high-resolution simulation, the scale height of the gas disc is marginally resolved and we set $f_t=1$, but for the low-resolution simulation this is not the case, implying that the mass distribution within the smoothing kernel is anisotropic and the kernel-averaged free-fall time overestimates the true free-fall time in the disc mid-plane. In addition, the intrinsic feedback energy of a single star particle in the low-resolution simulation is higher by a factor of 5 than that in the high-resolution simulation due to the difference in particle mass, causing feedback events from single particles (i.e.~0.5 per cent of a smoothing kernel) to be much more disruptive at fixed ambient gas pressure. This may lead to the destruction of the gas condensation by the first star particle that forms, whereas this need not be the case at higher resolution. Both of these effects mean that the expected gas lifetime is overestimated for the low-resolution simulation. We can correct for this problem by setting $f_t>1$. For the experiments based on the low-resolution simulation discussed in this section, a value $f_t=3$ provides an accurate match to the obtained results.

Using the maps at the fiducial densities (e.g.~experiment IDs 1--11) as an example, equation~(\ref{eq:tgasexp}) predicts that we should expect to find gas phase lifetimes of $\tgasexp=15~\myr$ for the low-resolution simulations and $\tgasexp=2.6~\myr$ for the high-resolution simulations. As stated previously, these values have no physical meaning, but act as a point of reference when interpreting the results of the experiments.

The expected duration of the overlap phase $\toverexp$ can be obtained analogously to the duration of the gas phase. Following \citet[equation~20]{kruijssen12d}, \citet[equation~4]{reinacampos17}, and \citet[equation~5]{pfeffer18}, we define the expected feedback time-scale as the time required for the feedback-driven energy density to overcome the ambient turbulent gas pressure, i.e.
\be
\label{eq:tfb}
t_{\rm fb}=\frac{\tsn}{2}\left(1+\sqrt{1+\frac{4\tdyn(\rhog)\sigma^2}{\chi_{{\rm fb},E}\phi_{\rm fb}\epsilon\tsn^2}}\right) ,
\ee
where $\tsn=3~\myr$ is the time after which the SN feedback is applied, $\sigma\approx15~\kms$ is the local turbulent velocity dispersion, and $\phi_{\rm fb}\approx3.2\times10^{-4}~{\rm m}^2~{\rm s}^{-3}\approx10^4~(\kms)^2~\myr^{-1}$ is a constant indicating the energy deposition rate per unit mass for a normal \citep[e.g.][]{chabrier03} stellar IMF, and $\chi_{{\rm fb},E}\approx1\times10^{-3}$ is the feedback-to-ISM coupling efficiency that we typically find for the experiments with minimum densities closest to the density threshold for star formation. As before, we account for star formation below the density threshold used to generate the gas maps, i.e.~stellar regions of which the gaseous progenitor would have been invisible at the adopted minimum volume density, by writing
\be
\label{eq:toverexp}
\toverexp=t_{\rm fb}(\rhomin)\frac{p_{\rm gas}(\rhomin)}{f_t} ,
\ee
with $p_{\rm gas}(\rhog)$ as defined in equation~(\ref{eq:pgas}). This final expression provides the expected measured duration of the overlap phase. For the maps at the fiducial densities (e.g.~experiment IDs 1--11), equation~(\ref{eq:toverexp}) predicts that we should expect to find overlap phase lifetimes of $\toverexp=1.7~\myr$ for the low-resolution simulations and $\toverexp=0.35~\myr$ for the high-resolution simulations.

Finally, the expected region separation length $\lambdaexp$ can be estimated by geometric considerations. For this, we require the SFR, the total duration of the evolutionary timeline, and the stellar mass formed per region to calculate the total number of regions, which in combination with the total area of the map gives an indication of how `crowded' the maps are and what the expected separation length is. First, we calculate the expected total number of regions across both the gas and the young stellar maps as
\be
\label{eq:nreg}
N_{\rm reg}=\frac{\tau_{\rm exp}\sfr}{N_{\rm part,sf}m_{\rm part}} ,
\ee
where we use $\sfr=0.3~\msun~\yr^{-1}$ (see Section~\ref{sec:discprops}) and set $m_{\rm part}$ according to \autoref{tab:ics} for the low-resolution and high-resolution simulations. This expression assumes conservation of numbers, such that a single stellar region is formed from a single gaseous region. The expected total duration of the evolutionary timeline $\tau_{\rm exp}$ is
\be
\label{eq:ttotalexp}
\tau_{\rm exp}=\tgasexp+\tstarref ,
\ee
which does not subtract $\toverexp$ (see below), because in these experiments we have set {\tt tstar\_incl}~$=0$ (see \autoref{tab:flags}) and hence $\tstarref$ represents the duration of the isolated stellar phase (see Section~\ref{sec:gasexp}). Finally, the number of star particles formed per region ($N_{\rm part,sf}$) is obtained by combining the typical star formation efficiency per star formation event that we find in Section~\ref{sec:derivephys} ($\sfe\approx0.01$) with the number of neighbours, i.e.
\be
\label{eq:npart}
N_{\rm part,sf}=\sfe N_{\rm ngb} ,
\ee
which results in $N_{\rm part,sf}=2$ and is consistent with the idea that the first 1--2 star particles formed in a region cause its destruction by feedback. Together, equations~(\ref{eq:nreg})--(\ref{eq:npart}) provide the total number of regions expected to be present in both maps.

With the number of regions in hand, their geometrically expected mean separation length follows as
\be
\label{eq:lambdageo}
\lambda_{\rm geo}=2f_{\rm struc}\sqrt{\frac{R_{\rm max}^2-R_{\rm min}^2}{N_{\rm reg}}} ,
\ee
where $R_{\rm min}$ and $R_{\rm max}$ reflect the minimum and maximum radii of the area considered (see \autoref{tab:input}), which for the experiments discussed here are set to $R_{\rm min}=0$ and $R_{\rm max}=\{10, 7\}~\kpc$ for the \{low, high\}-resolution simulations. The quantity $f_{\rm struc}$ in equation~(\ref{eq:lambdageo}) is a correction factor that is necessary to account for the substructure in the map. For a random, unclustered distribution of points, we have $f_{\rm struc}=1$. However, the simulated galaxy maps contain substructure and voids (see \autoref{fig:gasmaps}), which means that the particles cluster in space and the geometric average implied by setting $f_{\rm struc}=1$ overestimates the region separation length in the direct vicinity of emission peaks. Fortunately, the value of $f_{\rm struc}$ can be constrained using the quantities $\exc_{\rm star,glob}$ and $\exc_{\rm gas,glob}$ (see Section~\ref{sec:stepfluxratios}), which represent the flux surface density contrast between a size-scale $\lambda$ and the entire map, thus providing a measure of the degree of flux clustering. Assuming that changes in the flux density with spatial scale are driven by changes in filling factor, we expect that $f_{\rm struc}$ scales with $\exc_{\rm glob}$ as
\be
\label{eq:fstruc}
f_{\rm struc}\approx\exc_{\rm glob}^{-1/2} .
\ee
In the experiments considered here, we measure mean values of $\exc_{\rm glob}=6$--$12$ for the low-resolution simulation and $\exc_{\rm glob}=5$--$10$ for the high-resolution simulation, resulting in $f_{\rm struc}=0.29$--$0.41$ and $f_{\rm struc}=0.32$--$0.46$, respectively. Therefore, we adopt $f_{\rm struc}=0.35$ for the low-resolution simulation and $f_{\rm struc}=0.40$ for the high-resolution simulation.\footnote{We stress that these numbers are specific to the simulations used. It is unclear whether they would also provide a reasonable description of observed galaxies and it is undoubtedly a function of the galaxy morphology. In observational applications, the quantities $\exc_{\rm star,glob}$ and $\exc_{\rm gas,glob}$ can be used to infer the degree of spatial clustering and, through equation~(\ref{eq:fstruc}), the difference between the geometrically-expected value of the separation length for a random distribution of regions (corresponding to $f_{\rm struc}=1$) and the measured, typical separation length around emission peaks in a possibly substructured medium ($f_{\rm struc}<1$).}

For an infinitely large number of regions, $\lambda_{\rm geo}$ approaches zero. However, infinitesimally small separation lengths cannot be attained in practice due to limits imposed by the spatial resolution of the maps. As will be discussed in Section~\ref{sec:lapmin}, the retrieved separation length never drops below two resolution elements. We therefore expand the expression from equation~(\ref{eq:lambdageo}) to write the expected region separation length as
\be
\label{eq:lambdaexp}
\lambdaexp=\sqrt{\lambda_{\rm geo}^2+(2\lapmin)^2} ,
\ee
in which the first term represents the geometrically expected separation length and the second term reflects the minimum separation imposed by the spatial resolution of the maps. For the experiments of \autoref{tab:gasruns_in}, the range of expected region separation lengths is $\lambdaexp=200$--$1000~\pc$ for the low-resolution simulation and $\lambdaexp=100$--$400~\pc$ for the high-resolution simulation.

The expected values of $\tgasexp$, $\toverexp$, and $\lambdaexp$ from equations~(\ref{eq:tgasexp}), (\ref{eq:toverexp}), and~(\ref{eq:lambdaexp}) will be included in the figures discussed in Sections~\ref{sec:tstar}--\ref{sec:incl}. Note that they should only be compared to measurements using the extended emission maps (cf.~\autoref{fig:gasmaps}). Any comparison to the results for the point particle maps would not be insightful, because each of the expressions depends on kernel-averaged densities and the number of neighbours used to construct the kernel. As will be shown below, the method accurately retrieves the numbers expected from the above expressions when applied to the extended emission maps.

\subsubsection{Dependence on the reference time-scale} \label{sec:tstar}
\begin{figure*}
\includegraphics[width=\hsize]{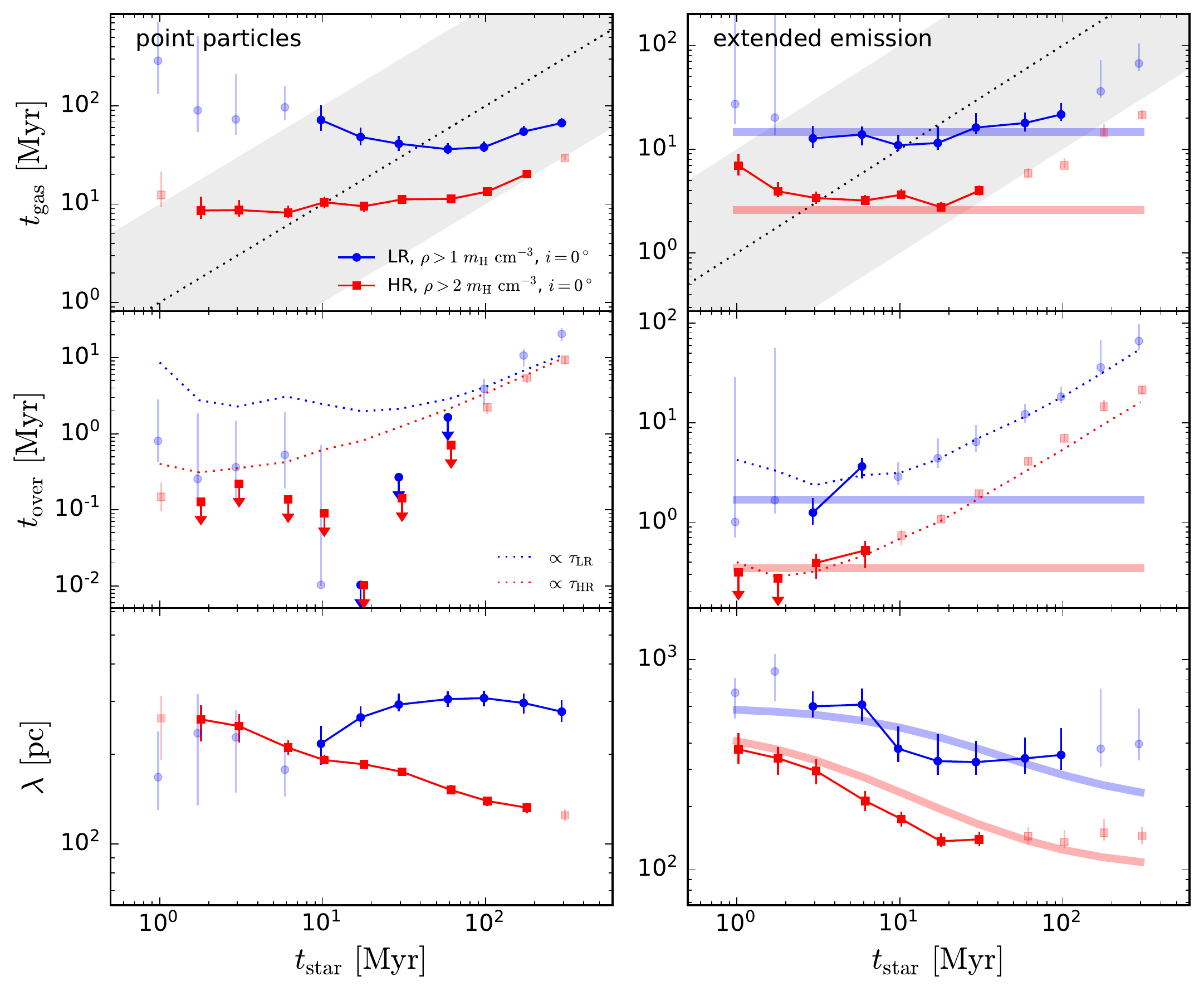}%
\vspace{-1mm}\caption{
\label{fig:tstar}
Influence of the choice of SFR tracer (captured by varying the reference time-scale $\tstar$) on the best-fitting values of the duration of the gas phase ($\tgas$, top), the duration of the overlap phase ($\tover$, middle), and the mean separation length between independent regions ($\lambda$, bottom), for experiments ID~1--11 in \autoref{tab:gasruns_in} using point particle maps (left panels) and extended emission maps (right panels). Transparent symbols represent experiments that do not satisfy each of the conditions for the reliable application of the method summarised in Section~\ref{sec:guide}. Non-transparent symbols are connected by lines as shown by the legend. In the right-hand panels, the thick blue (low resolution) and red (high resolution) lines show the expected values from Section~\ref{sec:expval}. In the top panel, the black dotted line indicates $\tgas=\tstar$ with the grey-shaded area indicating the area where both time-scales differ by less than a factor of 10 (i.e.~$|\log_{10}(\tstar/\tgas)|\leq1$). In the middle panel, downward arrows indicate upper limits (see Section~\ref{sec:guide}) and the blue (low resolution) and red (high resolution) dotted lines illustrate a proportionality to the total duration of the timeline $\tau$, as indicated by the legends. This figure shows that the (non-transparent) values of $\tgas$, $\tover$, and $\lambda$ behave as expected, in that they either show little variation with $\tstar$ or follow the predicted dependence, implying that the method can be applied reliably to constrain the cloud-scale physics of star formation and feedback using a variety of star formation tracers.\vspace{-1mm}
}
\end{figure*}
We first carry out a test using the top $4\times11=44$ experiments presented in Section~\ref{sec:gasexp} (i.e.~ID~1--11) to determine how the choice of the reference map (which sets the reference time-scale $\tstarref$) affects the constrained quantities $\tgas$, $\tover$, and $\lambda$. In Section~\ref{sec:validtgas}, we showed that the reference time-scale is important, because the method provides accurate constraints on cloud-scale star formation and feedback when $\tstar$ and $\tgas$ differ by at most an order of magnitude. Fortunately, real-Universe applications of our method have some freedom in making the choice of reference map. When using an SFR tracer to set the reference time-scale, $\halpha$, FUV, and NUV together cover characteristic time-scales from $\tstarref=5$--$50~\myr$ \citep{haydon18}. This provides a large dynamic range, implying that in many observational applications it is feasible to minimise the difference between $\tstar$ and $\tgas$. To quantify the stability of the constrained quantities across a wide dynamic range, we consider values of $\tstar=1$--$300~\myr$, which extends well beyond the limits of observational applications.

\autoref{fig:tstar} shows the best-fitting values of $\tgas$, $\tover$, and $\lambda$ for all 44 experiments considered here. As discussed in Section~\ref{sec:expval}, when calculating the expected values of the constrained quantities we make use of the particle smoothing kernels, which cover $N_{\rm ngb}=200$ neighbours. Therefore, the expected values are only shown in comparison to the extended emission experiments in the right-hand panels. Across the diagram, we see that the agreement with the expected values is good. The non-transparent symbols either show no dependence on $\tstar$ (for $\tgas$ and $\tover$) or the expected dependence (for $\lambda$). The first-order result of these experiments is therefore that the method can be applied reliably, irrespective of the choice of reference map, as long as the requirements from Section~\ref{sec:guide} below are satisfied.

Specifically, the top panels of \autoref{fig:tstar} confirm the finding of Section~\ref{sec:validtgas} that the duration of the gas phase is accurately constrained provided that $\tstar$ and $\tgas$ differ by at most an order of magnitude, which is indicated by the grey-shaded region in the figure. Experiments outside of this area have highly dissimilar durations of the (young) stellar and gas phases, which results in highly asymmetric tuning fork diagrams. In such cases, one of the two branches in the diagram closely approaches the $\bias=1$ line (corresponding to the galactic average gas-to-stellar flux ratio), where the shape of the diagram is much less sensitive to the underlying time-scales. As a result, the method becomes less accurate at constraining the time-scales and the measurements of $\tgas$ begin to deviate considerably from the expected values, in extreme cases even to the extent that $\tgas\propto\tstar$. However, as long as $|\log_{10}(\tstar/\tgas)|\leq1$, we find that $\tgas$ varies as a function of $\tstar$ by less than a factor of 2, and typically by less than 50~per cent. The most stable and reliable solutions are found when $\tstar$ and $\tgas$ differ by less than a factor of 4, or $|\log_{10}(\tstar/\tgas)|\leq0.6$. This means that, when combining the range of reference time-scales covered by $\halpha$ ($\tstarref\sim5~\myr$, \citealt{haydon18}), FUV, and NUV ($\tstarref=15$--$35~\myr$, \citealt{haydon18}) emission, the method can be applied at the highest accuracy for $\tgas=1$--$140~\myr$, which comfortably encompasses the (hotly debated) range of molecular cloud lifetimes considered in the literature \citep[cf.][]{dobbs14}.

Moving to the duration of the overlap phase ($\tover$) in the middle panels of \autoref{fig:tstar}, we see that the point particle experiments invariably provide upper limits on $\tover$. This occurs in cases where the retrieved duration occupies less than 5 per cent of the entire timeline (i.e.~$\tover/\tau<0.05$), because that percentage corresponds to the precision at which the method can constrain $\tover$ at the adopted critical region filling factor $\zeta$ (see Section~\ref{sec:validtover} and Appendix~\ref{sec:appblending}). Any value $\tover/\tau<0.05$ is therefore consistent with $\tover=0$. Indeed, the values shown in the left middle panel of the figure are very low ($\tover\sim0.1~\myr$), which is expected given that single gas and star particles can never coexist in the simulations.

The above behaviour changes when considering the extended emission experiments in the middle right panel of \autoref{fig:tstar}. In these cases, the retrieved durations of the overlap phase are longer (of the order $\tover\sim1~\myr$) and in reasonable agreement with the expected values from Section~\ref{sec:expval}. This change relative to the point particle experiments occurs because the extended emission maps are generated by smoothing the particle population over $N_{\rm ngb}=200$ neighbours, implying that we should start to see a physical time overlap between the gas and stellar phases. Interestingly, many of the symbols in this panel are transparent, particularly at $\tstar\geq10~\myr$. For these experiments, we find that the retrieved region filling factors $\zetastar$ and $\zetagas$ both exceed the critical value $\zetacrit=0.5$ above which the spatial blending of regions affects the accuracy of the $\tover$ measurement (see Section~\ref{sec:validtover}). When $\zeta>\zetacrit$, it becomes difficult (and sometimes impossible) to statistically distinguish between the time overlap of the stellar and gas phases in a single region and the spatial overlap between physically unrelated regions. The magnitude of this degeneracy is determined by the level of region crowding in the maps, which in turn is set by the total duration of the timeline $\tau$. For this reason, experiments that are strongly affected by blending result in retrieved values of $\tover$ that increase linearly with $\tstar$ and, given that these are found where $\tstar\gg\tgas$, also with $\tau$. This shows the importance of applying our recommended upper limit of $\zetacrit=0.5$. If observational applications of the method yield higher values of $\zeta$, it should be kept in mind that the uncertainty on $\tover$ will exceed $0.05\tau$, as shown in Appendix~\ref{sec:appblending}.

Finally, the constrained values of the region separation length $\lambda$ shown in the bottom panels of \autoref{fig:tstar} are in good agreement with the dependence on $\tstar$ expected from Section~\ref{sec:expval}. Overall, the separation length should decrease with increasing $\tstar$ for geometric reasons. A longer duration of the evolutionary timeline implies a larger number of regions, which in turn results in smaller separation lengths. While this behaviour is mostly followed by the experiments, there are two exceptions. The first occurs in the low-resolution point particle experiments, for which $\lambda$ increases with $\tstar$ at $\tstar<100~\myr$. This is caused by the fact that the young star particles in the low-resolution simulation are closely packed in spiral structures (see \autoref{fig:starmaps}) mirroring the gas structures from which they were born (see \autoref{fig:gasmaps}). Over time, the older star particles disperse and occupy the large voids outside the dense birth sites, increasing the distance to their neighbours. This results in an increasing trend of $\lambda$ with $\tstar$ that is only reversed when the voids get crowded and $\lambda$ decreases for geometric reasons, as expected. The high-resolution and extended emission experiments do not follow this same trend reversal, because either the star formation is much more distributed and the maps are lacking major voids (in the high-resolution simulation), or the closely-packed young star particles are merged to constitute single regions by the use of a smoothing kernel when generating the extended emission maps.

The second case of $\lambda$ (gently) increasing with $\tstar$ occurs in the extended emission experiments at large values of $\tstar$. This reflects a crowding of the regions. By increasing $\tstar$, the space between the regions has vanished and the overlapping envelopes of the regions give rise to a diffuse emission reservoir, within which single regions become increasingly hard to distinguish. As a result, the retrieved value of $\lambda$ increases and the accuracy of all constrained quantities is negatively affected. In Section~\ref{sec:diffuse}, we provide a quantitative criterion to identify applications of the method that are affected by diffuse emission reservoirs. Illustrative examples in \autoref{fig:tstar} are the two low-resolution, extended emission experiments at $\tstar>100~\myr$, which are therefore shown as transparent symbols.

In summary, the experiments summarised in \autoref{fig:tstar} show that the method can be reliably applied across a broad range of reference time-scales. Primarily, $\tstar$ and $\tgas$ should not differ by more than an order of magnitude due to the decreasing sensitivity of the method to the underlying timeline towards larger differences. It can be verified after applying the method whether this requirement is satisfied. For typical SFR tracers such as $\halpha$ ($\tstarref\sim5~\myr$, \citealt{haydon18}), FUV, and NUV ($\tstarref=15$--$35~\myr$, \citealt{haydon18}) emission, this requirement should not be problematic, as together the tracers cover a wide range of gas phase durations that would be deemed reliable, i.e.~$\tgas=0.5$--$50~\myr$ and $\tgas=1.5$--$350~\myr$, respectively. Next to this immediate requirement of the method that (somewhat) limits the range of $\tstar$, changing the duration of $\tstar$ can have secondary implications that negatively affect the results, mostly due to region crowding and blending. These effects are captured not in intrinsic limits on $\tstar$, but on other quantities, such as the region filling factor $\zeta$ (see Section~\ref{sec:validtover}) and the shape of the tuning fork diagram $\bias$ (see Section~\ref{sec:diffuse}). All requirements for reliable applications of the method are summarised in Section~\ref{sec:guide} below.

\subsubsection{Dependence on the gas density threshold} \label{sec:rhomin}
\begin{figure*}
\includegraphics[width=\hsize]{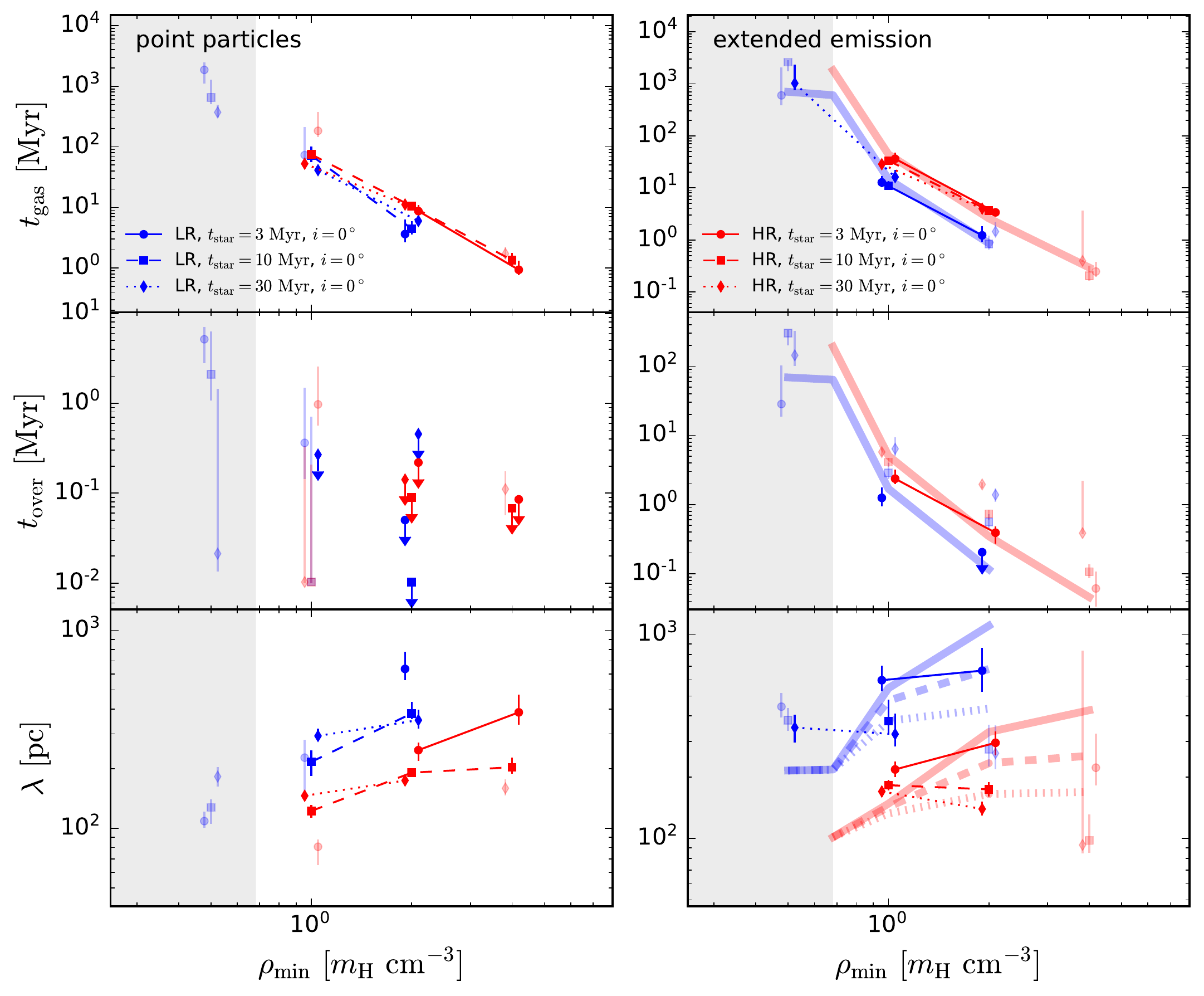}%
\vspace{-1mm}\caption{
\label{fig:densmin}
Influence of the choice of gas tracer (captured by varying the minimum gas volume density $\rhomin$) on the best-fitting values of the duration of the gas phase ($\tgas$, top), the duration of the overlap phase ($\tover$, middle), and the mean separation length between independent regions ($\lambda$, bottom), for experiments ID~3, 5, 7, and 12--17 in \autoref{tab:gasruns_in} using point particle maps (left panels) and extended emission maps (right panels). Transparent symbols represent experiments that do not satisfy each of the conditions for the reliable application of the method summarised in Section~\ref{sec:guide}. Non-transparent symbols are connected by lines as shown by the legend. The grey-shaded area indicates the density range where gas particles are ineligible for star formation (see Section~\ref{sec:models}). In the right-hand panels, the thick blue (low resolution) and red (high resolution) lines show the expected values from Section~\ref{sec:expval}, with line styles chosen to match the legend. In the middle panel, downward arrows indicate upper limits (see Section~\ref{sec:guide}). This figure illustrates how a combination of gas tracers covering a variety of critical densities can be used to probe the evolutionary time-scales at different stages of the cloud lifecycle.\vspace{-1mm}
}
\end{figure*}
The next test we carry out uses $4\times9=36$ experiments from \autoref{tab:gasruns_in}, with IDs 3, 5, 7, and 12--17, to determine how the choice of gas tracer (represented by the minimum gas volume density used to construct the gas map) affects the constrained quantities $\tgas$, $\tover$, and $\lambda$. The choice of gas tracer or its excitation density is expected to strongly influence the retrieved duration of the gas phase, because higher-density gas represents a more advanced stage of collapse towards star formation. More specifically, high-density gas is short-lived in any star formation theory in which star formation proceeds on the gas free-fall time \citep[e.g.][]{elmegreen02,krumholz05,padoan11,hennebelle11,federrath12}. In the simulations considered here, star formation does indeed proceed on a free-fall time, which will lead to a decrease of $\tgas$ towards higher values of $\rhomin$. Furthermore, we showed in Section~\ref{sec:expval} that the use of minimum gas volume densities above the star formation threshold enables star particles to form from gas that could not have been visible in the gas map. The absence of visible progenitors to young stellar emission peaks leads to an additional (and even stronger) decrease of $\tgas$ and $\tover$ with $\rhomin$, through the factor $p_{\rm gas}$ in equation~(\ref{eq:tgasexp}) and~(\ref{eq:toverexp}), which is propagated into the expected region separation length $\lambda$ through equations~(\ref{eq:nreg})--(\ref{eq:lambdaexp}). This strong variation of the constrained quantities as a function of the minimum gas density is physical. By varying the gas tracer such that it covers a range of different excitation densities, the method can thus be used to assign lifetimes to gas at these densities and, by inversion, to probe the gas density evolution of the gas towards star formation as a function of absolute time. In order to test this behaviour, we consider an order of magnitude range in $\rhomin$, by using $\rhominlr=0.5$--$2~{\rm m}_{\rm H}~\cmc$ and $\rhominhr=1$--$4~{\rm m}_{\rm H}~\cmc$.

In \autoref{fig:densmin}, we test if the dependence of the constrained quantities $\tgas$, $\tover$, and $\lambda$ on the minimum gas density of the gas map follows the relations predicted in Section~\ref{sec:expval}. As before in Section~\ref{sec:tstar}, we are only including the expected values in the right-hand panels showing the extended emission experiments. Across the three panels, we find remarkably good agreement between the values retrieved using the presented methods and the expected values. To first order, we see that the method performs well in tracing the gas density evolution as a function of absolute time. This is encouraging in the light of the wide variety densities probed by gas tracers that are readily detectable with ALMA in nearby galaxies \citep[e.g.][]{shirley15,leroy17}.

Specifically, the top panels of \autoref{fig:densmin} show that the expected, steep dependence of $\tgas$ on $\rhomin$ is quantitatively reproduced by the method. There is remarkably little difference between the point particle and extended emission experiments, which highlights that at the high densities probed here, the individual gas particles within a single region are so crowded (see \autoref{fig:gasmaps}) that they are not separated, but together constitute a region in the point particle maps. The small vertical offset between the low-resolution and high-resolution measurements is attributed to a numerical resolution effect in Section~\ref{sec:expval}.

The middle panels of \autoref{fig:densmin} exhibit contrasting trends. In the point particle experiments (middle left panel), the upper limits on the data points show that the duration of the overlap phase $\tover$ is very short ($\tover<1~\myr$), which may be expected given that a single star particle never coexists with its parent gas particle. However, the point-like nature of the emission peaks in these maps only fundamentally applies to the stellar maps (\autoref{fig:starmaps}), because the gas particles are so strongly spatially correlated that it is hard to discriminate individual gas particles in the gas maps (\autoref{fig:gasmaps}). It turns out that the point-like nature of the stellar maps is sufficient to maintain a short duration of the overlap phase.

The measurements of $\tover$ obtained in the extended emission experiments (middle right panel) behave quite differently to the point particle experiments, yet follow a relation with $\rhomin$ that is very similar to the gas phase lifetime in the top panels. As explained in the derivation of Section~\ref{sec:expval}, this is not surprising. Because the regions consist of large numbers of neighbours with long star formation time-scales, the expected duration of the overlap phase is long ($\sim100~\myr$) at the density threshold for star formation $\rhocrit$. At higher densities, the gaseous progenitors to young stellar regions may be lost to star formation before they reach sufficiently high densities to be visible in the gas maps. This decreases the inferred duration of the entire timeline and thus affects $\tgas$ and $\tover$ to an equal extent.

We remind the reader that, even though the steep dependence of $\tgas$ and $\tover$ on $\rhomin$ behaves as predicted in Section~\ref{sec:expval}, this is strictly a numerical effect due to the stochastic star formation algorithm used in the simulations. In observational applications of the method, any dependence of the evolutionary timeline on the gas tracer density should be considerably shallower. In real-Universe systems, gaseous regions attain higher densities as they evolve towards star formation. As long as the observational sensitivity is sufficiently high, the high-density gaseous progenitor to young stellar populations should always be detectable, because stars represent the evolutionary end state of the evolution towards increasing densities. This presents a strong contrast to the simulations, in which stars may form out of low-density gas, provided that it has a density in excess of the star formation threshold ($\rhog\geq\rhocrit$).

Despite the fundamentally different physics at play, the steep dependence of the retrieved time-scales on $\rhomin$ serves as both a warning and an exciting prospect for observational applications of the method. The warning is that the observations must be sufficiently sensitive to have been able to detect the gaseous progenitors to the young stellar emission peaks considered, as well as to have been able to detect the young stellar populations that may eventually emerge from the gas emission peaks hosted by the maps (also see the discussion in \citetalias{kruijssen14}). The exciting prospect of the density dependences shown in the top and middle panels of \autoref{fig:densmin} is that by probing gas of various densities, it is not only possible to reconstruct the density evolution as a function of time towards star formation, but also the dissociation of the dense gas during the stellar feedback phase.

Finally, the bottom two panels in \autoref{fig:densmin} show the region separation lengths retrieved by the method, which (in the bottom right panel) again show good agreement with the expected values. The naive expectation shown by the thick lines is that these should increase with $\rhomin$. At higher minimum gas densities, the emission peaks are shorter lived, implying that the gas maps should host fewer of them, in turn leading to larger separation lengths. This behaviour is indeed followed by the point particle experiments, but for the extended emission maps this expectation only applies to the experiments with short stellar reference time-scales ($\tstarref$). For $\tstarref\geq10~\myr$, the bottom right panel shows that $\lambda$ decreases with $\rhomin$. We find that this trend occurs due to the fact that higher-density gas more closely traces the spiral structure in the simulations than the low-density gas. As a result, the high-density gas is more spatially correlated, which may lead to smaller region separation lengths if the stellar regions follow the spiral structure too. Indeed they do, but the clustering of the stellar regions near spiral arms only becomes statistically noticeable if they are sufficiently numerous to sample the spiral structure well, which happens for $\tstarref\geq10~\myr$.

In summary, the experiments summarised in \autoref{fig:densmin} show that by varying the tracer (and specifically such that different characteristic excitation densities are probed), the physical dependence of the constrained quantities on the gas density can be probed. The presented examples cover an order of magnitude in gas density, which is considerably smaller than the $\sim3$ orders of magnitude in density that are plausibly spanned by observations of nearby galaxies with ALMA \citep[e.g.][]{leroy17}. Therefore, this test illustrates how the application of the method to well-chosen sets of gas tracers can provide even more insight in the cloud-scale star formation process.

\subsubsection{Dependence on the spatial resolution or distance} \label{sec:lapmin}
We now use $4\times18=72$ experiments from \autoref{tab:gasruns_in}, with IDs 3, 5, 7, and 18--32, to test how the spatial resolution of the maps (in units of the mean separation length) affects the constrained quantities $\tgas$, $\tover$, and $\lambda$. This is a critical test for future applications of the method, because any limitations placed on the spatial resolution will directly set the maximum distance up to which the method can be applied (also see Section~\ref{sec:dist} below). With high-sensitivity interferometers like ALMA being operational and 30-metre class optical telescopes with adaptive optics coming online in the near future, the application of the method out to large distances using high-resolution imaging presents a logical research avenue. To define a point of reference for these applications, we convolve the simulated maps to a wide range of spatial resolutions that are defined by the minimum aperture size, covering $\lapmin=25$--$800~\pc$ or $1.5~\dex$ in PSF size.

\begin{figure*}
\includegraphics[width=\hsize]{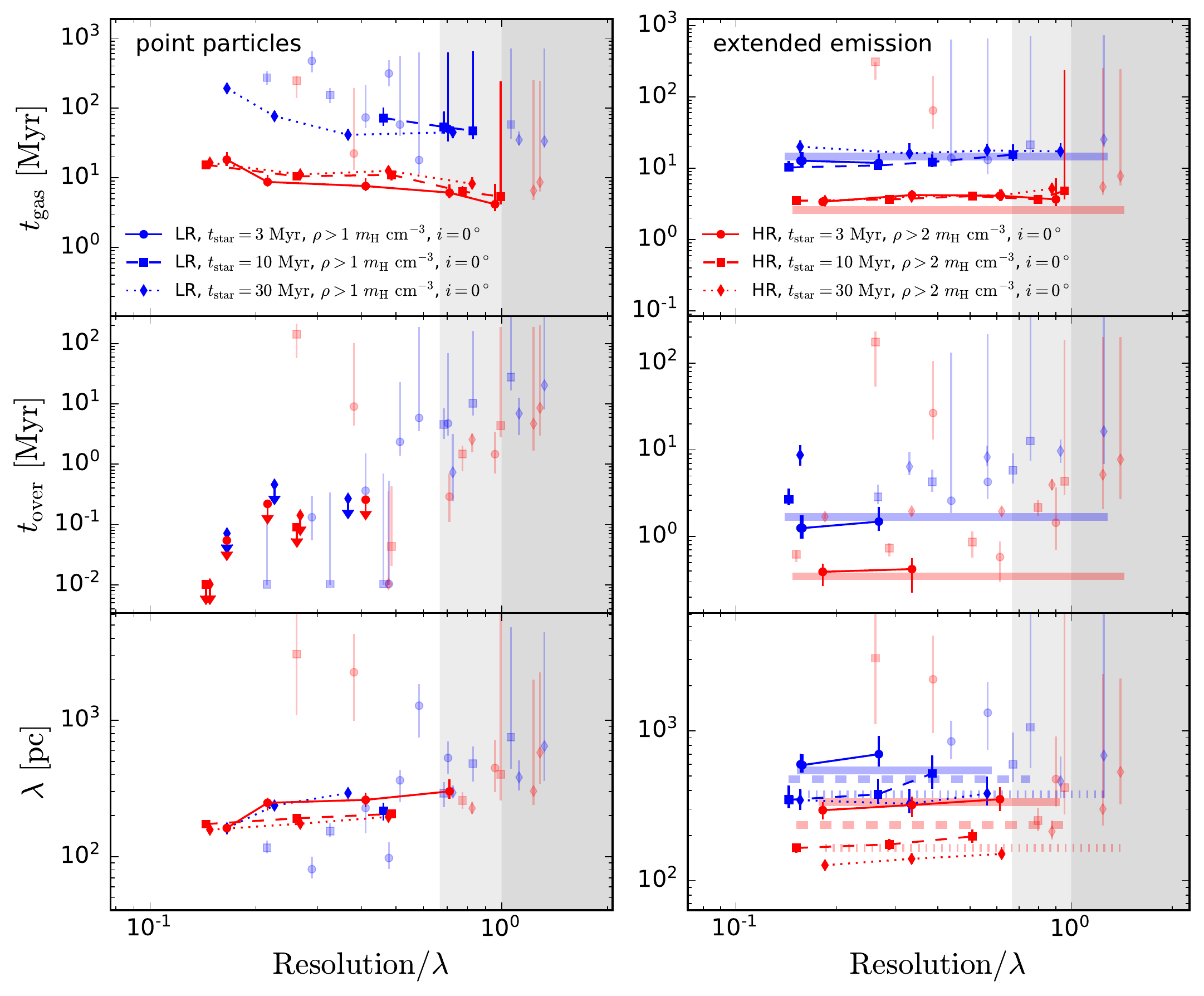}%
\vspace{-1mm}\caption{
\label{fig:minres}
Influence of the physical spatial resolution (FWHM or $\lapmin$) in units of the mean separation length on the best-fitting values of the duration of the gas phase ($\tgas$, top), the duration of the overlap phase ($\tover$, middle), and the mean separation length between independent regions ($\lambda$, bottom), for experiments ID~3, 5, 7, and 18--32 in \autoref{tab:gasruns_in} using point particle maps (left panels) and extended emission maps (right panels). Transparent symbols represent experiments that do not satisfy each of the conditions for the reliable application of the method summarised in Section~\ref{sec:guide}. Non-transparent symbols are connected by lines as shown by the legend. The light grey shaded region indicates resolutions worse than $0.67\lambda$ (or $\zeta>0.56$ for unresolved point sources), above which the measurements of $\tover$ and $\lambda$ start to show systematic deviations. The dark grey shaded regions indicate resolutions worse than $\lambda$ (or $\zeta>0.84$ for unresolved point sources), above which $\tgas$ shows minor deviations. In the right-hand panels, the thick blue (low resolution) and red (high resolution) lines show the expected values from Section~\ref{sec:expval}, with line styles chosen to match the legend. In the middle panel, downward arrows indicate upper limits (see Section~\ref{sec:guide}). This figure shows that (especially) the non-transparent retrieved quantities depend only weakly on the spatial resolution, implying that the method can be reliably applied as long as the region separation length $\lambda$ is resolved by at least one (for $\tgas$) or 1.5 (for $\tover$ and $\lambda$) resolution elements.\vspace{-1mm}
}
\end{figure*}
In \autoref{fig:minres}, we test the accuracy of the method across the above range of spatial resolutions, which for the purpose of the comparison here are normalised to the region separation length $\lambda$. This is a necessary normalisation to enable a quantitative comparison between all experiments and reflects the fact that the characteristic size scale of the gas-to-stellar flux ratio bias modelled in Section~\ref{sec:stepmodel} is set entirely by the region separation length (see the top right panel of \autoref{fig:stepmodel}). Again, we display the expected values from Section~\ref{sec:expval} in the right-hand panels showing the results from the extended emission experiments. Across the three panels, we find good agreement between the values retrieved using the presented methods and the expected values, as well as little dependence of the constrained quantities on the spatial resolution, provided that $\lapmin/\lambda\ll1$. Around $\lapmin/\lambda\sim1$, some of the measurements start to show deviations from both the higher-resolution measurements and the expected values. We see that all of the experiments have $\lapmin/\lambda<1.5$, despite the fact that the largest PSF size used ($\lapmin=800~\pc$) is an order of magnitude larger than the smallest region separation length ($\lambda\approx80~\pc$). This shows that convolving the maps with a PSF much larger than the region separation length will simply result in a larger retrieved value of $\lambda$, reflecting the unsurprising result that is not possible to retrieve sub-resolution separations. Therefore, the first-order result of \autoref{fig:minres} is that the constrained quantities $\tgas$, $\tover$, and $\lambda$ do not strongly depend on the spatial resolution, as long as the region separation length is well-resolved. Below, we quantify this statement further for each of the three quantities.

Another general observation that can be made for each of $\tgas$, $\tover$, and $\lambda$, is that the results of the point particle experiments approach those of the extended emission experiments towards large $\lapmin/\lambda$. Again, this is not necessarily surprising -- the lifecycle and separation of individual particles differs from those of concentrations of particles within a smoothing kernel, but the convolution with a large PSF to worsen the resolution of the maps obscures that difference. Indeed, for $\lapmin\gg h_{\rm smooth}$, the point particle maps are visually nearly indistinguishable from the extended emission maps. For the \{low, high\} resolution simulations, we find that the differences between the point particle and extended emission experiments start to vanish at aperture sizes of $\lapmin\geq400, 200\}~\pc$. Given that typical values of $\lambda$ are of the same order at high spatial resolutions (see Appendix~\ref{sec:appexp}), we expect the results of the point particle and extended emission experiments to yield similar results for $\lapmin/\lambda>0.5$. \autoref{fig:minres} shows that this is indeed the case.

If we now first focus on the top panels in \autoref{fig:minres}, it is clear that the retrieved durations of the gas phase closely match the expected value, independently of the spatial resolution. The point particle maps return longer lifetimes than the extended emission maps, which is to be expected given that individual particles should be longer-lived than collections of particles if these regions are destroyed by stellar feedback. The lack of correlation between $\tgas$ and $\lapmin/\lambda$ is encouraging for future application of the method. Only at $\lapmin/\lambda\ga1$ (highlighted by the dark grey area), some of the data points show large upward error bars or overestimate $\tgas$ relative to the expected value. The large upward error bars are easy to understand in the context of the tuning fork diagram of \autoref{fig:tuningfork} -- at low spatial resolution, the separation of the two branches is not well-resolved. If one of the branches does not deviate strongly from the galactic average (i.e.~it is consistent with $\bias=1$ at $1$--$2\sigma$), then the retrieved duration of the gas phase may be arbitrarily long-lived (up to the galaxy-wide gas depletion time). Likewise, the measurement of longer lifetimes at large $\lapmin/\lambda$ can be understood in terms of a smaller deviation from the galactic average gas-to-stellar flux ratio at larger aperture sizes, which results in more extended evolutionary timelines. This is an undesirable outcome and shows that measurements of $\tgas$ obtained with our method should only be used if $\lapmin<\lambda$.

In the middle panels of \autoref{fig:minres}, the retrieved durations of the overlap phase ($\tover$) behave roughly as expected. In the point particle experiments, we obtain very low ($\tover<1~\myr$) upper limits, which approach the higher values found for the extended emission experiments as $\lapmin/\lambda$ approaches unity. The results of the extended emission experiments are consistent with the expectations from Section~\ref{sec:expval}. Unfortunately, most of the experiments do not pass the requirements for reliably constraining $\tover$ listed in Section~\ref{sec:guide} below, which inhibits drawing firm conclusions regarding any trends of $\tover$ with the spatial resolution of the maps. The transparent symbols mostly do not satisfy these requirements because the retrieved region filling factors ($\zeta\equiv2r/\lambda$) exceed the critical value ($\zetacrit=0.5$), implying that spatial overlap between regions obstructs constraining their temporal overlap. However, if we focus on the trustworthy experiments (shown as the opaque data points) that reach the largest $\lapmin/\lambda$, i.e.~the experiments with $\tstar=3~\myr$ (filled circles on solid lines), then the middle right panel of \autoref{fig:minres} does show that following set of unreliable (transparent) experiments remain close to the results obtained at high spatial resolution until $\lambda$ is resolved by less than 1.5 resolution elements (highlighted by the light grey area). Therefore, we recommend to disregard measurements of $\tover$ resulting from applications for which $\lapmin/\lambda>0.67$.

Finally, the bottom panels of \autoref{fig:minres} show that the region separation length $\lambda$ is well-constrained until about $\lapmin/\lambda\sim0.67$, similarly to $\tover$. As discussed above, this is not surprising. The retrieved value of $\lambda$ cannot be much smaller than a resolution element, implying that applications of the method to low-resolution maps must overestimate the region separation length. In the extreme case of PSF sizes much larger than the sub-resolution region separation length, we expect that the measured separation length satisfies $\lambda\propto\lapmin$. Indeed, this behaviour can be seen in the bottom right panel of \autoref{fig:minres} near $\lapmin/\lambda=1$. This shows that region separation lengths measured with our method should only be used if they are sufficiently well-resolved, which we quantify as $\lapmin/\lambda<0.67$ (highlighted by the light grey area).

In summary, the spatial resolution of the map is a key quantity in assessing the accuracy of measurements obtained with the method. We require $\lapmin/\lambda<1$ when measuring $\tgas$ and $\lapmin/\lambda<0.67$ when measuring $\tover$ or $\lambda$. Fortunately, these are rather weak requirements. The typical separation lengths obtained in the presented experiments (see Appendix~\ref{sec:appexp}) and in the first observational applications that we are currently carrying out (\citealt{kruijssen18}; Hygate et al.~in prep.; Schruba et al.~in prep.; Chevance et al.~in prep.; Ward et al. in prep.) are all comfortably resolved out to tens of $\mpc$ with modern observatories (see Section~\ref{sec:dist}). This implies that our method enables the systematic characterisation of the evolutionary timeline of cloud-scale star formation across the galaxy population.

\subsubsection{Dependence on the galaxy inclination} \label{sec:incl}
We now use $4\times21=84$ experiments from \autoref{tab:gasruns_in}, with IDs 3, 5, 7, and 33--50, to test how the galaxy inclination affects the constrained quantities $\tgas$, $\tover$, and $\lambda$. While the inclination angles of galaxies are weakly correlated with the orientation of the cosmic web by which they have been fed and torqued \citep[e.g.][]{white84,dekel09b,foreroromero14}, their distribution of inclination angles can be assumed to be flat for the purposes in this paper. As a result, by determining the inclination angle above which the method no longer returns accurate results, we can effectively determine which fraction of the galaxy population should be avoided in observational applications. Because galaxies are found with all inclination angles, our experiments are chosen to cover a (nearly) complete range of $i=0.0$--$87.1^\circ$. At the maximum inclination angle, the projected diameter and thickness of the simulated galaxy discs are equal.

\begin{figure*}
\includegraphics[width=\hsize]{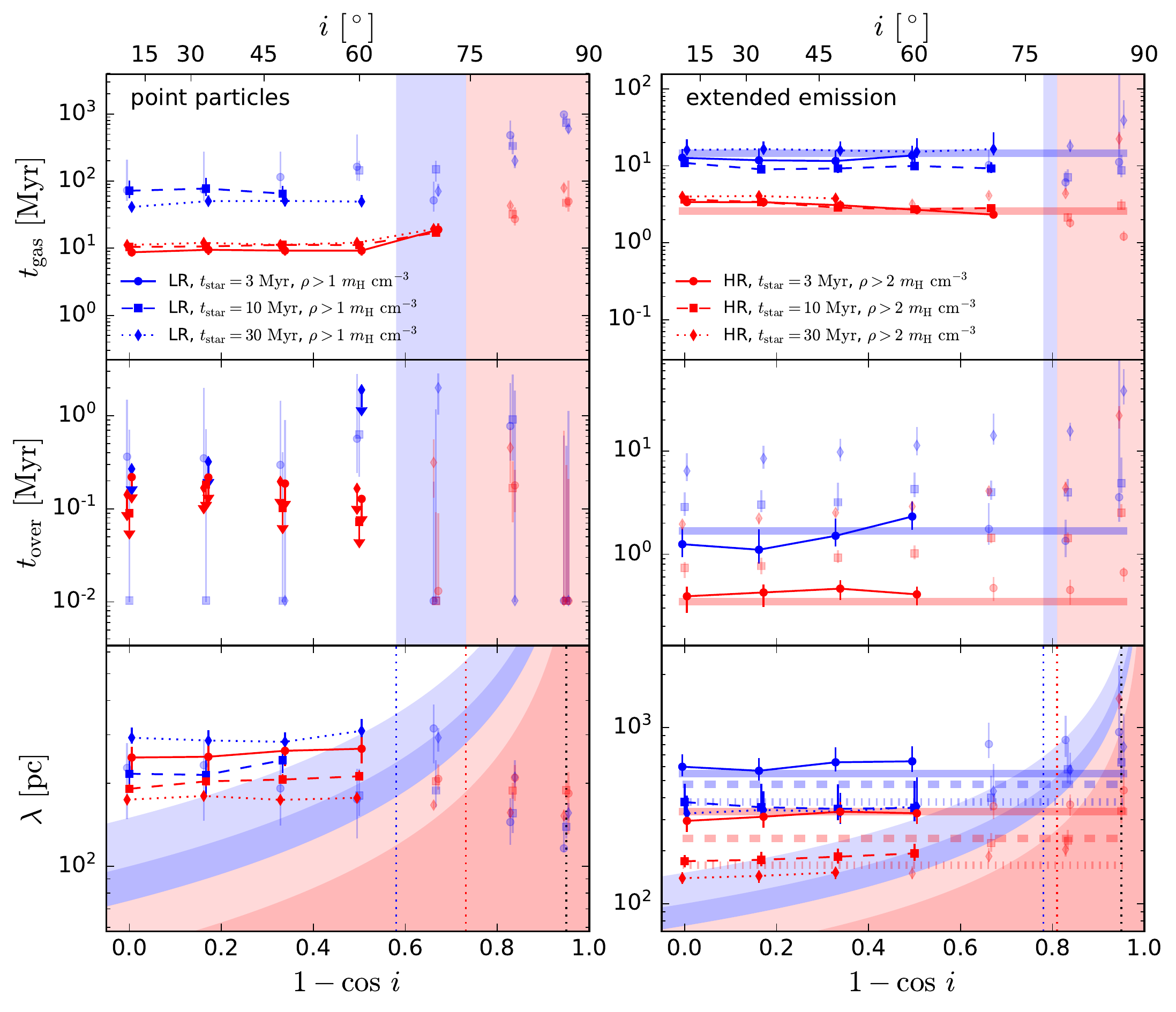}%
\vspace{-1mm}\caption{
\label{fig:incl}
Influence of the galaxy inclination angle $i$ on the best-fitting values of the duration of the gas phase ($\tgas$, top), the duration of the overlap phase ($\tover$, middle), and the mean separation length between independent regions ($\lambda$, bottom), for experiments ID~3, 5, 7, and 33--50 in \autoref{tab:gasruns_in} using point particle maps (left panels) and extended emission maps (right panels). Transparent symbols represent experiments that do not satisfy each of the conditions for the reliable application of the method summarised in Section~\ref{sec:guide}. Non-transparent symbols are connected by lines as shown by the legend. In the right-hand panels, the thick blue (low resolution) and red (high resolution) lines show the expected values from Section~\ref{sec:expval}, with line styles chosen to match the legend. In the bottom panel, the light blue (low-resolution) and red (high-resolution) shaded regions indicate resolutions worse than $0.67\lambda$ (or $\zeta>0.56$), above which the measurements of $\tover$ and $\lambda$ start to show systematic deviations. Likewise, the dark shaded regions indicate resolutions worse than $\lambda$ (or $\zeta>0.84$), above which $\tgas$ shows minor deviations. The coloured dotted vertical lines indicate the inclination angles above which the retrieved region separation length is no longer well-resolved, whereas the black dash-dotted vertical line marks the inclination angle at which the galaxy diameter and vertical thickness have the same projected dimension. In the top and middle panel, the shaded regions indicate the inclination angles for which the retrieved region separation lengths are ill-resolved. In the middle panel, downward arrows indicate upper limits. This figure shows that all of the retrieved quantities show only a weak dependence on the inclination angle, enabling reliable applications of the method up to $i\sim75^\circ$.\vspace{-1mm}
}
\end{figure*}
In \autoref{fig:incl}, we test the accuracy of the method across the above range of galaxy inclination angles, which for the purpose of the comparison here are expressed as $1-\cos{i}$. This is a useful representation, because the distances in the disc plane are scaled by a factor of $\cos{i}$ in projection. The term $\cos{i}$ is subtracted from unity to have the inclination increase towards the right (see the top axis). As before, we display the expected values from Section~\ref{sec:expval} in the right-hand panels showing the results from the extended emission experiments. Across the three panels, we find good agreement between the values retrieved using the presented methods and the expected values from Section~\ref{sec:expval}, as well as little dependence of the constrained quantities on the inclination.

This figure exhibits great similarity to \autoref{fig:minres}, in that the retrieved quantities are accurate until the galaxy is inclined so far that the region separation length is no longer resolved. In the figure, this regime is shown by the blue (low-resolution) and red (high-resolution) shaded areas -- in the bottom panels, the light shades apply to the requirement for $\tover$ and $\lambda$ to resolve the region separation length with 1.5 resolution elements, whereas the dark shades reflect the requirement for $\tgas$ that $\lambda$ is resolved with a single resolution element (see Section~\ref{sec:lapmin}). The critical inclination angles above which the projected region separation lengths are no longer resolved are $i=65$--$75^\circ$ for the point particle experiments and $i=75$--$80^\circ$ for the extended emission experiments.

In the top panels of \autoref{fig:incl}, we see that the duration of the gas phase does not depend on the inclination angle until $i>\cos^{-1}(\lapmin/\lambda)$ (indicated by the blue and red shaded areas). Above these inclinations, the scatter of the measurements increases and the retrieved values of $\tgas$ are slightly elevated relative to low inclinations. These effects are relatively minor, implying that measurements of the gas phase lifetime are robust against variations of the inclination angle. If real-Universe systems have similar region separation lengths as the experiments used here ($\lambda=100$--$400~\pc$) and are observed at a similar spatial resolution ($\lapmin=50~\pc$), then $\tgas$ can be reliably constrained up to $i=75^\circ$. More generally, we require $\lambda\geq\lapmin/\cos{i}$ when measuring $\tgas$, which generalises the result from Section~\ref{sec:lapmin} to arbitrary inclinations.

The middle panels of \autoref{fig:incl} show similar results as before in Section~\ref{sec:tstar}--\ref{sec:lapmin}. The point particle experiments return durations of the overlap phase that are short, as indicated by upper limits $\tover\la1~\myr$. Similarly to \autoref{fig:minres} showing the dependence of the constrained quantities on the spatial resolution, the opaque symbols (indicative of satisfying all requirements for the reliable application of the method listed in Section~\ref{sec:guide}) do not extend to high inclination angles in the measurements of $\tover$. Perhaps unsurprisingly, this occurs for the same reason as in \autoref{fig:minres}. Already at intermediate inclinations ($i\sim60^\circ$), the region filling factors exceed the critical value above which region blending prohibits placing accurate constraints on the duration of the overlap phase (i.e.~$2r/\lambda>0.5$). For the measurements that can be considered to be accurate, the middle right panel shows that the retrieved values of $\tover$ reproduce the expectation from Section~\ref{sec:expval}. At large inclinations, we observe a slight increase of the retrieved $\tover$, which mirrors the trend from \autoref{fig:minres}, as well as the behaviour of $\tgas$ in the top panels of \autoref{fig:incl}. Therefore, we maintain the requirement from Section~\ref{sec:lapmin} that the region separation length should be resolved by at least 1.5 resolution elements to place accurate constraints on $\tover$ and generalise it to $\lambda\geq1.5\lapmin/\cos{i}$. For $\lambda=200~\pc$ and $\lapmin=50~\pc$, this corresponds to $i\leq68^\circ$.

Finally, the bottom panels of \autoref{fig:incl} confirm the above results by demonstrating that the region separation length is insensitive to the inclination angle, provided that $\lambda$ is resolved by at least 1.5 resolution elements in projection (indicated with the light shaded blue and red areas). When this requirement is satisfied, the retrieved values of $\lambda$ are consistent with the expected values from Section~\ref{sec:expval}. In line with Section~\ref{sec:lapmin}, we thus again require $\lambda\geq1.5\lapmin/\cos{i}$, or $i\leq68^\circ$ for $\lambda=200~\pc$ and $\lapmin=50~\pc$. We note that this is a very conservative limit, because the bottom panels in \autoref{fig:incl} both do not show significant deviations from the low-inclination measurements for $i<75^\circ$.

In summary, the galaxy inclination angle only weakly affects the applicability of the method. For the dimensions and properties of the simulated galaxies considered here, we require $i\leq75^\circ$ (or generally $\lambda\geq\lapmin/\cos{i}$) when measuring $\tgas$ and $i\leq68^\circ$ (or generally $\lambda\geq1.5\lapmin/\cos{i}$) when measuring $\tover$ or $\lambda$. Assuming a flat distribution of inclination angles, this implies that future applications of our method are not noticeably affected by the galaxy inclination for $\sim80$~per cent of the galaxy population. In conjunction with the resolution requirements from Section~\ref{sec:lapmin}, this means that the method is readily applicable to a statistically significant galaxy sample. We quantify this statement in Section~\ref{sec:dist}.

\subsubsection{Dependence on the number of independent regions} \label{sec:npeak}
Finally, we use all $4\times50=200$ experiments from \autoref{tab:gasruns_in} to quantify how the number of identified emission peaks (related to the size of the maps in units of the region separation length) affects the error bars on the constrained quantities $\tgas$, $\tover$, and $\lambda$. At low numbers of emission peaks, the uncertainties are expected to increase due to Poisson noise. If we make the assumption that the peak identification algorithm roughly selects independent regions (see Section~\ref{sec:steppeaks}), then this test places a lower limit on the galaxy size for a given value of the region separation length. Specifically, if we define a circular, face-on disc galaxy with outer radius $R_{\rm max}$ that hosts a population of randomly distributed independent regions and require at least a number of emission peaks $N_{\rm min}$, this places a lower limit on the galaxy radius of
\be
\label{eq:rmax}
R_{\rm max}\geq\frac{\lambda}{2}\frac{\sqrt{N_{\rm min}}}{f_{\rm struc}} ,
\ee
where the factor $0<f_{\rm struc}\leq1$ accounts for deviations from a random distribution ($f_{\rm struc}=1$), as in equation~(\ref{eq:lambdageo}) of Section~\ref{sec:expval}.

Of course, the above assumptions only enable an order-of-magnitude estimate of $R_{\rm max}$. We have already seen in the preceding discussion that independent regions are not randomly distributed -- for the simulations considered here, we find $f_{\rm struc}\approx0.4$ (see Section~\ref{sec:expval}). In addition, one of the reasons for using the presented method rather than just counting emission peaks to derive their relative time-scales is that it allows us to be agnostic about what constitutes an independent region. After all, it is unknown a priori under which conditions the structures selected by peak identification algorithms correspond to independent regions. In spite of these caveats, equation~(\ref{eq:rmax}) serves as a rough indication of how identifying a minimum number of emission peaks may help select suitable target galaxies given an expected value of $\lambda$. More generally, it is of immediate empirical relevance to define a value of $N_{\rm min}$ above which the error bars on the constrained quantities are acceptably small, because a visual check of a possible target galaxy can already give an indication of the rough number emission peaks that may plausibly be identified. Across all 200 experiments in \autoref{tab:gasruns_in}, we define $N_{\rm min}$ for each experiment as
\be
\label{eq:nmin}
N_{\rm min}=\min{(N_{\rm peak,star},N_{\rm peak,gas})} ,
\ee
so that it reflects the smallest number of peaks identified between both maps. By using all experiments, we cover the largest possible range in $N_{\rm min}$, spanning nearly 2.5 orders of magnitude from $N_{\rm min}=7$--$1700$.

\begin{figure*}
\includegraphics[width=\hsize]{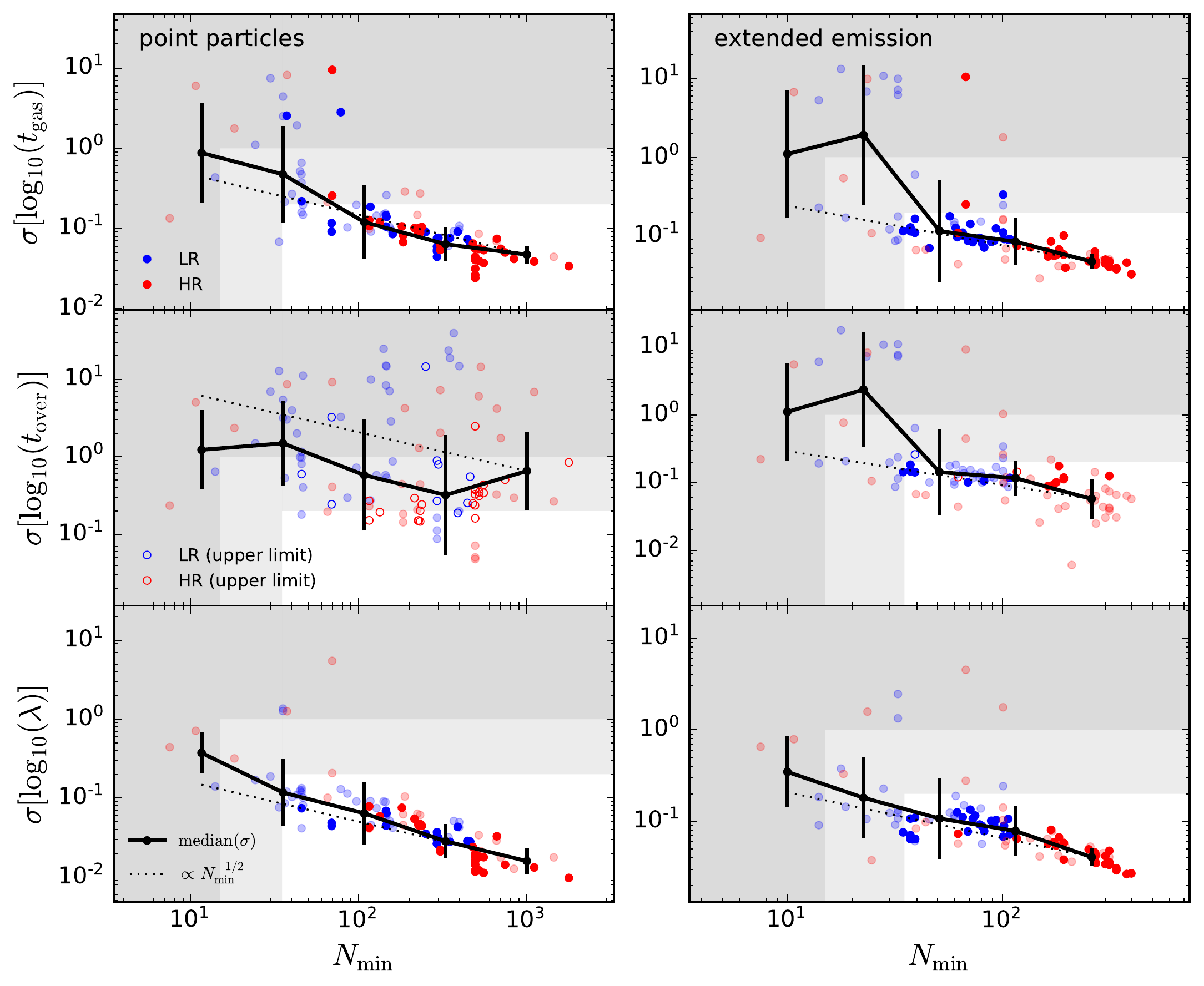}%
\vspace{-1mm}\caption{
\label{fig:npeak}
Influence of the smallest number of emission peaks in either of the two maps on the logarithmic uncertainties of the duration of the gas phase ($\tgas$, top), the duration of the overlap phase ($\tover$, middle), and the mean separation length between independent regions ($\lambda$, bottom), for all experiments in \autoref{tab:gasruns_in} using point particle maps (left panels) and extended emission maps (right panels). The uncertainties correspond to the logarithmic average of the upward and downward error for each experiment. Transparent symbols represent experiments that do not satisfy each of the conditions for the reliable application of the method summarised in Section~\ref{sec:guide}. Non-transparent symbols represent the low- and high-resolution simulations as indicated by the legends, with open symbols in the middle panels denoting upper limits. The thick black lines show the median uncertainties across five bins that are equally spaced in the logarithm of the number of peaks $N_{\rm min}$. For reference, the dotted line shows the expected trend for the standard deviation of the average, i.e.~$\propto N_{\rm min}^{-1/2}$. In the white region, the typical logarithmic uncertainties are $<0.2$ (i.e.~$<50$~per~cent), whereas the dark grey shaded area indicates where the typical uncertainty exceeds unity. The figure shows that meaningful quantitative measurements are obtained when each of the maps contains $N_{\rm min}>35$ emission peaks, whereas order-of-magnitude estimates are possible if $N_{\rm min}>15$.\vspace{-1mm}
}
\end{figure*}
In \autoref{fig:npeak}, we show the logarithmic (i.e.~relative) error bars on the retrieved quantities, taking the logarithmic average of the upward and downward error for each experiment, as a function of $N_{\rm min}$. Even if the figure can be used to determine the method's precision continuously as a function of $N_{\rm min}$, we consider two characteristic uncertainties of 50~per cent (i.e.~$0.2~\dex$, corresponding to the boundary between the white and light grey areas) and one order of magnitude (i.e.~$1~\dex$, corresponding to the boundary between the dark and light grey areas) to determine a minimum number of emission peaks. The first of these defines the number of peaks above which the evolutionary timeline of cloud-scale star formation and feedback is characterised at high precision, whereas the second limit represents a firm minimum -- while order-of-magnitude precision may be sufficient at high redshift to place first constraints on clump lifetimes, the diagnostic power of the method vanishes at even lower precision.

Across all panels in \autoref{fig:npeak} except the middle left (which is discussed below), the error bars roughly scale as the inverse square-root of the number of emission peaks $N_{\rm min}$, as expected for Poisson-dominated errors. Another similarity between the panels is the normalisation of this trend -- in nearly all panels, the median relation (thick black line) reaches the critical $0.2~\dex$ uncertainty at just below 50 emission peaks and the critical $1~\dex$ uncertainty at a little over 10 emission peaks. However, note that the median is calculated for all experiments. If we restrict ourselves to the experiments that satisfy all of the requirements listed in Section~\ref{sec:guide} (opaque symbols), we find that the number of emission peaks at which a $0.2~\dex$ uncertainty is reached can be constrained quite well, at $N_{\rm min}\approx35$. This is not possible for the $1~\dex$ uncertainty level, because none of the opaque data points have such large error bars. Using the transparent data points, we find that a $1~\dex$ uncertainty is typically reached at $N_{\rm min}\approx15$.

Exceptions exist to both of these rules of thumb. Most notably, there are four opaque data points with error bars well in excess of $1~\dex$  in the top panels. These experiments are easily identified as the data points with large error bars in the top panels of \autoref{fig:minres} and have in common that they barely resolve the region separation length. While this boosts their upward error bars, both their absolute values and downward error bars are not negatively affected by the poor resolution. The other main exception to the above minimum numbers of emission peaks is found in the middle left panel, showing the uncertainty on the duration of the overlap phase in the point particle experiments. Nearly all data points in this panel have large error bars, with a median error of about $1~\dex$. Again, this is not surprising. The retrieved values of $\tover$ in the point particle experiments are all close to zero and often represent upper limits. Physically, such an outcome is understandable, because individual gas particles in the simulation are instantaneously converted into stellar particles, without any time overlap. However, if we compare the retrieved small values of $\tover\sim0.1~\myr$ (see \autoref{fig:tstar}--\ref{fig:incl}) to its typical precision of $0.05\tau\ga1~\myr$, it is instantly clear that the typical uncertainty on $\tover$ must span an order of magnitude, even if the absolute uncertainty is still small. Therefore, the elevated error bars in the middle left panel are no fundamental shortcoming of the method, but correspond to an acceptable uncertainty.

In summary, \autoref{fig:npeak} shows that the method constrains the quantities $\tgas$, $\tover$, and $\lambda$ to a precision of $<0.2~\dex$ (or 50~per cent) if both maps contain at least $N_{\rm min}\geq35$ identified emission peaks. For order-of-magnitude estimates ($1~\dex$ precision), we require $N_{\rm min}>15$. Turning to equation~(\ref{eq:rmax}), these numbers of regions can be translated to minimum galaxy radii. We assume a region separation length of $\lambda=200~\pc$, which is reasonable in view of both the experiments carried out here and the first observational applications of the method (\citealt{kruijssen18}; Hygate et al.~in prep.; Schruba et al.~in prep.; Chevance et al.~in prep.; Ward et al.~in prep.), and adopt a typical correction factor for galactic structure of $f_{\rm struc}=0.4$ to find minimum galaxy radii of $R_{\rm max}\geq1.5~\kpc$ for $0.2~\dex$ precision and $R_{\rm max}\geq1~\kpc$ for order-of-magnitude estimates. If we characterise galaxies by their half-light radii, then the relation between stellar mass and half-light radius for local-Universe, late-type star-forming galaxies \citep{shen03} can be used to convert the above limits on galaxy radii to a limit on the galaxy stellar mass of $M_\star\ga10^9~\msun$. Practically, the above fields of view can also comfortably be covered, even with high-resolution (e.g.~$\lapmin=50~\pc$) observations on facilities like ALMA.

For a region separation length more appropriate for a high-redshift environment of $\lambda=1.5~\kpc$ (see Section~\ref{sec:dist}) and a galactic structure correction factor appropriate for galaxies without dominant spiral structure ($f_{\rm struc}=1$), we instead find minimum galaxy radii of $R_{\rm max}\geq4.4~\kpc$ for $0.2~\dex$ precision and $R_{\rm max}\geq2.9~\kpc$ for order-of-magnitude estimates. Again, such fields of view are easily achieved with current facilities. However, these numbers do show that some care should be taken in targeting sufficiently large galaxies (or galaxy samples) at high redshift. In view of the above estimates, we conclude that most galaxies of interest should host a sufficient number of emission peaks for the method to provide meaningful constraints on the cloud-scale physics of star formation and feedback.

\subsubsection{The influence of diffuse emission} \label{sec:diffuse}
Finally, we briefly comment on how the presence of a diffuse emission reservoir affects the derived evolutionary timeline. In this context, `diffuse' refers to emission on spatial scales larger than the region separation length $\lambda$. If such a reservoir is present in a given tracer, it decreases the rarity of the emission found in the identified peaks, resulting in different lifetimes for that tracer. Physically, this means that the diffuse phase is included in the timeline and therefore affects its duration. This is not necessarily incorrect, but does require that the diffuse and clumpy emission follow the same large-scale distribution throughout the maps. If diffuse emission is more prominent in one region and emission peaks can only be identified in another, the diffuse emission only contributes to the galactic average emission (obtained for large aperture sizes) and not to the peaks (which dominate at small aperture sizes). As a result, the presence of diffuse emission that is systematically displaced from emission peaks can lead to the counterintuitive situation in which focusing a small aperture on emission peaks of a certain tracer causes a flux {\it deficit} of that tracer relative to the galactic average gas-to-stellar flux ratio \citep{hygate18}. In the tuning fork diagram of \autoref{fig:tuningfork}, this would manifest itself as a branch crossing the galactic average ($\bias=1$) into the opposite half of the diagram. Therefore, quantitative applications of our method can be fundamentally obstructed by considerable amounts of diffuse emission.

To minimise the impact of diffuse emission on the constrained quantities, any tuning fork diagram showing signs of a substantial contribution from diffuse emission should be discarded. In the present work, we therefore rule out the extreme case in which focusing on an emission peak of a certain tracer results in a deficit of that tracer by more than $1\sigma$. Formally, we thus require
\be
\label{eq:diffusestar}
\max{(\bias_{\rm star}-\sigma_{\log_{10}\bias_{\rm star}})}\leq1 ,
\ee
and
\be
\label{eq:diffusegas}
\min{(\bias_{\rm gas}+\sigma_{\log_{10}\bias_{\rm gas}})}\geq1 .
\ee
These conditions can be satisfied in two ways. Firstly, there may not exist any substantial diffuse emission reservoirs in either map, which is the simplest situation and the one most conducive to the application of the method. Alternatively, diffuse emission reservoirs may exist, but they have similar contributions to both the young stellar and gas maps while having a similar spatial distribution to the clumpy emission. Such a configuration does not change the galactic average gas-to-stellar flux ratio relative to a strictly clumpy distribution and implies that the flux ratios around the identified peaks are representative for most of the emission in the map. As a result, the gas phase lifetime obtained from the tuning fork diagram is accurate, even if the presence of diffuse emission may increase the effective region filling factors $\zetastar$ and $\zetagas$ and thus lead to overestimated durations of the overlap phase. However, such cases are easily dealt with by following the guidelines for the permitted range of region filling factors presented in Section~\ref{sec:validtover}.

While the above conditions already provide a good way of avoiding any major uncertainties due to diffuse emission, we will quantify this further in \citet{hygate18} by testing how the accuracy of the quantities constrained by the method depends on the `diffuse fraction' of the maps, i.e.~the total fraction of the emission that arises on size scales larger than $\lambda$, obtained by analysing the two-dimensional Fourier transforms of the maps. We will use this to define a critical diffuse fraction above which the obtained results are no longer reliable. While the definition of such a critical value is useful for evaluating the accuracy of a measurement, it does not allow one to improve on a rejected experiment, because a map's suitability for the method is unchanged. However, a more active approach is to filter out the diffuse emission in a map and obtain the best-fitting quantities for the resulting map. This requires an iterative approach, in which the original map is used to derive $\lambda$, after which emission on scales $>\lambda$ is filtered out, and the analysis is repeated using the updated, filtered map, until convergence is reached. This process is expected to lead to convergence, because filtering a map should increase $\lambda$ and thus require less filtering during the next iteration. We have developed a module of \code that extends the method described in this paper by filtering out diffuse emission in Fourier space, enabling the application of the method to observed maps that contain substantial diffuse emission reservoirs. This optional module is described and validated in \citet{hygate18}.

\subsection{Summary: guidelines for observational applications} \label{sec:guide}
Sections~\ref{sec:starstar} and~\ref{sec:gasstar} report on how we have systematically tested the method presented in Section~\ref{sec:method} with 288 experiments, in which it has been applied to simulated galaxy maps. The goal of these experiments has been to push the method to its extremes and quantitatively assess under which circumstances the retrieved characterisation of the cloud lifecycle are still accurate. Across these tests, we have found that the resulting limits on the applicability of the method are encouraging, enabling systematic applications from the local Universe out to high redshift (see Section~\ref{sec:dist} below). We now summarise the requirements for obtaining accurate constraints on the three quantities that are fundamentally obtained with the method, i.e.~the duration of the gas phase $\tgas$, the duration of the overlap phase $\tover$, and the region separation length $\lambda$. For a detailed explanation of the origin and consequences of these requirements, we refer to the above tests. The accuracy of the additional, derived quantities is discussed in Section~\ref{sec:derivephys}.

We first list the requirements that apply to all three quantities:
\begin{enumerate}
\item[(i)]
$|\log_{10}(\tstar/\tgas)|\leq1$ (Sections~\ref{sec:validtgas} and~\ref{sec:tstar}): This condition states that the durations of the gas and stellar phases should not differ by more than an order of magnitude. For larger differences, the retrieved time-scales exhibit systematic biases.
\item[(ii)]
$\lambda\geq N_{\rm res}\lapmin/\cos{i}$ (Sections~\ref{sec:lapmin} and~\ref{sec:incl}): This condition states that the region separation length should be resolved in projection by at least $N_{\rm res}$ resolution elements. We require $N_{\rm res}=1$ for $\tgas$ and $N_{\rm res}=1.5$ for $\tover$ and $\lambda$. At insufficient resolution (or too small a separation length given the resolution), all three quantities exhibit systematic biases.
\item[(iii)]
$N_{\rm min}\geq\{15,35\}$ (Section~\ref{sec:npeak}): This condition states that the smallest number of emission peaks identified in either of the two maps should be at least 15 or 35. The identification of at least 15 emission peaks enables order-of-magnitude estimates of the constrained quantities, whereas a minimum of 35 peaks yield a precision better than $0.2~\dex$ (50 per cent). In Section~\ref{sec:npeak}, we convert the minimum number of peaks to a lower limit on the spatial extent of the target galaxy. Finally, we note that this is a soft requirement in the sense that the logarithmic precision scales continuously as $\sigma_{\log}\propto N_{\rm min}^{-1/2}$. A higher precision is achieved at a larger number of identified emission peaks.
\item[(iv)]
$\max{(\bias_{\rm star}-\sigma_{\log_{10}\bias_{\rm star}})}\leq1$ when focusing on stellar peaks and $\min{(\bias_{\rm gas}+\sigma_{\log_{10}\bias_{\rm gas}})}\geq1$ when focusing on gas peaks (Section~\ref{sec:diffuse}): This condition states that focusing on a stellar or gaseous emission peak should never lead to a deficit of that tracer relative to the galactic average. Such a deficit indicates the presence of a diffuse emission reservoir, which negatively impacts the accuracy of the derived quantities.
\item[(v)]
$|\log_{10}[\sfr(t\leq\tau)/\sfr(0)]|_{\mbox{${\rm d}t$}=\mbox{\{\tstar,\tgas\}}}\leq0.2$ (Section~\ref{sec:validtgas}): This condition states that, if available, the $\sfr$ as a function of age $t\leq\tau$ should not vary by more than $0.2~\dex$ when averaged over age intervals with a width of $\tstar$ or $\tgas$. This ensures that any bias of the retrieved $\tgas$ due to $\sfr$ variations is less than 50 per cent.
\item[(vi)]
Each independent region should be detectable in both tracer maps at some point in its lifecycle (Section~\ref{sec:rhomin}): This condition states that tracer pairs in which some regions in one tracer would never be visible in the other tracer should be avoided. This is unlikely to occur for the transition from gas to young stars traced by e.g.~CO and $\halpha$, as gas concentrations generally pass through a molecular phase before forming massive stars. However, other (rare) tracers may not be visible in all regions. In that case, the measured duration of that phase is decreased proportionally.
\end{enumerate}
Across all tests discussed in this section, we have removed any experiments that do not satisfy any of the conditions (i)--(iv). These experiments are shown as transparent symbols in the figures. In order for an experiment to be represented by a non-transparent symbol in any of the panels, it must satisfy all four conditions (i)--(iv).

For feedback-related quantities depending on $\tover$, we formulate an additional set of requirements:
\begin{enumerate}
\item[(vii)]
$\max{(\zetastar, \zetagas)}<\zeta(\delta\log_{10}{\f})$ (Section~\ref{sec:validtover} and Appendix~\ref{sec:appblending}): This condition states that the emission peak size in units of the region separation length should be small enough for the peak identification algorithm to identify adjacent peaks. If this condition is not satisfied, the value of $\tover$ is overestimated.
\item[(viii)]
$\max{(\zetastar, \zetagas)}<0.5$ (Section~\ref{sec:validtover} and Appendix~\ref{sec:appblending}): This condition states that the emission peak size in units of the region separation length should be small enough to limit the flux contamination from adjacent peaks to 5~per cent. For unresolved point sources, the peak size is set by the PSF (i.e.~$\lapmin$) and this requirement closely matches the condition of point~(ii). If this condition is not satisfied, the value of $\tover$ may be overestimated by more than the corresponding uncertainty of $0.05\tau$.
\item[(ix)]
$0.05<\tover/\tau<0.95$ (Sections~\ref{sec:validtover} and~\ref{sec:npeak}): This condition states that the duration of the overlap phase should not be too close to zero or the duration of the entire timeline. If this condition is not satisfied, then the retrieved value of $\tover$ is either an upper limit (if $\tover\tau<0.05$) or a lower limit (if $\tover/\tau>0.95$). These numbers are based on a fixed precision at which $\tover$ is measured of $0.05\tau$. However, this precision is set by $\zetastar$ and $\zetagas$ as summarised in point~(viii) and shown in Appendix~\ref{sec:appblending}, so any value of $\zeta>0.5$ will increase the 5~per cent systematic uncertainty range due to region blending.
\item[(x)]
$|\log_{10}[\sfr(t\leq\tau)/\sfr(0)]|_{\mbox{${\rm d}t$}=\tover}\leq0.2$ (Section~\ref{sec:validtover}): This condition states that, if available, the $\sfr$ as a function of age $t\leq\tau$ should not vary by more than $0.2~\dex$ when averaged over age intervals with a width of $\tover$. This ensures that any bias of the retrieved $\tover$ due to $\sfr$ variations is less than 50 per cent.
\item[(xi)]
A visual inspection of the maps does not reveal abundant region blending (Section~\ref{sec:validtover}): This condition accounts for the fact that, while the above quantitative requirements greatly increase the accuracy of $\tover$ measurements, a visual check of the maps is desirable to rule out region blending. Visibly isolated emission peaks rule out any contamination of the retrieved temporal overlap by spatial overlap.
\end{enumerate}
Across all tests discussed in this section that assess the accuracy of constraining $\tover$, we have removed any experiments not satisfying conditions (vii) and (viii), in addition to conditions (i)--(iv). These experiments are shown as transparent symbols in the figures. In order for an experiment to be represented by a non-transparent symbol in $\tover$-related panels, it must satisfy all six conditions (i)--(iv) and (vii)--(viii). If condition (ix) is not satisfied, then the quoted value of $\tover$ is considered to be a lower limit (if $\tover/\tau\leq0.05$) or an upper limit (if $\tover/\tau\geq0.95$). Throughout Section~\ref{sec:gasstar}, such cases are represented by symbols with arrows.

When all of the above conditions are satisfied, the {\it fitted} quantities ($\tgas$, $\tover$, $\lambda$) should be considered to be fully accurate for point particle maps and accurate to within $\sim0.1$~dex for extended emission maps, as shown by the various tests in this section. This refers to a systematic uncertainty and adds in quadrature to the formal logarithmic uncertainty returned by the method. The accuracy of the derived quantities varies (cf.~Section~\ref{sec:derivephys}) and is obtained through the formal error propagation of the systematic error according to the equations provided in Section~\ref{sec:stepderived}. Depending on the specific quantity, this can decrease the logarithmic accuracy by some factor $>1$. The most extreme case is the feedback energy efficiency of equations~(\ref{eq:chie}), which in the limit $\sfe\ll1$ effectively scales as $\chifbe\propto\lambda^2\tgas^{-1}\tover^{-3}$. If we assume the extreme case of zero covariance between the systematic uncertainties on $\tgas$, $\tover$, and $\lambda$, this boosts the systematic uncertainty by a factor of $\sqrt{6}\approx2.5$ relative to a single fitted quantity. We therefore conclude that, for extended emission maps, the systematic uncertainties of the {\it derived} quantities fall within a range of 0.1--0.3~dex.

The public release of \code will automatically verify that the guidelines listed in this section are satisfied upon the completion of the analysis. Any violations of these conditions are identified in the output. This evaluation also includes those guidelines for selecting appropriate input maps from Section~\ref{sec:stepreadmap} that are suitable for an automated quantitative assessment.

\section{Derived quantities describing cloud-scale star formation and feedback} \label{sec:derivephys}
The previous section showed that the three free parameters $\tgas$, $\tover$, and $\lambda$ are accurately retrieved by the method presented in Section~\ref{sec:method}. We now turn to a brief set of examples demonstrating that also the derived quantities describing cloud-scale star formation and feedback are well-constrained. This section serves as a proof of concept and is not as exhaustive as Section~\ref{sec:valid} -- rather than providing a discussion of all quantities listed in \autoref{tab:output}, we focus on five of these that we expect to be of immediate physical interest in future applications of the method. These are the radii of stellar regions $\rstar$, the radii of gas regions $\rgas$, the feedback outflow velocity $\vfb$, the star formation efficiency per star formation event $\sfe$, and the mass loading factor $\etaavgfb$. Firstly, we estimate the values that we expect to measure based on our knowledge of the simulations. Secondly, we carry out a quantitative comparison between these values and the retrieved quantities. This comparison exclusively uses the experiments combining the gas maps and young stellar maps from Section~\ref{sec:gasstar} and only considers those based on extended emission maps (cf.~\autoref{fig:gasmaps}). To avoid large spreads in the discussed quantities due to their likely physical dependence on the minimum gas density used to generate the maps, we also restrict ourselves to the maps with the fiducial values of $\rhomin$, i.e.~$\rhominlr=1~{\rm m}_{\rm H}~\cmc$ and $\rhominhr=2~{\rm m}_{\rm H}~\cmc$. This allows us to narrow down the range of expected and retrieved values and isolate the accuracy of the method in constraining the derived quantities. Finally, we only consider the experiments that satisfy the conditions for the reliable application of the method listed in Section~\ref{sec:guide}, i.e.~the opaque data points from Figures~\ref{fig:tstar}--\ref{fig:npeak}.

\subsection{Expected values of the derived quantities} \label{sec:expder}
Here, we briefly provide expressions for the expected values of the five derived quantities considered here. Analogously to the discussion in Section~\ref{sec:expval}, these cannot necessarily be obtained directly from the simulations themselves, because the minimum gas densities exceed the density threshold for star formation (i.e.~$\rhomin>\rhocrit$). As a result, only a fraction of the progenitors of the young stellar regions could have been visible in the gas maps, which changes the retrieved time-scales and other quantities from the true values in the simulation. This is strictly a numerical effect, caused by the stochastic model for star formation used in the simulations, in which stars can be spawned from low-density gas without ever going through a high-density phase. Real-Universe applications of the method will not suffer from this behaviour, because the collapse towards star formation eventually causes the gas to emit in any of the commonly-used gas tracers. Most importantly for the question at hand, we can easily accommodate how the above effect changes the retrieved quantities, such that their expected values can still be accurately predicted based on our knowledge of the simulations. In equations~(\ref{eq:tgasexp}) and (\ref{eq:toverexp}), this is done by including a factor $p_{\rm gas}(\rhog)$, which represents the fraction of gaseous regions that survives to a density $\rhog$ without forming stars. The values of $\tgasexp$, $\toverexp$, and $\lambdaexp$ that we thus predict to be retrieved using the method are accurate, as shown in Sections~\ref{sec:tstar}--\ref{sec:incl}.

The first two quantities that we consider are the radii of the young stellar and gaseous regions, denoted as $\rstar$ and $\rgas$, respectively. Out of the five quantities discussed in this section, these are the only ones that can be predicted directly from the simulations, without accounting for the numerical effects due to the stochastic star formation model. As evident from equation~(\ref{eq:prof3}), we define the region radius as its Gaussian dispersion, such that we can express the expected radius as
\be
\label{eq:rexp}
r_{\rm exp}=\frac{1}{2}\sqrt{\frac{f_{\rm Wend}^2\overline{h}_{\rm smooth}^2+\lapmin^2}{2\ln{2}}} ,
\ee
where the first term in the numerator represents the mean FWHM of the smoothing kernels of the particles from the simulation shown in the maps, the second term in the numerator represents the minimum aperture size, which sets the FWHM of the resolution at which the maps are analysed, and the denominator converts the FWHM to a Gaussian dispersion. For a Wendland $C^4$ kernel defined on the interval $0<r/h<1$, the FWHM spans a fraction $f_{\rm Wend}\equiv\Delta(r/h)=0.56$ of the smoothing length. The mean smoothing length $\overline{h}$ is determined by averaging over all gas particles with densities above the minimum density (i.e.~$\rhog\geq\rhominlr$ for the low-resolution simulation and $\rhog\geq\rhominlr$ for the high-resolution simulation), which is $h_{\rm smooth,LR}=267~\pc$ for the low-resolution simulation and $h_{\rm smooth,HR}=126~\pc$ for the high-resolution simulation. The minimum aperture size ranges from $\lapminlr=50$--$400~\pc$ for the low-resolution simulation and $\lapminhr=25$--$400~\pc$ for the high-resolution simulation. These minimum aperture sizes cover a wide range, because in Section~\ref{sec:lapmin} we changed the spatial resolution to quantify the requirements for the reliable application of the method. Together, the above numbers imply expected radii of $r_{\rm exp,LR}=67$--$181~\pc$ for the low-resolution simulation and $r_{\rm exp,HR}=32$--$172~\pc$ for the high-resolution simulation. Because the young stellar maps and the gas maps are both constructed using a Wendland $C^4$ smoothing kernel with 200 neighbours, we do not distinguish between the radii expected for both types of maps -- the retrieved values of $\rstar$ and $\rgas$ should be similar.

The third quantity that we consider is the feedback outflow velocity $\vfb$. As explained above, the expected value of this quantity cannot be estimated directly from the simulations, because the measured time-scales are affected by the stochastic star formation model used in the simulations. Fortunately, it is straightforward to express the value of $\vfb$ that we expect to retrieve with the method in terms of the expected duration of the overlap phase and the expected region separation length, i.e.
\be
\label{eq:vfbexp}
v_{\rm fb,exp}=\frac{\lambdaexp}{2\toverexp} .
\ee
Given the mean expected values of $\toverexp=\{1.7,0.35\}~\myr$ and $\lambdaexp=\{548,284\}~\pc$ for the experiments considered here (see Section~\ref{sec:expval}) based on the \{low, high\}-resolution simulations, we thus expect to measure typical feedback outflow velocities of $v_{\rm fb,exp}=\{162,408\}~\kms$, where we have taken the mean velocity over all experiments.

\begin{figure*}
\includegraphics[width=\hsize]{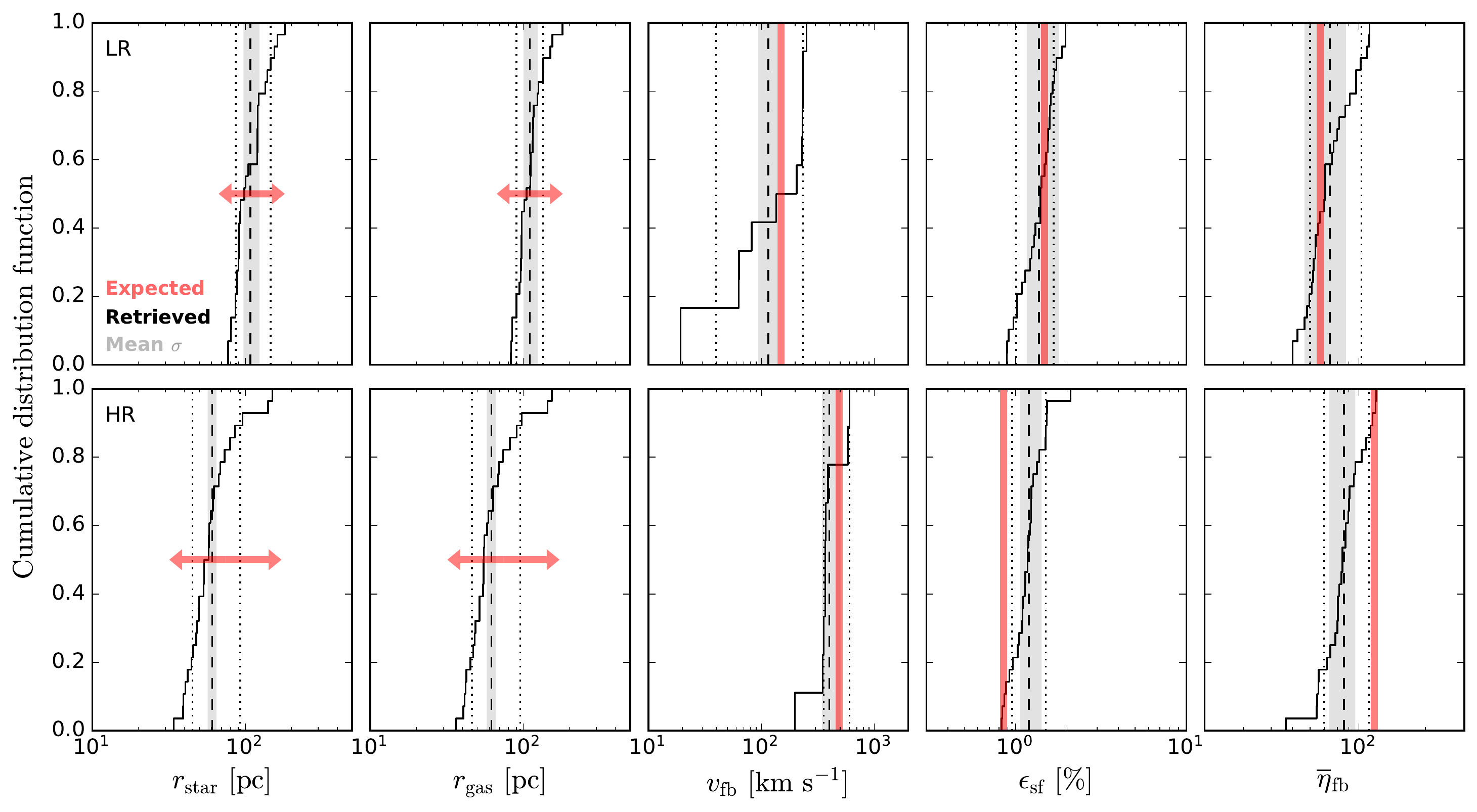}%
\vspace{-1mm}\caption{
\label{fig:derived}
Comparison between five `derived' quantities describing cloud-scale star formation and feedback and the expected values based on the simulations (as described in Section~\ref{sec:expder}). Shown are the cumulative distributions (black solid lines) of the measured quantities for the subset of experiments that satisfy the guidelines from Section~\ref{sec:guide} (see the text for details on the sample selection), with the median values, dispersions, and logarithmic mean measurement uncertainties highlighted by vertical dashed lines, dotted lines, and grey bands, respectively. The dispersions of the distributions are derived using the same approach as for the PDFs described in Section~\ref{sec:steppdfs}. For comparison, the (range of) expected values from the simulations as derived in the text are highlighted by thick red arrows or vertical lines. These results are shown for the low and high-resolution simulations (top and bottom panels, respectively). The figure shows that not just the three free parameters discussed in Section~\ref{sec:valid} are well-constrained, but also the derived quantities are accurately retrieved using the method presented in this paper.\vspace{-1mm}
}
\end{figure*}
The final two quantities discussed here are closely related. The first is the star formation efficiency per star formation event, which by definition follows from the star formation relation in equation~(\ref{eq:sfrelation}) as the ratio between the cloud lifetime (over which stars can form) and the galactic gas depletion time (reflecting how long it takes the galaxy to run out of gas at the current SFR), i.e.
\be
\label{eq:esfexp}
\epsilon_{\rm sf,exp}=\frac{\tgasexp}{\tdepl} .
\ee
For the \{low, high\}-resolution simulations, we find expected durations of the gas phase of $\tgasexp=\{15,2.6\}~\myr$ (see Section~\ref{sec:expval}) while for $\rhog\geq\{1,2\}~{\rm m}_{\rm H}~\cmc$, the simulations have gas depletion times of $\tdepl=\{1.0,0.3\}~\gyr$. After again taking the mean over all experiments, we thus expect to measure typical star formation efficiencies per event of $\epsilon_{\rm sf,exp}=\{1.5,0.85\}$~per~cent. The final considered quantity is the average mass loading factor, i.e.~the time-integrated mass outflow rate in units of the star formation rate, of which the expected value follows from $\epsilon_{\rm sf,exp}$ immediately as
\be
\label{eq:etaexp}
\overline{\eta}_{\rm fb,exp}=\frac{1-\epsilon_{\rm sf,exp}}{\epsilon_{\rm sf,exp}} ,
\ee
dividing the outflowing mass fraction by the star-forming mass fraction. For the above numbers, we expect to find typical mass loading factors of $\overline{\eta}_{\rm fb,exp}=\{67,117\}$, averaged over all experiments as before.

\subsection{Comparison to the measurements using the method} \label{sec:compder}
\autoref{fig:derived} shows the cumulative distributions of the five derived quantities considered in this section as retrieved by our method from the low and high-resolution, extended emission experiments that satisfy the conditions for the reliable application of the method (see Section~\ref{sec:guide}). For reference, the expected values based on the lines of reasoning presented in Section~\ref{sec:expder} are displayed in red. Across the ten panels in \autoref{fig:derived}, there is good agreement between the red and black lines. In the bottom-right two panels, the expected values fall marginally outside the $1\sigma$-equivalent percentiles (dotted lines), which is approximately the statistically expected number when considering ten distributions. Therefore, the main result of this comparison is that the method does not only accurately retrieve the free parameters $\tgas$, $\tover$, and $\lambda$, but also puts reliable constraints on the derived quantities describing cloud-scale star formation and feedback.

It is particularly worth commenting on the region radii $\rstar$ and $\rgas$, for which the horizontal arrows highlight ranges rather than typical values. For these quantities, we expect a range of values, because the spatial resolution of the experiments used to construct the distribution varies through the minimum aperture size ($\lapmin$). This is highlighted by the fact that the dispersion of the retrieved values (dotted lines in \autoref{fig:derived}) significantly exceeds the mean measurement uncertainty (indicated with the grey area), especially for the high-resolution simulation. Equation~(\ref{eq:rexp}) quantifies how we expect a change of $\lapmin$ to affect the retrieved region radii. While this is not directly shown by the cumulative distribution shown here, there is indeed a strong correlation between $\rstar$ or $\rgas$ and $\lapmin$, which quantitatively traces the prediction from equation~(\ref{eq:rexp}) with a mean difference of just 30~per cent. In fact, all experiments fall within a factor of two of the radius expected for their minimum aperture size.

The range of minimum aperture sizes is large enough to cover both the regimes where the smoothing kernels are resolved into several (1--10) resolution elements, which occurs at $\lapmin<\{150,70\}~\pc$ for the \{low, high\}-resolution simulations, and where they are smaller than one resolution element, which applies to larger values of $\lapmin$. Thanks to this dynamic range, the good agreement between the measured and expected region radii has two important implications. Firstly, it shows the obvious result that the radii (and hence densities) of clouds and star-forming regions can be constrained to high accuracy with the presented method. Secondly, the good performance at $\lapmin>\{150,70\}~\pc$ (for the \{low, high\}-resolution simulations) also illustrates that even sub-resolution sizes can be measured by subtracting the PSF dispersion in quadrature from the measured region radii, somewhat analogously to the early work on stellar clusters by \citet{larsen99}.

Across the remaining three variables, i.e.~the feedback velocity $\vfb$, the star formation efficiency $\sfe$, and the mass loading factor $\etaavgfb$, the dispersions of the retrieved values (dotted lines) largely match the mean measurement uncertainties (grey areas). This implies that the method obtains well-converged measurements, as is evidenced by the good agreement between the medians and the expected values shown by the red vertical lines in \autoref{fig:derived}. We also note that the dispersions are small in an absolute sense, at about 30~per cent (or 0.1~dex), which is similar to the mean difference between the measured region radii and their expected values. Only the feedback velocity of the low-resolution simulation seems to show a systematic variation, with a large (0.4~dex) dispersion that clearly exceeds the mean uncertainty. This discrepancy is largely driven by experiment IDs 20 and 23, which are run with a longer reference time-scale ($\tstar=30~\myr$) than the other experiments considered here ($\tstar=3$--$10~\myr$). Even though the stellar maps of these experiments do not have unacceptably large values of the region filling factor $\zetastar$, the relatively long reference time-scale does lead to visible region blending, because the emission peaks have relatively extended envelopes in the low-resolution simulation. As a result, the two experiments pass the quantitative criteria for reliable measurements of the duration of the overlap phase in Section~\ref{sec:guide}, but they do not pass the qualitative criterion that the maps should not exhibit any visual signs of region blending. In practical applications of the method, they would therefore have been considered unreliable. Quantitatively, the blending leads to overestimated durations of the overlap phase and, hence, underestimated feedback velocities. This behaviour does not occur for the high-resolution simulation, which better resolves the regions and does not show as high a degree of blending as the low-resolution simulation. Therefore, the dispersion of the retrieved feedback velocities in the high-resolution simulation matches the mean measurement uncertainty.

In summary, \autoref{fig:derived} demonstrates that the derived quantities describing cloud-scale star formation and feedback that are obtained using the method are accurate. While we have only considered five examples in this section to act as a proof of concept, the other quantities are derived using similarly simple and straightforward steps (see Section~\ref{sec:stepderived}). We thus conclude that the diagnostic power of the method is very promising, provided that it can be systematically applied to large samples of galaxies. In the next section, we quantitatively show that this can readily be achieved with current observational facilities.

\section{Applicability across cosmic history} \label{sec:dist}
\begin{figure*}
\includegraphics[width=\hsize]{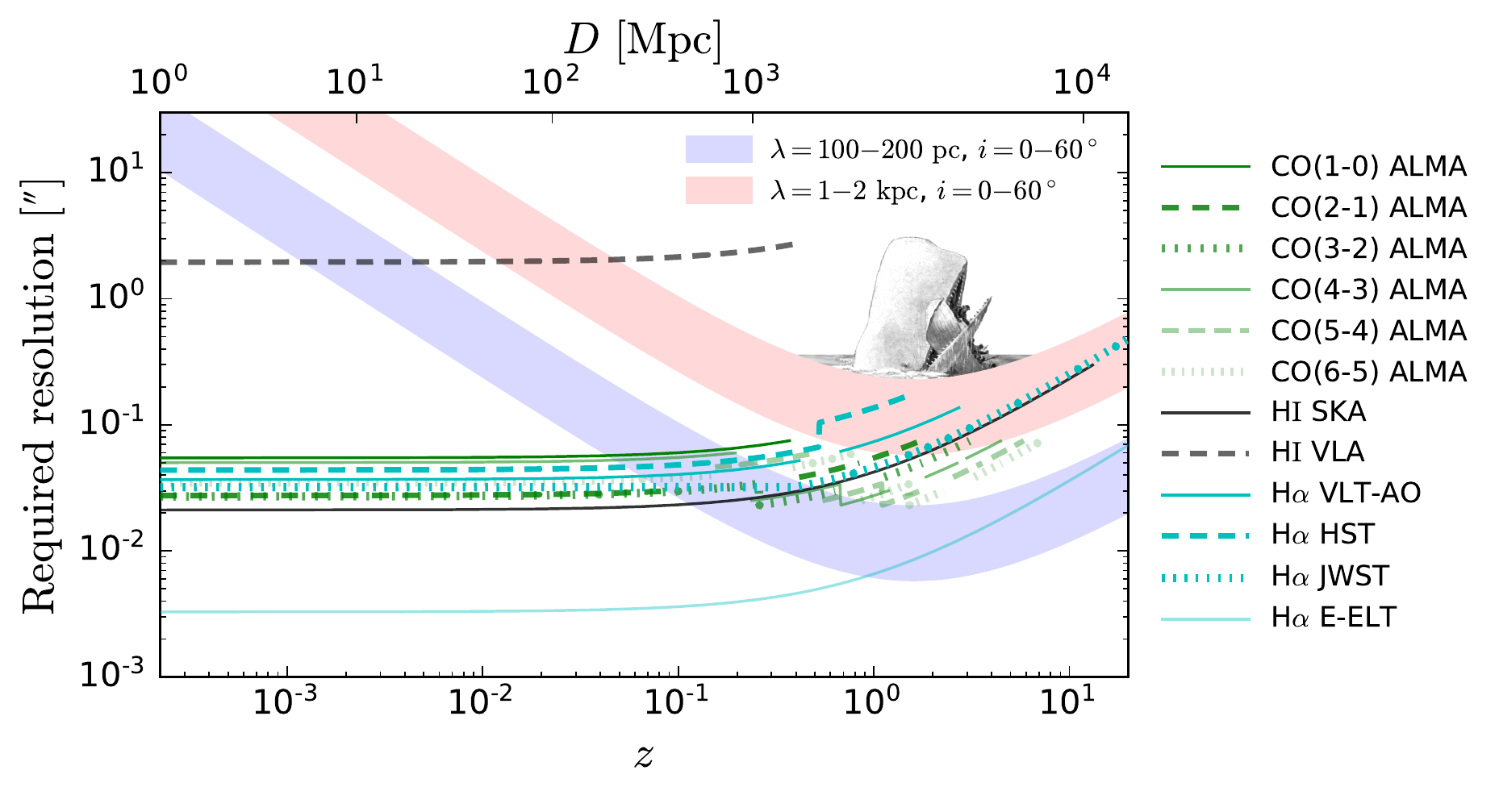}\vspace{-3mm}%
\vspace{-1mm}\caption{
\label{fig:redshift}
Comparison between the spatial resolution required for applying the method to galaxies with certain properties (shaded bands) and the resolution attained by current and upcoming observational facilities (lines) as a function of redshift (bottom $x$-axis) or distance (top $x$-axis). The blue-shaded band assumes mean separation lengths $100$--$200~\pc$, which is characteristic for local-Universe galaxies, whereas the red-shaded band assumes mean separation lengths $1$--$2~\kpc$, which is seen in high-redshift systems. Both shaded bands cover a range of inclination angles $i=0$--$60^\circ$. The spreads in $\lambda$ and $i$ contribute equally to the widths of the bands. As long as the lines are below or within the bands, the attainable resolution is sufficient for the application of our method. The figure shows that, in terms of spatial resolution and wavelength coverage, the method is comfortably applicable out to $z\sim4$. However, the rapid decrease of the sensitivity with redshift implies that observations at $z>1$ (marked by the white whale) require a major (but feasible) time investment (see the text).\vspace{-1mm}
}
\end{figure*}
In Section~\ref{sec:guide}, we have summarised the requirements for the successful application of the method detailed in Section~\ref{sec:method} to observed galaxy maps. Broadly speaking, these requirements can be divided into a `soft' and a `hard' category. By soft requirements, we refer to necessary target properties, which can be satisfied by picking the right galaxy. Examples are a low inclination, a sufficient number of independent regions (usually corresponding to a sufficient spatial extent of the galaxy), a relatively constant SFR over a time-scale $\tau$, and best-fitting time-scales $\tgas$ and $\tover$ in the range where they are considered reliable. Not all of these are necessarily known a priori, but when applying the method to a sample of sensibly-selected galaxies with a variety of gas and SFR tracers to choose from, these requirements should generally be satisfied for at least some (and usually most) of the targets.

By hard requirements we mean those that are technically restrictive when formulating observing strategies for applying the method. The two main examples of such requirements are the desired sensitivity [see requirement~(vi) in Section~\ref{sec:guide}] and spatial resolution [see requirement~(ii) in Section~\ref{sec:guide}]. In principle, any sensitivity can be obtained with infinite observing time, but the spatial resolution is fundamentally limited by the specifications of the observational facility. Therefore, resolution requirements present the most stringent condition for the observational application of our method. The method's validation in Section~\ref{sec:valid} demonstrates that accurate results are obtained for a minimum aperture size (i.e.~resolution) of at most $\lambda\cos{i}$ (for $\tgas$) or $\lambda\cos{i}/1.5$ (for $\tover$ and $\lambda$). In other words, a suitable observation must resolve the typical separation between independent regions along the minor axis of an inclined galaxy. This requirement can be evaluated for current and upcoming observational facilities with a high resolving power.

In \autoref{fig:redshift}, we show the spatial resolution required for the application of our method to a typical low-redshift star-forming galaxy (blue band) and a typical high-redshift star-forming galaxy (red band). In this context, a `galaxy' is characterised by a choice of the region separation length $\lambda$ and the galaxy inclination $i$. For local-Universe galaxies, it is possible to carry out a physically-motivated decomposition of position-velocity data cubes into coherent structures that are commonly referred to as clouds or H{\sc ii} regions \citep[e.g.][]{rosolowsky06,henshaw16}, which have typical separation lengths of the order $\lambda=100$--$200~\pc$ \citep[e.g.][]{colombo14,freeman17}. Early applications of our method yield broadly similar values of $\lambda$ (\citealt{kruijssen18}; Hygate et al.~in prep.; Schruba et al.~in prep.; Chevance et al.~in prep.; Ward et al. in prep.). We therefore characterise low-redshift star-forming galaxies with $\lambda=100$--$200~\pc$ in \autoref{fig:redshift}. By contrast, high-redshift galaxies exhibit clumpy morphologies on much larger scales, of order $1$--$2~\kpc$ \citep[e.g.][]{genzel11,genzel14,swinbank11,swinbank12,hodge12}. While observations using gravitational lensing \citep{dessauges17} and numerical simulations \citep[e.g.][]{behrendt16,oklopcic17} suggest that these clumps may fragment into smaller structures when observed at higher spatial resolution, it is important to consider that $\lambda$ represents the separation length between {\it independent} regions. The clumps seen in high-redshift galaxies are thought to represent the largest self-gravitating scale \citep{dekel09,reinacampos17}, which means that the evolutionary phases of the expected substructure within these clumps will be correlated rather than independent. We therefore adopt $\lambda=1$--$2~\kpc$ to characterise the separation length in high-redshift star-forming galaxies. Finally, we adopt a range of inclinations $i=0$--$60^\circ$, although we note that this is mainly for illustrative purposes. For these choices, the spreads of $\lambda$ and $i$ contribute equally to the width of each band in \autoref{fig:redshift}.

For comparison, the lines in \autoref{fig:redshift} show the spatial resolutions that can be attained as a function of redshift for a number of key spectral lines with current and upcoming observational facilities. The legend indicates which spectral lines are observed at the maximum resolution of which facility. We consider several CO transitions for tracing molecular gas, H{\sc i} for tracing atomic gas, and \halpha for tracing recent star formation. For the CO lines, we show the maximum attainable resolution with ALMA, the resolution in H{\sc i} is shown for the Karl G.~Jansky Very Large Array (JVLA) and the Square Kilometer Array (SKA), and the resolution in \halpha is shown for the Very Large Telescope (VLT; using the MUSE or SINFONI instruments with adaptive optics enabled), the Hubble Space Telescope (HST; using WFC3), the James Webb Space Telescope (JWST; using NIRCam or MIRI), and the European Extremely Large Telescope (E-ELT; using HARMONI, MICADO, or METIS). Some of the lines exhibit resolution jumps at certain redshifts, which is caused by the lines redshifting into a different band or instrument, leading to a change in the maximum attainable resolution. All resolutions are taken from the instruments' technical specifications prior to the submission of this paper. Whenever the documentation provides a single resolution per band, it is assumed that it applies to the centre of the central wavelength of that band $\lambda_{\rm band}$.\footnote{For the JVLA, we include an additional factor of 1.5 to represent natural weighting rather than uniform weighting.} The resolution at the wavelength of a line at a given redshift $\lambda_{\rm line}(z)$ is obtained by scaling the band's central resolution by a factor $\lambda_{\rm line}(z)/\lambda_{\rm band}=(1+z)\lambda_{\rm line}(0)/\lambda_{\rm band}$.

\autoref{fig:redshift} shows that currently available facilities (JVLA, ALMA, VLT, HST) allow the application of our method to typical low-redshift galaxies up to distances of $D\sim500~\mpc$ (or redshift $z\sim0.1$ for CO and \halpha, whereas for atomic hydrogen this can be achieved within $D<10~\mpc$ (with prohibitively long integration times already beyond a few $\mpc$). This implies that the lifecycle of molecular clouds and star-forming regions can be probed systematically using our method across hundreds of low-redshift galaxies covering a wide variety of properties. First efforts in this direction are currently ongoing (\citealt{kruijssen18}; Hygate et al.~in prep.; Schruba et al.~in prep.; Chevance et al.~in prep.; Ward et al.~in prep.). Upon the arrival of the SKA, we will be able to include atomic hydrogen across a distance range similar to the CO and \halpha, enabling the evolutionary analysis of molecular cloud condensation from the atomic phase.

Only a modest time investment is required for achieving the required spatial resolution and sensitivity across a representative sample of nearby galaxies. Observations of the molecular gas are generally the most prohibitive. We are currently undertaking a 75-hour ALMA Large Programme as part of the PHANGS collaboration, which targets all 80 massive ($10^{9.75}<M_\star/\msun<10^{11}$), not edge-on ($i<75\degr$), actively star-forming (${\rm SFR}/M_\star>10^{-2}~\gyr^{-1}$) galaxies at distances $D<17~\mpc$ (Leroy et al., in preparation), which is sensitive to individual molecular clouds of masses $M>10^5~\msun$ at a spatial resolution $1\arcsec$ (corresponding to $50$--$80~\pc$, depending on distance). Optical and UV observations of SFR tracers at matched spatial resolution can either be obtained with small, ground-based telescopes or have already been taken with HST. This shows that the observations required for the systematic application of our method can be attained with modern facilities.

In addition, \autoref{fig:redshift} demonstrates that the method can be applied to typical high-redshift galaxies out to $z\sim2.6$ using currently available facilities. This redshift range is limited by the accessibility of \halpha with the VLT and HST until the first light of the JWST and the E-ELT. Once these facilities are online, the redshift range will instead be limited by access to the peak of the CO spectral line energy distribution (SLED) with ALMA, which is expected to be around CO(5-4) in high-redshift galaxies \citep{carilli13}. For CO line observations up to the peak of the CO SLED, the method's range of application thus reaches out to $z=4$--$5$.

We do note that these kinds of high-redshift observations are highly time-consuming. At $z\sim1$, ALMA observations of galaxies with the highest CO surface brightnesses at a sensitivity of $0.1~{\rm mJy}$ and a required resolution of $<0.2"$ take $>20$ hours per galaxy. By $z\sim4$, reaching the required sensitivity at a similar resolution requires $>100$ hours (also see \citealt{hodge12}), which is not fundamentally prohibitive, but does imply that a large programme is required to apply the presented method at $z\ga2$.\footnote{Another way of gaining access to a larger galaxy sample may be to include high-resolution observations of gravitationally lensed galaxies \citep[e.g.][]{swinbank11,livermore15}. However, our method relies strongly on the spatial structure of the galaxy maps, implying that lensed galaxies will require an extremely high-precision lens model to obtain accurate images in the source plane.} Eventually, it will be possible to systematically characterise the lifecycle of molecular and ionised clumps in high-redshift galaxies for 10--20 galaxies at $z\sim1$ and a handful of targets at $z>2$. These numbers increase by a factor of several if the dust continuum is used to trace the molecular ISM. Such maps lack kinematic information, but require less integration time because the dust continuum is brighter than CO line emission. While this may not seem like a large number, we stress that the questions under consideration here previously represented Local Group or solar neighbourhood science. Being able to systematically probe the cloud-scale physics of star formation and feedback for any statistical sample of galaxies (let alone out to $z\sim4$) is an unprecedented and exciting prospect that is now within reach.

\section{Discussion} \label{sec:disc}
In Sections~\ref{sec:method}--\ref{sec:dist}, we have presented and validated a new method for probing the cloud-scale physics of star formation and feedback across cosmic history. The described elements of the method represent an optimised set of tools -- they have been carefully designed to yield the most accurate measurements of the constrained quantities. We now turn to a broader discussion of the results, paying particular attention to other approaches that we have tested and did not yield satisfactory results, as well as the limitations of the method as described in Section~\ref{sec:method}.

Specifically, the presented method is based on several choices of parameters and algorithmic approaches, most of which are either straightforward or carry a clear advantage relative to the alternatives. However, there are a number of these choices where (at face value) reasonable alternatives would have been possible. These warrant some further discussion. Below, we discuss the methods used for determining the overlap-to-isolated flux ratios $\betastar$ and $\betagas$, the way in which we account for the spatially extended nature of emission peaks, the influence of galactic morphology or substructure, and the general reliance of our method on the central limit theorem. Readers interested in specific, quantitative guidelines for observational applications are referred to Sections~\ref{sec:guide} and~\ref{sec:concl}.

\subsection{Unsuccessful alternative approaches} \label{sec:fails1}

\subsubsection{Methods for determining the overlap-to-isolated flux ratios $\betastar$ and $\betagas$}
The (possibly complex) time evolution of the young stellar or gas flux in a region can influence the best-fitting evolutionary timeline (see Section~\ref{sec:stepmodel}). In the context of our method, this time evolution is captured by the parameters $\betastar$ and $\betagas$, which represent the ratio between the flux regions residing in the overlap phase of \autoref{fig:tschem} relative to that of regions outside of that phase (i.e.~in the `isolated' phase). This simplistic representation of flux evolution is enabled by the fact that the method decomposes that evolution in just two phases per tracer. As discussed in Section~\ref{sec:stepfluxratios} and \autoref{fig:stepbeta}, $\betastar$ and $\betagas$ are determined by sorting the gas-to-stellar (stellar-to-gas) flux ratios of all stellar (gas) peaks and assigning the fraction $\tover/\tstar$ ($\tover/\tgas$) of the highest flux ratios to the overlap phase, whereas the remainder is taken to reside in the isolated phase. However, the best-fitting time-scales $\tstar$, $\tgas$, and $\tover$ are only obtained after fitting the model to the data in Section~\ref{sec:stepfit}. It is therefore necessary to calculate $\betastar$ and $\betagas$ as a function of $\tover/\tstar$ and $\tover/\tgas$, respectively, and let these two parameters vary with the time-scales during the fitting process.

As discussed in Section~\ref{sec:validbeta}, the above procedure results in accurate measurements of $\betastar$ and $\betagas$. However, it assumes that the identified peaks reasonably sample the underlying timeline. Even if the uncertainty resulting from this assumption is small (see Section~\ref{sec:stepfluxratios}), it is worth investigating if other approaches yield more accurate results. Therefore, we have evaluated two alternative methods for determining $\betastar$ and $\betagas$.

Firstly, one could attempt to define a critical gas-to-stellar flux ratio (or its contrast relative to the galactic average) for deciding whether an emission peak is considered to reside in the overlap phase. However, for gradually evolving flux levels of independent regions, defining a critical flux ratio is entirely ad hoc. In practice, it would also depend on the specific tracer used, as different tracers will exhibit different typical evolutionary histories. As demonstrated by \autoref{fig:stepbeta}, there are no obvious jumps near the best-fitting value of (in this example) $\betastar$ that would enable identifying a threshold value. The other experiments show similar behaviour. We have experimented with several different thresholds, such as requiring the gas-to-stellar flux ratio to reside within a certain factor of the galactic average to identify a region as `overlap', where we let that factor (or its inverse) vary from $0.05$--$0.5$. The resulting values of $\betastar$ and $\betagas$ vary greatly as a function of the adopted threshold value, even if the impact on the best-fitting time-scales is modest (as expected, see Section~\ref{sec:stepfluxratios}). Only for the subset of stellar point particle experiments discussed in Section~\ref{sec:validbeta} it is possible to define a threshold, but this is enabled exclusively by the imposed step function flux evolution of the particles in these experiments, which experience a flux change by a factor of 2 upon entering or emerging from the overlap phase (see Section~\ref{sec:starexp}). In other words, defining a critical flux ratio for deciding whether an emission peak resides in the overlap phase may only work with prior knowledge of the evolutionary timeline. This approach is therefore undesirable.

Secondly, one could try to iteratively constrain $\betastar$ and $\betagas$, i.e.~by applying the procedure of Section~\ref{sec:stepfluxratios}--\ref{sec:steppdfs} for some initial guess of these parameters to obtain the best-fitting $\tstar$, $\tgas$, and $\tover$ and use these best-fitting values to refine $\betastar$ and $\betagas$ analogously to how we currently use the time-scales to determine the fraction of peaks that should reside in the overlap phase. This can then be repeated until convergence is obtained. To some extent, this approach is a less elegant version of how we determine $\betastar$ and $\betagas$ in Section~\ref{sec:stepfluxratios}, because during a single iteration this version (incorrectly) assumes that the time-scales can be varied without changing the overlap-to-isolated flux ratios accordingly. In addition, it is computationally more expensive, because the procedure of Section~\ref{sec:stepfluxratios}--\ref{sec:steppdfs} needs to be repeated multiple times. Most importantly, an iterative approach ends up being highly unreliable -- tests of this approach often identify two `attractor' solutions that the iteration keeps alternating between, implying that convergence is never obtained.

In summary, the adopted approach for determining the overlap-to-isolated flux ratios $\betastar$ and $\betagas$ is the most accurate, most computationally efficient, and the most self-consistent of the options that we have considered. Due to a lack of feasible alternatives, we have not included any other (optional) approaches in the \code code.

\subsubsection{Accounting for the extended nature of emission peaks}
Contrary to the original model \citepalias{kruijssen14} for the `tuning fork diagram' of \autoref{fig:tuningfork}, the model described in Section~\ref{sec:stepmodel} does not assume that emission peaks are point sources, but accounts for their extended nature. This is a critical enhancement of the model -- in the original form, the flattening of the tuning fork diagram at small aperture sizes is unambiguously interpreted as being due to a non-zero overlap time-scale $\tover>0$, whereas such a flattening is also caused by extended emission peaks with finite central surface densities. As a result, distinguishing the influences of extended emission and the duration of the overlap phase on the flattening of the gas-to-stellar flux ratio at aperture sizes smaller than the region separation length (i.e.~$\lap<\lambda$) is critical for obtaining a non-degenerate characterisation of the evolutionary timeline. As evident by comparing panel~(b) with panels~(g) and~(h) in \autoref{fig:stepmodel} (in which ${\cal E}\rightarrow\infty$ corresponds to delta function-shaped peak profiles), avoiding this degeneracy may be difficult, but can be achieved using our method. However, this relies quite strongly on which functional form is assumed for the emission peak profiles. This represents the main uncertainty when using our method to constrain feedback-related physics, because it predominantly affects the measurement of the duration of the overlap phase $\tover$. We have therefore evaluated various profiles and their possible variation with tracer or evolutionary phase.

Practical applications of the method will often need to deal with objects near the resolution limit. Therefore, we make a strong case in Section~\ref{sec:stepmodel} that two-dimensional Gaussians are the preferred profile, because these represent both the observational PSF and the convolution kernel used when convolving the input maps to a common spatial resolution. However, \code also includes the option of using point (delta function) or constant surface density disc profiles. When applying these profiles to the extended emission experiments from Section~\ref{sec:valid}, the results become less accurate than when using two-dimensional Gaussians, as expected. Adopting point particle profiles invariably leads to overestimated overlap time-scales, because any flattening of the tuning fork caused by the peak profile is erroneously attributed to the duration of the overlap phase. A constant surface density disc generally does better, because its finite surface density at small aperture sizes appropriately decreases $\tover$ and could be preferred over a two-dimensional Gaussian, because it is `simple'. However, it also provides less accurate results, because a constant surface density does not properly describe the actual peak emission profiles. In the context of the tuning fork in \autoref{fig:tuningfork}, this profile requires the gas-to-stellar flux ratios to be constant once $\lap<2r$, which implies a sudden transition in the model tuning fork (i.e.~a discontinuity in its first derivative) as shown in panel~(i) of \autoref{fig:stepmodel} that is not present in the observed data points. This mismatch leads to an unsatisfactory recovery of the input time-scales in the experiments of Section~\ref{sec:starstar}. Therefore, we recommend using a two-dimensional Gaussian peak profile.

Even if a Gaussian profile is preferred, it remains to be decided how to set its width, which we express in terms of a dispersion radius $r$. In Section~\ref{sec:stepmodel}, we use the contrast between the flux level of the emission peaks and the average across the map to determine the ratio $\zeta\equiv2r/\lambda$. This is done separately for young stellar peaks and gas peaks, which is the sensible choice for peaks in the `isolated' phase, i.e.~that do not reside in the overlap phase. However, it is not obvious how to determine the peak radii for peaks residing in the overlap phase. Should the stellar and gas components of a single `overlap' peak be allowed to have different radii? If so, should these radii be assumed to be the same across the entire peak sample, irrespective of whether they reside in the isolated or overlap phase, or should a distinction be made between the evolutionary phases? Depending on these choices, it may be necessary to obtain up to four different peak radii.

We start by addressing the first of these questions with an example. If we select a stellar or gas peak that resides in the overlap phase and allow its radius to be different for the gas and stellar flux in the numerator and denominator of the gas-to-stellar flux ratio, then the extreme of the gas-to-stellar flux ratio bias may be reached at a radius larger than the smallest aperture size if the radius difference between both tracers is large enough. In other words, the `tuning fork' curve is no longer monotonically increasing (when focusing on a gas peak) or decreasing (when focusing on a young stellar peak) towards smaller aperture sizes, but exhibits a minimum or maximum at some finite aperture size. This happens when the peak that is being focused on is more extended in its primary tracer (by which it was selected) than in its secondary tracer (the presence of which makes the peak reside in the overlap phase). In that case, the primary tracer (which drives the bias of the gas-to-stellar flux ratio relative to the galactic average) reaches the central surface density at a larger size scale than the secondary tracer does, implying that at smaller apertures the secondary tracer contributes more of the flux density and the degree of flux ratio bias decreases.

The behaviour in which a secondary tracer is considerably more centrally concentrated than the primary tracer might manifest itself on the small scales of individual molecular clumps, which often exhibit a higher star formation efficiency in their central regions than globally \citep[e.g.][]{kruijssen12,longmore14,ginsburg16}. However, on the scales of entire molecular clouds (or `independent regions') considered here, the structure of the ISM is scale-free \citep[e.g.][]{elmegreen96,maclow04}, which implies that the distribution of star formation should roughly trace the distribution of mass in the cloud even if the stars themselves form at the small-scale density peaks. As a result, the trend reversal described above has not been observed in the several observational applications of our method that are currently underway (\citealt{kruijssen18}, Hygate et al.~in prep., Schruba et al.~in prep., Chevance et al.~in prep., Ward et al.~in prep.), which cover over a dozen different galaxies and tracer pairs. This suggests that using different radii for the stellar and gas emission of a peak in the overlap phase is not the right approach.

However, differences in peak radius are expected as a function of evolutionary stage. For instance, a molecular cloud may contract due to self-gravity or grow by accretion, whereas H{\sc ii} regions may expand before shrinking when the most massive stars have ended their lives. As a result, it is not obvious that the peak radii during the overlap phase should be the same as during the isolated stellar and gas phases. Indeed, tests using a description in which peaks in the isolated and overlap phases share the same radius (but still differ between gas and stars) show less accurate results than when the peak radius is allowed to vary with evolutionary phase for each tracer. Therefore, it is necessary to account for the evolution of the peak radii -- the flux contrast between the emission peaks and the map average should be determined separately for peaks in the isolated and overlap phases, as is done in Section~\ref{sec:stepmodel}.

In summary, exploring alternative descriptions for the radius evolution shows that the model presented in Section~\ref{sec:stepmodel} provides the best match to simulated and observed tuning fork diagrams. It thus most accurately constrains the free parameters.

\subsection{Current limitations of the method} \label{sec:fails2}
\subsubsection{The influence of galactic morphology or substructure} \label{sec:morphology}
\begin{figure}
\includegraphics[width=\hsize]{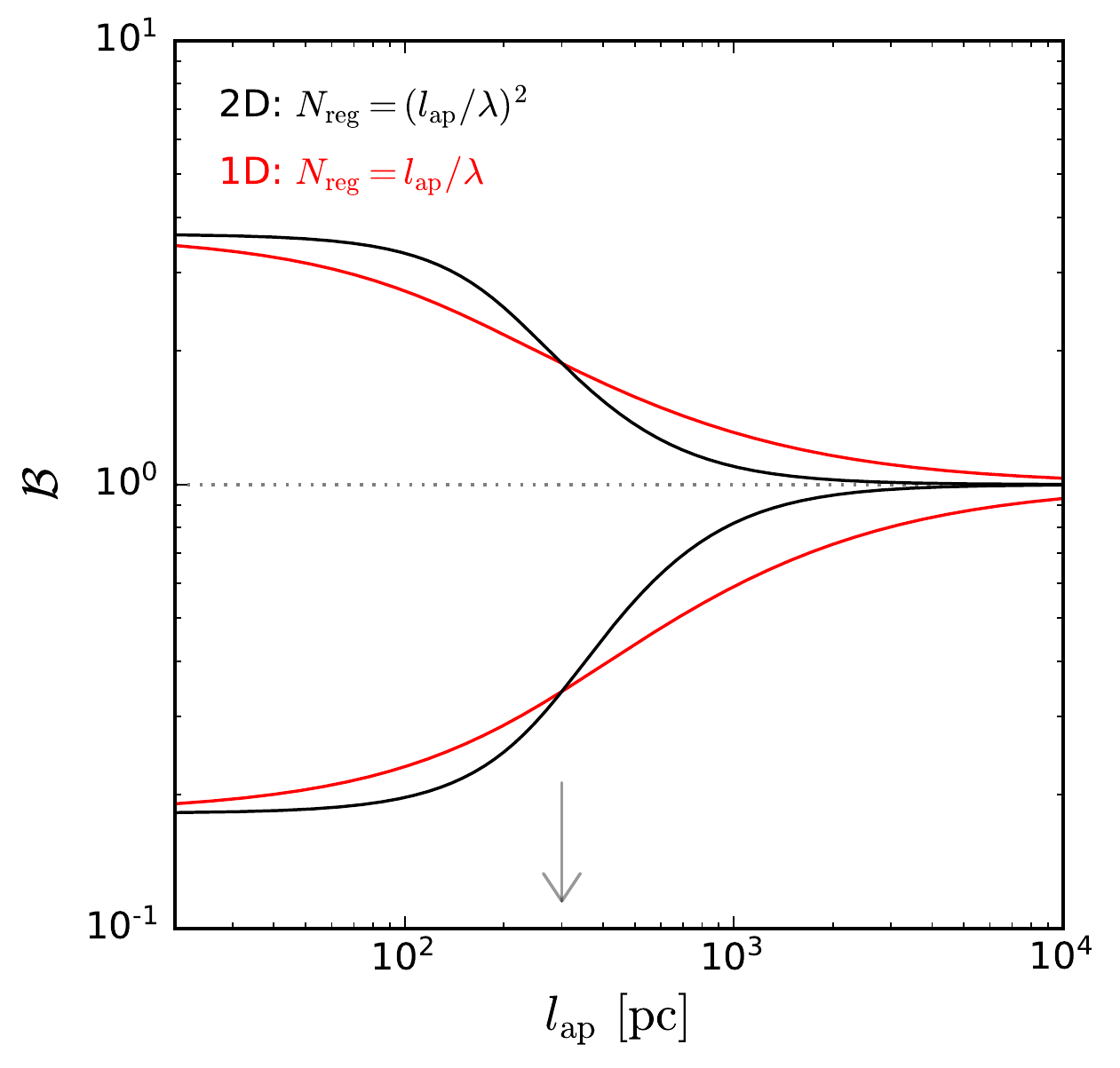}%
\vspace{-1mm}\caption{
\label{fig:morphology}
Influence of galactic morphology on the predicted gas-to-stellar flux ratio bias as a function of the aperture size when focusing apertures on stellar peaks (bottom branches) or gas peaks (top branches), with parameters corresponding to the fiducial model in \autoref{fig:stepmodel} (see \autoref{tab:model}). The black lines represent the random distribution of independent regions in two dimensions adopted in our model, whereas the red lines indicate the same model when the regions are randomly positioned along a line. These two extremes bracket the range of `tuning fork' shapes expected for real galaxies with prominent morphological features. The aperture size where $\lap=\lambda$ is highlighted by the grey arrow.\vspace{-1mm}
}
\end{figure}
The model that is fitted to observed `tuning fork' diagrams in Section~\ref{sec:stepfit} and is described in Section~\ref{sec:stepmodel} assumes that independent regions are randomly distributed in two dimensions. Specifically, this assumption implies that the number of regions enclosed by an aperture $N_{\rm reg}$ scales as $N_{\rm reg}=(\lap/\lambda)^2$, which determines how steeply the gas-to-stellar flux ratio bias (in small apertures) transitions to the galactic average (in large apertures) around a characteristic aperture size $\lap=\lambda$. By contrast, if the regions are not randomly distributed in two dimensions, but are situated along a line, then the number of regions enclosed by an aperture scales as $N_{\rm reg}=\lap/\lambda$. Real galaxies that exhibit some degree of substructure like spiral arms or a ring may have an $N_{\rm reg}$ scaling in between these two extremes.

\autoref{fig:morphology} illustrates the effect of morphology or substructure on the tuning fork diagram by considering both the two-dimensional distribution and the linear sequence of regions. The difference between both models is modest, especially when considering that the black and red lines represent morphological extremes. However, a characteristic distinction is that the linear model transitions more gradually between the single-peak and galactic-average regimes of the aperture size range. Based on these model predictions, it is clear that galactic morphology does not affect the measurement of $\lambda$ (as both models match at $\lap=\lambda$), but may influence the obtained evolutionary timeline, particularly if any morphological features are more prominent in one tracer than in the other (i.e.~one branch of the tuning fork follows a red line in \autoref{fig:morphology}, whereas the other follows a black line).

We have attempted to account for galactic morphology by letting the $(\lap/\lambda)^2$ terms in equations~(\ref{eq:rstar}) and~(\ref{eq:rgas}) vary according to the actual substructure in the considered maps by stacking the radial emission profiles around the identified emission peaks. However, this is a highly delicate exercise. As before, much of the uncertainty in accounting for morphology comes from the treatment of regions in the overlap phase, for which the spatial structure in both maps needs to be combined. In our experiments investigating this problem, we find that a morphology correction to the default $N_{\rm reg}=(\lap/\lambda)^2$ is very sensitive to the definition of which regions reside in the isolated or overlap phases. To remedy such a dependence, the present paper follows the conservative assumption that the regions are randomly distributed in two dimensions. The validity of this assumption is supported by the fact that the galaxy simulations on which the method is tested do exhibit some degree of substructure (see \autoref{fig:maps}, \ref{fig:starmaps}, and~\ref{fig:gasmaps}), yet the retrieved evolutionary timelines are in good agreement with the input values (see Section~\ref{sec:starstar}). This suggests that the influence of galactic morphology on the accuracy of the method is limited.

As shown by \autoref{fig:morphology}, our model for the gas-to-stellar flux ratio as a function of aperture size can provide predictions for the two bracketing cases of substructure, as well as morphologies in between. The challenge is to characterise the substructure in a way that provides a clean measurement of $N_{\rm reg}$ as a function of $\lap/\lambda$ for any pair of galaxy maps without relying too strongly on the definition of a `region'. This line of research is currently being continued and is planned to be addressed in a future paper. Until the completion of that work, we recommend applications of the method to galaxies dominated by strong spiral arms or rings to visually inspect how well the model matches the slope of the observed tuning fork near $\lap=\lambda$. A discrepancy between both slopes would indicate non-random morphological features. Such cases may be improved by the partial masking of the young stellar and gas maps used, or by filtering them at different emission levels to enhance or suppress their morphological features.

If the current version of the method is applied to a galaxy with dominant one-dimensional morphological features such as spiral arms, bars, or rings that propagate down to scales $\lap\approx\lambda$, the constrained quantities may be inaccurate. \autoref{fig:morphology} shows that $\lambda$ is unaffected, as the flux ratio bias at $\lap=\lambda$ is insensitive to the dimensionality of the emission structure. By contrast, the best-fitting value of $\tgas$ will be incorrect if the morphological features are present down to a scale $\lap\approx\lambda$ only (or mainly) in one of the maps. In that case, the tuning fork diagram becomes asymmetric, with one branch providing a bad match to the modelled two-dimensional morphology. This affects the accuracy of $\tgas$, because in our model it controls the asymmetry of the tuning fork diagram (see \autoref{fig:stepmodel}). In other words, varying $\tgas$ is somewhat degenerate with morphological differences between both maps. Fortunately, such cases are easily recognised, because the fitting procedure then often returns a best-fitting solution that reproduces one branch of the tuning fork diagram, but not the other. Finally, $\tover$ is relatively unaffected, because it is set on the smallest scales, where $\lap<\lambda$ and the $\lap/\lambda$ term in the model becomes unimportant.

In closing, it is important to emphasise that while our model assumes a random distribution of regions, this only affects the quality of the fit around $\lap=\lambda$. At aperture sizes $\lap\ll\lambda$, the gas-to-stellar flux ratio bias is set by the properties of the central peak, specifically the central flux density and the duration of the overlap phase. At aperture sizes $\lap\gg\lambda$, the gas-to-stellar flux ratio converges to the galactic average, meaning that large-scale morphological features do not affect the accuracy of the results. In an observational context, this means that spiral arms, bars, or rings can only affect the constrained quantities if they persist down to $\lap\approx\lambda$. In practice, we find that a reasonable degree of two-dimensional spatial isotropy is only needed for $\lap\leq2\lambda$. This is implies a much weaker requirement than ruling out any applications of the method to all galaxies with prominent morphological features. We plan to further investigate the applicability of the method to galaxies with pronounced morphological features in future work, as part of our ongoing observational surveys of the nearby galaxy population (see Section~\ref{sec:dist}).

\subsubsection{The method's reliance on the central limit theorem} \label{sec:limit}
A final point worth discussing is that our method relies quite strongly on the central limit theorem by attributing a common evolutionary timeline to all regions. This approach effectively assumes that region-to-region variations can be linearly combined to retrieve an intensity-weighted average gas-to-stellar flux ratio bias [see equations~(\ref{eq:rstar}) and~(\ref{eq:rgas}) in Section~\ref{sec:stepfluxratios}]. We demonstrated in \citetalias{kruijssen14}~(Figure~5) that this assumption is correct when the regions follow a mass spectrum of non-zero width. The only effect of such a mass spectrum is to increase the statistical uncertainty on the observed gas-to-stellar flux ratio bias, but it does not affect the absolute value of the bias.\footnote{This also applies to the environmental variation of other region properties, such as their sizes, expansion velocities, or emissivities. If a major variation of these properties is a concern (e.g.~due to a sharp radial metallicity gradient in the target galaxy), we recommend sub-dividing the galaxy into fields or radial annuli across which these properties are more homogeneous.} We note that this increased uncertainty due to the mass spectrum is accounted for in the derivation of the uncertainties in Section~\ref{sec:steperrors}. Even if regions of different masses experience different evolutionary timelines, our method of constraining a timeline is not incorrect as long as it is kept in mind that it represents a flux-weighted population average. In essence, it provides a characteristic evolutionary history for Lagrangian mass elements in a galaxy.

Our approach is less adequate when some subset of the independent regions never evolves through a phase in which it is bright in either of the tracers. Section~\ref{sec:rhomin} provides a numerical example based on the stochastic star formation model used in our simulations. In real galaxies, a subset of the molecular cloud population may be disrupted by galactic shear, without ever being associated with high-mass star formation. Alternatively, low-mass clouds may have too low star formation efficiencies to produce massive stars that are traced by \halpha, FUV, or NUV. In both cases, the rarity of the stellar phase is increased relative to the gas phase, implying that the cloud lifetime inferred with our method will be longer than the lifetime of the star-forming subset of clouds. Specifically, if a fraction $f_{\rm vis}$ of all clouds will at a later stage also be visible in the young stellar phase, then the observed lifetime will be a factor of $f_{\rm vis}^{-1}$ longer than the lifetime of (eventually) star-forming clouds. A good way of picturing this is that our method measures the statistical average of the total amount of time spent by a mass element in the gas phase before becoming associated with the young stellar phase. For instance, if a cloud is disrupted by shear before it forms massive stars, its lifetime is effectively added to that of a similar cloud that will become associated with star formation.

It depends on the science question at hand whether the above behaviour is desirable. For instance, the theory for cloud lifetimes by \citet{jeffreson18} follows a similar statistical approach by adding up cloud destruction rates due to various mechanisms and subtracting the mechanism that is not associated with star formation (i.e.~shear), such that it increases the effective cloud lifetime. In that case, theory and observation are dealing with the same quantities. However, if one is interested in the absolute lifetime of a certain ionised emission line (e.g.~H$\beta$, [O{\sc ii}], [O{\sc iii}], [N{\sc ii}], or [S{\sc ii}]), which is linked to a stellar evolutionary time-scale, it is undesirable to obtain the population-integrated lifetime. In such a case, an effort should be made to mask or filter out emission from regions that will never be visible in the other tracer. For gas maps, this can be achieved by filtering out emission from low-mass clouds (possibly in a diffuse form), which typically have higher virial parameters \citep{dobbs11} and are more prone to destruction by shear. Low-mass clouds are also statistically less likely to form high-mass stars due to the effects of IMF sampling, but this effect is accounted for in our calibrations of the lifetimes of young stellar tracers such as \halpha, FUV, and NUV, which decrease towards low region masses due to stochastic effects \citep{haydon18}. In practice, the impact of low-mass clouds and star-forming regions may be limited, because the luminosity functions (${\rm d}N/{\rm d}L\propto L^\gamma$) of both populations have slopes $\gamma>-2$, implying that most of the flux emerges from massive clouds and regions \citep[e.g.][]{elmegreen96,lee12}.

The examples considered above show that there are feasible ways of dealing with tracer pairs for which the population-averaged evolutionary timeline may not provide meaningful time-scales. However, there is no universal way of doing so. The best approach depends on the tracer pair and sometimes even on the particular data set. This highlights a fundamental point -- the method presented in this paper does not represent a `black box' and should not be used as such. A sound astrophysical interpretation of the measurements requires the user to carefully evaluate to what extent the constrained quantities match the physical quantities of interest. When such care is taken, the presented method can provide strong and powerful constraints on important physical quantities describing cloud-scale star formation and feedback across cosmic history.

\section{Conclusions} \label{sec:concl}
\subsection{Summary} \label{sec:summary}
We present a new method for measuring the key quantities describing the cloud-scale physics of star formation and feedback from high-resolution imaging of the ISM and star formation across galaxy discs. These quantities include the (molecular) cloud lifetime, the feedback time-scale, the mean separation length between independent regions, the region size, the star formation efficiency per star formation event, the feedback outflow velocity, the mass loading factor, the feedback-to-ISM coupling efficiency, as well as several other important quantities (\autoref{tab:output}). This paper explains the method in detail (Section~\ref{sec:method}) and demonstrates its accuracy using nearly 300 controlled applications to hydrodynamical simulations of star-forming galaxies (Sections~\ref{sec:valid} and~\ref{sec:derivephys}), finding that the method is suitable for systematic applications across statistically relevant samples of galaxies, from the nearby Universe out to high redshift (Section~\ref{sec:dist}). This extends the `Local Group science' of cloud-scale star formation and feedback into the realm of galaxy evolution across cosmic time. The main results of this work are as follows.
\begin{enumerate}

\item
{\it Section~\ref{sec:kl14}}: Star formation in galaxies can be described as an ensemble average over a population of `independent regions', i.e.~concentrations of gas or young stars that reside on a timeline of star formation in an evolutionary phase that is independent of their neighbours. The star formation relation between the gas mass (surface density) and the SFR (surface density) observed when averaging these quantities over large areas of galaxies breaks down below a few times the region separation length $\lambda$, where the discretisation of a galaxy into independent regions introduces significant departures from the galaxy-wide average relation due to region-to-region differences in evolutionary phase. These departures become systematic when {\it focusing} small ($\lap\lesssim2\lambda$) apertures on emission peaks of either gas or young stars, such that the gas-to-stellar flux ratio (tracing the gas mass per unit SFR or the gas depletion time) becomes biased to elevated or suppressed values, respectively, because specific evolutionary stages are preferentially selected. The magnitude of these biases directly probes the underlying evolutionary timeline of the regions.

\item
{\it Section~\ref{sec:method}}: We introduce the code \code, developed in \idl, which applies the above formalism to obtain quantitative measurements of the duration of each evolutionary phase and the region separation length from observed galaxy maps. The evolutionary phases are defined by the emission maps used, such that a timeline consists of two phases reflected by each map and a third `overlap' phase during which regions are visible in both maps. The method works by identifying emission peaks in the maps, measuring the flux ratio between both maps around the identified peaks as a function of the spatial averaging scale (or aperture size) in a `tuning fork diagram' (\autoref{fig:tuningfork}), and fitting a statistical model that predicts the flux ratio bias as a function of size scale and the underlying evolutionary timeline. This way, the method directly translates the observed flux ratio bias when focusing apertures on emission peaks in either map into the (relative) lifetimes of these peaks. Fundamentally, the method only measures relative lifetimes, but these can be converted into absolute time-scales by using a `reference time' if one of the tracer lifetimes is known.

In the context of star formation in molecular gas clouds and feedback from young stars, obvious emission tracers are CO and \halpha, although it is one of the method's main strengths that it can be applied to any pair of emission maps that represent different phases of an evolutionary progression. The choice of the SFR tracer is important, because it can be used to set the aforementioned `reference time'. For instance, \halpha has a lifetime of about $5~\myr$ \citep{leroy12,haydon18}. Because the method makes use of flux ratio biases, only relative flux differences are important. As a result, the measured time-scales and separation lengths are themselves independent of the (uncertain) conversion factors from flux to physical quantities such as gas masses or SFRs, provided that these conversion factors are roughly constant across the analysed field. However, by combining the measured evolutionary timeline with these physical quantities, a range of important derived quantities (such as the star formation efficiency and mass loading factor) can be constrained. The method employs Monte-Carlo error propagation, providing the complete PDF for each constrained quantity.

\item
{\it Section~\ref{sec:valid}}: The method is systematically tested and validated using simulated galaxy maps, with the goal of identifying exactly under which conditions it reliably characterises cloud-scale star formation and feedback. To this end, we performed hydrodynamical simulations of isolated, star-forming disc galaxies, which are used to generate gaseous and young stellar emission maps. The first set of 88 controlled experiments (described in Section~\ref{sec:starstar}) uses pairs of stellar maps showing only the stars in specific age bins, granting us complete control over the duration of each phase. These tests show that the obtained time-scales are accurate to within 30~per~cent or 0.11~dex. The largest source of uncertainty is the variation of the SFR across the duration of the timeline. We also find that the duration of the overlap phase can be obtained to an accuracy of $\sim5$~per~cent of the total duration of the evolutionary timeline jointly spanned by both maps. In order to accurately measure the duration of the overlap phase, it is important that the region diameters in both maps do not exceed half the region separation length to avoid the misidentification of spatial overlap as temporal overlap.

The second set of 200 controlled experiments (described in Section~\ref{sec:gasstar}) combines age-binned stellar maps with maps of the gas in the simulations, with the goal of addressing the applicability of the method as a function of observational conditions, such as the choice of SFR tracer and its characteristic lifetime, the choice of gas tracer and its characteristic gas volume density, the spatial resolution of the maps, the inclination angle of the galaxy, the number of independent regions per map, and the diffuse emission fraction. The main results of these tests are as follows.
\begin{enumerate}
\item
The lifetimes of the gas and SFR tracers should not differ by more than an order of magnitude. This means that the choice of SFR tracer (which sets the reference time-scale as described by \citealt{haydon18}) may need to be modified after applying the method. For an SFR tracer lifetime of $5~\myr$ (typical of e.g.~\halpha), accurate cloud lifetimes are obtained for $\tgas=0.5$--$50~\myr$. By including high-resolution FUV or NUV coverage, the upper limit of this range is increased to several $100~\myr$, implying that the method is not limited by the availability of suitable reference time-scales.
\item
The measured cloud lifetimes depend on the gas tracer used and the characteristic volume densities it traces. This dependence is physical in nature and opens up the exciting possibility that the density evolution of gas clouds towards star formation can be probed as a function of absolute time by combining different gas tracers.
\item
After correcting for galaxy inclination, the maps need to have a spatial resolution sufficient to resolve the region separation length $\lambda$ in order to accurately measure the tracer lifetimes. To obtain the duration of the overlap phase and $\lambda$ itself, we find that it is necessary to resolve $\lambda/1.5$. For cloud separation lengths that are characteristic of nearby galaxies ($\lambda\sim200~\pc$), this implies spatial resolutions of the order $2"$, whereas at clump separations seen in high-redshift galaxies ($\lambda\sim2~\kpc$), this implies spatial resolutions of the order $0.2"$.
\item
In terms of the galaxy inclination angle, the above requirements on the spatial resolution typically require $i<75^\circ$, even though this is somewhat dependent on galaxy properties. This implies that galaxy inclination does not significantly limit the method's applicability.
\item
To constrain the measured time-scales to better than 50~per~cent, at least 35 emission peaks per map are needed. Making order-of-magnitude estimates of the evolutionary timeline requires 15 emission peaks per map. For the aforementioned separation lengths and accounting for substructure, the requirement of 35 emission peaks translates to minimum galaxy radii of $R\geq1.5~\kpc$ in local galaxies and $R\geq4.4~\kpc$ at high redshift.
\item
The presence of a large diffuse emission reservoir can cause the selected tracer to be underemphasised around emission peaks relative to the galactic average. This manifests itself as a negative emission bias in the tuning fork diagram of \autoref{fig:tuningfork} and leads to inaccurate results. Applications showing signs of large diffuse emission fractions should be discarded or spatially filtered \citep[see][]{hygate18}.
\end{enumerate}
As discussed in Section~\ref{sec:valid}, the above conditions are comfortably satisfied with modern observatories such as ALMA and VLT/MUSE across the nearby galaxy population. The method is even readily applicable at high redshift (see below).

\item
{\it Section~\ref{sec:derivephys}}: The accuracy of the derived physical quantities is assessed by comparing the retrieved values to those expected from the numerical simulations. Specifically, we consider the region radii in both the gas and stellar maps, the feedback velocity, the star formation efficiency per star formation event, and the mass loading factor. We show that each of these is consistent with the values expected based on the included physics, usually to within the uncertainties, but universally to within 50~per~cent. This shows that the method provides an accurate way of quantitatively constraining the cloud-scale physics of star formation and feedback.

\item
{\it Section~\ref{sec:dist}}: We consider the spatial resolutions attainable by current observational facilities as a function of distance and redshift and compare these to the spatial resolutions required for applications of our method, assuming the aforementioned region separation lengths that are typical of galaxies in the local Universe and at high redshift. With currently available facilities (ALMA, VLT, HST), the method can be applied to typical low-redshift galaxies up to $D\sim500~\mpc$ or $z\sim0.1$ for molecular gas traced by CO and star formation traced by \halpha. The atomic gas phase can currently be included up to a few $\mpc$ with the JVLA before it becomes prohibitively time-consuming, but the arrival of the SKA will extend the method's applicability to H{\sc i} to a similar distance range as for CO. Across the wavelength range, the method can be applied to hundreds of nearby galaxies spanning a variety of masses, morphologies, and SFRs. At high redshifts ($z>1$), the method can be applied to CO line imaging of 10-20 galaxies at $z\sim1$ with a reasonable time investment and a handful of galaxies at $z>2$. When dust continuum is used to trace the gas phase, the number of galaxies that can be observed within a feasible time-scale increases considerably. These numbers mean that the lifecycle of molecular clouds and star-forming regions can be probed systematically for more than a hundred galaxies across cosmic time, spanning a representative range of cosmic environments. This is a major step relative to previous methods and facilities, for which studying the cloud-scale physics of star formation and feedback was restricted to the Local Group or the solar neighbourhood.

\item
{\it Section~\ref{sec:disc}}: The presented method is the end result of a wide variety of renditions that we have explored over the past years. We document the most important subset of these various approaches by discussing a number of key choices and current limitations of the method and comparing these to the alternatives. We show that our current method for quantifying the flux evolution of independent regions between the isolated and overlap phases (captured by the parameters $\betastar$ and $\betagas$) is accurate and performs better than reasonable alternatives. The same holds for the way in which we account for the extended density profiles of the emission peaks. However, extremely extended emission components (i.e.~diffuse emission) still negatively affect the obtained results. We present reasonable ways of dealing with this problem in Section~\ref{sec:diffuse}, whereas a new approach to systematically filter diffuse components from emission maps is presented by \citet{hygate18}.

Another current limitation of the method is that it assumes a random spatial distribution of independent regions on a size scale $\lap\approx\lambda$ and does not account for galactic structure such as dominant spiral arm patterns or bars, which could persist down to these small scales. We quantitatively show that this does not negatively impact the accuracy of $\tover$ and $\lambda$, but can prohibit accurate measurements of $\tgas$. Fortunately, such cases are easily recognised as a bad fit to the observed `tuning fork diagram', reproducing only one branch while poorly matching the other. We intend to explore this further in future work and currently list points of attention for applications to galaxies with prominent morphological features to ensure accurate results. Finally, we discuss how the method's reliance on the central limit theorem, in which regions are connected through an evolutionary timeline and the galactic average incidence ratio between their evolutionary phases emerges on scales much larger than the region separation length, can fail if some regions are never visible in one of the tracers considered, or appear multiple times in one tracer before becoming visible in the other. We present practical ways of interpreting the results in such cases.
\end{enumerate}

\subsection{Applicability and future work} \label{sec:outlook}
To conclude this paper, we present a brief outlook on how real-Universe applications of the method are expected to provide insight into the physics of star formation and feedback. Throughout this work, we have mostly presented examples in which our method is applied to a star formation rate tracer map and a (molecular) gas tracer map. However, it has also been reiterated that the method can be applied to any pair of tracer maps that share an evolutionary connection according to the schematic timeline of \autoref{fig:tschem}. For instance, the method can be applied to pairs of different gas tracer maps, pairs of different ionised emission line maps, or any combination of these. This enables a broad range of applications, including measurements of the molecular cloud lifetime, the cloud condensation time-scale, the feedback time-scale for cloud destruction, feedback outflow velocity, the mass loading factor, the feedback-ISM coupling efficiency, the star formation efficiency, the fragmentation length of galaxy discs, and many more. The systematic applicability to a broad wavelength and redshift range is key in achieving a census of cloud-scale star formation and feedback as a function of galactic environment.

Each of the above quantities can not only be obtained across a sample of different galaxies, but also as a function of galactocentric radius or local environment within single galaxies. Different tracers can be combined to obtain maps of derived physical quantities such as the density or electron temperature, which can then be used to mask the emission maps prior to applying the method, providing a way to measure the time-scales on which these quantities change. The generality of the method provides an enormous freedom in choosing such applications. Varying the gas tracer (e.g.~H{\sc i}, CO, HCN, HCO$^+$) constrains the absolute time-evolution towards increasing densities during cloud condensation, gravitational collapse, and star formation. When considering the physics of feedback, it may even be possible to choose different tracers to highlight individual feedback mechanisms. For instance, SN remnants can be traced by an elevated [S{\sc ii}]/\halpha ratio \citep[e.g.][]{kreckel17}, implying that their time evolution can be probed by combining a reference map with a young stellar map masked by a critical [S{\sc ii}]/\halpha line ratio. Similarly, the infrared photon processing that drives radiation pressure feedback may be traced with JWST in the near-infrared, whereas photoionization feedback is traced by $\halpha$. These examples show that a promising way forward is to simultaneously observe a variety of feedback tracers with MUSE on the VLT to measure the individual time-scales of each mechanism.

In addition to the observational applications of the presented method, it may be applied to numerical simulations of galaxy formation and evolution. Such applications can serve a variety of purposes. For instance, the method may be used to infer the lifecycle of clouds and star-forming regions in simulations with sparse output intervals, in which the evolution of individual structures cannot be followed between different snapshots. In addition, galaxy simulations may be used to test the method further by considering galaxies spanning a wider variety of physical properties than included in this work. By post-processing such simulations to generate synthetic observations of (molecular) gas or SFR tracers \citep[see e.g.][]{pawlik11,dasilva12,krumholz14c,haworth18}, it is possible to derive the tracer lifetimes and test how the method's performance is affected by environmental variations in tracer emissivity. First efforts in this direction are currently in progress.

This reference paper is accompanied by two companion papers. Firstly, in \citet{haydon18}, we apply the method to synthetic SFR tracer maps of simulated disc galaxies, to calibrate the reference time-scales of the main SFR tracers (e.g.~\halpha, FUV, NUV) as a function of the metallicity and SFR surface density. Both of these quantities are expected to change the SFR tracer lifetimes due to differences in massive star lifetimes and varying degrees of IMF sampling, respectively. The results presented by \citet{haydon18} systematically quantify these dependences and thus provide the reference time-scales needed for the systematic application of the method to large galaxy samples. Secondly, in \citet{hygate18}, we develop a new approach for quantifying the amount of diffuse emission in tracer maps and correcting for its presence. To this end, we extend the method with a module for filtering out the diffuse component. This enables the method's application to galaxies with high diffuse fractions, which generally have high surface densities. The approach presented by \citet{hygate18} thus greatly increases the method's range of possible applications. 

With the presented method at hand, it is possible to empirically constrain the main unknowns in galaxy formation simulations, such as the star formation time-scale, the star formation efficiency, the feedback outflow rate, and its coupling efficiency. We are currently carrying out systematic applications of the method to a large sample of nearby galaxies aimed at probing and understanding these physical quantities as a function of the galactic environment. Initially, this focuses on single-tracer observations of individual galaxies, such as NGC300 \citep{kruijssen18}, M33 (Hygate et al.~in prep.), and M31 (Schruba et al.~in prep.), enabling a detailed understanding of the interplay between galactic environment and the cloud lifecycle across the face of nearby galaxies, as well as providing a point of reference for comparison to previous work. In the intermediate term, we intend to capitalise on the method's straightforward applicability by extending these applications to large galaxy samples (Chevance et al.~in prep.) and panchromatic observations of many different pairs of emission maps, initially in the Large Magellanic Cloud (Ward et al.~in prep.), but eventually covering the nearby galaxy population. We anticipate that the resulting measurements will provide the most systematic and accurate constraints to date on the cloud-scale physics of star formation and feedback across the galaxy population, that way providing a critical point of reference for calibrating the sub-grid physics in the next generation of large-scale galaxy formation models.

\section*{Acknowledgements}
JMDK gratefully acknowledges funding from the German Research Foundation (DFG) in the form of an Emmy Noether Research Group (grant number KR4801/1-1, PI Kruijssen), from the European Research Council (ERC) under the European Union's Horizon 2020 research and innovation programme via the ERC Starting Grant MUSTANG (grant agreement number 714907, PI Kruijssen), and from Sonderforschungsbereich SFB 881 ``The Milky Way System'' (subproject P1) of the DFG.
APSH and DTH are fellows of the International Max Planck Research School for Astronomy and Cosmic Physics at the University of Heidelberg (IMPRS-HD).
The Flatiron Institute is supported by the Simons Foundation.
The numerical simulations were performed on the Odin cluster hosted by the Max Planck Computing \& Data Facility (\url{http://www.mpcdf.mpg.de/}).
This work has made use of Numpy \citep{vanderwalt11}, Scipy \citep{jones01}, Astropy \citep{astropy13}, and Pynbody \citep{pontzen13}. All figures in this paper were produced using the Python library Matplotlib \citep{hunter07}.
The image of the white whale in \autoref{fig:redshift} was adapted from Herman Melville's `Moby-Dick; Or, The Whale', 1851, page~510 of the 1892 edition (illustration by Augustus Burnham Shute).
JMDK thanks Simon White for many stimulating conversations that helped shape the results presented in this paper.
We thank Frank Bigiel, Andi Burkert, Daniela Calzetti, M\'{e}lanie Chevance, Bruce Elmegreen, Robert Feldmann, Simon Glover, Jonathan Henshaw, Annie Hughes, Sarah Jeffreson, Cliff Johnson, Rob Kennicutt, Ralf Klessen, Kathryn Kreckel, Sophie Kruijssen, Mark Krumholz, Charlie Lada, Adam Leroy, Sharon Meidt, Thorsten Naab, Eve Ostriker, Paolo Padoan, Eric Pellegrini, Marta Reina-Campos, Hans-Walter Rix, Joop Schaye, Eva Schinnerer, Andreas Schmidt, Linda Tacconi, Fabian Walter, and Jacob Ward for insightful comments and discussions during the development of this work.
We are greatly indebted to an anonymous referee for their careful reading of the manuscript and for their constructive feedback, which improved the presentation of this paper.

\bibliographystyle{mnras}
\bibliography{mybib}

\appendix

\section{Calculating the exact enclosed pixel area when intersected by a circular aperture} \label{sec:appcircseg}
As discussed in Section~\ref{sec:stepconvolve}, the numerical use of a tophat kernel convolution to determine all flux within an aperture of a certain size must account for the fact that the kernel boundary intersects with pixels, for which the kernel value is a real number between zero and unity, i.e.~$\{W_{ij}\in\mathbb{R}~|~0\leq W_{ij}\leq1\}$. This number represents the fraction of the pixel area enclosed by the aperture. In this appendix, we explain how the exact fraction of each intersected pixel's surface area for which $r\leq h$ is analytically calculated by dividing up the pixel area into square, rectangular, triangular, or circular segment shapes and adding or subtracting the appropriate parts of the pixel.

\begin{figure*}
\includegraphics[width=\hsize]{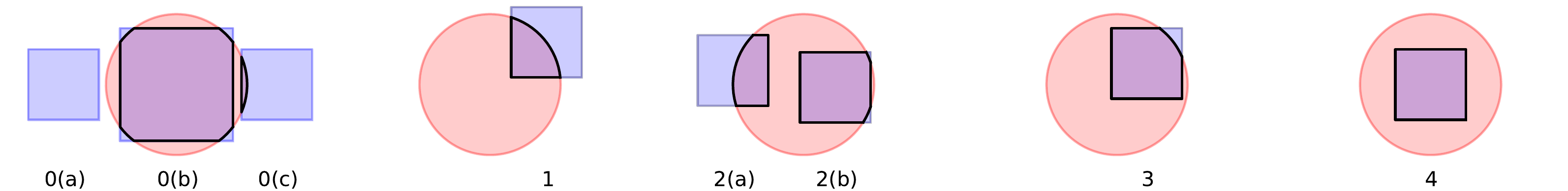}%
\vspace{-1mm}\caption{
\label{fig:circseg}
Illustration of all possible intersections (black lines enclosing purple-shaded areas) between circular apertures (red circles) and pixels (blue squares). From left to right, the panels show all cases of having $\{0,1,2,3,4\}$ corners of the pixel within the aperture. The numbers (with letters) along the bottom of the figure indicate the eight possible topologies. These illustrate that the surface areas of all intersections can be calculated through simple geometry by adding and subtracting squares, rectangles, triangles, and circular segments. For reference, the intersection of pixel 0(c) resembles a single circular segment. Note that the relative sizes and positions of the apertures and pixels are chosen to best illustrate the different topologies, but are not representative of the relative sizes (or positions) when calculating the total flux within apertures of various sizes in Section~\ref{sec:stepconvolve}. Typical aperture sizes exceed the pixel size by a factor $3$--$500$ (see Sections~\ref{sec:method} and~\ref{sec:valid}), implying that the curvature of the aperture boundary on the scale of a pixel varies as a function of aperture size from being highly significant to effectively unnoticeable.\vspace{-1mm}
}
\end{figure*}
To determine the area of an intersected pixel that falls within the aperture boundary, we first determine the coordinates of the pixel corners and consider the five separate cases of $N=\{0,1,2,3,4\}$ corners enclosed by the aperture. By considering these different cases, we can easily generalise the problem to dealing with aperture-pixel intersections having eight different topologies, which are shown in \autoref{fig:circseg}. Intersections~0(a) and~4 represent the trivial cases of pixels residing entirely outside or within the aperture, respectively. Each of the six remaining cases can be assembled by adding and subtracting squares, rectangles, triangles, and circular segments. Intersection~0(c) shows the shape of a single circular segment, which consists of a straight line and a circular arc. If we define the thickness of the circular segment $l_{\rm cs}$ as the maximum distance between the straight line and the arc, measured perpendicularly to the straight line, then the area under the circular segment is expressed analytically as
\be
\label{eq:circseg}
A_{\rm cs}=\frac{\lap^2}{4}\cos^{-1}\left(\frac{R-l_{\rm cs}}{R}\right)-(R-l_{\rm cs})\sqrt{2Rl_{\rm cs}-l_{\rm cs}^2} .
\ee
In combination with the familiar expressions for the areas of a square, rectangle, and triangle, this means that for each of the pixels that are intersected by the aperture boundary, it is straightforward to obtain the exact fraction of the pixel area enclosed by the aperture.

\section{Requirements for minimising the effects of blending from adjacent peaks} \label{sec:appblending}
Here we quantify the effects of the blending between adjacent peaks with two-dimensional Gaussian surface density profiles (see Section~\ref{sec:validtover}). Specifically, we focus on the identification of those peaks using the clump finding algorithm (see Section~\ref{sec:method}) and on the mutual contamination between these peaks.

\subsection{Peak identification}
To identify peaks, our method uses a commonly-used clump finding algorithm \citep{williams94} that selects peaks by looking for closed contours in the maps. By default, these contours are logarithmically spaced by a specified amount $\delta\log_{10}{\f}$. For the \{low, high\} resolution simulations, we use $\delta\log_{10}{\f}=\{0.25, 0.5\}$. In order to identify two adjacent peaks, the contrast between the logarithms of the peak flux and the minimum in between both peaks should be at least $\delta\log_{10}{\f}$. For a set of two Gaussian profiles with standard deviation $r$, centred on $(x, y)=(0, 0)$ and $(x, y)=(\lambda, 0)$, the flux in the two-dimensional plane is
\be
\label{eq:gauss}
\f(x,y) \propto {\rm e}^{-(x^2+y^2)/2r^2}+{\rm e}^{-[(x-\lambda)^2+y^2]/2r^2} ,
\ee
with the partial derivative along the line connecting both Gaussians ($y=0$) given by
\be
\label{eq:gaussder}
\frac{\partial\f(x,0)}{\partial x} \propto (\lambda-x){\rm e}^{-(x-\lambda)^2/2r^2}-x{\rm e}^{-x^2/2r^2} .
\ee
We numerically solve $\partial\f(x, 0)/\partial x=0$ to locate the peaks and the minimum between both peaks. The profile only has a minimum for separation lengths $\lambda>2r$ (or filling factors $\zeta\equiv2r/\lambda<1$), because for smaller separations (or larger filling factors) the Gaussians blend together into a single peak. For $\lambda>2r$, the positions of both peaks are shifted inwards relative to the centres of the isolated Gaussians, due to the asymmetry of having a neighbour contributing emission from only one direction. The observed peak separation is therefore smaller than the true separation between the underlying Gaussians and the measured filling factor $\zeta$ exceeds the true filling factor. However, the difference is minor -- for $\zeta\leq0.6$, the true and observed filling factor are effectively indistinguishable.

Having located the locations of the blended minimum and maxima in the double-Gaussian profile, it is trivial to evaluate the combined flux at each of these positions and determine the flux contrast between maximum and minimum as a function of $\zeta$. This is shown in the top panel of \autoref{fig:blending}. We highlight the flux contrasts used to identify the peaks in the simulated maps as horizontal and vertical dotted lines, which define the maximum values $\zetacrit$ below which the contrast is large enough to identify peaks. We discard all experiments with $\zeta>\zetacrit$, where for the \{low, high\} resolution simulations we obtain $\zetacrit=\{0.64, 0.52\}$.

\subsection{Contamination}
Given two adjacent Gaussians, the integrated flux under each of these includes some contribution from the other peak. For our method to accurately determine the fraction of time for which both tracers coexist (i.e.~the duration of the overlap phase $\tover$), we wish to limit the contamination by adjacent peaks. To determine this contamination as a function of the filling factor $\zeta$, we determine the fraction of the integrated flux on one side of the equidistance line ($x=\lambda/2$) between both Gaussians that is contributed by the peak on the other side of that line. Due to symmetry, the total integrated flux on each side of the line must equal to the integrated flux of a single Gaussian. If we normalise each (two-dimensional) Gaussian to unity, the contamination fraction $f_{\rm cont}$ reduces to the integral of a single two-dimensional Gaussian with standard deviation $r$ over $x\geq\lambda/2$. It is straightforward to show that $f_{\rm cont}$ then follows as
\be
\label{eq:fcont}
f_{\rm cont}=\int_{\lambda/2}^{\infty} \frac{1}{\pi}\arccos{\left(\frac{\lambda}{2x}\right)}x{\rm e}^{-x^2/2r^2}{\rm d}x ,
\ee
which is a one-dimensional function of $\zeta\equiv2r/\lambda$.

\begin{figure}
\includegraphics[width=\hsize]{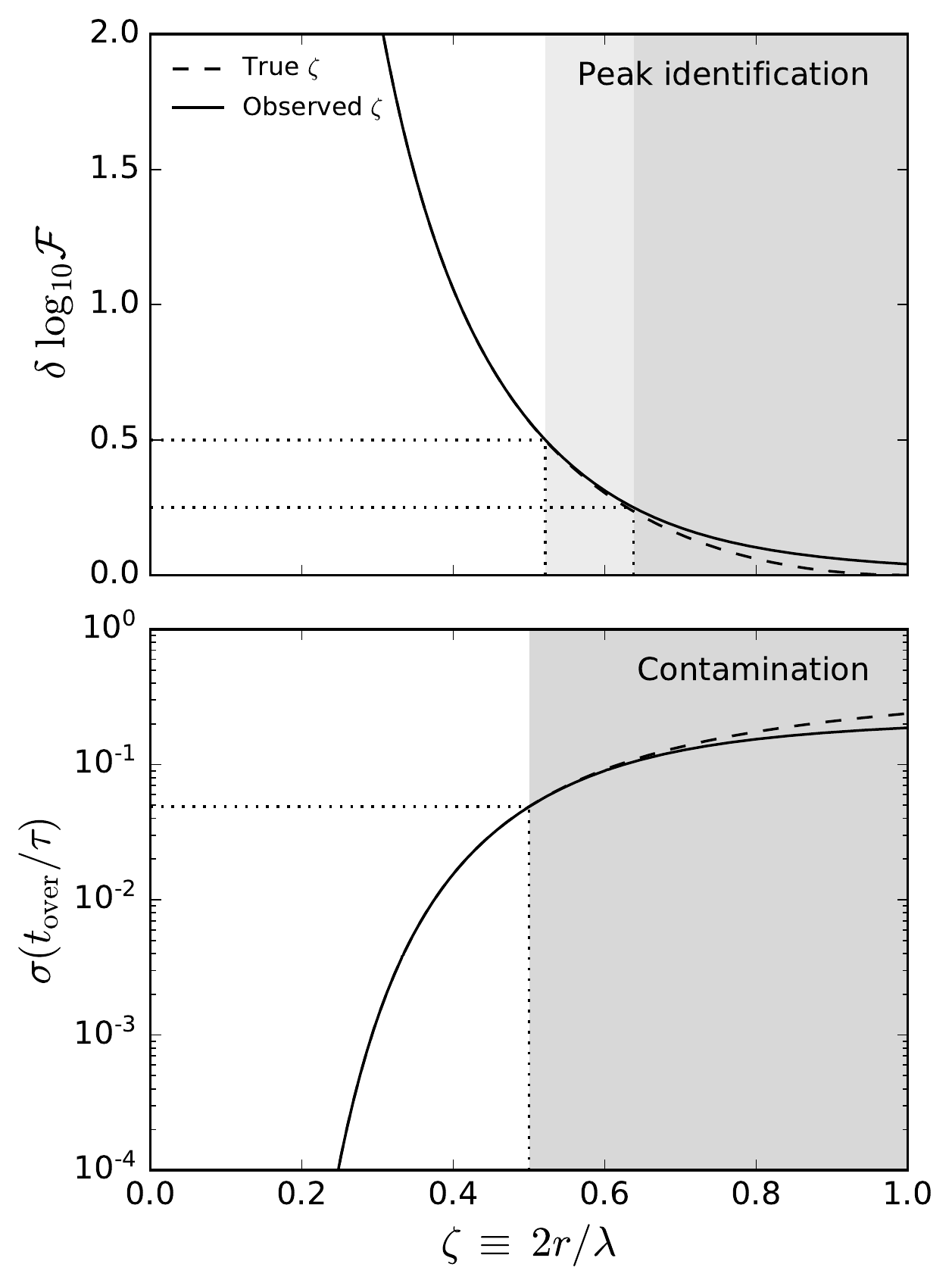}%
\vspace{-1mm}\caption{
\label{fig:blending}
Effects of Gaussian blending on the identification of peaks and on the contamination of regions by neighbouring peaks. Top panel: logarithmic density contrast $\delta\log_{10}{\f}$ between the peak flux and the minimum in between two adjacent Gaussian profiles with standard deviation $r$ and separation $\lambda$, as a function of the filling factor $\zeta\equiv2r/\lambda$. The dotted lines indicate the contrasts used to identify peaks in the simulated maps of Section~\ref{sec:valid}. The grey-shaded areas indicate the range of filling factors where blending inhibits the identification of the peaks. Bottom panel: systematic uncertainty of the relative duration of the overlap phase due to blending as a function of the filling factor $\zeta$. This uncertainty is equal to the fraction of the integrated flux on one side of the equidistance line between two Gaussians ($x=\lambda/2$) that is contributed by the peak on the other side of that line. The dotted line indicates the desired maximum level of such contamination (see the text) and the grey-shaded area indicates the range of filling factors where this maximum is exceeded. In both panels, the dashed line uses the true separation between both Gaussians ($\lambda$), whereas the solid line accounts for the blending-induced shift of both peaks towards one another.\vspace{-1mm}
}
\end{figure}
Our method determines the duration of the overlap phase by measuring the correlation between both tracers in excess of the statistical correlation expected for a random distribution in the plane. This measurement assumes a linear relation between the population-integrated photon flux and the mean lifetime. Therefore, the amount by which blending affects the relative duration of the overlap phase $\tover/\tau$ is equal to the contamination fraction from equation~(\ref{eq:fcont}). If we express this amount as a blending-induced uncertainty, we can thus define $\sigma(\tover/\tau)\equiv f_{\rm cont}$. This uncertainty is shown as a function of $\zeta$ in the bottom panel of \autoref{fig:blending}. Given that our method is able to determine $\tover/\tau$ at an accuracy of $\sim5$~per~cent (see Section~\ref{sec:validtover}), we require that $\sigma(\tover/\tau)<0.05$. This condition is satisfied for $\zeta<0.50$, which corresponds to the requirement that regions are separated by at least $1.69$ times their FWHM. We impose this condition to guarantee the reliability of the measured values of $\tover$, resulting in an average blending-induced uncertainty of $\sigma(\tover/\tau)<0.05$. Of course, it is possible to impose a higher upper limit on $\zeta$, as long as \autoref{fig:blending} is used to quantify the associated blending-induced uncertainty on (or possible overestimation of) the duration of the overlap phase.

\section{The probability of gas reaching high densities prior to star formation} \label{sec:apppgas}
As discussed in Section~\ref{sec:expval}, the duration of the gas phase retrieved from the simulations is affected by the fact that some regions never reach the gas densities necessary for being shown in the gas map before being destroyed by feedback. Because star-forming regions in the real Universe do not form stars stochastically like our simulations do, this is not expected to influence observational applications of our method. However, for interpreting the results of our experiments, it is important to derive the fraction of gas particles that is capable of reaching the minimum density shown in the gas map $\rhomin$ without being affected by star formation.

The formation of star particles within a given smoothing kernel in the simulation is a Poisson process with an occurrence rate $\lambda_p$ set by
\be
\label{eq:lambdap}
\lambda_p=\frac{\epsilon N_{\rm ngb}}{\tdyn} ,
\ee
where the factor $N_{\rm ngb}=200$ represents the number of neighbours and accounts for the fact that any star formation event within the smoothing kernel will associate that kernel with star formation. For a constant dynamical time-scale, the probability of $k$ particles being turned into stars after a time $t$ is given by
\be
\label{eq:poisson}
p(k)=\frac{(\lambda_pt)^k}{k!}{\rm e}^{-\lambda_pt} ,
\ee
such that the probability of a gas concentration to be unaffected by star formation (i.e.~$k=0$) for at least a time $t$ is
\be
\label{eq:pgas1}
p_{\rm gas}={\rm e}^{-\lambda_pt} .
\ee
However, the dynamical time-scale on which the particles are spawned is time-dependent, because it is set by the kernel-averaged gas density. We therefore use a more general expression
\be
\label{eq:pgas2}
p_{\rm gas}(t)={\rm e}^{-\tilde{\lambda}_p(t)} ,
\ee
where we define
\be
\label{eq:lambdap2}
\tilde{\lambda}_p(t)=\int_0^t\lambda_p(t'){\rm d}t' ,
\ee
as the integral over a time-dependent occurrence rate $\lambda[\tdyn(t)]$.

While equation~(\ref{eq:lambdap2}) accounts for the time-dependence of $\lambda_p(t)$, we are not interested in obtaining $p_{\rm gas}$ at a given time, but at a given gas density. In order to obtain such an expression, we need to describe how the density changes with time. For this purpose, we define a dimensionless quantity
\be
\label{eq:xirho}
\xi_\rho \equiv \left(\frac{\rhocrit}{\rhog}\right)^{1/3} ,
\ee
that reflects the relative density increase beyond the star formation threshold $\rhocrit$. This will allow us to derive a relation $\xi_\rho(t)$ and rewrite equation~(\ref{eq:lambdap2}) as
\be
\label{eq:lambdap3}
\tilde{\lambda}_p(\xi_\rho)=\int_1^{\xi_\rho}\lambda_p(\xi_\rho')\frac{{\rm d}t}{{\rm d}\xi_\rho'}{\rm d}\xi_\rho' ,
\ee
so that we obtain $p_{\rm gas}(\xi_\rho)$ and, hence, $p_{\rm gas}(\rhog)$.

The key missing ingredient allowing us to derive an expression for $p_{\rm gas}(\xi_\rho)$ is the time-evolution of the gas density, or $\xi_\rho(t)$, so that we can specify its (inverse) time-derivative ${\rm d}t/{\rm d}\xi_\rho$ in equation~(\ref{eq:lambdap3}). For this purpose, we describe the star-forming gas by the gravitational free-fall of a homogeneous sphere. Defining the initial and current radii of the sphere as $R$ and $r$, respectively, and its mass as $M$, the textbook rate of collapse is
\be
\label{eq:drdt}
\frac{{\rm d}r}{{\rm d}t}=-\sqrt{2GM\left(\frac{1}{r}-\frac{1}{R}\right)} ,
\ee
which can be rewritten as
\be
\label{eq:dtdxi}
\frac{{\rm d}t}{{\rm d}\xi_\rho}=-\left(\frac{3}{8\pi G\rhocrit}\right)^{1/2}\left(\frac{\xi_\rho}{1-\xi_\rho}\right)^{1/2} ,
\ee
where we have set the initial density of the homogeneous sphere equal to the critical density for star formation in the simulations, i.e.~$\rhocrit=3M/4\pi R^3$.

We now combine equation~(\ref{eq:dtdxi}) with the density-dependent occurrence rate, which follows from equation~(\ref{eq:lambdap}) as
\be
\label{eq:lambdap4}
\lambda_p(\xi_\rho)=\epsilon N_{\rm ngb}(4\pi G\rhocrit)^{1/2}\xi_\rho^{-3/2} ,
\ee
upon which substitution into equation~(\ref{eq:lambdap3}) results in an integral that can be solved analytically (recall that $0<\xi_\rho\leq1$):
\be
\label{eq:lambdap5}
\begin{aligned}
\tilde{\lambda}_p(\xi_\rho)=&-\left(\frac{3}{2}\right)^{1/2}\epsilon N_{\rm ngb}\int_1^{\xi_\rho} \left({\xi_\rho'}^2-{\xi_\rho'}^3\right)^{-1/2}{\rm d}\xi_\rho' \\
=& \left(\frac{3}{2}\right)^{1/2}\epsilon N_{\rm ngb}\left[2\tanh^{-1}\left(\sqrt{1-\xi_\rho'}\right)\right]_{\xi_\rho'=1}^{\xi_\rho'=\xi_\rho}\\
=& \left(\frac{3}{2}\right)^{1/2}\epsilon N_{\rm ngb}\ln{\left(\frac{1+\sqrt{1-\xi_\rho}}{1-\sqrt{1-\xi_\rho}}\right)} .
\end{aligned}
\ee
This provides the required expression of the integrated density-dependent occurrence rate. Substitution into equation~(\ref{eq:pgas2}) then yields the probability that a gas concentration is unaffected by star formation until at least a density $\rhog$ has been reached as
\be
\label{eq:pgas3}
p_{\rm gas}(\xi_\rho)=\left(\frac{1+\sqrt{1-\xi_\rho}}{1-\sqrt{1-\xi_\rho}}\right)^{-\sqrt{3/2}\epsilon N_{\rm ngb}} ,
\ee
or, in terms of the volume density, as
\be
\label{eq:pgas4}
p_{\rm gas}(\rhog)=\left[\frac{1+\sqrt{1-(\rhocrit/\rhog)^{1/3}}}{1-\sqrt{1-(\rhocrit/\rhog)^{1/3}}}\right]^{-\sqrt{3/2}\epsilon N_{\rm ngb}} ,
\ee
which is valid for $\rhog\geq\rhocrit$. For $\rhog<\rhocrit$, we set $p_{\rm gas}=1$.

\section{Complete results of the performed experiments} \label{sec:appexp}
Here we tabulate the best-fitting solutions from all experiments discussed in Section~\ref{sec:valid}, including some of the derived quantities.

\subsection{Age-binned stellar maps} \label{sec:appstarstar}
Section~\ref{sec:starstar} discusses 88 different experiments using the star particles in the numerical simulations described in Section~\ref{sec:models}. These are constituted by a $2\times2$ matrix of the 22 experiments listed in \autoref{tab:starruns_in}, using maps of \{point particles, extended emission\} in the \{low, high\} resolution galaxy simulations. The output is listed in \autoref{tab:starruns_p_lr_out}--\ref{tab:starruns_hr_out}, which represent the low-resolution point particle experiments, the high-resolution point particle experiments, the low-resolution extended emission experiments, and the high-resolution extended emission experiments, respectively.
\begin{table*}
 \centering
 \begin{minipage}{\hsize}
  \caption{Best-fitting solutions for age-binned stellar maps (low resolution, point particles)}\label{tab:starruns_p_lr_out}\vspace{-1mm}
  \begin{tabular}{c c c c c c c c c c c c}
   \hline
   ID & $N_{\rm star}$ & $N_{\rm gas}$ & $t_{\rm gas}$ & $t_{\rm over}$ & $\lambda$ & $\beta_{\rm star}$ & $\beta_{\rm gas}$& $\zeta_{\rm star}$ & $\zeta_{\rm gas}$ & $\chi_{\rm red}^2$ & Symbol\\ 
   \hline
   $1$ & $57$ & $49$ & $1.91_{-0.44}^{+0.98}$ & $0.01_{-0.00}^{+0.01}$ & $907_{-73}^{+88}$ & $0.49_{-0.00}^{+0.00}$ & $1.00_{-0.00}^{+0.00}$ & $0.09_{-0.01}^{+0.01}$ & $0.10_{-0.01}^{+0.01}$ & 5.38 & $\color{black}\blacksquare$ \\ 
   $2$ & $158$ & $157$ & $10.85_{-1.89}^{+3.80}$ & $0.01_{-0.00}^{+0.09}$ & $579_{-33}^{+46}$ & $0.48_{-0.00}^{+0.15}$ & $0.99_{-0.00}^{+0.00}$ & $0.16_{-0.01}^{+0.01}$ & $0.17_{-0.01}^{+0.01}$ & 0.78 & $\color{black}\blacksquare$ \\ 
   $3$ & $420$ & $426$ & $24.53_{-3.88}^{+4.62}$ & $0.04_{-0.02}^{+0.22}$ & $513_{-20}^{+41}$ & $1.57_{-0.30}^{+0.00}$ & $0.67_{-0.00}^{+0.16}$ & $0.19_{-0.01}^{+0.01}$ & $0.21_{-0.01}^{+0.01}$ & 0.55 & $\color{black}\blacksquare$ \\ 
   $4$ & $977$ & $1219$ & $135.56_{-12.52}^{+15.20}$ & $0.01_{-0.00}^{+1.58}$ & $300_{-6}^{+19}$ & $0.96_{-0.02}^{+0.03}$ & $0.87_{-0.00}^{+0.14}$ & $0.32_{-0.02}^{+0.01}$ & $0.31_{-0.01}^{+0.01}$ & 0.49 & $\color{black}\blacksquare$ \\[1.5ex] 
   $5$ & $26$ & $49$ & $1.69_{-0.45}^{+1.07}$ & $0.01_{-0.00}^{+0.01}$ & $1226_{-159}^{+134}$ & $0.49_{-0.00}^{+0.00}$ & $0.90_{-0.00}^{+0.01}$ & $0.07_{-0.01}^{+0.01}$ & $0.07_{-0.01}^{+0.01}$ & 4.31 & $\color{blue}\rotatebox[origin=c]{90}{$\blacktriangle$}$ \\ 
   $6$ & $59$ & $15$ & $0.62_{-0.17}^{+0.51}$ & $0.07_{-0.02}^{+0.07}$ & $1134_{-180}^{+414}$ & $1.04_{-0.05}^{+0.00}$ & $1.01_{-0.00}^{+0.02}$ & $0.08_{-0.02}^{+0.01}$ & $0.08_{-0.02}^{+0.02}$ & 1.24 & $\color{blue}\rotatebox[origin=c]{270}{$\blacktriangle$}$ \\ 
   $7$ & $174$ & $426$ & $18.99_{-3.68}^{+4.70}$ & $0.03_{-0.01}^{+0.13}$ & $543_{-29}^{+52}$ & $1.13_{-0.07}^{+0.00}$ & $0.91_{-0.01}^{+0.19}$ & $0.18_{-0.01}^{+0.01}$ & $0.19_{-0.01}^{+0.01}$ & 0.42 & $\color{blue}\rotatebox[origin=c]{90}{$\blacktriangle$}$ \\ 
   $8$ & $452$ & $157$ & $10.79_{-1.50}^{+3.58}$ & $0.01_{-0.00}^{+0.20}$ & $488_{-22}^{+54}$ & $0.48_{-0.00}^{+0.24}$ & $0.99_{-0.02}^{+0.00}$ & $0.19_{-0.01}^{+0.01}$ & $0.21_{-0.02}^{+0.01}$ & 0.57 & $\color{blue}\rotatebox[origin=c]{270}{$\blacktriangle$}$ \\ 
   $9$ & $15$ & $157$ & $19.89_{-6.61}^{+20.73}$ & $0.08_{-0.04}^{+0.12}$ & $561_{-164}^{+147}$ & $0.53_{-0.03}^{+0.10}$ & $0.95_{-0.00}^{+0.09}$ & $0.17_{-0.03}^{+0.07}$ & $0.15_{-0.02}^{+0.04}$ & 2.79 & $\color{blue}\blacktriangledown$ \\ 
   $10$ & $156$ & $15$ & $0.82_{-0.23}^{+0.82}$ & $0.11_{-0.04}^{+0.12}$ & $695_{-112}^{+322}$ & $1.03_{-0.01}^{+0.01}$ & $1.01_{-0.01}^{+0.01}$ & $0.13_{-0.03}^{+0.02}$ & $0.14_{-0.05}^{+0.03}$ & 2.45 & $\color{blue}\blacktriangle$ \\ 
   $11$ & $61$ & $426$ & $12.88_{-2.71}^{+6.29}$ & $0.14_{-0.06}^{+0.06}$ & $704_{-120}^{+89}$ & $1.07_{-0.01}^{+0.02}$ & $1.04_{-0.14}^{+0.46}$ & $0.14_{-0.02}^{+0.03}$ & $0.16_{-0.01}^{+0.02}$ & 0.43 & $\color{blue}\blacktriangledown$ \\ 
   $12$ & $442$ & $49$ & $3.15_{-0.60}^{+1.46}$ & $0.01_{-0.00}^{+0.05}$ & $528_{-43}^{+90}$ & $0.48_{-0.00}^{+0.08}$ & $1.02_{-0.00}^{+0.00}$ & $0.18_{-0.01}^{+0.01}$ & $0.19_{-0.03}^{+0.02}$ & 1.94 & $\color{blue}\blacktriangle$ \\[1.5ex] 
   $13$ & $148$ & $157$ & $12.03_{-2.03}^{+3.08}$ & $1.05_{-0.20}^{+0.21}$ & $584_{-72}^{+72}$ & $0.92_{-0.02}^{+0.05}$ & $0.93_{-0.00}^{+0.01}$ & $0.16_{-0.01}^{+0.02}$ & $0.16_{-0.01}^{+0.02}$ & 0.31 & $\color{red}\blacksquare$ \\ 
   $14$ & $149$ & $157$ & $10.80_{-1.19}^{+1.68}$ & $3.77_{-0.33}^{+0.38}$ & $577_{-97}^{+115}$ & $0.99_{-0.02}^{+0.03}$ & $0.93_{-0.01}^{+0.03}$ & $0.16_{-0.02}^{+0.03}$ & $0.16_{-0.02}^{+0.02}$ & 0.26 & $\color{red}\blacksquare$ \\ 
   $15$ & $148$ & $157$ & $11.05_{-0.69}^{+0.96}$ & $8.32_{-0.31}^{+0.43}$ & $603_{-184}^{+438}$ & $1.04_{-0.05}^{+0.01}$ & $0.87_{-0.00}^{+0.02}$ & $0.15_{-0.05}^{+0.05}$ & $0.15_{-0.05}^{+0.05}$ & 0.36 & $\color{red}\blacksquare$ \\[1.5ex] 
   $16$ & $176$ & $426$ & $23.53_{-3.07}^{+4.03}$ & $3.17_{-0.29}^{+0.31}$ & $582_{-87}^{+96}$ & $1.02_{-0.00}^{+0.01}$ & $1.01_{-0.08}^{+0.09}$ & $0.17_{-0.02}^{+0.02}$ & $0.18_{-0.01}^{+0.02}$ & 0.42 & $\color{green}\rotatebox[origin=c]{90}{$\blacktriangle$}$ \\ 
   $17$ & $452$ & $157$ & $9.23_{-1.25}^{+1.53}$ & $3.43_{-0.43}^{+0.43}$ & $479_{-62}^{+72}$ & $0.96_{-0.02}^{+0.03}$ & $0.96_{-0.04}^{+0.00}$ & $0.19_{-0.01}^{+0.02}$ & $0.21_{-0.02}^{+0.02}$ & 0.67 & $\color{green}\rotatebox[origin=c]{270}{$\blacktriangle$}$ \\[1.5ex] 
   $18$ & $148$ & $157$ & $10.64_{-1.20}^{+1.49}$ & $3.65_{-0.33}^{+0.22}$ & $573_{-100}^{+97}$ & $0.52_{-0.02}^{+0.05}$ & $0.92_{-0.00}^{+0.01}$ & $0.16_{-0.02}^{+0.03}$ & $0.16_{-0.02}^{+0.03}$ & 0.35 & $\color{cyan}\blacksquare$ \\ 
   $19$ & $149$ & $157$ & $10.84_{-1.05}^{+2.22}$ & $3.71_{-0.36}^{+0.44}$ & $581_{-105}^{+90}$ & $0.99_{-0.02}^{+0.02}$ & $0.47_{-0.01}^{+0.02}$ & $0.16_{-0.02}^{+0.03}$ & $0.16_{-0.02}^{+0.03}$ & 0.17 & $\color{cyan}\blacksquare$ \\ 
   $20$ & $148$ & $157$ & $10.95_{-1.13}^{+1.94}$ & $3.61_{-0.39}^{+0.25}$ & $565_{-95}^{+82}$ & $0.51_{-0.01}^{+0.04}$ & $0.48_{-0.00}^{+0.02}$ & $0.16_{-0.02}^{+0.03}$ & $0.16_{-0.02}^{+0.02}$ & 0.25 & $\color{cyan}\blacksquare$ \\[1.5ex] 
   $21$ & $163$ & $348$ & $20.91_{-2.21}^{+3.12}$ & $3.87_{-0.32}^{+0.35}$ & $643_{-95}^{+105}$ & $1.00_{-0.00}^{+0.00}$ & $0.98_{-0.02}^{+0.02}$ & $0.14_{-0.02}^{+0.02}$ & $0.14_{-0.02}^{+0.02}$ & 0.31 & $\color{magenta}\rotatebox[origin=c]{90}{$\blacktriangle$}$ \\ 
   $22$ & $336$ & $58$ & $4.84_{-1.03}^{+3.17}$ & $0.02_{-0.01}^{+0.12}$ & $707_{-61}^{+153}$ & $1.00_{-0.00}^{+0.73}$ & $1.29_{-0.00}^{+0.00}$ & $0.13_{-0.01}^{+0.01}$ & $0.14_{-0.02}^{+0.01}$ & 0.29 & $\color{magenta}\blacktriangle$ \\ 
   \hline
  \end{tabular} \\ 
  All time-scales are in units of Myr. All length-scales are in units of pc.\vspace{-1mm}
 \end{minipage}
\end{table*}

\begin{table*}
 \centering
 \begin{minipage}{\hsize}
  \caption{Best-fitting solutions for age-binned stellar maps (high resolution, point particles)}\label{tab:starruns_p_hr_out}\vspace{-1mm}
  \begin{tabular}{c c c c c c c c c c c c}
   \hline
   ID & $N_{\rm star}$ & $N_{\rm gas}$ & $t_{\rm gas}$ & $t_{\rm over}$ & $\lambda$ & $\beta_{\rm star}$ & $\beta_{\rm gas}$& $\zeta_{\rm star}$ & $\zeta_{\rm gas}$ & $\chi_{\rm red}^2$ & Symbol\\ 
   \hline
   $1$ & $321$ & $309$ & $3.30_{-0.44}^{+0.61}$ & $0.01_{-0.00}^{+0.01}$ & $460_{-18}^{+21}$ & $1.00_{-0.00}^{+0.01}$ & $0.97_{-0.00}^{+0.00}$ & $0.10_{-0.00}^{+0.00}$ & $0.10_{-0.00}^{+0.00}$ & 0.85 & $\color{black}\blacksquare$ \\ 
   $2$ & $956$ & $838$ & $8.48_{-0.74}^{+0.89}$ & $0.01_{-0.00}^{+0.01}$ & $328_{-7}^{+9}$ & $0.94_{-0.00}^{+0.01}$ & $0.99_{-0.01}^{+0.00}$ & $0.15_{-0.00}^{+0.00}$ & $0.15_{-0.00}^{+0.00}$ & 3.60 & $\color{black}\blacksquare$ \\ 
   $3$ & $2302$ & $2195$ & $27.84_{-1.54}^{+1.89}$ & $0.01_{-0.00}^{+0.07}$ & $208_{-3}^{+5}$ & $0.97_{-0.00}^{+0.01}$ & $1.09_{-0.00}^{+0.31}$ & $0.22_{-0.00}^{+0.00}$ & $0.23_{-0.00}^{+0.00}$ & 1.16 & $\color{black}\blacksquare$ \\ 
   $4$ & $5668$ & $5998$ & $111.20_{-3.55}^{+3.62}$ & $0.01_{-0.00}^{+0.11}$ & $126_{-1}^{+1}$ & $1.02_{-0.05}^{+0.07}$ & $0.91_{-0.03}^{+0.00}$ & $0.37_{-0.00}^{+0.00}$ & $0.37_{-0.00}^{+0.00}$ & 10.34 & $\color{black}\blacksquare$ \\[1.5ex] 
   $5$ & $126$ & $309$ & $3.25_{-0.55}^{+0.85}$ & $0.01_{-0.00}^{+0.00}$ & $573_{-44}^{+41}$ & $1.01_{-0.00}^{+0.00}$ & $1.00_{-0.00}^{+0.00}$ & $0.08_{-0.01}^{+0.01}$ & $0.08_{-0.00}^{+0.00}$ & 1.66 & $\color{blue}\rotatebox[origin=c]{90}{$\blacktriangle$}$ \\ 
   $6$ & $345$ & $89$ & $0.87_{-0.14}^{+0.30}$ & $0.01_{-0.00}^{+0.01}$ & $588_{-36}^{+66}$ & $0.99_{-0.00}^{+0.02}$ & $1.02_{-0.00}^{+0.01}$ & $0.08_{-0.00}^{+0.00}$ & $0.08_{-0.01}^{+0.01}$ & 0.31 & $\color{blue}\rotatebox[origin=c]{270}{$\blacktriangle$}$ \\ 
   $7$ & $786$ & $2195$ & $27.92_{-2.31}^{+2.54}$ & $0.01_{-0.00}^{+0.03}$ & $255_{-6}^{+9}$ & $0.96_{-0.00}^{+0.04}$ & $0.97_{-0.03}^{+0.00}$ & $0.19_{-0.01}^{+0.00}$ & $0.19_{-0.00}^{+0.00}$ & 0.73 & $\color{blue}\rotatebox[origin=c]{90}{$\blacktriangle$}$ \\ 
   $8$ & $2178$ & $838$ & $11.91_{-0.91}^{+1.13}$ & $0.01_{-0.00}^{+0.02}$ & $268_{-6}^{+8}$ & $0.95_{-0.00}^{+0.02}$ & $0.93_{-0.00}^{+0.00}$ & $0.18_{-0.00}^{+0.00}$ & $0.19_{-0.00}^{+0.00}$ & 2.10 & $\color{blue}\rotatebox[origin=c]{270}{$\blacktriangle$}$ \\ 
   $9$ & $108$ & $838$ & $8.72_{-1.52}^{+2.05}$ & $0.02_{-0.00}^{+0.01}$ & $370_{-32}^{+34}$ & $0.99_{-0.00}^{+0.01}$ & $0.93_{-0.00}^{+0.05}$ & $0.13_{-0.01}^{+0.01}$ & $0.12_{-0.01}^{+0.01}$ & 0.45 & $\color{blue}\blacktriangledown$ \\ 
   $10$ & $852$ & $89$ & $0.97_{-0.14}^{+0.29}$ & $0.01_{-0.00}^{+0.01}$ & $364_{-23}^{+47}$ & $1.03_{-0.00}^{+0.00}$ & $1.02_{-0.01}^{+0.00}$ & $0.12_{-0.01}^{+0.01}$ & $0.13_{-0.02}^{+0.01}$ & 0.58 & $\color{blue}\blacktriangle$ \\ 
   $11$ & $241$ & $2195$ & $28.88_{-4.00}^{+4.46}$ & $0.02_{-0.01}^{+0.04}$ & $273_{-16}^{+21}$ & $1.04_{-0.04}^{+0.00}$ & $0.94_{-0.01}^{+0.03}$ & $0.18_{-0.01}^{+0.01}$ & $0.17_{-0.01}^{+0.01}$ & 0.34 & $\color{blue}\blacktriangledown$ \\ 
   $12$ & $2158$ & $309$ & $4.18_{-0.43}^{+0.58}$ & $0.01_{-0.00}^{+0.02}$ & $267_{-11}^{+15}$ & $0.93_{-0.00}^{+0.05}$ & $0.98_{-0.00}^{+0.03}$ & $0.18_{-0.00}^{+0.00}$ & $0.18_{-0.01}^{+0.01}$ & 1.42 & $\color{blue}\blacktriangle$ \\[1.5ex] 
   $13$ & $944$ & $838$ & $8.73_{-0.61}^{+0.70}$ & $1.28_{-0.09}^{+0.08}$ & $313_{-16}^{+15}$ & $0.98_{-0.00}^{+0.00}$ & $0.92_{-0.00}^{+0.00}$ & $0.15_{-0.01}^{+0.01}$ & $0.15_{-0.01}^{+0.01}$ & 0.38 & $\color{red}\blacksquare$ \\ 
   $14$ & $916$ & $838$ & $8.92_{-0.47}^{+0.53}$ & $2.66_{-0.13}^{+0.11}$ & $318_{-21}^{+21}$ & $0.98_{-0.00}^{+0.00}$ & $0.93_{-0.01}^{+0.03}$ & $0.15_{-0.01}^{+0.01}$ & $0.15_{-0.01}^{+0.01}$ & 0.26 & $\color{red}\blacksquare$ \\ 
   $15$ & $858$ & $838$ & $9.59_{-0.25}^{+0.19}$ & $7.54_{-0.15}^{+0.10}$ & $318_{-41}^{+50}$ & $1.01_{-0.00}^{+0.01}$ & $0.93_{-0.01}^{+0.01}$ & $0.14_{-0.01}^{+0.02}$ & $0.14_{-0.01}^{+0.02}$ & 0.35 & $\color{red}\blacksquare$ \\[1.5ex] 
   $16$ & $773$ & $2195$ & $33.13_{-1.98}^{+2.40}$ & $3.61_{-0.17}^{+0.15}$ & $255_{-16}^{+15}$ & $0.98_{-0.00}^{+0.01}$ & $0.95_{-0.01}^{+0.02}$ & $0.19_{-0.01}^{+0.01}$ & $0.18_{-0.01}^{+0.01}$ & 0.32 & $\color{green}\rotatebox[origin=c]{90}{$\blacktriangle$}$ \\ 
   $17$ & $2124$ & $838$ & $9.73_{-0.59}^{+0.71}$ & $2.93_{-0.17}^{+0.18}$ & $243_{-12}^{+14}$ & $0.95_{-0.00}^{+0.01}$ & $0.92_{-0.00}^{+0.01}$ & $0.19_{-0.01}^{+0.01}$ & $0.20_{-0.01}^{+0.01}$ & 0.56 & $\color{green}\rotatebox[origin=c]{270}{$\blacktriangle$}$ \\[1.5ex] 
   $18$ & $926$ & $838$ & $8.72_{-0.52}^{+0.59}$ & $2.56_{-0.15}^{+0.11}$ & $315_{-19}^{+17}$ & $0.50_{-0.00}^{+0.01}$ & $0.92_{-0.00}^{+0.01}$ & $0.15_{-0.01}^{+0.01}$ & $0.15_{-0.01}^{+0.01}$ & 0.28 & $\color{cyan}\blacksquare$ \\ 
   $19$ & $916$ & $849$ & $9.01_{-0.44}^{+0.65}$ & $2.61_{-0.14}^{+0.14}$ & $314_{-20}^{+18}$ & $0.98_{-0.00}^{+0.00}$ & $0.47_{-0.01}^{+0.02}$ & $0.15_{-0.01}^{+0.01}$ & $0.15_{-0.01}^{+0.01}$ & 0.18 & $\color{cyan}\blacksquare$ \\ 
   $20$ & $926$ & $849$ & $8.83_{-0.48}^{+0.65}$ & $2.53_{-0.16}^{+0.13}$ & $315_{-18}^{+15}$ & $0.50_{-0.00}^{+0.00}$ & $0.46_{-0.00}^{+0.01}$ & $0.15_{-0.00}^{+0.01}$ & $0.15_{-0.01}^{+0.01}$ & 0.20 & $\color{cyan}\blacksquare$ \\[1.5ex] 
   $21$ & $1075$ & $2458$ & $26.50_{-1.26}^{+1.56}$ & $4.26_{-0.14}^{+0.14}$ & $249_{-14}^{+12}$ & $0.98_{-0.01}^{+0.01}$ & $0.96_{-0.00}^{+0.00}$ & $0.19_{-0.01}^{+0.01}$ & $0.18_{-0.01}^{+0.01}$ & 0.64 & $\color{magenta}\rotatebox[origin=c]{90}{$\blacktriangle$}$ \\ 
   $22$ & $2400$ & $320$ & $3.91_{-0.38}^{+0.54}$ & $0.01_{-0.00}^{+0.01}$ & $297_{-12}^{+16}$ & $0.96_{-0.00}^{+0.01}$ & $0.98_{-0.00}^{+0.00}$ & $0.16_{-0.00}^{+0.00}$ & $0.16_{-0.01}^{+0.01}$ & 1.74 & $\color{magenta}\blacktriangle$ \\ 
   \hline
  \end{tabular} \\ 
  All time-scales are in units of Myr. All length-scales are in units of pc. \vspace{-1mm}
 \end{minipage}
\end{table*}

\begin{table*}
 \centering
 \begin{minipage}{\hsize}
  \caption{Best-fitting solutions for age-binned stellar maps (low resolution, extended emission)}\label{tab:starruns_lr_out}\vspace{-1mm}
  \begin{tabular}{c c c c c c c c c c c c}
   \hline
   ID & $N_{\rm star}$ & $N_{\rm gas}$ & $t_{\rm gas}$ & $t_{\rm over}$ & $\lambda$ & $\beta_{\rm star}$ & $\beta_{\rm gas}$& $\zeta_{\rm star}$ & $\zeta_{\rm gas}$ & $\chi_{\rm red}^2$ & Symbol\\ 
   \hline
   $1$ & $49$ & $43$ & $2.44_{-0.68}^{+1.15}$ & $0.18_{-0.07}^{+0.16}$ & $908_{-129}^{+225}$ & $1.35_{-0.17}^{+0.16}$ & $0.89_{-0.22}^{+0.12}$ & $0.30_{-0.05}^{+0.04}$ & $0.31_{-0.05}^{+0.04}$ & 0.36 & $\color{black}\blacksquare$ \\ 
   $2$ & $87$ & $87$ & $20.31_{-3.83}^{+5.75}$ & $2.04_{-0.71}^{+0.76}$ & $533_{-81}^{+80}$ & $0.58_{-0.02}^{+0.02}$ & $1.58_{-0.12}^{+0.10}$ & $0.44_{-0.03}^{+0.04}$ & $0.44_{-0.03}^{+0.04}$ & 0.76 & $\color{black}\blacksquare$ \\ 
   $3$ & $117$ & $105$ & $16.41_{-3.15}^{+7.49}$ & $4.08_{-0.77}^{+2.42}$ & $700_{-95}^{+246}$ & $1.94_{-0.44}^{+0.28}$ & $1.62_{-0.04}^{+0.13}$ & $0.44_{-0.05}^{+0.02}$ & $0.44_{-0.06}^{+0.03}$ & 0.10 & $\color{black}\blacksquare$ \\ 
   $4$ & $145$ & $114$ & $180.02_{-43.37}^{+54.06}$ & $46.33_{-8.42}^{+9.52}$ & $547_{-130}^{+199}$ & $1.23_{-0.03}^{+0.15}$ & $1.95_{-0.13}^{+0.08}$ & $0.49_{-0.06}^{+0.06}$ & $0.47_{-0.04}^{+0.05}$ & 0.65 & $\color{black}\blacksquare$ \\[1.5ex] 
   $5$ & $22$ & $43$ & $2.37_{-0.65}^{+1.35}$ & $0.12_{-0.05}^{+0.07}$ & $1273_{-267}^{+303}$ & $0.92_{-0.04}^{+0.09}$ & $0.64_{-0.01}^{+0.04}$ & $0.26_{-0.05}^{+0.07}$ & $0.22_{-0.03}^{+0.04}$ & 0.16 & $\color{blue}\rotatebox[origin=c]{90}{$\blacktriangle$}$ \\ 
   $6$ & $46$ & $15$ & $0.54_{-0.13}^{+0.48}$ & $0.16_{-0.04}^{+0.09}$ & $1292_{-253}^{+647}$ & $0.90_{-0.05}^{+0.06}$ & $0.87_{-0.01}^{+0.11}$ & $0.22_{-0.05}^{+0.04}$ & $0.26_{-0.08}^{+0.06}$ & 0.28 & $\color{blue}\rotatebox[origin=c]{270}{$\blacktriangle$}$ \\ 
   $7$ & $94$ & $105$ & $12.67_{-2.66}^{+3.92}$ & $1.88_{-0.43}^{+0.85}$ & $773_{-161}^{+268}$ & $1.50_{-0.25}^{+0.20}$ & $1.66_{-0.91}^{+0.62}$ & $0.39_{-0.06}^{+0.05}$ & $0.41_{-0.06}^{+0.04}$ & 0.19 & $\color{blue}\rotatebox[origin=c]{90}{$\blacktriangle$}$ \\ 
   $8$ & $111$ & $87$ & $15.81_{-2.06}^{+5.46}$ & $4.88_{-0.75}^{+1.75}$ & $616_{-81}^{+138}$ & $0.92_{-0.06}^{+0.01}$ & $0.95_{-0.01}^{+0.04}$ & $0.42_{-0.03}^{+0.03}$ & $0.43_{-0.04}^{+0.03}$ & 0.43 & $\color{blue}\rotatebox[origin=c]{270}{$\blacktriangle$}$ \\ 
   $9$ & $11$ & $87$ & $17.60_{-5.42}^{+6.14}$ & $0.63_{-0.11}^{+0.19}$ & $875_{-180}^{+317}$ & $0.98_{-0.25}^{+0.01}$ & $0.82_{-0.06}^{+0.07}$ & $0.39_{-0.09}^{+0.07}$ & $0.31_{-0.06}^{+0.05}$ & 0.18 & $\color{blue}\blacktriangledown$ \\ 
   $10$ & $81$ & $15$ & $1.39_{-0.34}^{+1.45}$ & $0.36_{-0.08}^{+0.33}$ & $1078_{-179}^{+594}$ & $0.91_{-0.10}^{+0.01}$ & $1.06_{-0.13}^{+0.01}$ & $0.28_{-0.06}^{+0.03}$ & $0.31_{-0.11}^{+0.06}$ & 0.39 & $\color{blue}\blacktriangle$ \\ 
   $11$ & $50$ & $105$ & $8.31_{-2.55}^{+4.17}$ & $0.88_{-0.35}^{+0.25}$ & $977_{-240}^{+202}$ & $1.38_{-0.08}^{+0.65}$ & $1.16_{-0.24}^{+1.02}$ & $0.32_{-0.04}^{+0.08}$ & $0.34_{-0.03}^{+0.05}$ & 0.23 & $\color{blue}\blacktriangledown$ \\ 
   $12$ & $101$ & $43$ & $4.93_{-0.89}^{+1.78}$ & $1.62_{-0.23}^{+0.47}$ & $574_{-82}^{+193}$ & $0.59_{-0.25}^{+0.21}$ & $0.79_{-0.02}^{+0.03}$ & $0.42_{-0.04}^{+0.03}$ & $0.45_{-0.06}^{+0.04}$ & 0.42 & $\color{blue}\blacktriangle$ \\[1.5ex] 
   $13$ & $85$ & $87$ & $21.49_{-3.76}^{+2.87}$ & $3.39_{-0.61}^{+0.30}$ & $564_{-94}^{+116}$ & $0.68_{-0.04}^{+0.05}$ & $1.28_{-0.06}^{+0.14}$ & $0.44_{-0.04}^{+0.04}$ & $0.42_{-0.04}^{+0.04}$ & 0.54 & $\color{red}\blacksquare$ \\ 
   $14$ & $81$ & $87$ & $13.49_{-2.20}^{+2.70}$ & $5.53_{-0.71}^{+0.24}$ & $569_{-146}^{+173}$ & $0.68_{-0.02}^{+0.01}$ & $1.10_{-0.07}^{+0.17}$ & $0.42_{-0.05}^{+0.07}$ & $0.42_{-0.05}^{+0.08}$ & 0.38 & $\color{red}\blacksquare$ \\ 
   $15$ & $77$ & $87$ & $11.39_{-0.85}^{+1.97}$ & $8.46_{-0.08}^{+0.82}$ & $931_{-423}^{+722}$ & $1.44_{-0.23}^{+0.42}$ & $1.33_{-0.11}^{+0.03}$ & $0.29_{-0.09}^{+0.12}$ & $0.31_{-0.10}^{+0.13}$ & 0.17 & $\color{red}\blacksquare$ \\[1.5ex] 
   $16$ & $98$ & $105$ & $13.24_{-1.58}^{+2.78}$ & $3.92_{-0.62}^{+0.51}$ & $848_{-213}^{+213}$ & $1.13_{-0.06}^{+0.05}$ & $1.09_{-0.17}^{+0.99}$ & $0.38_{-0.05}^{+0.06}$ & $0.38_{-0.04}^{+0.05}$ & 0.13 & $\color{green}\rotatebox[origin=c]{90}{$\blacktriangle$}$ \\ 
   $17$ & $114$ & $87$ & $17.31_{-2.57}^{+3.52}$ & $8.05_{-1.28}^{+0.73}$ & $674_{-137}^{+130}$ & $1.02_{-0.12}^{+0.11}$ & $1.04_{-0.04}^{+0.14}$ & $0.41_{-0.03}^{+0.04}$ & $0.41_{-0.04}^{+0.05}$ & 0.25 & $\color{green}\rotatebox[origin=c]{270}{$\blacktriangle$}$ \\[1.5ex] 
   $18$ & $82$ & $87$ & $15.18_{-3.18}^{+2.63}$ & $5.20_{-0.72}^{+0.26}$ & $612_{-174}^{+227}$ & $0.55_{-0.04}^{+0.12}$ & $1.28_{-0.10}^{+0.06}$ & $0.40_{-0.06}^{+0.07}$ & $0.40_{-0.06}^{+0.08}$ & 0.48 & $\color{cyan}\blacksquare$ \\ 
   $19$ & $81$ & $84$ & $13.06_{-1.90}^{+1.84}$ & $4.70_{-0.78}^{+0.34}$ & $574_{-139}^{+163}$ & $0.69_{-0.04}^{+0.52}$ & $0.83_{-0.02}^{+0.03}$ & $0.42_{-0.05}^{+0.06}$ & $0.41_{-0.05}^{+0.07}$ & 0.25 & $\color{cyan}\blacksquare$ \\ 
   $20$ & $82$ & $84$ & $14.36_{-2.21}^{+2.83}$ & $4.41_{-0.65}^{+0.50}$ & $574_{-136}^{+164}$ & $0.54_{-0.05}^{+0.05}$ & $0.91_{-0.03}^{+0.04}$ & $0.41_{-0.05}^{+0.06}$ & $0.41_{-0.05}^{+0.07}$ & 0.38 & $\color{cyan}\blacksquare$ \\[1.5ex] 
   $21$ & $110$ & $141$ & $16.03_{-2.61}^{+2.60}$ & $5.55_{-0.68}^{+0.68}$ & $776_{-178}^{+207}$ & $1.03_{-0.10}^{+0.08}$ & $0.69_{-0.10}^{+0.03}$ & $0.39_{-0.06}^{+0.06}$ & $0.34_{-0.04}^{+0.05}$ & 0.19 & $\color{magenta}\rotatebox[origin=c]{90}{$\blacktriangle$}$ \\ 
   $22$ & $145$ & $48$ & $8.68_{-2.41}^{+11.11}$ & $1.91_{-0.32}^{+0.37}$ & $1119_{-233}^{+507}$ & $1.83_{-1.15}^{+2.47}$ & $2.45_{-0.41}^{+2.18}$ & $0.29_{-0.04}^{+0.04}$ & $0.27_{-0.06}^{+0.05}$ & 0.18 & $\color{magenta}\blacktriangle$ \\ 
   \hline
  \end{tabular} \\ 
  All time-scales are in units of Myr. All length-scales are in units of pc. \vspace{-1mm}
 \end{minipage}
\end{table*}

\begin{table*}
 \centering
 \begin{minipage}{\hsize}
  \caption{Best-fitting solutions for age-binned stellar maps (high resolution, extended emission)}\label{tab:starruns_hr_out}\vspace{-1mm}
  \begin{tabular}{c c c c c c c c c c c c}
   \hline
   ID & $N_{\rm star}$ & $N_{\rm gas}$ & $t_{\rm gas}$ & $t_{\rm over}$ & $\lambda$ & $\beta_{\rm star}$ & $\beta_{\rm gas}$& $\zeta_{\rm star}$ & $\zeta_{\rm gas}$ & $\chi_{\rm red}^2$ & Symbol\\ 
   \hline
   $1$ & $248$ & $230$ & $3.31_{-0.38}^{+0.53}$ & $0.10_{-0.04}^{+0.07}$ & $455_{-28}^{+36}$ & $0.89_{-0.08}^{+0.00}$ & $0.95_{-0.12}^{+0.14}$ & $0.31_{-0.02}^{+0.02}$ & $0.30_{-0.02}^{+0.01}$ & 0.22 & $\color{black}\blacksquare$ \\ 
   $2$ & $447$ & $435$ & $8.42_{-0.70}^{+0.73}$ & $0.32_{-0.17}^{+0.13}$ & $283_{-14}^{+15}$ & $0.64_{-0.17}^{+0.09}$ & $0.69_{-0.09}^{+0.05}$ & $0.44_{-0.01}^{+0.01}$ & $0.44_{-0.01}^{+0.01}$ & 0.47 & $\color{black}\blacksquare$ \\ 
   $3$ & $567$ & $514$ & $23.37_{-1.49}^{+1.70}$ & $4.87_{-0.62}^{+0.58}$ & $253_{-19}^{+19}$ & $0.85_{-0.05}^{+0.03}$ & $0.85_{-0.03}^{+0.02}$ & $0.49_{-0.02}^{+0.02}$ & $0.51_{-0.01}^{+0.02}$ & 0.21 & $\color{black}\blacksquare$ \\ 
   $4$ & $454$ & $396$ & $123.83_{-7.22}^{+7.34}$ & $39.27_{-2.74}^{+3.13}$ & $198_{-16}^{+19}$ & $0.83_{-0.01}^{+0.02}$ & $0.98_{-0.03}^{+0.03}$ & $0.59_{-0.02}^{+0.01}$ & $0.58_{-0.02}^{+0.01}$ & 0.56 & $\color{black}\blacksquare$ \\[1.5ex] 
   $5$ & $114$ & $230$ & $3.00_{-0.44}^{+0.71}$ & $0.07_{-0.03}^{+0.03}$ & $593_{-64}^{+55}$ & $1.13_{-0.15}^{+0.17}$ & $0.79_{-0.02}^{+0.09}$ & $0.27_{-0.02}^{+0.03}$ & $0.23_{-0.01}^{+0.02}$ & 0.34 & $\color{blue}\rotatebox[origin=c]{90}{$\blacktriangle$}$ \\ 
   $6$ & $250$ & $83$ & $0.99_{-0.13}^{+0.29}$ & $0.01_{-0.00}^{+0.02}$ & $588_{-27}^{+61}$ & $0.46_{-0.01}^{+0.22}$ & $0.77_{-0.00}^{+0.09}$ & $0.24_{-0.02}^{+0.01}$ & $0.25_{-0.02}^{+0.01}$ & 0.40 & $\color{blue}\rotatebox[origin=c]{270}{$\blacktriangle$}$ \\ 
   $7$ & $449$ & $514$ & $20.11_{-1.65}^{+1.87}$ & $1.73_{-0.30}^{+0.16}$ & $298_{-26}^{+23}$ & $0.87_{-0.02}^{+0.01}$ & $0.84_{-0.13}^{+0.26}$ & $0.44_{-0.02}^{+0.02}$ & $0.46_{-0.02}^{+0.02}$ & 0.31 & $\color{blue}\rotatebox[origin=c]{90}{$\blacktriangle$}$ \\ 
   $8$ & $527$ & $435$ & $14.65_{-1.16}^{+1.47}$ & $1.38_{-0.34}^{+0.38}$ & $267_{-18}^{+19}$ & $1.02_{-0.22}^{+0.05}$ & $0.92_{-0.03}^{+0.01}$ & $0.47_{-0.01}^{+0.02}$ & $0.47_{-0.02}^{+0.02}$ & 0.43 & $\color{blue}\rotatebox[origin=c]{270}{$\blacktriangle$}$ \\ 
   $9$ & $99$ & $435$ & $8.07_{-1.31}^{+1.34}$ & $0.08_{-0.03}^{+0.03}$ & $398_{-35}^{+47}$ & $0.79_{-0.14}^{+0.12}$ & $0.82_{-0.01}^{+0.08}$ & $0.38_{-0.04}^{+0.02}$ & $0.33_{-0.02}^{+0.02}$ & 0.39 & $\color{blue}\blacktriangledown$ \\ 
   $10$ & $432$ & $83$ & $1.39_{-0.20}^{+0.34}$ & $0.08_{-0.04}^{+0.05}$ & $426_{-33}^{+53}$ & $0.52_{-0.07}^{+0.29}$ & $0.63_{-0.05}^{+0.18}$ & $0.32_{-0.02}^{+0.02}$ & $0.34_{-0.04}^{+0.03}$ & 0.36 & $\color{blue}\blacktriangle$ \\ 
   $11$ & $201$ & $514$ & $17.79_{-1.68}^{+2.72}$ & $0.40_{-0.09}^{+0.09}$ & $336_{-30}^{+24}$ & $0.98_{-0.15}^{+0.04}$ & $0.71_{-0.02}^{+0.06}$ & $0.43_{-0.02}^{+0.02}$ & $0.43_{-0.01}^{+0.02}$ & 0.24 & $\color{blue}\blacktriangledown$ \\ 
   $12$ & $524$ & $230$ & $5.97_{-0.61}^{+0.80}$ & $0.60_{-0.14}^{+0.19}$ & $290_{-17}^{+24}$ & $0.82_{-0.30}^{+0.10}$ & $0.80_{-0.15}^{+0.03}$ & $0.44_{-0.01}^{+0.01}$ & $0.44_{-0.02}^{+0.02}$ & 0.50 & $\color{blue}\blacktriangle$ \\[1.5ex] 
   $13$ & $429$ & $435$ & $10.02_{-0.66}^{+0.86}$ & $1.62_{-0.21}^{+0.19}$ & $294_{-20}^{+25}$ & $0.93_{-0.00}^{+0.02}$ & $0.84_{-0.02}^{+0.06}$ & $0.44_{-0.02}^{+0.02}$ & $0.42_{-0.02}^{+0.02}$ & 0.53 & $\color{red}\blacksquare$ \\ 
   $14$ & $423$ & $435$ & $9.98_{-0.58}^{+0.68}$ & $2.92_{-0.18}^{+0.20}$ & $312_{-25}^{+29}$ & $1.11_{-0.03}^{+0.02}$ & $0.96_{-0.03}^{+0.02}$ & $0.42_{-0.02}^{+0.02}$ & $0.41_{-0.02}^{+0.02}$ & 0.44 & $\color{red}\blacksquare$ \\ 
   $15$ & $433$ & $435$ & $10.25_{-0.30}^{+0.27}$ & $7.92_{-0.16}^{+0.15}$ & $323_{-55}^{+86}$ & $1.11_{-0.06}^{+0.05}$ & $0.88_{-0.03}^{+0.04}$ & $0.39_{-0.05}^{+0.05}$ & $0.38_{-0.05}^{+0.04}$ & 0.11 & $\color{red}\blacksquare$ \\[1.5ex] 
   $16$ & $432$ & $514$ & $26.49_{-2.13}^{+2.49}$ & $4.25_{-0.24}^{+0.36}$ & $287_{-26}^{+29}$ & $1.11_{-0.02}^{+0.02}$ & $1.05_{-0.03}^{+0.02}$ & $0.46_{-0.02}^{+0.02}$ & $0.46_{-0.02}^{+0.02}$ & 0.37 & $\color{green}\rotatebox[origin=c]{90}{$\blacktriangle$}$ \\ 
   $17$ & $503$ & $435$ & $14.95_{-1.19}^{+1.27}$ & $4.41_{-0.43}^{+0.36}$ & $274_{-19}^{+22}$ & $1.01_{-0.01}^{+0.04}$ & $0.91_{-0.03}^{+0.02}$ & $0.46_{-0.02}^{+0.02}$ & $0.46_{-0.02}^{+0.02}$ & 0.39 & $\color{green}\rotatebox[origin=c]{270}{$\blacktriangle$}$ \\[1.5ex] 
   $18$ & $424$ & $435$ & $10.97_{-0.82}^{+0.98}$ & $2.18_{-0.15}^{+0.23}$ & $301_{-20}^{+26}$ & $0.81_{-0.03}^{+0.01}$ & $0.90_{-0.01}^{+0.01}$ & $0.43_{-0.02}^{+0.02}$ & $0.41_{-0.02}^{+0.02}$ & 0.59 & $\color{cyan}\blacksquare$ \\ 
   $19$ & $423$ & $423$ & $9.23_{-0.72}^{+0.75}$ & $2.22_{-0.23}^{+0.24}$ & $309_{-19}^{+25}$ & $0.95_{-0.01}^{+0.04}$ & $0.68_{-0.02}^{+0.02}$ & $0.42_{-0.02}^{+0.02}$ & $0.41_{-0.02}^{+0.02}$ & 0.37 & $\color{cyan}\blacksquare$ \\ 
   $20$ & $424$ & $423$ & $9.68_{-0.59}^{+0.71}$ & $1.82_{-0.21}^{+0.20}$ & $307_{-25}^{+25}$ & $0.66_{-0.05}^{+0.05}$ & $0.56_{-0.05}^{+0.06}$ & $0.42_{-0.02}^{+0.02}$ & $0.42_{-0.02}^{+0.02}$ & 0.42 & $\color{cyan}\blacksquare$ \\[1.5ex] 
   $21$ & $607$ & $675$ & $17.92_{-0.97}^{+1.10}$ & $4.98_{-0.16}^{+0.23}$ & $281_{-22}^{+24}$ & $0.96_{-0.05}^{+0.02}$ & $0.85_{-0.02}^{+0.01}$ & $0.45_{-0.02}^{+0.02}$ & $0.46_{-0.02}^{+0.02}$ & 0.58 & $\color{magenta}\rotatebox[origin=c]{90}{$\blacktriangle$}$ \\ 
   $22$ & $651$ & $276$ & $4.65_{-0.40}^{+0.57}$ & $0.54_{-0.11}^{+0.13}$ & $296_{-16}^{+21}$ & $0.98_{-0.11}^{+0.03}$ & $0.77_{-0.02}^{+0.05}$ & $0.45_{-0.01}^{+0.01}$ & $0.45_{-0.02}^{+0.02}$ & 0.29 & $\color{magenta}\blacktriangle$ \\ 
   \hline
  \end{tabular} \\ 
  All time-scales are in units of Myr. All length-scales are in units of pc. \vspace{-1mm}
 \end{minipage}
\end{table*}

\subsection{Gas and stellar maps} \label{sec:appgasstar}
Section~\ref{sec:gasstar} discusses 200 different experiments using the gas and star particles in the numerical simulations described in Section~\ref{sec:models}. These are constituted by a $2\times2$ matrix of the 50 experiments listed in \autoref{tab:gasruns_in}, using maps of \{point particles, extended emission\} in the \{low, high\} resolution galaxy simulations. The output is listed in \autoref{tab:gasruns_p_lr_out}--\ref{tab:gasruns_hr_out}, which represent the low-resolution point particle experiments, the high-resolution point particle experiments, the low-resolution extended emission experiments, and the high-resolution extended emission experiments, respectively.
\begin{table*}
 \centering
 \begin{minipage}{\hsize}
  \caption{Best-fitting solutions for gas maps (low resolution, point particles)}\label{tab:gasruns_p_lr_out}\vspace{-1mm}
  \begin{tabular}{c c c c c c c c c c c}
   \hline
   ID & $N_{\rm star}$ & $N_{\rm gas}$ & $t_{\rm gas}$ & $t_{\rm over}$ & $\lambda$ & $\beta_{\rm star}$ & $\beta_{\rm gas}$& $\zeta_{\rm star}$ & $\zeta_{\rm gas}$ & $\chi_{\rm red}^2$\\ 
   \hline
   $1$ & $15$ & $314$ & $287.61_{-156.25}^{+418.18}$ & $0.81_{-0.38}^{+2.01}$ & $167_{-37}^{+71}$ & $1.00_{-0.01}^{+0.04}$ & $1.48_{-0.00}^{+0.15}$ & $0.53_{-0.12}^{+0.08}$ & $0.59_{-0.08}^{+0.05}$ & 1.23 \\ 
   $2$ & $26$ & $314$ & $90.02_{-35.90}^{+421.01}$ & $0.25_{-0.14}^{+1.61}$ & $235_{-100}^{+83}$ & $0.96_{-0.02}^{+0.13}$ & $1.48_{-0.00}^{+0.19}$ & $0.41_{-0.10}^{+0.19}$ & $0.51_{-0.07}^{+0.13}$ & 1.17 \\ 
   $3$ & $49$ & $314$ & $72.98_{-21.91}^{+137.29}$ & $0.36_{-0.22}^{+1.13}$ & $227_{-79}^{+53}$ & $0.98_{-0.02}^{+0.04}$ & $0.65_{-0.00}^{+0.24}$ & $0.42_{-0.08}^{+0.15}$ & $0.52_{-0.05}^{+0.09}$ & 0.57 \\ 
   $4$ & $104$ & $314$ & $96.86_{-25.44}^{+62.39}$ & $0.53_{-0.34}^{+1.43}$ & $177_{-33}^{+41}$ & $0.97_{-0.00}^{+0.02}$ & $0.14_{-0.00}^{+0.45}$ & $0.51_{-0.07}^{+0.06}$ & $0.58_{-0.05}^{+0.04}$ & 0.90 \\ 
   $5$ & $157$ & $314$ & $71.79_{-16.06}^{+30.17}$ & $0.01_{-0.00}^{+0.70}$ & $217_{-32}^{+32}$ & $0.97_{-0.00}^{+0.02}$ & $0.14_{-0.00}^{+0.38}$ & $0.44_{-0.06}^{+0.06}$ & $0.54_{-0.04}^{+0.04}$ & 0.72 \\ 
   $6$ & $269$ & $314$ & $48.27_{-8.52}^{+11.57}$ & $0.01_{-0.00}^{+0.69}$ & $265_{-19}^{+24}$ & $0.00_{-0.00}^{+0.79}$ & $0.21_{-0.01}^{+0.32}$ & $0.35_{-0.03}^{+0.03}$ & $0.50_{-0.02}^{+0.02}$ & 0.70 \\ 
   $7$ & $426$ & $314$ & $41.18_{-6.27}^{+8.08}$ & $0.27_{-0.18}^{+0.93}$ & $293_{-14}^{+26}$ & $0.71_{-0.27}^{+0.18}$ & $0.21_{-0.00}^{+0.61}$ & $0.32_{-0.02}^{+0.02}$ & $0.49_{-0.02}^{+0.01}$ & 0.23 \\ 
   $8$ & $803$ & $314$ & $36.21_{-4.18}^{+5.73}$ & $1.64_{-0.96}^{+1.08}$ & $305_{-17}^{+20}$ & $0.96_{-0.01}^{+0.01}$ & $0.68_{-0.51}^{+0.09}$ & $0.29_{-0.02}^{+0.02}$ & $0.49_{-0.01}^{+0.01}$ & 0.78 \\ 
   $9$ & $1219$ & $314$ & $37.96_{-3.98}^{+5.46}$ & $3.89_{-1.50}^{+1.43}$ & $308_{-19}^{+19}$ & $0.97_{-0.01}^{+0.00}$ & $0.72_{-0.03}^{+0.06}$ & $0.29_{-0.01}^{+0.02}$ & $0.50_{-0.01}^{+0.01}$ & 1.49 \\ 
   $10$ & $1714$ & $314$ & $54.78_{-5.40}^{+7.66}$ & $10.70_{-3.06}^{+2.47}$ & $296_{-24}^{+22}$ & $0.95_{-0.03}^{+0.03}$ & $0.78_{-0.08}^{+0.03}$ & $0.31_{-0.02}^{+0.02}$ & $0.51_{-0.01}^{+0.02}$ & 2.02 \\ 
   $11$ & $2587$ & $314$ & $66.96_{-5.70}^{+7.95}$ & $20.54_{-4.19}^{+4.10}$ & $277_{-21}^{+26}$ & $0.98_{-0.02}^{+0.02}$ & $0.77_{-0.01}^{+0.05}$ & $0.35_{-0.02}^{+0.02}$ & $0.53_{-0.02}^{+0.02}$ & 2.61 \\[1.5ex] 
   $12$ & $49$ & $1230$ & $1857.33_{-754.59}^{+603.85}$ & $5.16_{-2.35}^{+1.93}$ & $109_{-8}^{+12}$ & $0.94_{-0.02}^{+0.01}$ & $1.26_{-0.01}^{+0.03}$ & $0.67_{-0.04}^{+0.03}$ & $0.69_{-0.02}^{+0.02}$ & 1.47 \\ 
   $13$ & $157$ & $1230$ & $646.41_{-146.66}^{+630.03}$ & $2.10_{-1.03}^{+4.20}$ & $128_{-22}^{+13}$ & $0.99_{-0.02}^{+0.00}$ & $1.14_{-0.01}^{+0.15}$ & $0.62_{-0.03}^{+0.07}$ & $0.66_{-0.02}^{+0.04}$ & 1.58 \\ 
   $14$ & $426$ & $1230$ & $370.96_{-65.21}^{+118.86}$ & $0.02_{-0.01}^{+1.44}$ & $183_{-20}^{+21}$ & $0.03_{-0.00}^{+0.83}$ & $0.91_{-0.00}^{+0.69}$ & $0.51_{-0.03}^{+0.03}$ & $0.61_{-0.02}^{+0.02}$ & 0.77 \\ 
   $15$ & $49$ & $74$ & $3.64_{-0.96}^{+2.70}$ & $0.05_{-0.03}^{+0.11}$ & $638_{-77}^{+141}$ & $1.01_{-0.02}^{+0.00}$ & $1.32_{-0.00}^{+0.12}$ & $0.14_{-0.02}^{+0.02}$ & $0.24_{-0.03}^{+0.02}$ & 0.52 \\ 
   $16$ & $157$ & $74$ & $4.46_{-0.88}^{+1.51}$ & $0.01_{-0.00}^{+0.15}$ & $380_{-23}^{+54}$ & $1.00_{-0.00}^{+0.02}$ & $0.59_{-0.00}^{+0.56}$ & $0.22_{-0.02}^{+0.01}$ & $0.37_{-0.03}^{+0.02}$ & 0.67 \\ 
   $17$ & $426$ & $74$ & $5.98_{-0.97}^{+1.54}$ & $0.45_{-0.22}^{+0.29}$ & $352_{-32}^{+46}$ & $0.98_{-0.03}^{+0.01}$ & $1.20_{-0.09}^{+0.05}$ & $0.24_{-0.02}^{+0.02}$ & $0.40_{-0.03}^{+0.03}$ & 0.64 \\[1.5ex] 
   $18$ & $50$ & $7541$ & $471.59_{-148.21}^{+173.05}$ & $0.13_{-0.08}^{+0.17}$ & $81_{-12}^{+19}$ & $1.02_{-0.01}^{+0.00}$ & $0.75_{-0.01}^{+0.00}$ & $0.29_{-0.05}^{+0.05}$ & $0.35_{-0.03}^{+0.03}$ & 1.10 \\ 
   $19$ & $167$ & $7541$ & $271.56_{-58.19}^{+62.26}$ & $0.01_{-0.00}^{+0.13}$ & $116_{-10}^{+15}$ & $1.01_{-0.00}^{+0.01}$ & $0.76_{-0.01}^{+0.08}$ & $0.20_{-0.02}^{+0.02}$ & $0.31_{-0.01}^{+0.01}$ & 2.20 \\ 
   $20$ & $495$ & $7541$ & $190.93_{-30.26}^{+27.21}$ & $0.07_{-0.04}^{+0.14}$ & $162_{-9}^{+12}$ & $0.56_{-0.00}^{+0.47}$ & $0.74_{-0.00}^{+0.07}$ & $0.14_{-0.01}^{+0.01}$ & $0.29_{-0.00}^{+0.00}$ & 8.00 \\ 
   $21$ & $50$ & $1579$ & $312.08_{-109.06}^{+173.40}$ & $0.01_{-0.00}^{+0.53}$ & $98_{-16}^{+29}$ & $1.00_{-0.01}^{+0.01}$ & $0.34_{-0.00}^{+0.28}$ & $0.48_{-0.11}^{+0.05}$ & $0.52_{-0.06}^{+0.04}$ & 0.82 \\ 
   $22$ & $164$ & $1579$ & $153.08_{-34.84}^{+38.62}$ & $0.01_{-0.00}^{+0.33}$ & $154_{-15}^{+22}$ & $1.03_{-0.02}^{+0.01}$ & $0.43_{-0.00}^{+0.06}$ & $0.31_{-0.04}^{+0.03}$ & $0.43_{-0.02}^{+0.02}$ & 1.21 \\ 
   $23$ & $476$ & $1579$ & $76.22_{-10.82}^{+14.91}$ & $0.46_{-0.27}^{+0.27}$ & $237_{-16}^{+15}$ & $0.92_{-0.18}^{+0.16}$ & $0.42_{-0.12}^{+0.10}$ & $0.20_{-0.01}^{+0.01}$ & $0.38_{-0.01}^{+0.01}$ & 2.20 \\ 
   $24$ & $46$ & $171$ & $57.83_{-18.01}^{+497.53}$ & $2.34_{-0.96}^{+20.68}$ & $363_{-113}^{+71}$ & $0.91_{-0.00}^{+0.17}$ & $0.58_{-0.23}^{+0.26}$ & $0.51_{-0.06}^{+0.13}$ & $0.57_{-0.05}^{+0.08}$ & 0.24 \\ 
   $25$ & $126$ & $171$ & $53.51_{-10.89}^{+35.02}$ & $4.55_{-1.93}^{+3.81}$ & $292_{-62}^{+61}$ & $1.04_{-0.03}^{+0.05}$ & $0.28_{-0.10}^{+0.29}$ & $0.56_{-0.06}^{+0.09}$ & $0.62_{-0.03}^{+0.05}$ & 0.23 \\ 
   $26$ & $274$ & $171$ & $44.78_{-7.83}^{+9.73}$ & $0.73_{-0.49}^{+2.46}$ & $295_{-20}^{+37}$ & $0.55_{-0.15}^{+0.33}$ & $0.04_{-0.03}^{+0.14}$ & $0.56_{-0.04}^{+0.02}$ & $0.63_{-0.02}^{+0.01}$ & 0.41 \\ 
   $27$ & $40$ & $92$ & $50.42_{-17.15}^{+575.57}$ & $4.74_{-1.78}^{+64.32}$ & $530_{-124}^{+169}$ & $0.98_{-0.10}^{+0.07}$ & $0.76_{-0.19}^{+0.21}$ & $0.60_{-0.08}^{+0.10}$ & $0.61_{-0.07}^{+0.10}$ & 0.14 \\ 
   $28$ & $84$ & $92$ & $47.08_{-11.45}^{+600.70}$ & $10.28_{-3.85}^{+150.84}$ & $482_{-126}^{+162}$ & $1.16_{-0.10}^{+0.18}$ & $0.70_{-0.20}^{+0.17}$ & $0.63_{-0.07}^{+0.07}$ & $0.64_{-0.07}^{+0.06}$ & 0.31 \\ 
   $29$ & $130$ & $92$ & $34.88_{-5.46}^{+11.09}$ & $6.92_{-3.89}^{+5.59}$ & $382_{-77}^{+124}$ & $1.17_{-0.32}^{+0.33}$ & $0.27_{-0.11}^{+0.24}$ & $0.71_{-0.08}^{+0.00}$ & $0.71_{-0.07}^{+0.00}$ & 0.50 \\ 
   $30$ & $32$ & $38$ & $17.91_{-6.48}^{+609.93}$ & $5.84_{-2.32}^{+183.83}$ & $1281_{-533}^{+573}$ & $1.01_{-0.09}^{+0.12}$ & $0.68_{-0.37}^{+0.17}$ & $0.53_{-0.10}^{+0.18}$ & $0.57_{-0.08}^{+0.14}$ & 0.27 \\ 
   $31$ & $45$ & $38$ & $57.94_{-21.91}^{+649.02}$ & $27.65_{-10.87}^{+392.63}$ & $753_{-313}^{+4079}$ & $1.69_{-0.39}^{+0.14}$ & $0.51_{-0.12}^{+0.63}$ & $0.71_{-0.45}^{+0.00}$ & $0.71_{-0.34}^{+0.00}$ & 0.41 \\ 
   $32$ & $53$ & $38$ & $33.41_{-6.48}^{+673.41}$ & $20.36_{-12.19}^{+496.81}$ & $647_{-288}^{+3767}$ & $2.29_{-0.80}^{+2.37}$ & $0.36_{-0.21}^{+0.14}$ & $0.71_{-0.39}^{+0.00}$ & $0.71_{-0.38}^{+0.00}$ & 0.69 \\[1.5ex] 
   $33$ & $49$ & $320$ & $74.19_{-23.57}^{+201.30}$ & $0.35_{-0.21}^{+1.65}$ & $233_{-87}^{+56}$ & $0.98_{-0.02}^{+0.01}$ & $0.84_{-0.03}^{+0.38}$ & $0.41_{-0.08}^{+0.16}$ & $0.53_{-0.05}^{+0.10}$ & 0.46 \\ 
   $34$ & $156$ & $320$ & $77.41_{-17.21}^{+35.13}$ & $0.01_{-0.00}^{+0.71}$ & $215_{-34}^{+34}$ & $1.08_{-0.05}^{+0.05}$ & $0.97_{-0.07}^{+0.12}$ & $0.44_{-0.06}^{+0.06}$ & $0.55_{-0.04}^{+0.04}$ & 0.57 \\ 
   $35$ & $421$ & $320$ & $50.27_{-7.21}^{+10.45}$ & $0.32_{-0.21}^{+0.97}$ & $286_{-16}^{+26}$ & $0.81_{-0.30}^{+0.09}$ & $0.31_{-0.01}^{+0.76}$ & $0.33_{-0.03}^{+0.02}$ & $0.50_{-0.02}^{+0.01}$ & 0.46 \\ 
   $36$ & $49$ & $375$ & $114.89_{-37.73}^{+159.95}$ & $0.30_{-0.18}^{+1.15}$ & $192_{-51}^{+52}$ & $0.79_{-0.24}^{+0.13}$ & $0.98_{-0.00}^{+0.12}$ & $0.49_{-0.10}^{+0.09}$ & $0.57_{-0.06}^{+0.06}$ & 0.52 \\ 
   $37$ & $156$ & $375$ & $65.14_{-12.18}^{+19.48}$ & $0.01_{-0.00}^{+0.39}$ & $243_{-24}^{+25}$ & $0.06_{-0.00}^{+0.59}$ & $0.36_{-0.00}^{+0.15}$ & $0.40_{-0.04}^{+0.04}$ & $0.52_{-0.02}^{+0.02}$ & 1.14 \\ 
   $38$ & $414$ & $375$ & $50.66_{-8.22}^{+9.47}$ & $0.01_{-0.00}^{+0.89}$ & $283_{-14}^{+23}$ & $0.55_{-0.00}^{+0.38}$ & $0.43_{-0.04}^{+0.11}$ & $0.33_{-0.03}^{+0.02}$ & $0.51_{-0.01}^{+0.01}$ & 0.41 \\ 
   $39$ & $48$ & $417$ & $163.41_{-61.16}^{+328.18}$ & $0.57_{-0.32}^{+2.25}$ & $175_{-49}^{+51}$ & $0.92_{-0.10}^{+0.06}$ & $0.55_{-0.00}^{+0.17}$ & $0.51_{-0.10}^{+0.11}$ & $0.58_{-0.05}^{+0.07}$ & 0.37 \\ 
   $40$ & $155$ & $417$ & $144.62_{-44.57}^{+58.09}$ & $0.63_{-0.41}^{+1.28}$ & $180_{-28}^{+46}$ & $0.61_{-0.52}^{+0.26}$ & $0.18_{-0.00}^{+0.21}$ & $0.50_{-0.08}^{+0.05}$ & $0.58_{-0.05}^{+0.03}$ & 0.61 \\ 
   $41$ & $424$ & $417$ & $49.46_{-7.58}^{+13.11}$ & $1.90_{-0.81}^{+0.83}$ & $310_{-29}^{+32}$ & $0.98_{-0.23}^{+0.01}$ & $0.40_{-0.11}^{+0.38}$ & $0.30_{-0.03}^{+0.03}$ & $0.49_{-0.01}^{+0.01}$ & 0.70 \\ 
   $42$ & $43$ & $563$ & $51.66_{-16.70}^{+47.38}$ & $0.01_{-0.00}^{+0.19}$ & $316_{-80}^{+70}$ & $1.01_{-0.00}^{+0.01}$ & $0.77_{-0.00}^{+0.33}$ & $0.31_{-0.06}^{+0.10}$ & $0.46_{-0.03}^{+0.05}$ & 0.44 \\ 
   $43$ & $151$ & $563$ & $149.92_{-50.87}^{+51.85}$ & $0.01_{-0.00}^{+1.17}$ & $188_{-25}^{+54}$ & $0.65_{-0.10}^{+0.28}$ & $0.31_{-0.00}^{+0.11}$ & $0.49_{-0.10}^{+0.04}$ & $0.57_{-0.05}^{+0.03}$ & 0.67 \\ 
   $44$ & $425$ & $563$ & $70.45_{-11.05}^{+20.46}$ & $1.99_{-0.97}^{+0.86}$ & $293_{-32}^{+25}$ & $0.92_{-0.03}^{+0.02}$ & $0.43_{-0.15}^{+0.12}$ & $0.32_{-0.03}^{+0.04}$ & $0.49_{-0.01}^{+0.02}$ & 0.63 \\ 
   $45$ & $39$ & $1052$ & $482.90_{-157.14}^{+319.90}$ & $0.77_{-0.45}^{+1.46}$ & $145_{-25}^{+32}$ & $0.90_{-0.11}^{+0.05}$ & $0.62_{-0.04}^{+0.13}$ & $0.57_{-0.06}^{+0.07}$ & $0.60_{-0.05}^{+0.05}$ & 0.17 \\ 
   $46$ & $137$ & $1052$ & $333.58_{-80.62}^{+141.32}$ & $0.91_{-0.58}^{+1.85}$ & $156_{-23}^{+30}$ & $0.89_{-0.19}^{+0.09}$ & $0.36_{-0.05}^{+0.14}$ & $0.55_{-0.05}^{+0.05}$ & $0.58_{-0.04}^{+0.04}$ & 0.30 \\ 
   $47$ & $395$ & $1052$ & $200.92_{-44.20}^{+41.01}$ & $0.01_{-0.00}^{+1.87}$ & $210_{-17}^{+32}$ & $0.53_{-0.00}^{+0.41}$ & $0.27_{-0.00}^{+0.17}$ & $0.45_{-0.06}^{+0.03}$ & $0.53_{-0.03}^{+0.02}$ & 1.42 \\ 
   $48$ & $36$ & $2656$ & $981.95_{-281.42}^{+26.15}$ & $0.01_{-0.00}^{+0.61}$ & $116_{-4}^{+37}$ & $1.02_{-0.02}^{+0.00}$ & $0.94_{-0.02}^{+0.06}$ & $0.63_{-0.12}^{+0.01}$ & $0.63_{-0.10}^{+0.01}$ & 0.16 \\ 
   $49$ & $127$ & $2656$ & $738.60_{-131.87}^{+178.95}$ & $0.01_{-0.00}^{+0.47}$ & $139_{-13}^{+16}$ & $1.14_{-0.13}^{+0.00}$ & $0.66_{-0.00}^{+0.16}$ & $0.56_{-0.04}^{+0.04}$ & $0.56_{-0.04}^{+0.04}$ & 1.39 \\ 
   $50$ & $368$ & $2656$ & $601.10_{-76.62}^{+91.14}$ & $0.01_{-0.00}^{+1.11}$ & $156_{-10}^{+13}$ & $1.05_{-0.08}^{+0.00}$ & $0.51_{-0.00}^{+0.08}$ & $0.51_{-0.03}^{+0.02}$ & $0.52_{-0.02}^{+0.02}$ & 1.12 \\ 
   \hline
  \end{tabular} \\ 
  All time-scales are in units of Myr. All length-scales are in units of pc. \vspace{-1mm}
 \end{minipage}
\end{table*}

\begin{table*}
 \centering
 \begin{minipage}{\hsize}
  \caption{Best-fitting solutions for gas maps (high resolution, point particles)}\label{tab:gasruns_p_hr_out}\vspace{-1mm}
  \begin{tabular}{c c c c c c c c c c c}
   \hline
   ID & $N_{\rm star}$ & $N_{\rm gas}$ & $t_{\rm gas}$ & $t_{\rm over}$ & $\lambda$ & $\beta_{\rm star}$ & $\beta_{\rm gas}$& $\zeta_{\rm star}$ & $\zeta_{\rm gas}$ & $\chi_{\rm red}^2$\\ 
   \hline
   $1$ & $61$ & $462$ & $12.37_{-3.03}^{+9.18}$ & $0.15_{-0.05}^{+0.08}$ & $263_{-72}^{+51}$ & $0.99_{-0.00}^{+0.02}$ & $1.05_{-0.43}^{+0.23}$ & $0.18_{-0.03}^{+0.07}$ & $0.30_{-0.03}^{+0.07}$ & 0.60 \\ 
   $2$ & $125$ & $462$ & $8.61_{-1.52}^{+3.25}$ & $0.13_{-0.05}^{+0.06}$ & $261_{-41}^{+30}$ & $1.02_{-0.01}^{+0.01}$ & $0.70_{-0.48}^{+0.26}$ & $0.18_{-0.02}^{+0.03}$ & $0.31_{-0.02}^{+0.03}$ & 0.16 \\ 
   $3$ & $217$ & $462$ & $8.71_{-1.27}^{+2.30}$ & $0.22_{-0.08}^{+0.07}$ & $248_{-30}^{+24}$ & $1.01_{-0.01}^{+0.00}$ & $0.68_{-0.41}^{+0.19}$ & $0.18_{-0.02}^{+0.03}$ & $0.33_{-0.02}^{+0.03}$ & 0.32 \\ 
   $4$ & $448$ & $462$ & $8.17_{-0.99}^{+1.45}$ & $0.14_{-0.07}^{+0.09}$ & $210_{-11}^{+12}$ & $1.01_{-0.01}^{+0.01}$ & $0.16_{-0.03}^{+0.28}$ & $0.21_{-0.01}^{+0.01}$ & $0.38_{-0.01}^{+0.01}$ & 0.42 \\ 
   $5$ & $622$ & $462$ & $10.45_{-1.34}^{+1.40}$ & $0.09_{-0.05}^{+0.11}$ & $191_{-6}^{+8}$ & $1.01_{-0.02}^{+0.05}$ & $0.13_{-0.00}^{+0.04}$ & $0.23_{-0.01}^{+0.01}$ & $0.41_{-0.01}^{+0.01}$ & 0.44 \\ 
   $6$ & $1131$ & $462$ & $9.51_{-1.11}^{+0.83}$ & $0.01_{-0.00}^{+0.12}$ & $185_{-4}^{+6}$ & $1.08_{-0.04}^{+0.01}$ & $0.26_{-0.05}^{+0.09}$ & $0.23_{-0.01}^{+0.00}$ & $0.43_{-0.01}^{+0.01}$ & 0.66 \\ 
   $7$ & $1603$ & $462$ & $11.17_{-1.05}^{+1.03}$ & $0.14_{-0.08}^{+0.16}$ & $174_{-4}^{+5}$ & $1.26_{-0.11}^{+0.09}$ & $0.17_{-0.00}^{+0.06}$ & $0.25_{-0.01}^{+0.01}$ & $0.45_{-0.01}^{+0.01}$ & 0.83 \\ 
   $8$ & $2860$ & $462$ & $11.32_{-0.74}^{+0.90}$ & $0.71_{-0.24}^{+0.29}$ & $152_{-4}^{+6}$ & $0.98_{-0.00}^{+0.03}$ & $0.30_{-0.08}^{+0.20}$ & $0.29_{-0.01}^{+0.01}$ & $0.49_{-0.01}^{+0.01}$ & 1.45 \\ 
   $9$ & $4070$ & $462$ & $13.40_{-0.74}^{+0.88}$ & $2.23_{-0.40}^{+0.33}$ & $139_{-5}^{+5}$ & $0.96_{-0.03}^{+0.02}$ & $0.55_{-0.06}^{+0.06}$ & $0.32_{-0.01}^{+0.01}$ & $0.52_{-0.01}^{+0.01}$ & 2.79 \\ 
   $10$ & $5250$ & $462$ & $20.21_{-1.09}^{+1.16}$ & $5.42_{-0.70}^{+0.57}$ & $132_{-6}^{+5}$ & $1.04_{-0.01}^{+0.01}$ & $0.59_{-0.06}^{+0.02}$ & $0.36_{-0.01}^{+0.01}$ & $0.53_{-0.01}^{+0.01}$ & 3.80 \\ 
   $11$ & $6021$ & $462$ & $29.49_{-1.53}^{+1.81}$ & $9.34_{-1.04}^{+1.03}$ & $125_{-6}^{+6}$ & $1.14_{-0.03}^{+0.03}$ & $0.50_{-0.05}^{+0.03}$ & $0.41_{-0.01}^{+0.01}$ & $0.55_{-0.01}^{+0.01}$ & 4.20 \\[1.5ex] 
   $12$ & $217$ & $1037$ & $183.35_{-39.14}^{+191.23}$ & $0.97_{-0.41}^{+1.58}$ & $81_{-15}^{+7}$ & $1.01_{-0.01}^{+0.02}$ & $0.52_{-0.05}^{+0.18}$ & $0.54_{-0.03}^{+0.07}$ & $0.61_{-0.01}^{+0.04}$ & 0.62 \\ 
   $13$ & $622$ & $1037$ & $74.85_{-10.95}^{+14.60}$ & $0.01_{-0.00}^{+0.20}$ & $122_{-9}^{+9}$ & $0.99_{-0.00}^{+0.03}$ & $0.27_{-0.00}^{+0.08}$ & $0.39_{-0.03}^{+0.03}$ & $0.53_{-0.01}^{+0.02}$ & 3.65 \\ 
   $14$ & $1603$ & $1037$ & $52.66_{-4.82}^{+4.57}$ & $0.01_{-0.00}^{+0.32}$ & $146_{-4}^{+5}$ & $0.03_{-0.00}^{+0.90}$ & $0.27_{-0.00}^{+0.19}$ & $0.32_{-0.01}^{+0.01}$ & $0.51_{-0.01}^{+0.00}$ & 6.46 \\ 
   $15$ & $217$ & $108$ & $0.94_{-0.17}^{+0.38}$ & $0.09_{-0.02}^{+0.04}$ & $385_{-52}^{+87}$ & $1.05_{-0.02}^{+0.00}$ & $0.91_{-0.11}^{+0.31}$ & $0.10_{-0.01}^{+0.02}$ & $0.20_{-0.03}^{+0.03}$ & 0.62 \\ 
   $16$ & $622$ & $108$ & $1.35_{-0.24}^{+0.42}$ & $0.07_{-0.03}^{+0.05}$ & $203_{-15}^{+24}$ & $1.12_{-0.00}^{+0.04}$ & $0.38_{-0.04}^{+0.04}$ & $0.19_{-0.01}^{+0.01}$ & $0.35_{-0.03}^{+0.02}$ & 0.70 \\ 
   $17$ & $1603$ & $108$ & $1.73_{-0.32}^{+0.47}$ & $0.11_{-0.05}^{+0.06}$ & $159_{-13}^{+17}$ & $1.50_{-0.08}^{+0.16}$ & $0.30_{-0.01}^{+0.02}$ & $0.25_{-0.02}^{+0.02}$ & $0.42_{-0.03}^{+0.03}$ & 0.68 \\[1.5ex] 
   $18$ & $222$ & $1667$ & $18.14_{-3.39}^{+5.32}$ & $0.05_{-0.03}^{+0.04}$ & $162_{-17}^{+16}$ & $1.01_{-0.01}^{+0.02}$ & $0.52_{-0.04}^{+0.09}$ & $0.14_{-0.01}^{+0.02}$ & $0.31_{-0.01}^{+0.01}$ & 3.62 \\ 
   $19$ & $696$ & $1667$ & $15.34_{-1.73}^{+1.82}$ & $0.01_{-0.00}^{+0.02}$ & $173_{-5}^{+6}$ & $1.02_{-0.00}^{+0.02}$ & $0.47_{-0.00}^{+0.10}$ & $0.12_{-0.00}^{+0.00}$ & $0.32_{-0.00}^{+0.00}$ & 4.66 \\ 
   $20$ & $2013$ & $1667$ & $16.42_{-1.41}^{+1.15}$ & $0.01_{-0.00}^{+0.04}$ & $158_{-3}^{+4}$ & $1.12_{-0.00}^{+0.22}$ & $0.54_{-0.04}^{+0.13}$ & $0.13_{-0.00}^{+0.00}$ & $0.35_{-0.00}^{+0.00}$ & 7.49 \\ 
   $21$ & $201$ & $286$ & $7.60_{-1.21}^{+2.32}$ & $0.26_{-0.15}^{+0.19}$ & $261_{-31}^{+35}$ & $1.09_{-0.07}^{+0.01}$ & $0.32_{-0.20}^{+0.21}$ & $0.35_{-0.04}^{+0.05}$ & $0.44_{-0.03}^{+0.04}$ & 0.04 \\ 
   $22$ & $489$ & $286$ & $11.02_{-1.88}^{+1.29}$ & $0.04_{-0.02}^{+0.38}$ & $206_{-6}^{+15}$ & $0.47_{-0.02}^{+0.51}$ & $0.08_{-0.00}^{+0.03}$ & $0.44_{-0.03}^{+0.01}$ & $0.54_{-0.02}^{+0.01}$ & 0.83 \\ 
   $23$ & $905$ & $286$ & $12.55_{-1.75}^{+1.37}$ & $0.01_{-0.00}^{+0.34}$ & $196_{-8}^{+11}$ & $0.82_{-0.00}^{+0.54}$ & $0.02_{-0.00}^{+0.07}$ & $0.46_{-0.02}^{+0.01}$ & $0.57_{-0.02}^{+0.01}$ & 1.80 \\ 
   $24$ & $169$ & $172$ & $6.12_{-1.07}^{+1.92}$ & $0.29_{-0.18}^{+0.41}$ & $301_{-35}^{+70}$ & $1.01_{-0.11}^{+0.07}$ & $0.14_{-0.08}^{+0.50}$ & $0.54_{-0.07}^{+0.04}$ & $0.58_{-0.05}^{+0.03}$ & 0.21 \\ 
   $25$ & $264$ & $172$ & $6.35_{-0.71}^{+1.26}$ & $1.48_{-0.72}^{+0.53}$ & $258_{-34}^{+40}$ & $1.18_{-0.14}^{+0.09}$ & $0.31_{-0.22}^{+0.15}$ & $0.59_{-0.05}^{+0.06}$ & $0.63_{-0.03}^{+0.04}$ & 0.37 \\ 
   $26$ & $321$ & $172$ & $8.19_{-1.09}^{+1.99}$ & $2.57_{-1.00}^{+0.70}$ & $226_{-21}^{+26}$ & $1.69_{-0.07}^{+0.08}$ & $0.22_{-0.13}^{+0.10}$ & $0.66_{-0.03}^{+0.03}$ & $0.67_{-0.03}^{+0.03}$ & 0.39 \\ 
   $27$ & $94$ & $65$ & $4.17_{-0.86}^{+4.07}$ & $1.46_{-0.77}^{+1.99}$ & $447_{-151}^{+275}$ & $1.17_{-0.14}^{+0.31}$ & $0.38_{-0.11}^{+0.19}$ & $0.65_{-0.14}^{+0.06}$ & $0.66_{-0.13}^{+0.04}$ & 0.26 \\ 
   $28$ & $94$ & $65$ & $5.37_{-1.22}^{+233.79}$ & $4.35_{-1.58}^{+182.55}$ & $402_{-143}^{+9995}$ & $1.43_{-0.05}^{+1.29}$ & $0.42_{-0.01}^{+0.04}$ & $0.69_{-0.42}^{+0.02}$ & $0.70_{-0.47}^{+0.01}$ & 0.28 \\ 
   $29$ & $35$ & $65$ & $6.52_{-1.70}^{+244.80}$ & $4.69_{-3.02}^{+182.89}$ & $303_{-64}^{+1691}$ & $3.65_{-1.04}^{+2.99}$ & $0.29_{-0.01}^{+0.03}$ & $0.71_{-0.28}^{+0.00}$ & $0.71_{-0.41}^{+0.00}$ & 0.60 \\ 
   $30$ & $17$ & $20$ & $22.22_{-9.37}^{+171.97}$ & $9.01_{-4.66}^{+92.86}$ & $2255_{-1260}^{+2037}$ & $3.26_{-2.43}^{+0.38}$ & $0.46_{-0.09}^{+0.01}$ & $0.46_{-0.04}^{+0.15}$ & $0.39_{-0.09}^{+0.23}$ & 0.08 \\ 
   $31$ & $7$ & $20$ & $244.49_{-105.33}^{+45.77}$ & $143.26_{-85.76}^{+69.35}$ & $3063_{-1969}^{+4255}$ & $1.54_{-0.00}^{+1.18}$ & $0.35_{-0.15}^{+0.08}$ & $0.40_{-0.01}^{+0.06}$ & $0.33_{-0.06}^{+0.25}$ & 0.01 \\ 
   $32$ & $10$ & $20$ & $8.63_{-2.31}^{+236.94}$ & $8.57_{-5.60}^{+193.38}$ & $580_{-236}^{+1664}$ & $3.48_{-1.17}^{+0.12}$ & $0.27_{-0.00}^{+0.05}$ & $0.71_{-0.17}^{+0.00}$ & $0.71_{-0.33}^{+0.00}$ & 0.62 \\[1.5ex] 
   $33$ & $211$ & $461$ & $9.47_{-1.53}^{+2.40}$ & $0.22_{-0.07}^{+0.08}$ & $250_{-29}^{+25}$ & $0.99_{-0.01}^{+0.05}$ & $0.75_{-0.28}^{+0.15}$ & $0.18_{-0.02}^{+0.02}$ & $0.33_{-0.02}^{+0.03}$ & 0.24 \\ 
   $34$ & $613$ & $461$ & $10.64_{-1.15}^{+1.60}$ & $0.18_{-0.10}^{+0.10}$ & $203_{-9}^{+10}$ & $1.02_{-0.02}^{+0.06}$ & $0.33_{-0.17}^{+0.20}$ & $0.21_{-0.01}^{+0.01}$ & $0.40_{-0.01}^{+0.01}$ & 0.58 \\ 
   $35$ & $1548$ & $461$ & $12.03_{-1.11}^{+1.19}$ & $0.17_{-0.09}^{+0.16}$ & $180_{-5}^{+5}$ & $1.21_{-0.11}^{+0.07}$ & $0.25_{-0.05}^{+0.10}$ & $0.24_{-0.01}^{+0.01}$ & $0.45_{-0.01}^{+0.01}$ & 1.10 \\ 
   $36$ & $213$ & $488$ & $9.20_{-1.35}^{+2.70}$ & $0.19_{-0.07}^{+0.06}$ & $263_{-31}^{+23}$ & $0.99_{-0.00}^{+0.00}$ & $0.84_{-0.38}^{+0.16}$ & $0.17_{-0.01}^{+0.02}$ & $0.33_{-0.01}^{+0.02}$ & 0.54 \\ 
   $37$ & $611$ & $488$ & $11.24_{-1.33}^{+1.54}$ & $0.10_{-0.06}^{+0.11}$ & $206_{-7}^{+10}$ & $0.95_{-0.00}^{+0.09}$ & $0.26_{-0.10}^{+0.20}$ & $0.21_{-0.01}^{+0.01}$ & $0.40_{-0.01}^{+0.01}$ & 0.69 \\ 
   $38$ & $1776$ & $488$ & $11.23_{-0.94}^{+1.09}$ & $0.20_{-0.11}^{+0.17}$ & $174_{-4}^{+5}$ & $1.09_{-0.04}^{+0.18}$ & $0.28_{-0.01}^{+0.10}$ & $0.25_{-0.01}^{+0.01}$ & $0.46_{-0.01}^{+0.01}$ & 0.82 \\ 
   $39$ & $217$ & $519$ & $9.22_{-1.44}^{+2.71}$ & $0.13_{-0.06}^{+0.06}$ & $267_{-30}^{+27}$ & $0.89_{-0.08}^{+0.03}$ & $0.90_{-0.40}^{+0.35}$ & $0.17_{-0.02}^{+0.02}$ & $0.35_{-0.02}^{+0.02}$ & 0.31 \\ 
   $40$ & $627$ & $519$ & $11.15_{-1.29}^{+1.46}$ & $0.07_{-0.04}^{+0.10}$ & $212_{-7}^{+10}$ & $0.99_{-0.02}^{+0.04}$ & $0.22_{-0.02}^{+0.32}$ & $0.21_{-0.01}^{+0.01}$ & $0.42_{-0.01}^{+0.01}$ & 0.65 \\ 
   $41$ & $1679$ & $519$ & $12.09_{-1.00}^{+1.06}$ & $0.17_{-0.10}^{+0.16}$ & $177_{-4}^{+5}$ & $1.01_{-0.01}^{+0.04}$ & $0.34_{-0.16}^{+0.11}$ & $0.25_{-0.01}^{+0.01}$ & $0.48_{-0.01}^{+0.00}$ & 1.04 \\ 
   $42$ & $214$ & $786$ & $18.94_{-4.26}^{+4.61}$ & $0.01_{-0.00}^{+0.08}$ & $208_{-18}^{+26}$ & $1.02_{-0.02}^{+0.00}$ & $0.27_{-0.00}^{+0.29}$ & $0.23_{-0.03}^{+0.02}$ & $0.42_{-0.02}^{+0.01}$ & 1.32 \\ 
   $43$ & $656$ & $786$ & $17.10_{-2.29}^{+2.10}$ & $0.01_{-0.00}^{+0.08}$ & $203_{-7}^{+10}$ & $1.00_{-0.00}^{+0.00}$ & $0.27_{-0.03}^{+0.07}$ & $0.22_{-0.01}^{+0.01}$ & $0.44_{-0.01}^{+0.00}$ & 1.81 \\ 
   $44$ & $1865$ & $786$ & $19.33_{-1.76}^{+1.96}$ & $0.31_{-0.18}^{+0.25}$ & $167_{-4}^{+5}$ & $1.05_{-0.05}^{+0.03}$ & $0.26_{-0.04}^{+0.16}$ & $0.27_{-0.01}^{+0.01}$ & $0.49_{-0.00}^{+0.00}$ & 2.24 \\ 
   $45$ & $209$ & $1356$ & $27.22_{-5.33}^{+13.17}$ & $0.18_{-0.09}^{+0.08}$ & $210_{-37}^{+24}$ & $1.03_{-0.01}^{+0.00}$ & $0.54_{-0.20}^{+0.10}$ & $0.18_{-0.02}^{+0.04}$ & $0.34_{-0.01}^{+0.02}$ & 2.18 \\ 
   $46$ & $619$ & $1356$ & $32.18_{-4.31}^{+5.96}$ & $0.17_{-0.10}^{+0.16}$ & $188_{-12}^{+13}$ & $1.03_{-0.01}^{+0.01}$ & $0.46_{-0.15}^{+0.24}$ & $0.20_{-0.01}^{+0.02}$ & $0.36_{-0.01}^{+0.01}$ & 4.27 \\ 
   $47$ & $1710$ & $1356$ & $43.00_{-3.83}^{+4.94}$ & $0.45_{-0.27}^{+0.29}$ & $157_{-6}^{+7}$ & $1.04_{-0.01}^{+0.03}$ & $0.48_{-0.16}^{+0.22}$ & $0.24_{-0.01}^{+0.01}$ & $0.40_{-0.00}^{+0.00}$ & 9.27 \\ 
   $48$ & $176$ & $499$ & $49.87_{-14.52}^{+51.85}$ & $0.01_{-0.00}^{+0.20}$ & $184_{-53}^{+36}$ & $1.06_{-0.04}^{+0.01}$ & $0.74_{-0.00}^{+0.05}$ & $0.38_{-0.06}^{+0.16}$ & $0.48_{-0.03}^{+0.11}$ & 3.54 \\ 
   $49$ & $485$ & $499$ & $47.56_{-7.34}^{+11.39}$ & $0.01_{-0.00}^{+0.28}$ & $189_{-15}^{+15}$ & $1.07_{-0.07}^{+0.00}$ & $0.26_{-0.00}^{+0.33}$ & $0.36_{-0.03}^{+0.03}$ & $0.48_{-0.01}^{+0.02}$ & 6.19 \\ 
   $50$ & $1147$ & $499$ & $79.12_{-9.60}^{+11.28}$ & $0.01_{-0.00}^{+0.68}$ & $152_{-6}^{+9}$ & $0.89_{-0.00}^{+0.15}$ & $0.26_{-0.00}^{+0.21}$ & $0.45_{-0.03}^{+0.02}$ & $0.55_{-0.01}^{+0.01}$ & 12.44 \\ 
   \hline
  \end{tabular} \\ 
  All time-scales are in units of Myr. All length-scales are in units of pc. \vspace{-1mm}
 \end{minipage}
\end{table*}

\begin{table*}
 \centering
 \begin{minipage}{\hsize}
  \caption{Best-fitting solutions for gas maps (low resolution, extended emission)}\label{tab:gasruns_lr_out}\vspace{-1mm}
  \begin{tabular}{c c c c c c c c c c c}
   \hline
   ID & $N_{\rm star}$ & $N_{\rm gas}$ & $t_{\rm gas}$ & $t_{\rm over}$ & $\lambda$ & $\beta_{\rm star}$ & $\beta_{\rm gas}$& $\zeta_{\rm star}$ & $\zeta_{\rm gas}$ & $\chi_{\rm red}^2$\\ 
   \hline
   $1$ & $15$ & $137$ & $27.21_{-9.80}^{+653.94}$ & $1.01_{-0.30}^{+27.84}$ & $695_{-171}^{+121}$ & $1.54_{-0.22}^{+0.31}$ & $1.31_{-0.11}^{+0.18}$ & $0.46_{-0.04}^{+0.08}$ & $0.42_{-0.03}^{+0.07}$ & 0.38 \\ 
   $2$ & $25$ & $137$ & $20.13_{-6.50}^{+625.50}$ & $1.67_{-0.44}^{+55.31}$ & $878_{-247}^{+184}$ & $1.25_{-0.03}^{+1.79}$ & $1.06_{-0.17}^{+0.09}$ & $0.38_{-0.05}^{+0.10}$ & $0.37_{-0.04}^{+0.08}$ & 0.50 \\ 
   $3$ & $42$ & $169$ & $12.72_{-2.51}^{+3.99}$ & $1.25_{-0.31}^{+0.51}$ & $598_{-70}^{+107}$ & $0.66_{-0.02}^{+0.05}$ & $1.03_{-0.09}^{+0.11}$ & $0.40_{-0.03}^{+0.03}$ & $0.37_{-0.03}^{+0.03}$ & 0.85 \\ 
   $4$ & $74$ & $169$ & $13.85_{-2.92}^{+2.67}$ & $3.64_{-0.87}^{+0.82}$ & $610_{-101}^{+117}$ & $0.87_{-0.01}^{+0.04}$ & $0.97_{-0.07}^{+0.03}$ & $0.39_{-0.03}^{+0.04}$ & $0.38_{-0.04}^{+0.05}$ & 1.25 \\ 
   $5$ & $86$ & $169$ & $10.92_{-1.35}^{+2.76}$ & $2.87_{-0.51}^{+1.13}$ & $377_{-52}^{+103}$ & $0.92_{-0.02}^{+0.09}$ & $0.83_{-0.12}^{+0.08}$ & $0.52_{-0.06}^{+0.02}$ & $0.52_{-0.07}^{+0.03}$ & 0.66 \\ 
   $6$ & $100$ & $169$ & $11.47_{-1.55}^{+5.06}$ & $4.40_{-0.87}^{+2.57}$ & $329_{-45}^{+112}$ & $0.99_{-0.05}^{+0.01}$ & $0.81_{-0.10}^{+0.15}$ & $0.52_{-0.05}^{+0.03}$ & $0.55_{-0.07}^{+0.03}$ & 0.37 \\ 
   $7$ & $108$ & $169$ & $16.14_{-2.19}^{+6.10}$ & $6.40_{-1.29}^{+3.07}$ & $325_{-43}^{+86}$ & $1.06_{-0.07}^{+0.07}$ & $0.62_{-0.05}^{+0.09}$ & $0.55_{-0.03}^{+0.02}$ & $0.56_{-0.04}^{+0.03}$ & 0.20 \\ 
   $8$ & $109$ & $169$ & $17.87_{-2.96}^{+4.60}$ & $12.20_{-2.23}^{+3.23}$ & $339_{-52}^{+88}$ & $0.96_{-0.03}^{+0.10}$ & $0.71_{-0.02}^{+0.02}$ & $0.54_{-0.03}^{+0.03}$ & $0.56_{-0.05}^{+0.03}$ & 0.30 \\ 
   $9$ & $114$ & $169$ & $21.59_{-2.75}^{+6.21}$ & $18.26_{-2.57}^{+4.69}$ & $352_{-52}^{+120}$ & $1.04_{-0.08}^{+0.20}$ & $0.59_{-0.02}^{+0.05}$ & $0.53_{-0.05}^{+0.03}$ & $0.56_{-0.07}^{+0.03}$ & 0.29 \\ 
   $10$ & $108$ & $169$ & $36.08_{-5.00}^{+36.07}$ & $35.98_{-5.82}^{+32.34}$ & $376_{-68}^{+352}$ & $1.08_{-0.09}^{+0.14}$ & $1.74_{-1.11}^{+0.00}$ & $0.52_{-0.11}^{+0.03}$ & $0.55_{-0.17}^{+0.03}$ & 0.90 \\ 
   $11$ & $85$ & $169$ & $66.92_{-9.73}^{+37.49}$ & $66.15_{-11.96}^{+30.82}$ & $397_{-65}^{+189}$ & $0.83_{-0.04}^{+0.11}$ & $0.68_{-0.13}^{+0.05}$ & $0.51_{-0.07}^{+0.03}$ & $0.54_{-0.11}^{+0.03}$ & 0.65 \\[1.5ex] 
   $12$ & $42$ & $206$ & $598.74_{-210.80}^{+1452.38}$ & $28.35_{-9.67}^{+73.57}$ & $444_{-52}^{+73}$ & $0.22_{-0.07}^{+0.03}$ & $1.59_{-0.09}^{+0.02}$ & $0.50_{-0.03}^{+0.03}$ & $0.53_{-0.03}^{+0.03}$ & 0.70 \\ 
   $13$ & $86$ & $206$ & $2608.31_{-872.57}^{+222.74}$ & $301.92_{-102.52}^{+37.24}$ & $381_{-43}^{+57}$ & $0.39_{-0.10}^{+0.13}$ & $1.30_{-0.03}^{+0.03}$ & $0.56_{-0.03}^{+0.02}$ & $0.56_{-0.03}^{+0.02}$ & 0.94 \\ 
   $14$ & $108$ & $206$ & $1022.09_{-280.18}^{+1303.32}$ & $143.67_{-43.61}^{+181.92}$ & $350_{-54}^{+56}$ & $1.09_{-0.52}^{+1.23}$ & $1.30_{-0.02}^{+0.06}$ & $0.58_{-0.02}^{+0.02}$ & $0.57_{-0.02}^{+0.03}$ & 0.58 \\ 
   $15$ & $42$ & $64$ & $1.22_{-0.30}^{+0.63}$ & $0.20_{-0.12}^{+0.13}$ & $666_{-143}^{+194}$ & $0.93_{-0.01}^{+0.02}$ & $0.45_{-0.26}^{+0.24}$ & $0.34_{-0.05}^{+0.05}$ & $0.29_{-0.05}^{+0.06}$ & 0.46 \\ 
   $16$ & $86$ & $64$ & $0.84_{-0.18}^{+0.36}$ & $0.56_{-0.11}^{+0.16}$ & $274_{-48}^{+88}$ & $1.36_{-0.10}^{+0.08}$ & $1.28_{-0.42}^{+0.20}$ & $0.56_{-0.05}^{+0.04}$ & $0.54_{-0.07}^{+0.06}$ & 0.24 \\ 
   $17$ & $108$ & $64$ & $1.44_{-0.25}^{+0.76}$ & $1.38_{-0.29}^{+0.31}$ & $262_{-43}^{+98}$ & $2.08_{-0.01}^{+0.09}$ & $1.18_{-0.42}^{+0.10}$ & $0.57_{-0.05}^{+0.03}$ & $0.55_{-0.08}^{+0.05}$ & 0.20 \\[1.5ex] 
   $18$ & $42$ & $169$ & $12.72_{-2.52}^{+3.96}$ & $1.25_{-0.31}^{+0.51}$ & $598_{-70}^{+109}$ & $0.66_{-0.02}^{+0.05}$ & $1.03_{-0.09}^{+0.11}$ & $0.40_{-0.03}^{+0.03}$ & $0.37_{-0.03}^{+0.03}$ & 0.85 \\ 
   $19$ & $88$ & $198$ & $10.25_{-1.05}^{+2.35}$ & $2.68_{-0.38}^{+0.92}$ & $346_{-43}^{+83}$ & $0.85_{-0.03}^{+0.00}$ & $0.75_{-0.13}^{+0.12}$ & $0.47_{-0.03}^{+0.02}$ & $0.48_{-0.04}^{+0.03}$ & 0.99 \\ 
   $20$ & $116$ & $198$ & $20.01_{-3.76}^{+4.68}$ & $8.72_{-2.16}^{+2.55}$ & $345_{-49}^{+65}$ & $0.77_{-0.02}^{+0.07}$ & $0.67_{-0.06}^{+0.06}$ & $0.45_{-0.02}^{+0.02}$ & $0.49_{-0.03}^{+0.03}$ & 0.38 \\ 
   $21$ & $42$ & $169$ & $12.72_{-2.51}^{+3.98}$ & $1.25_{-0.31}^{+0.51}$ & $591_{-66}^{+109}$ & $0.66_{-0.02}^{+0.05}$ & $1.03_{-0.09}^{+0.11}$ & $0.41_{-0.03}^{+0.03}$ & $0.38_{-0.03}^{+0.03}$ & 0.85 \\ 
   $22$ & $88$ & $198$ & $10.25_{-1.05}^{+2.35}$ & $2.68_{-0.38}^{+0.92}$ & $348_{-43}^{+83}$ & $0.85_{-0.03}^{+0.00}$ & $0.75_{-0.13}^{+0.12}$ & $0.46_{-0.03}^{+0.02}$ & $0.48_{-0.04}^{+0.03}$ & 0.97 \\ 
   $23$ & $116$ & $198$ & $20.01_{-3.76}^{+4.69}$ & $8.72_{-2.16}^{+2.55}$ & $345_{-49}^{+66}$ & $0.77_{-0.02}^{+0.07}$ & $0.67_{-0.05}^{+0.06}$ & $0.45_{-0.02}^{+0.02}$ & $0.49_{-0.03}^{+0.03}$ & 0.39 \\ 
   $24$ & $40$ & $137$ & $11.86_{-2.11}^{+4.19}$ & $1.49_{-0.32}^{+0.71}$ & $699_{-120}^{+227}$ & $1.04_{-0.11}^{+0.32}$ & $1.00_{-0.17}^{+0.18}$ & $0.46_{-0.08}^{+0.05}$ & $0.43_{-0.06}^{+0.05}$ & 0.33 \\ 
   $25$ & $80$ & $137$ & $12.21_{-1.85}^{+3.39}$ & $4.29_{-1.01}^{+1.62}$ & $519_{-108}^{+169}$ & $1.28_{-0.10}^{+0.06}$ & $0.90_{-0.13}^{+0.11}$ & $0.54_{-0.07}^{+0.07}$ & $0.52_{-0.07}^{+0.07}$ & 0.58 \\ 
   $26$ & $95$ & $137$ & $17.70_{-2.57}^{+4.29}$ & $8.26_{-1.67}^{+2.96}$ & $381_{-63}^{+116}$ & $1.11_{-0.08}^{+0.17}$ & $0.58_{-0.06}^{+0.04}$ & $0.63_{-0.06}^{+0.02}$ & $0.61_{-0.07}^{+0.03}$ & 0.24 \\ 
   $27$ & $35$ & $84$ & $13.81_{-3.13}^{+626.59}$ & $2.59_{-0.71}^{+130.64}$ & $850_{-206}^{+314}$ & $1.32_{-0.14}^{+1.12}$ & $0.90_{-0.12}^{+0.27}$ & $0.52_{-0.09}^{+0.08}$ & $0.51_{-0.09}^{+0.07}$ & 0.26 \\ 
   $28$ & $65$ & $84$ & $15.49_{-2.89}^{+6.21}$ & $5.80_{-1.68}^{+3.24}$ & $597_{-142}^{+380}$ & $1.22_{-0.14}^{+0.28}$ & $1.04_{-0.06}^{+0.11}$ & $0.61_{-0.11}^{+0.06}$ & $0.61_{-0.12}^{+0.07}$ & 0.60 \\ 
   $29$ & $71$ & $84$ & $17.25_{-2.93}^{+5.06}$ & $9.70_{-2.62}^{+3.54}$ & $461_{-106}^{+213}$ & $1.46_{-0.17}^{+0.30}$ & $0.64_{-0.06}^{+0.07}$ & $0.67_{-0.06}^{+0.03}$ & $0.67_{-0.07}^{+0.04}$ & 0.43 \\ 
   $30$ & $30$ & $35$ & $13.03_{-4.81}^{+641.06}$ & $4.27_{-1.55}^{+211.70}$ & $1324_{-574}^{+814}$ & $1.18_{-0.16}^{+6.36}$ & $1.04_{-0.41}^{+0.47}$ & $0.54_{-0.13}^{+0.17}$ & $0.56_{-0.12}^{+0.15}$ & 0.20 \\ 
   $31$ & $44$ & $35$ & $21.18_{-4.90}^{+692.71}$ & $12.67_{-5.14}^{+442.59}$ & $1058_{-495}^{+5997}$ & $1.87_{-0.34}^{+4.25}$ & $0.69_{-0.09}^{+0.42}$ & $0.62_{-0.33}^{+0.09}$ & $0.62_{-0.34}^{+0.09}$ & 0.49 \\ 
   $32$ & $37$ & $35$ & $25.26_{-5.54}^{+713.28}$ & $16.33_{-9.51}^{+539.97}$ & $685_{-328}^{+7399}$ & $2.09_{-0.93}^{+6.93}$ & $0.60_{-0.04}^{+0.10}$ & $0.71_{-0.43}^{+0.00}$ & $0.71_{-0.47}^{+0.00}$ & 0.71 \\[1.5ex] 
   $33$ & $40$ & $172$ & $11.82_{-2.08}^{+4.95}$ & $1.11_{-0.29}^{+0.64}$ & $568_{-67}^{+101}$ & $0.70_{-0.01}^{+0.07}$ & $0.85_{-0.07}^{+0.19}$ & $0.43_{-0.03}^{+0.03}$ & $0.41_{-0.04}^{+0.03}$ & 0.90 \\ 
   $34$ & $78$ & $172$ & $9.01_{-1.07}^{+2.41}$ & $3.02_{-0.47}^{+1.15}$ & $352_{-48}^{+126}$ & $0.79_{-0.02}^{+0.09}$ & $0.91_{-0.04}^{+0.09}$ & $0.53_{-0.06}^{+0.02}$ & $0.55_{-0.07}^{+0.03}$ & 0.32 \\ 
   $35$ & $98$ & $172$ & $16.50_{-2.35}^{+4.26}$ & $8.45_{-1.71}^{+2.81}$ & $340_{-50}^{+94}$ & $0.86_{-0.02}^{+0.11}$ & $0.65_{-0.02}^{+0.04}$ & $0.52_{-0.04}^{+0.03}$ & $0.57_{-0.05}^{+0.03}$ & 0.11 \\ 
   $36$ & $40$ & $153$ & $11.56_{-2.07}^{+4.31}$ & $1.51_{-0.32}^{+0.69}$ & $634_{-70}^{+136}$ & $0.73_{-0.04}^{+0.06}$ & $0.96_{-0.05}^{+0.15}$ & $0.43_{-0.03}^{+0.02}$ & $0.40_{-0.04}^{+0.03}$ & 1.12 \\ 
   $37$ & $70$ & $153$ & $9.24_{-1.15}^{+3.57}$ & $3.19_{-0.54}^{+1.73}$ & $345_{-48}^{+129}$ & $0.76_{-0.02}^{+0.04}$ & $0.89_{-0.09}^{+0.13}$ & $0.52_{-0.05}^{+0.03}$ & $0.57_{-0.07}^{+0.02}$ & 0.40 \\ 
   $38$ & $84$ & $153$ & $15.90_{-2.19}^{+4.77}$ & $9.75_{-1.84}^{+3.38}$ & $332_{-48}^{+94}$ & $0.77_{-0.03}^{+0.09}$ & $0.66_{-0.01}^{+0.04}$ & $0.52_{-0.04}^{+0.03}$ & $0.58_{-0.05}^{+0.02}$ & 0.10 \\ 
   $39$ & $37$ & $139$ & $13.62_{-2.76}^{+4.52}$ & $2.32_{-0.60}^{+0.93}$ & $642_{-83}^{+142}$ & $0.52_{-0.02}^{+0.05}$ & $0.84_{-0.03}^{+0.07}$ & $0.46_{-0.03}^{+0.03}$ & $0.42_{-0.04}^{+0.03}$ & 0.72 \\ 
   $40$ & $66$ & $139$ & $9.98_{-1.23}^{+3.24}$ & $4.27_{-0.69}^{+1.93}$ & $350_{-55}^{+147}$ & $0.57_{-0.03}^{+0.07}$ & $0.83_{-0.13}^{+0.07}$ & $0.52_{-0.06}^{+0.03}$ & $0.58_{-0.08}^{+0.02}$ & 0.30 \\ 
   $41$ & $76$ & $139$ & $15.25_{-2.55}^{+7.47}$ & $11.30_{-2.22}^{+5.77}$ & $356_{-57}^{+164}$ & $0.78_{-0.01}^{+0.05}$ & $0.82_{-0.11}^{+0.17}$ & $0.51_{-0.05}^{+0.03}$ & $0.59_{-0.09}^{+0.02}$ & 0.14 \\ 
   $42$ & $35$ & $90$ & $10.25_{-2.36}^{+6.04}$ & $1.76_{-0.53}^{+1.38}$ & $804_{-136}^{+263}$ & $0.64_{-0.01}^{+0.10}$ & $1.05_{-0.08}^{+0.14}$ & $0.48_{-0.06}^{+0.04}$ & $0.45_{-0.08}^{+0.04}$ & 0.46 \\ 
   $43$ & $49$ & $120$ & $9.26_{-1.18}^{+1.82}$ & $4.00_{-0.65}^{+1.20}$ & $400_{-61}^{+133}$ & $0.64_{-0.01}^{+0.07}$ & $0.90_{-0.12}^{+0.06}$ & $0.52_{-0.06}^{+0.03}$ & $0.58_{-0.07}^{+0.02}$ & 0.35 \\ 
   $44$ & $61$ & $120$ & $16.49_{-2.87}^{+10.73}$ & $14.10_{-2.63}^{+8.81}$ & $435_{-66}^{+185}$ & $0.67_{-0.07}^{+0.05}$ & $1.06_{-0.27}^{+0.40}$ & $0.46_{-0.03}^{+0.02}$ & $0.57_{-0.08}^{+0.03}$ & 0.37 \\ 
   $45$ & $32$ & $78$ & $6.10_{-1.13}^{+2.29}$ & $1.35_{-0.41}^{+0.81}$ & $849_{-169}^{+315}$ & $0.50_{-0.07}^{+0.17}$ & $0.72_{-0.10}^{+0.20}$ & $0.47_{-0.06}^{+0.05}$ & $0.51_{-0.08}^{+0.05}$ & 0.27 \\ 
   $46$ & $35$ & $78$ & $7.09_{-1.02}^{+1.93}$ & $3.97_{-0.71}^{+1.38}$ & $579_{-96}^{+236}$ & $0.75_{-0.01}^{+0.03}$ & $0.96_{-0.06}^{+0.41}$ & $0.53_{-0.06}^{+0.04}$ & $0.60_{-0.06}^{+0.03}$ & 0.12 \\ 
   $47$ & $34$ & $82$ & $18.08_{-3.06}^{+4.17}$ & $15.56_{-3.06}^{+3.14}$ & $573_{-98}^{+145}$ & $0.65_{-0.01}^{+0.04}$ & $1.27_{-0.32}^{+0.51}$ & $0.45_{-0.04}^{+0.04}$ & $0.57_{-0.03}^{+0.03}$ & 0.18 \\ 
   $48$ & $19$ & $33$ & $11.19_{-3.34}^{+675.54}$ & $3.57_{-1.50}^{+290.54}$ & $940_{-320}^{+1310}$ & $0.37_{-0.06}^{+9.10}$ & $0.82_{-0.08}^{+0.10}$ & $0.58_{-0.18}^{+0.07}$ & $0.60_{-0.21}^{+0.06}$ & 0.15 \\ 
   $49$ & $20$ & $41$ & $8.70_{-1.59}^{+5.29}$ & $4.86_{-0.91}^{+3.75}$ & $632_{-127}^{+294}$ & $0.84_{-0.26}^{+0.08}$ & $0.92_{-0.11}^{+0.07}$ & $0.57_{-0.07}^{+0.03}$ & $0.64_{-0.05}^{+0.02}$ & 0.18 \\ 
   $50$ & $15$ & $41$ & $39.00_{-8.42}^{+32.87}$ & $38.18_{-9.98}^{+23.76}$ & $776_{-249}^{+411}$ & $0.37_{-0.03}^{+0.06}$ & $0.73_{-0.02}^{+0.10}$ & $0.44_{-0.04}^{+0.09}$ & $0.61_{-0.08}^{+0.03}$ & 0.21 \\ 
   \hline
  \end{tabular} \\ 
  All time-scales are in units of Myr. All length-scales are in units of pc. \vspace{-1mm}
 \end{minipage}
\end{table*}

\begin{table*}
 \centering
 \begin{minipage}{\hsize}
  \caption{Best-fitting solutions for gas maps (high resolution, extended emission)}\label{tab:gasruns_hr_out}\vspace{-1mm}
  \begin{tabular}{c c c c c c c c c c c}
   \hline
   ID & $N_{\rm star}$ & $N_{\rm gas}$ & $t_{\rm gas}$ & $t_{\rm over}$ & $\lambda$ & $\beta_{\rm star}$ & $\beta_{\rm gas}$& $\zeta_{\rm star}$ & $\zeta_{\rm gas}$ & $\chi_{\rm red}^2$\\ 
   \hline
   $1$ & $58$ & $293$ & $6.95_{-1.39}^{+2.13}$ & $0.31_{-0.06}^{+0.11}$ & $374_{-54}^{+73}$ & $1.13_{-0.14}^{+0.16}$ & $0.80_{-0.10}^{+0.12}$ & $0.33_{-0.05}^{+0.04}$ & $0.29_{-0.03}^{+0.03}$ & 0.21 \\ 
   $2$ & $109$ & $293$ & $3.93_{-0.49}^{+0.87}$ & $0.27_{-0.09}^{+0.09}$ & $339_{-55}^{+46}$ & $1.31_{-0.16}^{+0.02}$ & $0.60_{-0.23}^{+0.13}$ & $0.34_{-0.03}^{+0.04}$ & $0.33_{-0.03}^{+0.04}$ & 0.38 \\ 
   $3$ & $180$ & $293$ & $3.38_{-0.40}^{+0.51}$ & $0.39_{-0.12}^{+0.09}$ & $295_{-40}^{+40}$ & $1.08_{-0.08}^{+0.09}$ & $0.57_{-0.16}^{+0.12}$ & $0.37_{-0.03}^{+0.04}$ & $0.37_{-0.03}^{+0.04}$ & 0.61 \\ 
   $4$ & $297$ & $293$ & $3.21_{-0.27}^{+0.42}$ & $0.52_{-0.18}^{+0.12}$ & $212_{-23}^{+24}$ & $1.01_{-0.12}^{+0.12}$ & $0.38_{-0.16}^{+0.12}$ & $0.47_{-0.03}^{+0.03}$ & $0.49_{-0.03}^{+0.03}$ & 0.33 \\ 
   $5$ & $353$ & $293$ & $3.64_{-0.30}^{+0.49}$ & $0.74_{-0.15}^{+0.10}$ & $174_{-14}^{+15}$ & $1.17_{-0.12}^{+0.20}$ & $0.32_{-0.04}^{+0.11}$ & $0.52_{-0.01}^{+0.02}$ & $0.54_{-0.01}^{+0.02}$ & 0.37 \\ 
   $6$ & $414$ & $293$ & $2.76_{-0.18}^{+0.33}$ & $1.07_{-0.09}^{+0.12}$ & $137_{-9}^{+13}$ & $1.41_{-0.19}^{+0.03}$ & $0.46_{-0.05}^{+0.08}$ & $0.58_{-0.02}^{+0.01}$ & $0.59_{-0.02}^{+0.01}$ & 0.20 \\ 
   $7$ & $409$ & $293$ & $3.99_{-0.40}^{+0.48}$ & $1.96_{-0.23}^{+0.15}$ & $140_{-10}^{+12}$ & $1.43_{-0.01}^{+0.07}$ & $0.54_{-0.05}^{+0.04}$ & $0.58_{-0.01}^{+0.01}$ & $0.59_{-0.01}^{+0.01}$ & 0.32 \\ 
   $8$ & $351$ & $293$ & $5.84_{-0.58}^{+0.78}$ & $4.11_{-0.40}^{+0.33}$ & $144_{-12}^{+15}$ & $1.39_{-0.08}^{+0.12}$ & $0.49_{-0.03}^{+0.06}$ & $0.57_{-0.02}^{+0.02}$ & $0.59_{-0.02}^{+0.02}$ & 0.29 \\ 
   $9$ & $274$ & $293$ & $6.99_{-0.42}^{+1.18}$ & $6.96_{-0.51}^{+0.89}$ & $136_{-9}^{+19}$ & $1.45_{-0.03}^{+0.01}$ & $0.46_{-0.00}^{+0.17}$ & $0.60_{-0.02}^{+0.01}$ & $0.60_{-0.02}^{+0.01}$ & 1.59 \\ 
   $10$ & $239$ & $293$ & $14.48_{-1.07}^{+2.90}$ & $14.46_{-1.24}^{+2.38}$ & $150_{-12}^{+25}$ & $1.26_{-0.04}^{+0.01}$ & $0.70_{-0.22}^{+0.00}$ & $0.60_{-0.02}^{+0.01}$ & $0.58_{-0.03}^{+0.02}$ & 0.63 \\ 
   $11$ & $170$ & $293$ & $21.31_{-1.82}^{+2.27}$ & $21.31_{-2.05}^{+1.98}$ & $145_{-13}^{+15}$ & $1.51_{-0.00}^{+0.01}$ & $0.38_{-0.00}^{+0.12}$ & $0.60_{-0.01}^{+0.01}$ & $0.59_{-0.02}^{+0.02}$ & 2.99 \\[1.5ex] 
   $12$ & $180$ & $371$ & $36.17_{-5.75}^{+11.18}$ & $2.37_{-0.38}^{+0.84}$ & $218_{-18}^{+21}$ & $0.86_{-0.01}^{+0.02}$ & $1.03_{-0.01}^{+0.02}$ & $0.49_{-0.03}^{+0.03}$ & $0.50_{-0.02}^{+0.02}$ & 0.80 \\ 
   $13$ & $353$ & $371$ & $33.33_{-3.28}^{+3.81}$ & $4.11_{-0.57}^{+0.64}$ & $183_{-11}^{+11}$ & $0.88_{-0.01}^{+0.04}$ & $0.75_{-0.08}^{+0.04}$ & $0.55_{-0.01}^{+0.01}$ & $0.55_{-0.01}^{+0.01}$ & 0.95 \\ 
   $14$ & $409$ & $371$ & $28.50_{-1.91}^{+2.42}$ & $5.75_{-0.78}^{+0.74}$ & $170_{-11}^{+11}$ & $1.02_{-0.01}^{+0.05}$ & $0.73_{-0.01}^{+0.02}$ & $0.57_{-0.01}^{+0.01}$ & $0.58_{-0.01}^{+0.01}$ & 1.11 \\ 
   $15$ & $180$ & $94$ & $0.25_{-0.05}^{+0.13}$ & $0.06_{-0.03}^{+0.05}$ & $223_{-40}^{+105}$ & $1.17_{-0.12}^{+0.23}$ & $0.61_{-0.29}^{+0.22}$ & $0.41_{-0.10}^{+0.06}$ & $0.40_{-0.12}^{+0.06}$ & 0.46 \\ 
   $16$ & $353$ & $94$ & $0.20_{-0.04}^{+0.12}$ & $0.11_{-0.02}^{+0.03}$ & $98_{-13}^{+34}$ & $2.90_{-0.05}^{+0.24}$ & $0.56_{-0.22}^{+0.10}$ & $0.64_{-0.07}^{+0.01}$ & $0.63_{-0.08}^{+0.02}$ & 0.33 \\ 
   $17$ & $409$ & $94$ & $0.39_{-0.01}^{+3.25}$ & $0.39_{-0.01}^{+1.82}$ & $93_{-8}^{+743}$ & $1.93_{-0.38}^{+0.75}$ & $0.88_{-0.34}^{+0.16}$ & $0.65_{-0.36}^{+0.01}$ & $0.63_{-0.49}^{+0.01}$ & 1.17 \\[1.5ex] 
   $18$ & $180$ & $293$ & $3.38_{-0.40}^{+0.51}$ & $0.39_{-0.12}^{+0.09}$ & $295_{-40}^{+39}$ & $1.08_{-0.08}^{+0.09}$ & $0.57_{-0.16}^{+0.12}$ & $0.37_{-0.03}^{+0.04}$ & $0.37_{-0.03}^{+0.04}$ & 0.61 \\ 
   $19$ & $361$ & $316$ & $3.50_{-0.29}^{+0.33}$ & $0.62_{-0.11}^{+0.07}$ & $165_{-12}^{+12}$ & $0.95_{-0.05}^{+0.05}$ & $0.43_{-0.08}^{+0.02}$ & $0.48_{-0.01}^{+0.01}$ & $0.51_{-0.01}^{+0.01}$ & 0.40 \\ 
   $20$ & $467$ & $316$ & $3.49_{-0.26}^{+0.37}$ & $1.70_{-0.09}^{+0.15}$ & $127_{-7}^{+10}$ & $1.31_{-0.05}^{+0.03}$ & $0.61_{-0.02}^{+0.03}$ & $0.54_{-0.01}^{+0.01}$ & $0.57_{-0.02}^{+0.01}$ & 0.29 \\ 
   $21$ & $171$ & $251$ & $4.21_{-0.57}^{+0.75}$ & $0.42_{-0.20}^{+0.14}$ & $320_{-54}^{+45}$ & $1.21_{-0.25}^{+0.14}$ & $0.56_{-0.38}^{+0.07}$ & $0.43_{-0.03}^{+0.06}$ & $0.43_{-0.03}^{+0.05}$ & 0.47 \\ 
   $22$ & $294$ & $251$ & $4.06_{-0.39}^{+0.66}$ & $0.87_{-0.29}^{+0.28}$ & $197_{-18}^{+23}$ & $1.30_{-0.11}^{+0.03}$ & $0.27_{-0.14}^{+0.13}$ & $0.59_{-0.04}^{+0.02}$ & $0.60_{-0.03}^{+0.01}$ & 0.25 \\ 
   $23$ & $298$ & $251$ & $4.19_{-0.51}^{+0.72}$ & $1.95_{-0.25}^{+0.24}$ & $150_{-12}^{+16}$ & $1.98_{-0.06}^{+0.03}$ & $0.41_{-0.12}^{+0.02}$ & $0.64_{-0.01}^{+0.01}$ & $0.65_{-0.01}^{+0.01}$ & 0.43 \\ 
   $24$ & $157$ & $159$ & $4.11_{-0.67}^{+0.95}$ & $0.58_{-0.28}^{+0.30}$ & $349_{-58}^{+72}$ & $1.13_{-0.05}^{+0.16}$ & $0.47_{-0.35}^{+0.04}$ & $0.55_{-0.06}^{+0.05}$ & $0.56_{-0.05}^{+0.04}$ & 0.12 \\ 
   $25$ & $198$ & $159$ & $3.66_{-0.44}^{+0.57}$ & $2.17_{-0.43}^{+0.45}$ & $251_{-38}^{+53}$ & $1.28_{-0.03}^{+0.06}$ & $0.55_{-0.09}^{+0.03}$ & $0.64_{-0.05}^{+0.05}$ & $0.65_{-0.04}^{+0.04}$ & 0.37 \\ 
   $26$ & $151$ & $159$ & $5.17_{-0.59}^{+0.92}$ & $3.96_{-0.53}^{+0.65}$ & $213_{-25}^{+38}$ & $1.76_{-0.08}^{+0.06}$ & $0.55_{-0.16}^{+0.05}$ & $0.69_{-0.03}^{+0.02}$ & $0.69_{-0.04}^{+0.02}$ & 0.52 \\ 
   $27$ & $90$ & $63$ & $3.66_{-0.71}^{+3.56}$ & $1.44_{-0.74}^{+2.22}$ & $476_{-163}^{+446}$ & $1.30_{-0.15}^{+0.21}$ & $0.43_{-0.24}^{+0.16}$ & $0.63_{-0.17}^{+0.07}$ & $0.65_{-0.16}^{+0.06}$ & 0.28 \\ 
   $28$ & $81$ & $63$ & $4.83_{-1.17}^{+231.65}$ & $4.35_{-1.37}^{+183.22}$ & $418_{-141}^{+8529}$ & $1.46_{-0.02}^{+1.61}$ & $0.80_{-0.37}^{+0.08}$ & $0.67_{-0.40}^{+0.03}$ & $0.69_{-0.46}^{+0.02}$ & 0.21 \\ 
   $29$ & $22$ & $63$ & $5.46_{-1.25}^{+246.66}$ & $5.18_{-3.10}^{+194.68}$ & $300_{-67}^{+2109}$ & $3.43_{-0.97}^{+1.00}$ & $0.51_{-0.22}^{+0.02}$ & $0.71_{-0.31}^{+0.00}$ & $0.71_{-0.43}^{+0.00}$ & 0.55 \\ 
   $30$ & $17$ & $20$ & $64.46_{-28.57}^{+132.87}$ & $26.60_{-13.42}^{+80.29}$ & $2212_{-1248}^{+2111}$ & $1.63_{-0.84}^{+1.18}$ & $0.47_{-0.09}^{+0.01}$ & $0.47_{-0.05}^{+0.16}$ & $0.39_{-0.09}^{+0.23}$ & 0.07 \\ 
   $31$ & $7$ & $20$ & $309.14_{-135.03}^{+0.00}$ & $175.19_{-121.53}^{+56.84}$ & $3065_{-1956}^{+7254}$ & $1.56_{-0.00}^{+1.18}$ & $0.34_{-0.16}^{+0.07}$ & $0.41_{-0.01}^{+0.06}$ & $0.33_{-0.08}^{+0.24}$ & 0.00 \\ 
   $32$ & $10$ & $20$ & $7.76_{-2.02}^{+238.08}$ & $7.75_{-5.06}^{+192.65}$ & $532_{-210}^{+1719}$ & $3.65_{-1.32}^{+0.66}$ & $0.27_{-0.00}^{+0.05}$ & $0.71_{-0.16}^{+0.00}$ & $0.71_{-0.34}^{+0.00}$ & 0.65 \\[1.5ex] 
   $33$ & $166$ & $281$ & $3.37_{-0.42}^{+0.47}$ & $0.43_{-0.12}^{+0.08}$ & $311_{-42}^{+36}$ & $1.04_{-0.08}^{+0.03}$ & $0.65_{-0.16}^{+0.04}$ & $0.37_{-0.02}^{+0.04}$ & $0.37_{-0.02}^{+0.04}$ & 0.61 \\ 
   $34$ & $318$ & $281$ & $3.38_{-0.30}^{+0.40}$ & $0.77_{-0.13}^{+0.16}$ & $177_{-15}^{+19}$ & $1.30_{-0.03}^{+0.02}$ & $0.38_{-0.11}^{+0.05}$ & $0.54_{-0.02}^{+0.01}$ & $0.54_{-0.02}^{+0.02}$ & 0.27 \\ 
   $35$ & $329$ & $281$ & $4.06_{-0.45}^{+0.50}$ & $2.23_{-0.28}^{+0.03}$ & $144_{-12}^{+10}$ & $1.44_{-0.01}^{+0.08}$ & $0.52_{-0.04}^{+0.04}$ & $0.58_{-0.01}^{+0.01}$ & $0.59_{-0.01}^{+0.02}$ & 0.13 \\ 
   $36$ & $162$ & $254$ & $3.07_{-0.33}^{+0.47}$ & $0.46_{-0.10}^{+0.10}$ & $333_{-45}^{+46}$ & $1.06_{-0.07}^{+0.06}$ & $0.62_{-0.13}^{+0.08}$ & $0.38_{-0.03}^{+0.03}$ & $0.38_{-0.03}^{+0.04}$ & 0.41 \\ 
   $37$ & $281$ & $254$ & $2.86_{-0.23}^{+0.35}$ & $0.93_{-0.10}^{+0.17}$ & $185_{-16}^{+21}$ & $1.34_{-0.05}^{+0.00}$ & $0.49_{-0.06}^{+0.03}$ & $0.54_{-0.02}^{+0.01}$ & $0.56_{-0.02}^{+0.01}$ & 0.19 \\ 
   $38$ & $259$ & $254$ & $3.76_{-0.39}^{+0.43}$ & $2.52_{-0.20}^{+0.09}$ & $150_{-13}^{+12}$ & $1.52_{-0.03}^{+0.01}$ & $0.64_{-0.10}^{+0.02}$ & $0.59_{-0.01}^{+0.01}$ & $0.60_{-0.01}^{+0.01}$ & 0.12 \\ 
   $39$ & $152$ & $235$ & $2.68_{-0.33}^{+0.36}$ & $0.41_{-0.09}^{+0.08}$ & $326_{-42}^{+43}$ & $1.19_{-0.10}^{+0.09}$ & $0.64_{-0.12}^{+0.07}$ & $0.41_{-0.02}^{+0.03}$ & $0.42_{-0.02}^{+0.03}$ & 0.45 \\ 
   $40$ & $236$ & $235$ & $2.72_{-0.24}^{+0.36}$ & $1.02_{-0.13}^{+0.20}$ & $193_{-18}^{+26}$ & $1.39_{-0.03}^{+0.01}$ & $0.52_{-0.04}^{+0.07}$ & $0.56_{-0.03}^{+0.01}$ & $0.58_{-0.03}^{+0.01}$ & 0.22 \\ 
   $41$ & $196$ & $235$ & $3.18_{-0.17}^{+0.44}$ & $2.90_{-0.05}^{+0.03}$ & $148_{-11}^{+14}$ & $1.76_{-0.24}^{+0.12}$ & $0.56_{-0.02}^{+0.05}$ & $0.60_{-0.01}^{+0.01}$ & $0.62_{-0.01}^{+0.01}$ & 0.16 \\ 
   $42$ & $126$ & $164$ & $2.32_{-0.27}^{+0.50}$ & $0.47_{-0.12}^{+0.13}$ & $357_{-55}^{+53}$ & $1.35_{-0.23}^{+0.06}$ & $0.52_{-0.04}^{+0.04}$ & $0.48_{-0.03}^{+0.04}$ & $0.51_{-0.03}^{+0.03}$ & 0.41 \\ 
   $43$ & $183$ & $187$ & $2.83_{-0.25}^{+0.26}$ & $1.44_{-0.16}^{+0.14}$ & $221_{-24}^{+31}$ & $1.07_{-0.03}^{+0.16}$ & $0.56_{-0.04}^{+0.02}$ & $0.54_{-0.03}^{+0.03}$ & $0.57_{-0.03}^{+0.03}$ & 0.17 \\ 
   $44$ & $139$ & $187$ & $4.11_{-0.19}^{+0.36}$ & $4.10_{-0.26}^{+0.23}$ & $186_{-14}^{+22}$ & $1.31_{-0.00}^{+0.02}$ & $0.58_{-0.07}^{+0.08}$ & $0.58_{-0.02}^{+0.01}$ & $0.61_{-0.02}^{+0.01}$ & 0.45 \\ 
   $45$ & $97$ & $121$ & $1.81_{-0.23}^{+0.35}$ & $0.45_{-0.13}^{+0.12}$ & $365_{-50}^{+52}$ & $1.10_{-0.06}^{+0.03}$ & $0.42_{-0.03}^{+0.06}$ & $0.52_{-0.03}^{+0.03}$ & $0.56_{-0.02}^{+0.02}$ & 0.44 \\ 
   $46$ & $96$ & $138$ & $2.13_{-0.21}^{+0.29}$ & $1.43_{-0.16}^{+0.20}$ & $227_{-23}^{+34}$ & $1.39_{-0.02}^{+0.06}$ & $0.48_{-0.02}^{+0.01}$ & $0.56_{-0.02}^{+0.02}$ & $0.61_{-0.02}^{+0.02}$ & 0.25 \\ 
   $47$ & $58$ & $138$ & $4.37_{-0.18}^{+0.71}$ & $4.37_{-0.24}^{+0.65}$ & $203_{-18}^{+36}$ & $1.36_{-0.04}^{+0.00}$ & $0.81_{-0.27}^{+0.00}$ & $0.54_{-0.02}^{+0.01}$ & $0.63_{-0.03}^{+0.01}$ & 1.16 \\ 
   $48$ & $41$ & $45$ & $1.20_{-0.15}^{+0.22}$ & $0.67_{-0.12}^{+0.08}$ & $440_{-94}^{+106}$ & $1.24_{-0.26}^{+0.32}$ & $0.50_{-0.05}^{+0.07}$ & $0.57_{-0.05}^{+0.05}$ & $0.63_{-0.03}^{+0.03}$ & 0.28 \\ 
   $49$ & $37$ & $48$ & $3.04_{-0.33}^{+0.61}$ & $2.52_{-0.26}^{+0.53}$ & $336_{-51}^{+79}$ & $1.42_{-0.11}^{+0.04}$ & $0.55_{-0.07}^{+0.02}$ & $0.58_{-0.03}^{+0.02}$ & $0.62_{-0.03}^{+0.02}$ & 0.12 \\ 
   $50$ & $23$ & $48$ & $22.36_{-5.88}^{+5.34}$ & $22.01_{-5.56}^{+5.15}$ & $1456_{-186}^{+67}$ & $2.14_{-0.14}^{+0.06}$ & $0.20_{-0.00}^{+0.06}$ & $0.44_{-0.01}^{+0.02}$ & $0.36_{-0.01}^{+0.03}$ & 4.31 \\ 
   \hline
  \end{tabular} \\ 
  All time-scales are in units of Myr. All length-scales are in units of pc. \vspace{-1mm}
 \end{minipage}
\end{table*}

\bsp

\label{lastpage}

\end{document}